\documentclass[final,1p,times,sort]{elsarticle}
\pdfoutput=1
\journal{Nuclear Physics A}
\usepackage{graphicx}
\usepackage{amsmath}
\usepackage{amssymb}
\usepackage{cancel}
\usepackage{booktabs}
\usepackage{array}
\usepackage{caption}
\usepackage{hyperref}

\hypersetup{
    colorlinks=true,       % false: boxed links; true: colored links
    linkcolor=blue,        % Color of internal links
    citecolor=blue,        % Color of citation links
    filecolor=magenta,     % Color of file links
    urlcolor=magenta          % Color of external links
}

\def\ba{\begin{eqnarray}}
\def\ea{\end{eqnarray}}

\def\be{\begin{equation}}
\def\ee{\end{equation}}

\renewcommand{\Bar}{\overline}
\DeclareMathOperator{\arctanh}{arctanh}

\begin{document}

\begin{frontmatter}

\title{Far-from-equilibrium attractors in kinetic theory for a mixture of quark and gluon fluids}

\author[a]{Ferdinando Frasc\`{a}}
\author[a]{Andrea Beraudo}
\author{and Michael Strickland}

\affiliation[a]{INFN, Sezione di Torino, via Pietro Giuria 1, I-10125 Torino}
%\affiliation[b]{Department of Physics, Kent State University, Kent, OH 44242 United States}

%\emailAdd{ferdinando.frasca@to.infn.it}
%\emailAdd{beraudo@to.infn.it}
%\emailAdd{michael.strickland@gmail.com}

\begin{abstract}
    We exactly solve an RTA-Boltzmann equation that describes the dynamics of coupled massless quark and gluon fluids undergoing transversally homogeneous longitudinal boost-invariant expansion. We include a fugacity parameter that allows quarks to be out of chemical equilibrium and we account for the different collision rates of quarks and gluons, which are related by Casimir scaling.  Based on these assumptions, we numerically determine the evolution of a large set of moments of the quark and gluon distribution functions and reconstruct their entire distribution functions. We find that both late and early-time attractors exist for all moments of the distribution functions containing more than one power of the squared longitudinal momentum.  These attractors emerge long before the system reaches the regime where hydrodynamic approximations apply. In addition, we discuss how the shear viscous corrections and entropy density of the fluid mixture evolve and consider the properties of their respective attractors. Finally, the entropy production is also investigated for different initial values of momentum anisotropy and quark abundance.
\end{abstract}


\end{frontmatter}

%%%%%%%%%%%%%%%%%%%%%%%%%%%%%%%%%%%%%%%%%%%%%%%%%%%%%%%%%%%%%
\section{Introduction}
%%%%%%%%%%%%%%%%%%%%%%%%%%%%%%%%%%%%%%%%%%%%%%%%%%%%%%%%%%%%%

The main goal of ultra-relativistic heavy-ion collisions (HICs) experiments being performed at CERN's Large Hadron Collider (LHC) and Brookhaven National Laboratory's Relativistic Heavy Ion Collider (RHIC) is to produce and study a primordial state of matter called the quark-gluon plasma (QGP) \cite{Gross:2022hyw}.  At the highest accessible center-of-mass energies the matter created in such collisions is gluon dominated, with quarks and antiquarks being pair produced only afterwards.  As a consequence, the QGP generated in such events has approximately zero baryochemical potential. However, since the state initially consists primarily of overoccupied gluons and few quarks/antiquarks, the QGP is not in chemical equilibrium \cite{Biro:1993qt,Dumitru:1993vz,Strickland:1994rf}.  It takes some time for the system to approach chemical equilibrium and the time scale for chemical equilibration is in general distinct from the timescale for kinetic equilibration, entailing in particular local pressure isotropization \cite{Du:2020dvp,Du:2020zqg}.  As a consequence, one must examine the non-equilibrium dynamics of systems that are far from both chemical and kinetic equilibrium.

In recent years, substantial progress has been made in understanding the far-from-equilibrium dynamics relevant to HICs; for recent reviews see Refs.~\cite{Florkowski:2017olj,Romatschke:2017ejr,Berges:2020fwq,Strickland:2024moq}. This progress has included both strongly-coupled~\cite{Chesler:2010bi,Heller:2011ju,Heller:2012km,vanderSchee:2013pia,Heller:2013oxa,Casalderrey-Solana:2013aba,Chesler:2013lia,Chesler:2015wra,Keegan:2015avk,Spalinski:2017mel} and and weakly-coupled scenarios~\cite{Denicol:2014xca,Denicol:2014tha,Kurkela:2015qoa,Keegan:2015avk,Bazow:2015dha,Denicol:2016bjh,Heller:2016rtz,Romatschke:2017vte,Strickland:2017kux,Blaizot:2017ucy,Strickland:2018ayk,Kurkela:2018wud,Kurkela:2018vqr,Kurkela:2019set,Strickland:2019hff,Blaizot:2019scw,Denicol:2019lio,Brewer:2019oha,Almaalol:2020rnu,Du:2020dvp,Du:2020zqg,Ambrus:2021fej,Blaizot:2021cdv,Alalawi:2022pmg,Jaiswal:2022udf,Mullins:2022fbx,Ambrus:2022qya,Ambrus:2022koq,Rocha:2022ind,Du:2022bel}.  One of the key findings is that there exist non-equilibrium hydrodynamic attractors that make it possible to describe the evolution of the system using dissipative relativistic hydrodynamics on sub-fm/$c$ timescales \cite{Heller:2015dha}.  At these early times in the QGP evolution, the system is not in isotropic thermal equilibrium, but nonetheless can be well described by hydrodynamical approaches including standard second viscous hydrodynamics and schemes that do not truncate the evolution equations in the inverse Reynolds number, such as anisotropic hydrodynamics \cite{Florkowski:2010cf,Martinez:2010sc,Ryblewski:2010ch,Martinez:2012tu,Ryblewski:2012rr,Bazow:2013ifa,Tinti:2013vba,Nopoush:2014pfa,Tinti:2015xwa,Bazow:2015cha,Strickland:2015utc,Alqahtani:2015qja,Alqahtani:2016rth,Bluhm:2015raa,Bluhm:2015bzi,Molnar:2016vvu,Molnar:2016gwq,Florkowski:2017ovw,Alqahtani:2017jwl,Alqahtani:2017tnq,Alalawi:2020zbx,Alalawi:2021jwn,Arslandok:2023utm}.

With a few exceptions \cite{Florkowski:2012as,Florkowski:2013uqa,Florkowski:2015cba,Florkowski:2017ovw,Strickland:2019hff,Almaalol:2018jmz,Du:2020dvp,Du:2020zqg} the dynamics of the QGP out of chemical and kinetic equilibrium has received less attention.  In particular, it has not been established whether or not there are early-time dynamical attractors when one considers two component systems consisting of quarks and gluons.  Such studies are critical to understanding the early-time non-equilibrium dynamics of the QGP and its subsequent description using dissipative relativistic hydrodynamics.  In Ref.~\cite{Florkowski:2017jnz}, the authors studied the case of a two-component system using exact solutions of the boost-invariant and transversally-homogeneous Boltzmann equation in relaxation time approximation, however assuming the same constant relaxation time for quarks and gluons. After varying the initial condition at fixed initialization time a forward attractor  was found for the pressure anisotropy. Despite this early finding, many open questions remain.  

In this paper we consider the question of whether or not dynamical attractors exist if there are separate time-dependent relaxation times for quarks and gluons.  We address the existence of both forward and pullback attractors, with the latter being associated with variation of the initialization time while holding the initial conditions fixed.  We also go beyond previous studies by considering not only the pressure anisotropy, but a large set of integral moments constructed from the quark and gluon one-particle distribution functions.  To accomplish this, we include a fugacity parameter that allows quarks to be out of chemical equilibrium and allow for different collision rates for quarks and gluons, which are related using Casimir scaling.  We then derive coupled Boltzmann equations that are obeyed by all moments of the distribution functions.  We will demonstrate that both forward and pullback attractors exist for all moments of the distribution functions containing more than one power of the squared longitudinal momentum.  These attractors emerge long before the system reaches the regime where hydrodynamic approximations apply. Finally, we discuss how the shear viscous corrections and scaled entropy density of the fluid mixture evolve and consider the properties of their respective attractors.

The structure of our paper is as follows. In Sec.~\ref{ex_2rta}, we introduce the general setup used to describe a two-component QGP within kinetic theory and relate this to the usual hydrodynamic variables, including dissipative terms.  In Sec.~\ref{ex_num}, we present our numerical results for a large set of scaled moments obtained from the quark and gluon distribution functions.  We also present associated quantities such as the effective temperature of the system and the scaled shear corrections to the energy-momentum tensor.  We then show our results for the evolution of the quark and gluon distribution functions obtained from the exact solution to the two-component Boltzmann equation.  Finally, we study the evolution of the quark fugacity and of the entropy density for our mixture of quarks and gluons. In Sec.~\ref{conclusions}, we present our conclusions together with an outlook for future developments and applications.  We collect some technical details concerning the calculations presented in the body of the text in~\ref{Milne} and~\ref{strick_proof}.

%%%%%%%%%%%%%%%%%%%%%%%%%%%%%%%%%%%%%%%%%%%%%%%%%%%%%%%%%%%%%
\section{General setup for a two-component mixture}
\label{ex_2rta}
%%%%%%%%%%%%%%%%%%%%%%%%%%%%%%%%%%%%%%%%%%%%%%%%%%%%%%%%%%%%%

{In order to give a more realistic description, within kinetic theory, of the fireball formed in heavy-ion collisions we model it as a mixture of two different fluids, one for gluons and one for quarks (and antiquarks), sharing however the same collective four-velocity $u^\mu$, that we will define in the following}. Within the framework of anisotropic hydrodynamics (aHydro) this was done, for instance, in Refs.~\cite{Florkowski:2012as, Florkowski:2017ovw, Florkowski:2013uqa, Maksymiuk:2017szr}.

In order to distinguish the species in the system we introduce the label $a = q$ (quarks + antiquarks) and $g$ (gluons). Moreover, we admit that quarks and gluons in the plasma could be, in principle, characterized by two different relaxation times $\tau_{{\rm eq},q}$ and $\tau_{{\rm eq},g}$. In addition, we consider the case where the net baryon number is zero, i.e. the number of quarks equals the number of antiquarks. This leads also to a vanishing baryo-chemical potential, {so that antiparticles will simply enter as a further factor 2 in the counting of internal fermionic degrees of freedom}. The latter situation is typically achieved in collisions at very high center-of-mass energy, with negligible stopping of the incoming baryonic matter~\cite{Dainese:2016dea,Arslandok:2023utm}.

In our treatment we assume that {initially quarks and gluons -- due to the violent longitudinal expansion of the fireball -- are described by a common off-equilibrium anisotropic momentum distribution. Furthermore we assume that initially the fireball is also out of chemical equilibirium, the system being gluon-dominated with occupation of fermionic modes strongly suppressed~\cite{Busza_2018, Biro:1981zi, Kurkela:2018oqw, Kurkela:2018xxd, Dumitru:1993vz,Strickland:1994rf}. An effective temperature $T$ for this} out-of-equilibrium mixture can be defined by using a generalized Landau matching condition~\cite{Florkowski:2012as, Bhadury:2020ngq}, as we will see in Sec.~\ref{Sec:Landau}.

The stress-energy tensor and the energy density of the fluid mixture are given by the sum of the contributions from the quark and gluon species:
\begin{equation}
    T^{\mu \nu} \equiv \sum_a T^{\mu \nu}_a \hspace{0.3cm},\hspace{0.3cm} \varepsilon \equiv \sum_a \varepsilon_a \hspace{0.3cm}, \text{ \small{with}} \hspace{0.2cm} a = q, g\,.\label{eq:Tmunu-sum}
\end{equation}
{We work in the Landau frame~\cite{Landau1987Fluid}, in which the fluid four-velocity refers to the energy transport and is defined as} the eigenvector of the energy-momentum tensor with corresponding eigenvalue given by the energy density $\varepsilon$~\cite{Gavin:1985ph, Romatschke:2009im}, {i.e. $u_\mu \hspace{0.04cm} T^{\mu \nu} = \varepsilon \hspace{0.04cm} u^{\nu}$. Consistently with the normalization $u^2=1$ of the common fluid velocity, arising from the $(+,-,-,-)$ signature of the metric, one has for the energy density of each component} 
\begin{equation}
  \varepsilon_a=u_\mu u_\nu \hspace{0.04cm}T^{\mu\nu}_a\,.\label{eq:edens-def}  
\end{equation}
%%%%%%%%%%%%%%%%%%%%%%%%%%%%%%%%%%%%%%%%%%%%%%%%%%%%%%%%%%%%%%%%%%%%%
\subsection{The Boltzmann equation for the quark and gluon distribution functions}
%%%%%%%%%%%%%%%%%%%%%%%%%%%%%%%%%%%%%%%%%%%%%%%%%%%%%%%%%%%%%%%%%%%%%

{In order to establish the link between the hydrodynamic and kinetic description of our system the starting point is the relativistic} Boltzmann equation (BE, see Refs.~\cite{DEBBASCH20091079,DEBBASCH20091818}) in a flat spacetime (see \ref{Milne}), which describes the evolution of the quark and gluon distribution functions, {here written in Relaxation-Time Approximation} (RTA) as follows~\cite{2021mfrf.book.....D, Florkowski:2012as, Florkowski:2017ovw, PhysRev.94.511,ANDERSON1974466}:
\begin{equation}
\label{rta5}
    p^{\mu} \hspace{0.07cm} \partial_{\mu} \hspace{0.07cm} f_a (x;p) \simeq - \frac{(p \cdot u)}{\tau_{{\rm eq}, a}(x)} \Bigl[ f_a(x;p) - f_{a}^{\rm eq}(x;p)  \Bigr] \hspace{0.3cm},\text{ \small{with}} \hspace{0.2cm} a = q, g\,.
\end{equation}
{Within this approximation} the distribution function of the species $a$ tends to relax to its equilibrium configuration in a typical time-scale given by the relaxation time $\tau_{{\rm eq},a}$ (no distinction is made for elastic or inelastic processes), {which depends on the spacetime point $x$ through the local effective temperature (common to all particle species), but for which one ignores any possible momentum dependence}. 
For a comparison of RTA results with the ones obtained with the collision integral provided by effective kinetic theory (EKT, see Refs.~\cite{Arnold:2002zm, Kurkela:2015qoa}) we refer the reader to Ref.~\cite{Almaalol:2020rnu}.
In principle the relaxation times could be different for quarks and gluons in the plasma, due to their different color charge~\cite{Maksymiuk:2017szr, Florkowski:2013uqa}, {and this is the case (in the following referred to as 2RTA) explored in our study.} Notice that chemical and kinetic equilibration can occur over different timescales. Here, for the sake of simplicity, at variance with more microscopic calculations (see e.g. Ref.~\cite{Kurkela:2018xxd}) we neglect such a possibility taking in Eq.~(\ref{rta5}) a unique relaxation time for elastic and inelastic processes.

{In the following we are going to consider the lowest-order moments of Eq.~(\ref{rta5}), which are relevant to evaluate the number density of the different species in the medium (and hence the deviation from chemical equilibrium), to ensure energy-momentum conservation and to define an effective temperature of the system even out of kinetic equilibrium through a generalized Landau matching condition}.

%------------------------------------------------------------
\subsubsection{Lowest-order moments of the Boltzmann equation}\label{Sec:Landau}
%------------------------------------------------------------

The zeroth moment of the BE in RTA can be found by integrating Eq.~(\ref{rta5}) over momentum space~\cite{Maksymiuk:2017szr, Strickland:2014pga}, getting
\begin{equation}
    \int d \chi \hspace{0.07cm} p^{\mu} \hspace{0.07cm} \partial_{\mu} \hspace{0.07cm} f_a (x;p) = - \frac{u_{\mu}}{\tau_{{\rm eq},a}} \int d \chi \hspace{0.07cm} p^{\mu} \hspace{0.07cm} \Bigl[ f_a(x;p) - f_{a}^{\rm eq}(x;p)  \Bigr] \hspace{0.3cm},\text{ \small{with}} \hspace{0.2cm} d \chi \equiv \frac{d^3 p}{(2 \pi)^3 \hspace{0.07cm} p^0}\,.
\end{equation}
{In the following we will take into account a system of on-shell massless particles, for which $p^0 = |\Vec{p}\hspace{0.04cm}|$, which has to be substituted into the invariant integration measure $d\chi$.} 
In the LHS one can move the spacetime derivative out of the momentum integral. One then defines the following {particle-number currents}:
\begin{equation}
\label{goggo}
    N_{a}^{\mu} \equiv \int d \chi \hspace{0.07cm} p^{\mu} \hspace{0.07cm} f_a(x;p)\quad{\rm and}\quad
    N_{a,{\rm eq}}^{\mu} \equiv \int d \chi \hspace{0.07cm} p^{\mu} \hspace{0.07cm} f_a^{\rm eq}(x;p)\,.    
\end{equation}
In terms of the above currents the equation for the zeroth moment of the BE in RTA becomes:
\begin{equation}
    \label{rta6}
    \partial_{\mu} \hspace{0.07cm} N_{a}^{\mu} = -\frac{u_{\mu}}{\tau_{{\rm eq},a}} \left( N_a^\mu - N_{a, {\rm eq}}^{\mu}\right)\,.
\end{equation}
{Notice that the quark current refers to the sum of the quark and antiquark contributions (the baryon-number current, involving the difference of the two~\cite{Florkowski:2015cba, Florkowski:2017ovw, Florkowski:2013uqa}, is not relevant for our study since it identically vanishes at any time). Of course, as evident from Eq.~(\ref{rta6}), these are not conserved currents: no conservation law prevents gluons from being radiated/absorbed or $q\bar q$ pairs from being created/annihilated~\cite{Bhadury:2020ngq, Kurkela:2018oqw, Kurkela:2018xxd}. Our study actually addresses the case of a system in which quarks (and antiquarks) are initially underpopulated.} 
{It is possible to show that, in the case of Bjorken flow addressed in the present work, the particle-number currents have the following simple expression~\cite{Florkowski:2012as}}

\begin{equation}
    N_{a}^{\mu} \equiv n_a \hspace{0.07cm} u^{\mu},\quad
    N_{a,{\rm eq}}^{\mu} \equiv n_{a,{\rm eq}} \hspace{0.07cm} u^{\mu}\,,
\end{equation}
{\rm with no contribution orthogonal to the fluid four-velocity. Hence, particle densities can be calculated from}
\begin{equation}
\label{roeo}
    n_a \equiv N^{\mu}_a \hspace{0.07cm} u_{\mu} = \int d \chi \hspace{0.07cm} (p \cdot u) \hspace{0.07cm} f_a(x;p) \hspace{0.3cm}, \text{ \small{with}} \hspace{0.2cm} a = q, g\,.
\end{equation}
{Eventually, quarks and gluons will reach their} local equilibrium densities given by
\begin{equation}
\label{eq}
    n_{g,{\rm eq}} = \frac{g_g}{\pi^2} \hspace{0.07cm} T^3 \hspace{0.3cm},\hspace{0.3cm} n_{q,{\rm eq}} = \frac{2 \hspace{0.07cm} g_q}{\pi^2} \hspace{0.07cm} T^3\,,
\end{equation}
coming from inserting into Eq.~(\ref{roeo}) the corresponding relativistic Boltzmann distribution for massless particles. {In the above, the degeneracy factors $g_a$ counts the internal relativistic degrees of freedom. In our numerical calculations we will employ the values $g_g=2\times 8$ for the gluons (polarization and color) and $g_q=2\times 3\times 3$ for the light quarks (spin, color and flavor).} The further factor 2 in $n_{q,{\rm eq}}$ comes from the choice of including both quarks and antiquarks into the distribution function $f_q$.

{We now consider the first moment of} the BE in RTA, obtained by multiplying both sides of Eq.~(\ref{rta5}) by $p^{\nu}$ before integrating over the phase space~\cite{Strickland:2014pga}:
\begin{equation*}
    \int d \chi \hspace{0.07cm} p^{\mu} p^{\nu} \hspace{0.07cm} \partial_{\mu} \hspace{0.07cm} f_a (x;p) = - \frac{u_{\mu}}{\tau_{{\rm eq},a}} \hspace{0.07cm} \int d \chi \hspace{0.07cm} p^{\mu} p^{\nu} \hspace{0.07cm} \Bigl[ f_a(x;p) - f_a^{\rm eq}(x;p)  \Bigr]\,.
\end{equation*}
 {After introducing the definition of the} energy-momentum tensor for each fluid component {in terms of the corresponding single-particle distribution}
\begin{equation}
\label{baba}
    T^{\mu \nu}_a \equiv \int d \chi \hspace{0.07cm} p^{\mu} p^{\nu} \hspace{0.07cm} f_a (x;p) \hspace{0.3cm},\text{ \small{with}} \hspace{0.2cm} a = q, g\,
\end{equation}
{one gets the} differential equation
\begin{equation}
\label{rta14}
    \partial_{\mu} \hspace{0.07cm} T^{\mu \nu}_a = -\frac{1}{\tau_{{\rm eq},a}} \left( T^{\mu \nu}_a-T^{\mu \nu}_{a,{\rm eq}}\right) u_{\mu} = -\frac{1}{\tau_{{\rm eq},a}} \left(\varepsilon_a-\varepsilon_{a,{\rm eq}}\right) u^{\nu} \neq 0\,.
\end{equation}
The last equality follows from the choice of the Landau frame~\cite{Landau1987Fluid}, in which the eigenvalue equation $u_\mu \hspace{0.04cm} T^{\mu \nu}_a = \varepsilon_a \hspace{0.04cm} u^{\nu}$ holds for each particle species $a = q, g$. 
We note that the RHS of Eq.~(\ref{rta14}) does not vanish in general. This means that collisions occurring in the plasma {allow energy and momentum to be transferred from one species to another during the system evolution.}
However, we must require that the total energy-momentum tensor $T^{\mu \nu} \equiv \sum_a T^{\mu \nu}_a$ satisfies the continuity equation
\begin{equation}
\label{rta15}
    \partial_{\mu} \hspace{0.07cm} T^{\mu \nu} = -\sum_a \Bigl[ \frac{1}{\tau_{{\rm eq},a}} \left(\varepsilon_a- \varepsilon_{a,{\rm eq}}\right) \Bigr] u^{\nu} \equiv 0\,,
\end{equation}
since the globally conserved quantities are the energy and the momentum of the sum of all the different species in the mixture. It follows that
\begin{equation}
\label{rta16}
    \sum_a \Bigl[ \frac{1}{\tau_{{\rm eq},a}} \left(\varepsilon_a- \varepsilon_{a,{\rm eq}}\right) \Bigr] = 0\,.
\end{equation}
This is nothing but the generalization of the usual Landau matching condition (which would simply entail $\varepsilon_a=\varepsilon_{a,{\rm eq}}$) to the case of a fluid made of particles with two different relaxation times. {Its physical meaning can be better appreciated by rewriting the last equation as follows:}
\begin{equation}
\sum_a \frac{\varepsilon_a}{\tau_{{\rm eq},a}}= \sum_a \frac{\varepsilon_{a,{\rm eq}}}{\tau_{{\rm eq},a}}\,.\label{Eq:Landau}      
\end{equation}
{Since $\varepsilon_{a,{\rm eq}}\sim T^4$, Eq.~(\ref{Eq:Landau}), is a way of defining a unique common effective temperature of the system through a weighted average of the energy densities of its different components, in which particles with a shorter relaxation time provide a greater contribution to the sum. In the case of a system of quarks and gluons}, we can rewrite the above relation by introducing the {ratio of the two different relaxation times}
\begin{equation}
    \label{rta17}
    C_R \equiv \frac{\tau_{{\rm eq},g}}{\tau_{{\rm eq},q}} = \frac{C_F}{C_A} = \frac{4}{9}\,,
\end{equation}
whose value is suggested by Casimir scaling~\cite{Grandou:2014yfa}. {In the above, $C_F$ and $C_A$ represent the eigenvalue of the Casimir operator in the fundamental and adjoint representations of $SU(3)$, respectively.}
{According to Casimir scaling, the ratio between the gluon and quark mean free path} or -- equivalently, being the particles ultrarelativistic -- their respective relaxation times, should be equal to $C_F/C_A$. {Within a perturbative approach this arises from the color factors of the $2 \rightarrow 2$ QCD scattering cross-sections involving quarks and gluons.}
More generally, numerous lattice-QCD calculations have provided evidence that the Casimir scaling holds also in a non-perturbative domain~\cite{Bali:2000un,Mykkanen:2012ri,Semay:2004br}. {Under this assumption}, Eq.~(\ref{Eq:Landau}) can be put in the following form~\cite{Florkowski:2012as}:
\begin{equation}
    \label{eq:LandauMC}
    \varepsilon_q + \frac{\varepsilon_g}{C_R} = \varepsilon_{q, {\rm eq}} + \frac{\varepsilon_{g, {\rm eq}}}{C_R}\,,
\end{equation}
where
\begin{equation}
    \label{eq:LandauMC2}
    \varepsilon_{g, {\rm eq}} = \frac{3 \hspace{0.07cm} g_g}{\pi^2} \hspace{0.07cm} T^4 \hspace{0.3cm},\hspace{0.3cm}  \varepsilon_{q, {\rm eq}} = \frac{6 \hspace{0.07cm} g_q}{\pi^2} \hspace{0.07cm} T^4\,. 
\end{equation}
This represents the Landau matching condition {for our mixture made of quarks and gluons}, which relates the energy density of the two species in and out-of equilibrium and allows us to define an effective temperature for the two-component system.

Notice that, asymptotically, the system will reach equipartition of energy. In this case the equality in Eq.~(\ref{rta16}) will hold separately for each term of the sum, i.e. $\varepsilon_{a}=\varepsilon_{a,{\rm eq}}$, and the relative contribution of each species to the total energy will simply depend on the number of internal degress of freedom. Notice also that $\varepsilon_{a}=\varepsilon_{a,{\rm eq}}$ would be a trivial solution of Eq.~(\ref{rta16}) also before reaching thermodynamic equilibrium; this however would lead to defining two different temperatures for the two fluid components, preventing -- within our setup -- the asymptotic relaxation of the system to a common temperature.
%%%%%%%%%%%%%%%%%%%%%%%%%%%%%%%%%%%%%%%%%%%%%%%%%%%%%%%%%%%%%%%%%%%%%
\subsubsection{Exact solution for the single-particle distribution functions}
%%%%%%%%%%%%%%%%%%%%%%%%%%%%%%%%%%%%%%%%%%%%%%%%%%%%%%%%%%%%%%%%%%%%%

Thus far the setup was quite general. We now restrict our analysis to the simpler case of Bjorken flow, in which the medium undergoes a one-dimensional longitudinal boost-invariant expansion with fluid four-velocity $u^\mu\equiv(t/\tau,0,0,z/\tau)$.  In this case all physical quantities depend only on the longitudinal proper-time $\tau\equiv\sqrt{t^2-z^2}$. We begin by summarizing the main kinetic-theory results for this simplified scenario, referring the reader to \ref{Milne} for details.  The boost-invariant single-particle distribution functions of the two species are more conveniently written as functions of boost-invariant quantities only~\cite{Bialas:1984wv,Bialas:1987en}, $f_a\equiv f_a(\tau;w,p_T)$, where $p_T$ is the transverse momentum of the particle and
\begin{equation}
    w \equiv t \hspace{0.07cm} p_z - z \hspace{0.07cm} p^0\,.\label{eq:w-def}
\end{equation}
In this case the BE reduces to a much simpler form (see \ref{Milne} for details),
\begin{equation}
    \partial_{\tau} \hspace{0.05cm} f_a = -\frac{f_a-f_{{\rm eq}, a}}{\tau_{{\rm eq}, a}} \hspace{0.5cm} \text{ \small{with}} \hspace{0.5cm} a = q, g\,,
\end{equation}
which, similarly to the case of a single relaxation time, can be solved exactly, leading to the following result for the one-particle distribution~\cite{Strickland:2018ayk, Florkowski:2013lya}:
\begin{equation}
    \label{BF12}
    f_a (\tau; w, p_T) = D_a (\tau, \tau_0) \hspace{0.07cm} f_{0, a} (w, p_T) + \int_{\tau_0}^{\tau} \frac{d \tau'}{\tau_{{\rm eq}, a}(\tau')} \hspace{0.07cm} D_a (\tau, \tau') \hspace{0.07cm} f_{{\rm eq}, a} (\tau'; w, p_T)\,.
\end{equation}
In the above the damping function
\begin{equation}
\label{damp}
    D_a(\tau_2,\tau_1) \equiv \exp \Biggl[ - \int_{\tau_1}^{\tau_2} \frac{d \tau'}{\tau_{{\rm eq}, a}(\tau')} \Biggr]\,,
\end{equation}
was introduced, which, physically, quantifies the fraction of particles which have not undergone any interaction during the time interval from $\tau_1$ to $\tau_2$.
The damping function $D_a(\tau_2, \tau_1)$ has the following properties: $D_a(\tau, \tau ) = 1$, $D_a(\tau_3, \tau_2) \hspace{0.07cm} D_a(\tau_2, \tau_1) = D_a(\tau_3, \tau_1)$, and also~\cite{Soloviev:2021lhs}
\begin{equation}
\label{damp_prop}
    \frac{\partial}{\partial \tau_2}D_a (\tau_2, \tau_1)\!=\! - \frac{D_a (\tau_2, \tau_1)}{\tau_{{\rm eq}, a}(\tau_2)}\,,\quad
\frac{\partial}{\partial \tau_1} D_a (\tau_2, \tau_1)\! =\!  \frac{D_a (\tau_2, \tau_1)}{\tau_{{\rm eq}, a}(\tau_1)}\,,    
\quad \text{ \small{where}} \hspace{0.5cm} a = q, g\,.
\end{equation}
Moreover, $f_{{\rm eq},a}$ is the usual relativistic Boltzmann distribution, while $f_{0, a}$ represents the initial {condition at proper time $\tau_0$ for the} distribution function of the species $a$, which, {in the realistic case, is expected to display sizable deviations from thermodynamic equilibrium~\cite{Strickland:2018ayk, Strickland:2014pga}.}

{In order to allow an easier comparison with previous results, in our analysis we choose the well-known} Strickland-Romatschke {form for the initial} distribution of the two species~\cite{Romatschke:2003ms, Alqahtani:2017mhy}:
\begin{equation}
    \label{rta3}
    f_{0,g}(x;p) \equiv g_g \hspace{0.07cm} \exp \Biggl[ - \frac{\sqrt{(p \cdot u)^2 + \xi_0 \hspace{0.07cm} (p \cdot z)^2}}{\Lambda_0}  \Biggr]\,,
\end{equation}
\begin{equation}
\label{rta4}
    f_{0,q}(x;p) \equiv 2 g_q \hspace{0.07cm}  \gamma_{q,0} \hspace{0.07cm} \exp \Biggl[ - \frac{\sqrt{(p \cdot u)^2 + \xi_0 \hspace{0.07cm} (p \cdot z)^2}}{\Lambda_0}  \Biggr]\,,
\end{equation}
where, as already mentioned, the factor 2 in Eq.~(\ref{rta4}) {accounts for the particle-antiparticle degeneracy. In the above $z^\mu\equiv(z/\tau,0,0,t/\tau)$ is a space-like unit vector ($z^2\!=\!-1$), which in the local rest frame (LRF) of the fluid reduces to $z^\mu_{\rm LRF}=(0,0,0,1)$~\cite{Strickland:2018ayk, Florkowski:2013lya}.} 
For sake of simplicity, we assumed that {the initial} quark and gluon {kinematic distributions in Eqs.~(\ref{rta3}) and~(\ref{rta4})} are characterized by the same off-equilibrium parameters $\xi_{0,g} = \xi_{0,q} \equiv \xi_0$ and $\Lambda_{0,g} = \Lambda_{0,q} \equiv \Lambda_0$. {Their physical meaning can be better appreciated in the fluid LRF in which, for massless particles, one has
\begin{equation*}
    f_0\sim\exp\left[-\frac{\sqrt{p_T^2+(1+\xi_0) \hspace{0.07cm}p_z^2}}{\Lambda_0}\hspace{0.07cm}\right]\,.
\end{equation*}
Hence, $\xi_0\in(-1,+\infty)$ quantifies the momentum anisotropy in the LRF of the system, while $\Lambda_0$ sets the typical transverse-momentum scale of the particles~\cite{Florkowski:2012as, Strickland:2024moq, Strickland:2017kux}. In solving the kinetic equations the angular asymmetry is more conveniently parametrized through the coefficient
\begin{equation}
    \alpha_0\equiv(1+\xi_0)^{-1/2}\,.\label{eq:alpha0def}
\end{equation}
Since in the early stages of the evolution the fast longitudinal expansion of the system strongly suppresses the occupation of high-$p_z$ modes, in most cases in our numerical calculations we will set $\xi_0>0$ and hence $\alpha_0\in(0,1)$.}
Furthermore, we also introduced a fugacity parameter $\gamma_{q,0}$ to {let the system evolution start out of chemical equilibrium, as expected in the realistic scenario in which quarks are strongly under-populated~\cite{Kurkela:2018oqw,Kurkela:2018xxd}. Accordingly, in the following we will refer to the case where $\gamma_{q,0}\ll 1$ as the one of physical relevance.}
{Asymptotically, one expects the system to reach full thermodynamic equilibrium, in which one has $\gamma_q\!=\!1$ (chemical equilibrium), $\xi\!=\!0$ and $\Lambda$ can be identified with the temperature (kinetic equilibrium)}, retrieving a relativistic Boltzmann distribution for both species~\cite{Alqahtani:2017mhy, Strickland:2014pga}.

{The approach to chemical and kinetic equilibrium is usually modeled through $2\leftrightarrow 2$ scatterings (including $gg\leftrightarrow q\overline q$ processes) and effective $1\leftrightarrow 2$ splittings, according to the effective kinetic theory developed in Ref.~\cite{Arnold:2002zm} and employed in Refs.~\cite{Kurkela:2018oqw,Kurkela:2018xxd}. Rather than exploiting the results of the full weak-coupling calculation presented in Ref.~\cite{Arnold:2002zm}, here we follow the simpler strategy of solving Eq.~(\ref{rta5}) for each particle species. The two equations for quarks and gluons are coupled through the Landau matching condition in Eq.~(\ref{eq:LandauMC}).}

{The above reactions are balanced at equilibrium, but this occurs only asymptotically. During most of the system's evolution they occur more efficiently in one of the two directions, both due to the initial underpopulation of quarks, which makes $q\overline q$ annihilation unlikely, and due to high expansion rate of the fireball, which prevents interactions from reaching detailed balance. In the following the value of the $\gamma_q$ parameter, which weights the quark contribution to thermodynamic observables, will be used to quantify the departure of the fireball from chemical equilibrium.}

The equilibration time in our approach may be an arbitrary function of the proper time, namely $\tau_{{\rm eq}, a} = \tau_{{\rm eq}, a}(\tau)$. {However, in order to establish a connection between the microscopic kinetic approach and viscous hydrodynamics (vHydro)} we use the following relations:
\begin{equation}
\label{eq2}
    \tau_{{\rm eq}, g} \equiv \tau_{\rm eq} = \frac{5 \hspace{0.07cm} \eta/s}{T(\tau)} \hspace{0.5cm} \text{\small{and}} \hspace{0.5cm} \tau_{{\rm eq}, q} \equiv \frac{\tau_{\rm eq}}{C_R}
\end{equation}
provided by conformal kinetic theory~\cite{Strickland:2018ayk, Romatschke:2017ejr, Denicol:2010xn, Denicol:2011fa,Jaiswal:2013npa,Jaiswal:2013vta} and Casimir scaling, respectively.
In the above, the $\eta/s$ coefficient, assumed to be time-independent, can be simply taken as {a parameter fixing the relation between the effective temperature of the system and the gluon relaxation-time}. However, {considering its physical meaning}, the $\eta/s$ ratio in Eq.~(\ref{eq2}) represents the specific (shear) viscosity of gluons, since it appears in the equation for the relaxation time $\tau_{{\rm eq}, g}$. The {actual} specific viscosity of {our fluid mixture} receives contributions from both quarks and gluons, resulting in {\rm the following weighted average:}
\begin{equation}
(\eta/s)_{\rm QGP}=\frac{g_g \hspace{0.07cm} (\eta/s)_g + 2g_q \hspace{0.07cm} (\eta/s)_q}{g_g + 2g_q}\,,\quad \text{\small{with}}\quad \eta_a=\frac{4}{5}\,\tau_{{\rm eq},a}\,P_{a,{\rm eq}}\;\longrightarrow\;(\eta/s)_a \equiv \frac{1}{5} \hspace{0.07cm} \tau_{{\rm eq}, a} \hspace{0.07cm} T\,,\label{eq:etastot}
\end{equation}
which arises from {the first-order Chapman-Enskog expansion} of the Boltzmann equation~\cite{1970mtnu.book.....C, Gavin:1985ph}.
Because of Casimir scaling the specific viscosities of gluons and quarks are related by $(\eta/s)_g = C_R \hspace{0.07cm} (\eta/s)_q$.

{It is interesting to provide a physical interpretation of the two terms contributing to the exact solution of the BE in Eq.~(\ref{BF12}).} The first term in the RHS of Eq.~(\ref{BF12}), in which the initial distribution function expressed in terms of boost-invariant variables does not change during the evolution of the fluid, represents the so-called ``free-streaming term'' {and arises from} the fraction of particles which have not suffered any collision~\cite{Broniowski:2008qk, Giacalone:2019ldn}; on the contrary the second term is called ``evolution term'' {and is responsible for the approach to equilibrium}~\cite{Strickland:2018ayk, Florkowski:2013lya}.
We note that the first term in Eq.~(\ref{BF12}) carries {full information on} the initial condition of the system, but at times greater than $\tau_{{\rm eq},a}$ this first contribution decays exponentially fast {so that}, at sufficiently late times, the information about the initial condition is completely lost~\cite{Strickland:2018ayk, Strickland:2024moq}. Thus, the late-time behavior of the system is {essentially described by the second term}~\cite{Soloviev:2021lhs}.

The property {of the information on} the initial condition of being completely lost during the fluid evolution is related to the concept of hydrodynamic attractor~\cite{Romatschke:2017vte, Strickland:2017kux, Jankowski:2023fdz, Jaiswal:2019cju, Spalinski:2022cgj}, which is then expected to exist also in the case of two different relaxation times for quarks and gluons {addressed here. In the following we will verify this occurrence through the numerical integration of the BE for this fluid mixture}.  As we will see in Sec.~\ref{ex_num}, {this universal behavior} can occur both at early times and/or at late times, depending on the considered thermodynamic variable, and can be attributed either to the expansion rate or to the interaction rate of the system. This can lead to the identification of universal observables -- typically related to moments of the phase-space particle distributions -- describing the behavior of the QGP formed in heavy-ion collisions.
%%%%%%%%%%%%%%%%%%%%%%%%%%%%%%%%%%%%%%%%%%%%%%%%%%%%%%%%%%
\subsection{Exact solution for the general moments of the particle distributions}
%%%%%%%%%%%%%%%%%%%%%%%%%%%%%%%%%%%%%%%%%%%%%%%%%%%%%%%%%%
{Important} quantities {relevant to describe the fluid evolution} correspond to moments of the particle distribution functions. A convenient definition for an arbitrary moment of the one-particle distribution function of species $a$, in the case of a purely longitudinal expansion, is given by~\cite{Strickland:2018ayk, Strickland:2019hff}:
\begin{equation}
\label{robo}
    M_a^{nm} \equiv \int d \chi \hspace{0.07cm} ( p \cdot u)^n \hspace{0.07cm} ( p \cdot z)^{2m} \hspace{0.07cm} f_a( \tau; w, p_T)\,. 
\end{equation}
{In the above the invariant integration measure for a system of massless particles can be rewritten as follows}
\begin{equation*}
 d \chi \equiv  \frac{d^2 p_T\, dw}{(2\pi)^3 \,v}\,,\quad{\rm with}\quad  v = \sqrt{w^2 + p^2_T \hspace{0.07cm} \tau^2}\,.
\end{equation*}
The lowest-order moments {can be given a clear physical meaning.}
Thus, for the species $a$, we have that $M^{20}_a = \varepsilon_a$ (energy density), $M^{10}_a = n_a$ (particle-number density) and $M^{01}_a = P_{L,a}$ (longitudinal pressure). These relations {follows from Eqs.~(\ref{eq:edens-def}) and (\ref{roeo}).}

{Being interested in studying the evolution of the system as a whole, in analogy to what done for the energy-momentum tensor, we define the generic total moment}
\begin{equation}
    M^{nm} \equiv \sum_{a = q, g} M_a^{nm}
\end{equation}
{as the sum of the quark and gluon contributions.}

In principle,  the definition of a general moment of the particle distribution in Eq.~(\ref{robo}) {could also contain powers of $p_T^{2}$}. However, such moments can be expressed as a linear combination of the two-index moments appearing above using {the mass-shell condition} $p^2 = 0$ to write $p_T^{2l} = [(p \cdot u)^2 - (p \cdot z)^2]^l$. This, for a conformal system, allows one to use $\varepsilon = 2 P_T + P_L$ to determine the transverse pressure~\cite{Strickland:2024moq, Alqahtani:2017mhy}. 

After a pretty long calculation and some non-trivial algebra, one eventually gets {an} integral equation (the full derivation is performed in \ref{strick_proof}) {describing the evolution of $M^{nm}$, which is more conveniently expressed as the sum of two contibutions}:
\begin{equation}
    \label{eq:mom-split}
        M^{nm}(\tau) = M^{nm}_{0}(\tau) + M^{nm}_{\rm coll}(\tau)\,.
\end{equation}
The first term comes from the free-streaming contribution to the exact distribution function in Eq.~(\ref{BF12}) {and reads}
\begin{equation}
\label{eq:split-fs}  
M^{nm}_{0}(\tau) \equiv A \hspace{0.07cm} \left\{D(\tau, \tau_0) \hspace{0.07cm} \biggl[ r + 2 \hspace{0.07cm} \gamma_{q, 0} \hspace{0.07cm} \biggl( D(\tau,\tau_0)\biggr)^{C_R - 1} \biggr] \hspace{0.07cm} \biggl( \frac{2 \hspace{0.07cm} \left(2 + \Bar{r}\right)}{2 \hspace{0.07cm} \gamma_{q, 0} + \Bar{r}} \biggr)^\frac{n + 2m + 2}{4} \hspace{0.07cm} T_{0}^{\hspace{0.07cm} n +2m +2} \hspace{0.1cm} \frac{H^{nm}\left( \alpha_0 \hspace{0.07cm} \frac{\tau_0}{\tau} \right)}{\Bigl[H(\alpha_0) \Bigr]^{\frac{n +2m +2}{4}}} \right\}\,,
\end{equation}
%\begin{multline}\label{eq:split-fs}    
    %M^{nm}_{0}(\tau) \equiv A \hspace{0.07cm}\left\{D(\tau, \tau_0) \hspace{0.07cm} \biggl[ r + 2 \hspace{0.07cm} \gamma_{q, 0} \hspace{0.07cm} \biggl( D(\tau,\tau_0)\biggr)^{C_R - 1} \biggr]\times\right.\\
    %\left.\times\hspace{0.07cm} \biggl( \frac{2 \hspace{0.07cm} \left(2 + \Bar{r}\right)}{2 \hspace{0.07cm} \gamma_{q, 0} + \Bar{r}} \biggr)^\frac{n + 2m + 2}{4} \hspace{0.07cm} T_{0}^{\hspace{0.07cm} n +2m +2} \hspace{0.1cm} \frac{H^{nm}\left( \alpha_0 \hspace{0.07cm} \frac{\tau_0}{\tau} \right)}{\Bigl[H(\alpha_0) \Bigr]^{\frac{n +2m +2}{4}}} \right\}\,,
%\end{multline}    
while the second one contains the contribution of collisions between partons:
\begin{align}
\begin{split}\label{eq:split-coll}
    M^{nm}_{\rm coll}(\tau) \equiv A \hspace{0.07cm} \Biggl\{C_R \hspace{0.07cm} \int_{\tau_0}^{\tau} \frac{d \tau'}{\tau_{\rm eq}(\tau')} \hspace{0.07cm} D(\tau, \tau') \hspace{0.07cm} \biggl[ \Bar{r} + 2 \hspace{0.07cm} \biggl( D(\tau,\tau')\biggr)^{C_R - 1} \biggr] \hspace{0.07cm} T^{\hspace{0.07cm} n +2m +2}(\tau') \hspace{0.07cm} H^{nm} \left( \frac{\tau'}{\tau} \right) \Biggr\} \,.
\end{split}
\end{align}
{In the above} we introduced two new constants. The first one is given by:
\begin{equation}
\label{nonna}
    \Bar{r} \equiv \frac{r}{C_R} \,,\quad \text{ \small{with}} \quad r \equiv \frac{g_g}{g_q} = \frac{8}{9}\,,
\end{equation}
{coming from counting the} internal degrees of freedom of gluons $g_g = 16$ and quarks $g_q = 18$. The {second overall} constant $A$ reads:
\begin{equation}
    A \equiv g_q \hspace{0.07cm} \frac{\Gamma(n + 2m +2)}{(2\pi)^2}\,.
\end{equation}
In the previous integral equation only the damping function associated to gluons $D(a,b)\equiv D_g(a,b) $ appears, containing the relaxation time $\tau_{\rm eq}\equiv\tau_{{\rm eq}, g}$.
This represents one of the main theoretical results of this work and provides a generalization {of Eq.~(3.6) of Ref.~\cite{Strickland:2018ayk} to} the case of two different relaxation times for quarks (which, {furthermore}, can be out of chemical equilibrium) and gluons. In fact, by studying this equation {in the $C_R\!=\!1$, $\gamma_{q,0}\!=\!1$ limit (1RTA case in the following)}, we retrieve the well-known result
\begin{align}
    \begin{split}
        M^{nm}(\tau) = &\hspace{0.1cm} (g_g+2g_q)\hspace{0.07cm} \frac{\Gamma(n + 2m +2)}{(2\pi)^2} \biggl\{ D(\tau, \tau_0) \hspace{0.07cm} 2^{\frac{n + 2m +2}{4}} \hspace{0.07cm} T_{0}^{\hspace{0.07cm} n +2m +2} \hspace{0.07cm} \frac{H^{nm}\left( \alpha_0 \hspace{0.07cm} \frac{\tau_0}{\tau} \right)}{\Bigl[H(\alpha_0) \Bigr]^{\frac{n +2m +2}{4}}} \hspace{0.1cm}+ \\
       & + \int_{\tau_0}^{\tau} \frac{d \tau'}{\tau_{\rm eq}(\tau')} \hspace{0.07cm} D(\tau, \tau') \hspace{0.07cm} T^{\hspace{0.07cm} n +2m +2}(\tau') \hspace{0.07cm} H^{nm} \left( \frac{\tau'}{\tau} \right) \biggr\} \,,
       \end{split}
    \end{align}
valid for a system composed of three {different parton species} ($g, q, \Bar{q}$) with one single relaxation time~\cite{Strickland:2018ayk, Florkowski:2013lya}, where, {furthermore,} chemical equilibrium is assumed during the entire evolution.

In addition, from Eqs.~(\ref{eq:split-fs}) and (\ref{eq:split-coll}) it is evident that, in order to compute the moments $M^{nm}$, one needs to determine how the effective temperature varies during the longitudinal expansion of the system.
{This follows from the Landau matching condition in Eqs.~(\ref{eq:LandauMC}) and~(\ref{eq:LandauMC2}). One gets} the following integral equation (see \ref{strick_proof} {for more details}):
\begin{align}
    \begin{split}
    \label{T_ex}
        T^4(\tau) &=  D(\tau, \tau_0) \hspace{0.07cm} \biggl[ \Bar{r} + 2 \hspace{0.07cm} \gamma_{q, 0} \hspace{0.07cm} \biggl( D(\tau,\tau_0)\biggr)^{C_R - 1} \biggr] \hspace{0.07cm} \bigl( 2 \hspace{0.07cm} \gamma_{q,0} + \Bar{r} \bigr)^{-1} \hspace{0.07cm} T_0^4 \hspace{0.07cm} \frac{H \left( \alpha_0 \hspace{0.07cm} \frac{\tau_0}{\tau} \right)}{H(\alpha_0)} \hspace{0.1cm}+ \\
        &+ \frac{C_R}{2 + \Bar{r}} \hspace{0.07cm} \int_{\tau_0}^{\tau} \frac{d \tau'}{2 \hspace{0.07cm} \tau_{\rm eq}(\tau')} \hspace{0.07cm} D(\tau, \tau') \hspace{0.07cm} \biggl[ \frac{\Bar{r}}{C_R} + 2 \hspace{0.07cm} \biggl( D(\tau,\tau')\biggr)^{C_R - 1} \biggr] \hspace{0.07cm} T^4(\tau') \hspace{0.07cm} H \left( \frac{\tau'}{\tau} \right)\,.
    \end{split}
    \end{align}
{In the 1RTA limit this reduces to}
\begin{equation}
        T^4(\tau) =  D(\tau, \tau_0) \hspace{0.07cm} T_0^4 \hspace{0.07cm} \frac{H \left( \alpha_0 \hspace{0.07cm} \frac{\tau_0}{\tau} \right)}{H(\alpha_0)} + \int_{\tau_0}^{\tau} \frac{d \tau'}{2 \hspace{0.07cm} \tau_{\rm eq}(\tau')} \hspace{0.07cm} D(\tau, \tau') \hspace{0.07cm} T^4(\tau') \hspace{0.07cm} H \left( \frac{\tau'}{\tau} \right)\,,
    \end{equation} 
which corresponds to {Eq.~(2.19) of} Ref.~\cite{Strickland:2018ayk}, as expected.
Both in Eqs.~(\ref{eq:split-fs}) and~(\ref{T_ex}) the parameter $\alpha_0 \equiv \left( 1 + \xi_0 \right)^{-1/2}$ appears. This quantity is related to the initial anisotropy parameter $\xi_0$ and it can take values from $0$ to $\infty$, in principle.
{For the case of a medium undergoing a violent longitudinal expansion, as in the initial stages of a heavy-ion collisions, one expects $\xi_0\in(0,+\infty)$ and hence $\alpha_0\in(0,1)$. However, in the following we will provide results also for $\alpha_0 >1$}, just to be as general as possible.
In Eqs.~(\ref{eq:split-fs}) and~(\ref{eq:split-coll}) the same special function as in Ref.~\cite{Strickland:2018ayk} has been used,
\begin{equation}
\label{BF17}
    H^{nm}(y) \equiv \frac{2}{2m + 1} \hspace{0.07cm} y^{2m + 1} \hspace{0.07cm} _2F_1 \left(m + \frac{1}{2}, \frac{1 - n}{2}, m + \frac{3}{2}; 1 - y^2 \right)\,,
\end{equation}
where $_2F_1$ is a Gaussian hypergeometric function.
In particular, when $n = 2$ and $m = 0$ the special function $H^{nm}$ reduces to~\cite{Florkowski:2013lya}:
\begin{equation}
\label{uno}
    H^{20} \equiv H(y) = y \int_0^{\pi} d \theta \hspace{0.07cm} \sin \theta \hspace{0.07cm} \sqrt{y^2 \hspace{0.07cm} \cos^2 \theta + \sin^2 \theta}\,.
\end{equation}
Instead, for $n = 0$ and $m = 1$ we find $H^{01} \equiv H_L$, whose integral representation is given by~\cite{Florkowski:2013lya}:
\begin{equation}
\label{BF20}
    H^{01} \equiv H_L (y) = y^3 \hspace{0.07cm} \int_0^{\pi} d \theta \hspace{0.07cm} \frac{\sin \theta \hspace{0.07cm} \cos^2 \theta}{\sqrt{y^2 \hspace{0.07cm} \cos^2 \theta + \sin^2 \theta}}\,.
\end{equation}
In the {numerical} results {presented in} Sec.~\ref{ex_num}, {the moments} $M^{nm}(\tau)$ {will be conveniently rescaled} by their corresponding value at equilibrium,
\begin{equation}
    \overline{M}^{\hspace{0.05cm} nm}(\tau) \equiv \frac{M^{nm}(\tau)}{M^{nm}_{\rm eq}(\tau)}\,,\label{eq:red-mom}
\end{equation}
{in order to appreciate the asymptotic equilibration of the system. In the above}
\begin{equation}
    \label{BF18}
    M_{\rm eq}^{nm}(\tau) = g_q \hspace{0.07cm} \frac{\Gamma(n + 2m +2)}{(2 \pi)^2} \hspace{0.07cm} \frac{2 \hspace{0.07cm} \bigl( 2 + r \bigr)}{2 \hspace{0.07cm} m + 1} \hspace{0.07cm} T^{n + 2m + 2}(\tau)\,.
\end{equation}
The coupled integral equations for the effective temperature and the general moments of the system {can now be solved}. This will be {the topic of Sec.}~\ref{ex_num}, where {the existence of early and late-time attractors in kinetic theory will be also scrutinized.} {We remind the reader that} the case of one single relaxation time has been already discussed in detail in many previous works, like e.g. Refs.~\cite{Alalawi:2020zbx, Strickland:2018ayk, Soloviev:2021lhs,Strickland:2019hff, Florkowski:2013lya}.

%%%%%%%%%%%%%%%%%%%%%%%%%%%%%%%%%%%%%%%%%%%%%%%%%%%%%%%%%%%%%%%%%%%
\subsection{Shear stress tensor and inverse Reynolds number in 2RTA}
%%%%%%%%%%%%%%%%%%%%%%%%%%%%%%%%%%%%%%%%%%%%%%%%%%%%%%%%%%%%%%%%%%%
{Also in the case of a two-fluid mixture the energy-momentum tensor of the system, receiving contribution -- see Eq.~(\ref{eq:Tmunu-sum}) -- from both particle species, can be written in the following form~\cite{Jaiswal:2013npa, Jaiswal:2013vta, 2021mfrf.book.....D, Baier:2006gy, DelZanna:2013eua, Romatschke:2017ejr}:}
\begin{equation}
\label{tuono}
    T^{\mu \nu} \equiv \varepsilon \hspace{0.07cm} u^{\mu} u^{\nu} - P \hspace{0.07cm} \Delta^{\mu \nu} + \pi^{\mu \nu}\,,
\end{equation}
which holds {since we are working in the conformal limit (no bulk viscosity) and in the Landau frame (no heat conduction)~\cite{Landau1987Fluid}}. In the above equation {$\Delta_{\mu \nu} \equiv g_{\mu \nu} - u_{\mu} u_{\nu}$ is the usual transverse projector and} $\varepsilon$ and $P$ represent the total out-of-equilibrium energy density and pressure of the mixture, which, in general, differ from their equilibrium counterparts due to the Landau matching condition in Eq.~(\ref{eq:LandauMC}): $\varepsilon\!\equiv\!\varepsilon_g\,+\,\varepsilon_q\,\neq\,\varepsilon_{g,{\rm eq}}+\varepsilon_{q,{\rm eq}}$. 

Finally, {$\pi^{\mu \nu}$ represents the shear stress tensor, traceless ($\pi^{\mu}_{\hspace{0.25cm} \mu} \!=\! 0$) and transverse to the fluid velocity ($\pi^{\mu \nu} \hspace{0.07cm} u_{\nu}\!=\!0$)~\cite{Romatschke:2017ejr}.
The above conditions, in the case of Bjorken flow, allow one to write the energy momentum tensor of the fluid in its local rest frame (LRF) as follows~\cite{Jaiswal:2013npa}:}
    \begin{equation}
T^{\mu\nu}_{\rm LRF}=    \begin{pmatrix}
      \epsilon & 0 & 0 & 0\\
      0 & P+\phi/2 & 0 & 0\\
      0 & 0 & P+\phi/2 & 0\\
      0 & 0 & 0& P-\phi 
\end{pmatrix}\equiv
 \begin{pmatrix}
      \epsilon & 0 & 0 & 0\\
      0 & P_T & 0 & 0\\
      0 & 0 & P_T & 0\\
      0 & 0 & 0& P_L 
\end{pmatrix}\,.
    \end{equation}
{The relative size of the shear-stress correction to the energy-momentum tensor of a conformal relativistic fluid is quantified by the inverse Reynolds number~\cite{Strickland:2014pga, Alqahtani:2017mhy, Denicol:2012cn, Strickland:2017kux}
\begin{equation}
    {\rm Re}^{-1}\equiv\frac{\displaystyle{\sqrt{\pi^{\mu\nu}\pi_{\mu\nu}}}}{P_{\rm eq}}=\sqrt{\frac{3}{2}}\frac{|\phi|}{P_{\rm eq}}\equiv \sqrt{\frac{3}{2}}\,|\overline\phi|\,,\label{eq:Reynolds}   
\end{equation}
where the viscous contribution has been normalized to the equilibrium pressure, $\overline{\phi}\!\equiv\!\phi/P_{\rm eq}$. 
}
{Starting from this definition one can express $\overline\phi$ as}
\begin{equation}
    \label{visc}
    \overline{\phi} = \overline{M}^{\hspace{0.05cm} 20} - \overline{M}^{\hspace{0.05cm} 01}\,,
\end{equation}
{which follows from $\phi\!=\!P\!-\!P_L$, from the $P\!=\!\varepsilon/3$ Equation of State (EoS) of a conformal ($T^\mu_{\hspace{0.25cm}\mu}\!=\!0$) system which must hold also out of equilibrium~\cite{Romatschke:2009im, Baier:2007ix, Romatschke:2017ejr}, and from expressing the relevant thermodynamic quantities in terms of the normalized moments of the particle distributions: $\overline M^{\hspace{0.05cm} 01}\!=\! P_L/P_{\rm eq}$ and $\overline M^{\hspace{0.05cm} 20}\!=\!\varepsilon/\varepsilon_{\rm eq}$.}
From Eq.~(\ref{visc}) we can conclude that if the scaled moments $\overline{M}^{\hspace{0.05cm} 20}$ and $\overline{M}^{\hspace{0.05cm} 01}$ possess an attractor, then also $\overline{\phi}$ will exhibit a universal behavior. This consideration is true only in the conformal case, while for non-conformal systems it is not guaranteed due to bulk-viscosity effects, as emphasized in Refs.~\cite{Alalawi:2022pmg,Chattopadhyay:2021ive,Jaiswal:2021uvv}. In the limiting case of 1RTA, for massless particles, the scaled shear correction $\overline{\phi}$ in Eq.~(\ref{visc}) consistently reduces to the expression written in Refs.~\cite{Strickland:2017kux, Jaiswal:2019cju} and it is characterized by an attractor behavior both at early and late times.

{Coming back to Eq.~(\ref{eq:Reynolds}) we notice that vHydro is not suited to deal with situations arbitrarily far from local thermodynamic equilibrium and requires a sufficiently low value of the inverse Reynolds number (hence $|\overline\phi|\ll 1$) to work~\cite{Jaiswal:2019cju, Martinez:2010sc, Jaiswal:2013npa, Muronga:2003ta, Strickland:2014pga}. However, in heavy-ion collisions}, especially in the first stages of the longitudinal expansion, {the momentum distribution of the particles is highly anisotropic} and the viscous correction $\phi$ can {be comparable to} the pressure. {This is a regime in which kinetic theory represents a better approach. In the case of our fluid mixture, solving the 2RTA-BE will allow us to face situations in which vHydro equations would lead to inconsistencies, like e.g. negative longitudinal pressures~\cite{Strickland:2014pga, Martinez:2010sc}.} This topic will be discussed in more detail {while presenting our numerical} results in Sec.~\ref{visc_numerical}. 

{In the following, 2RTA-BE results will be compared with
first-order viscous hydrodynamics predictions (often referred to as relativistic Navier-Stokes theory)}, where the shear tensor is given by~\cite{Muronga:2003ta, Romatschke:2017ejr, Song:2007ux}:
\begin{equation}
    \pi^{\mu \nu}_{\rm NS} \equiv 2 \hspace{0.07cm} \eta \hspace{0.07cm} \sigma^{\mu \nu} \hspace{0.3cm},\hspace{0.3cm} \text{\small{with}} \hspace{0.3cm} \sigma^{\mu \nu} = \frac{1}{2} \hspace{0.07cm} \left( \nabla^{\mu} u^{\nu} + \nabla^{\nu} u^{\mu}  \right) - \frac{1}{3} \hspace{0.07cm} \Delta^{\mu \nu} \hspace{0.07cm} \Theta\,.\label{eq:NS}
\end{equation}
In the above $\eta \equiv \eta_g + \eta_q$ corresponds to the shear viscosity of the mixture of (anti-)quarks and gluons, $\Delta^{\mu} \equiv \Delta^{\mu \nu} \hspace{0.07cm} \partial_{\nu}$ is the transverse gradient and $\Theta\!\equiv\!\nabla_\mu u^\mu $ is the expansion rate of the system. {In the case of Bjorken flow $\theta\!=\!1/\tau$ and, using Eq.~(\ref{eq:NS}), one gets:}
\begin{equation}
\phi_{\rm NS}= \frac{4}{3}\frac{\eta}{\tau}=\frac{4}{3}\frac{\displaystyle{\frac{4}{5}\frac{T^4}{\pi^2}\left(g_g\tau_{{\rm eq},g}+2g_q\tau_{{\rm eq},q}\right)}}{\tau}\,.  
\end{equation}
{After rescaling by the equilibrium pressure $P_{\rm eq}$ one eventually obtains as a result}
\begin{equation}
    \label{navier}
    \Bar{\phi}_{\rm NS} = \frac{16}{15} \hspace{0.07cm} \frac{r + \frac{2}{C_R}}{r + 2} \hspace{0.07cm} \frac{1}{\Bar{\omega}}\,,
\end{equation}
{which depends only on the dimensionless scaling variable} $\Bar{\omega}\,\equiv\,\tau/\tau_{\rm eq}(\tau)$.

%%%%%%%%%%%%%%%%%%%%%%%%%%%%%%%%%%%%%%%%%%%%%%%%%%%%%%%%%%%%%%%%%%%%%%%%
\subsection{Entropy density for a two-component fluid undergoing Bjorken flow}
%%%%%%%%%%%%%%%%%%%%%%%%%%%%%%%%%%%%%%%%%%%%%%%%%%%%%%%%%%%%%%%%%%%%%%%%
Once the exact distribution function for quarks and gluons is known, one can compute the entropy density of the system through the following integral in momentum space~\cite{Florkowski:2013lya, Hosoya:1983xm}:
\begin{equation}
    \label{entropy}
    s(\tau) = - \sum_{a = g, q} G_a(\tau) \hspace{0.07cm} \int d\chi \hspace{0.07cm} (p \cdot u) \hspace{0.07cm} \widehat{f}_a(\tau; w , p_T) \hspace{0.07cm} \biggl\{ \ln \Bigl[ \widehat{f}_a(\tau; w , p_T) \Bigr] - 1 \biggr\}\,,
\end{equation}
which is valid for classical particles. Here $\widehat{f}$ represents the functional part of the exact distribution function in Eq.~(\ref{BF12}), {the one carrying the dependence on the particle momentum}, without {the number of active} the degrees of freedom, {which} are factorized out of the integral.
One has to pay attention to the fact that, {for a mixture of different parton species},
\begin{equation*}
G_a(\tau)\equiv\widetilde g_a \, \gamma_a(\tau) 
\end{equation*}
{can vary} during the expansion of the {system, since, beside counting through the $\widetilde g_a$ coefficient the internal degrees of freedom (color, spin, flavor) which are fixed, it also includes a fugacity factor $\gamma_a$ accounting for possible deviations from chemical equilibrium, which will be studied in detail}, in Sec.~\ref{distrib_num}.
{In the above, the $\widetilde g_a$ coefficient ( $\widetilde g_g\equiv g_g$ and $\widetilde g_q\equiv 2 g_q$) in the case of quarks includes a factor 2 accounting for particle-antiparticle degeneracy; this allows a unified treatment of quarks and gluons.

The starting point to study the evolution of $\gamma_a(\tau)$ is the exact solution of the RTA-BE for the single-particle distribution at zero momentum ($w = p_T = 0$), namely:
\begin{equation}
    f_a (\tau; 0, 0) = D_a (\tau, \tau_0) \hspace{0.07cm} f_{0, a} (0, 0) + \int_{\tau_0}^{\tau} \frac{d \tau'}{\tau_{{\rm eq}, a}(\tau')} \hspace{0.07cm} D_a (\tau, \tau') \hspace{0.07cm} f_{{\rm eq}, a} (\tau'; 0, 0)\,.\label{eq:p=0}
\end{equation}}
{Inserting the definition 
\begin{equation*}
f_a (\tau;w,p_T)\equiv G_a(\tau) \, \widehat{f}_a(\tau; w , p_T)\,,\quad{\rm with}\quad   \widehat{f}_a(\tau;0,0)=1\,, 
\end{equation*}
into Eq.~(\ref{eq:p=0}), taking into account that $\gamma_{a,{\rm eq}}\!=\!1$ and exploiting Eq.~(\ref{damp_prop}) one gets
\begin{equation}
\label{gamma_q}
\gamma_a(\tau)=D_a(\tau,\tau_0) \, \gamma_{a,0}+\Bigl[1-D_a(\tau,\tau_0)\Bigr]\,. 
\end{equation}
As expected, independently from the initial condition (no matter how small $\gamma_{a,0}$ is), asymptotically all particle species approach chemical equilibrium. Notice that, when a particle species is initially at chemical equilibrium, i.e. $\gamma_{a,0}\!=\!1$, the latter is preserved during the entire system's evolution, as will be confirmed also by our the numerical results in Sec.~\ref{distrib_num}. Even if the latter formalism is pretty general, in the following we will mainly address the case in which gluons are in chemical equilibrium, while (anti-)quarks are initially under-populated, trying to  ``disentagle" the entropy production associated to the system's isotropization from the one related to $q\overline q$ production (see Sec.~\ref{entropy_num}).} 

{It will be interesting to perform the comparison between the kinetic-theory result for the entropy density provided by Eq.~(\ref{entropy}) and the corresponding value for a system in thermodynamic equilibrium at the same temperature~\cite{Florkowski:2013lya, Romatschke:2009im}:}
\begin{equation}
    \label{entropy_eq}
    s_{\rm eq}(\tau) = 4 \, n_{\rm eq}(\tau) = \frac{4}{\pi^2}\left(g_g+2g_q \right) \hspace{0.07cm} T^3(\tau)\,,
\end{equation}
where the effective temperature corresponds to the solution of Eq.~(\ref{T_ex}). {This will represent also a consistency check of our numerical calculations, since one must have $s(\tau)\underset{\tau\to\infty}{\sim}s_{\rm eq}(\tau)$, with $s(\tau)$ approaching $s_{\rm eq}(\tau)$ from below, the equilibrium state being the one which maximizes the entropy.}

{Interestingly, starting with the SR distributions in Eqs.~(\ref{rta3}) and~(\ref{rta4}) one can find an analytic expression for the initial entropy density:}
\begin{equation*}
    s_0 = \frac{4}{\pi^2} \hspace{0.07cm} g_q \hspace{0.07cm} \left( 2 \hspace{0.07cm} \gamma_{q,0} + r \right) \hspace{0.07cm} \alpha_0 \hspace{0.07cm} \Biggl[\frac{2}{H(\alpha_0)} \hspace{0.07cm} \frac{2 + \Bar{r}}{2 \hspace{0.07cm} \gamma_{q, 0} + \Bar{r}} \Biggr]^{3/4} \hspace{0.07cm} T_0^3
\end{equation*}
which consistently reduces to the ideal expression when the system is initialized {at kinetic, $\alpha_0 = 1$, and chemical, $\gamma_{q, 0} = 1$, equilibrium}.

%%%%%%%%%%%%%%%%%%%%%%%%%%%%%%%%%%%%%%%%%%%%%%%%%%%%%%%%%%%%%
\section{Numerical results from 2RTA kinetic theory in the Bjorken-flow regime}
\label{ex_num}
%%%%%%%%%%%%%%%%%%%%%%%%%%%%%%%%%%%%%%%%%%%%%%%%%%%%%%%%%%%%%

Here we show the main numerical results obtained {in our 2RTA-BE description of a quark-gluon mixture}. This section is structured as follows. In Sec.~\ref{mom_num} we study the moments {of the single-particle} distribution function -- see Eq.~(\ref{robo}) -- for $n$, $m \in$ $[0,2]$, highlighting the analogies and differences among these cases.

{Then}, in Sec.~\ref{visc_numerical}, we address the evolution of the scaled viscous correction to the pressure {given by} Eq.~(\ref{visc}). Furthermore, in Sec.~\ref{distrib_num} we investigate the evolution of the distribution function itself, written in Eq.~(\ref{BF12}), {looking at differences in the isotropization of quarks and gluons and studying the approach of quarks to chemical equilibrium} during the system's longitudinal expansion.
In conclusion, in Sec.~\ref{entropy_num} we {display results for} the scaled entropy density -- see Eq.~(\ref{entropy}) -- and for the entropy production.

 \begin{figure}[!hbt]
    \centering
    \includegraphics[width=0.495\textwidth] {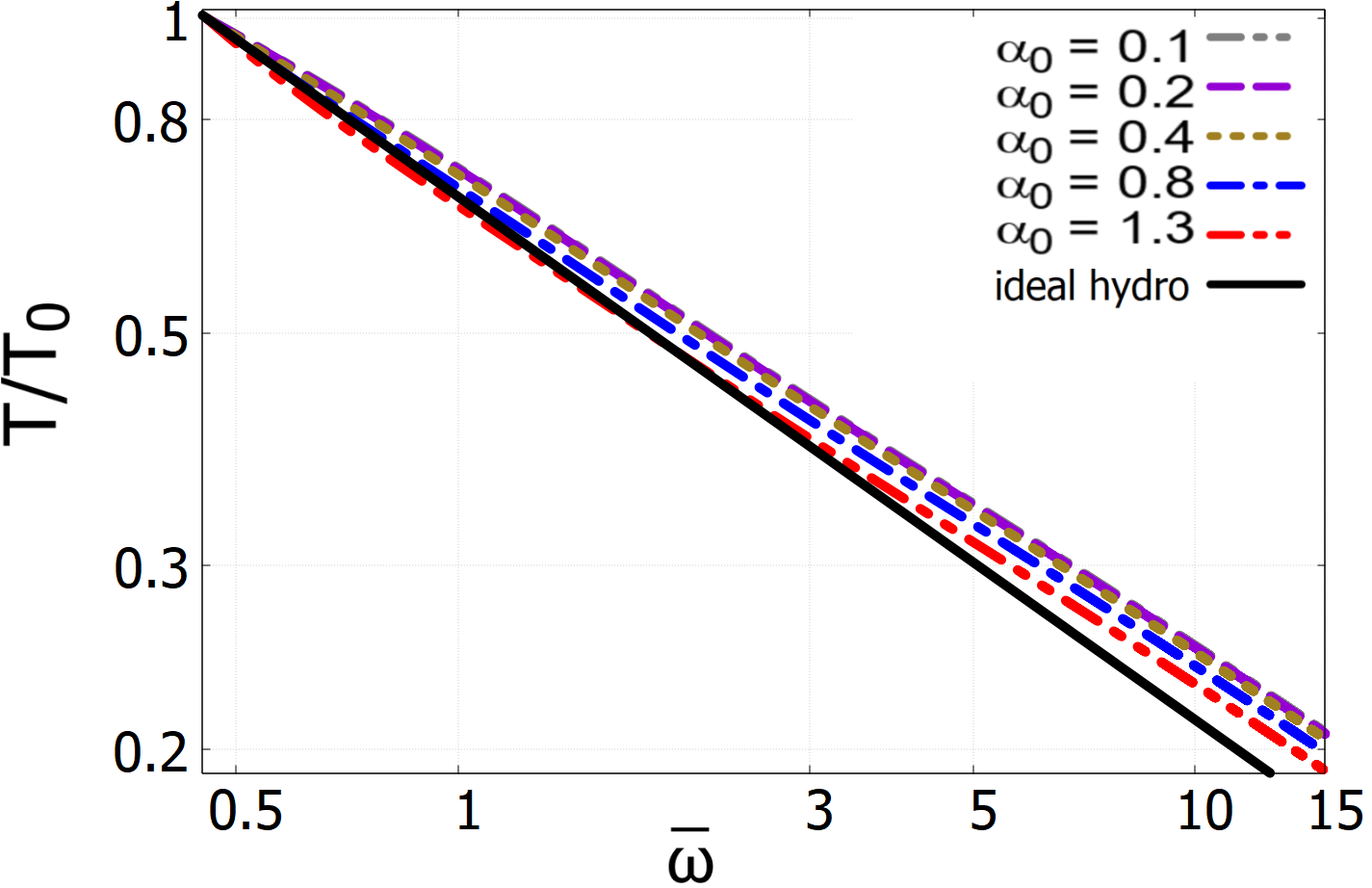}
    \includegraphics[width=0.495\textwidth] {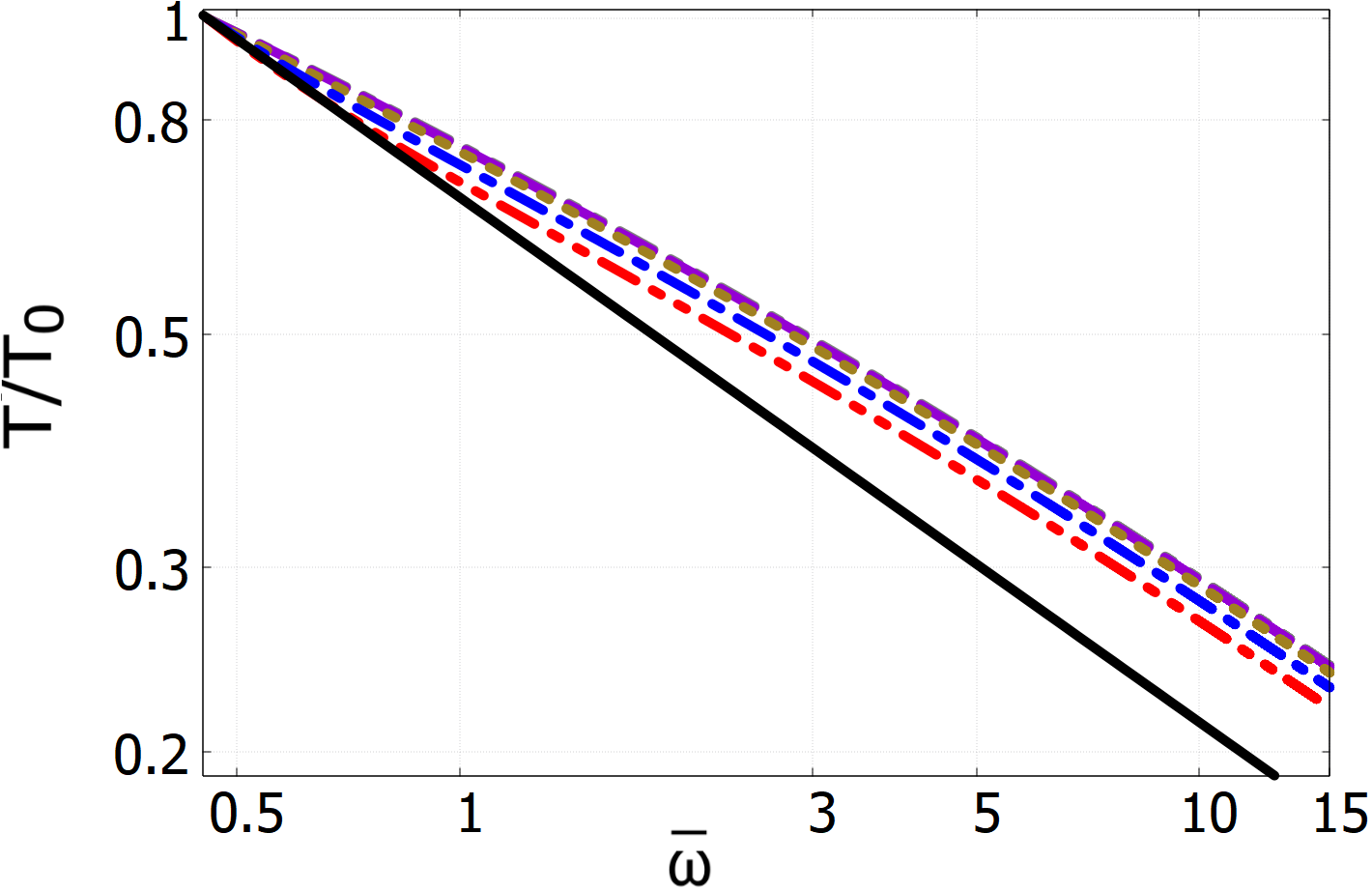} \\ 
    \caption{{Evolution of the} effective temperature of a mixture of quarks and gluons as a function of the scaled time $\overline{\omega}\equiv\tau/\tau_{\rm eq}(\tau)$. {The system's evolution is} initialized at $\tau_0 = 0.15$ fm/c and $T_0 = 600$ MeV. {The displayed curves refer to a specific gluon viscosity $\eta/s = 0.2$ and different values of the} initial anisotropy parameter $\alpha_0$. A logarithmic scale has been adopted for both axes {in order to emphasize deviations from the ideal power-law behavior $T\!\sim\!\tau^{-1/3}$}. {Results in the left panel refer to an initial under-population of quarks,} $\gamma_{q, 0} = 0.1$, {while the right panel refers to the chemical equilibrium case}, $\gamma_{q, 0} = 1.0$.}
    \label{temp}
\end{figure}
{In the following, all quantities will be studied as functions} of the {dimensionless scaling variable} $\Bar{\omega} \equiv \tau/\tau_{\rm eq}(\tau)$, where the expression for $\tau_{\rm eq}(\tau)$ is provided by Eq.~(\ref{eq2}). {The starting point of our numerical study is then the self-consistent solution of the integral equation (\ref{T_ex}) for the temperature evolution of the system, which is obtained through subsequent iterations of a secant root solver applied to a discretized version of Eq.~(\ref{T_ex}), using a trapezoidal quadrature rule, until convergence to the desired numerical precision is achieved for all values of $\tau$ (tolerance is set to $10^{-8}$). Our results are shown}  in Fig.~\ref{temp} for different initializations of the {momentum-space} anisotropy {parameter} $\alpha_0$ and {of the quark fugacity} $\gamma_{q, 0}$ {and compared to the ideal behaviour $T\,\sim\,\tau^{-1/3}$}.
{Our results} are qualitatively similar to those presented in Refs.~\cite{Florkowski:2013lya, Jaiswal:2013vta}, obtained in 1RTA.

{Since we are considering a dissipative medium with a finite relaxation time, due to entropy production, during the system expansion the temperature decreases more slowly than in the ideal case. Comparing the two panels in Fig.~\ref{temp} one observes that starting with an initial under-population of quarks (left panel) the temperature drops faster than in the chemical-equilibrium case (right panel). In the second case entropy is in fact produced essentially only in the form of heat, taming the decrease of the average energy per particle during the collective expansion of the medium, while starting with $\gamma_{q,0}\!\ll\! 1$ entropy production occurs also through the excitation of quark degrees of freedom, initially under-populated.}

From Fig.~\ref{temp} it is also evident that there is no attractor behavior for $T/T_0$, even when plotted as a function of {the scaling variable} $\Bar{\omega}$.
Since for a conformal system {$M^{nm}\!\sim\!T^{\hspace{0.05cm} n+2m+2}$ it will be important to bear in mind this fact in considering}
the potential forward-attractor property of the scaled moments $\Bar{M}^{\hspace{0.05cm}nm} \equiv M^{nm}/M^{nm}_{\rm eq}$ {which, if present, holds only because the explicit dependence on the temperature cancels in the ratio}. {This issue will be addressed in the following subsection.}
 
%%%%%%%%%%%%%%%%%%%%%%%%%%%%%%%%%%%%%%%%%%%%%%%%%%%%%%%%%%%%%%%%%%%%%%%
\subsection{Numerical solutions for the scaled moments}
\label{mom_num}
%%%%%%%%%%%%%%%%%%%%%%%%%%%%%%%%%%%%%%%%%%%%%%%%%%%%%%%%%%%%%%%%%%%%%%%

{We start this section displaying our results for} the normalized general moments $\overline{M}^{\hspace{0.05cm} nm}$, {plotted as functions} of the dimensionless scaled time $\Bar{\omega}$.

All the following plots {refer} to a specific viscosity $\eta/s = 0.2$ {for the gluon component -- see the discussion after Eq.~(\ref{eq:etastot})} --  which represents a reasonable value for the QGP {produced in relativistic HIC's collisions, close to the universal lower bound provided by} the AdS/CFT correspondence~\cite{Busza_2018, Casalderrey-Solana:2011dxg, Schenke:2011bn, Romatschke:2007mq, Bhalerao:2014owa, Song:2007ux}.
{However, one can easily verify that, due to conformal symmetry, indepentently from the value of $\eta/s$ all solutions of the kinetic equations for $\overline{M}^{\hspace{0.05cm} nm}$ tend to the same attractor once plotted as functions of the dimensionless time $\Bar{\omega}$}.
{Employing a spheroidal form - see Eqs.~(\ref{rta3}) and~(\ref{rta4}) - for the initial distribution function~\cite{Strickland:2014pga, Florkowski:2013lya, Alqahtani:2017mhy}, in all our plots} we compare the attractor solution (solid black line) for each scaled moment with a set of representative solutions {of kinetic equations} (colored dashed lines) {referring to different} initial momentum-space anisotropy {coefficient} $\alpha_0$ defined in Eq.~(\ref{eq:alpha0def}).

{In the following we will explore the values}
$ \alpha_0= \{0.1,0.2, 0.4, 0.8\}$, {referring to an oblate initial momentum distribution ($\xi_0>0$, $\langle p_z^2\rangle< \langle p_{x}^2\rangle,\langle p_y^2\rangle$, as expected in HIC's due to the rapid initial longitudinal expansion) and $\alpha_0=1.3$ representative of the prolate case ($-1<\xi_0<0$, $\langle p_z^2\rangle> \langle p_{x}^2\rangle,\langle p_y^2\rangle$)}. 
The initial {effective} temperature {is set to} $T_0 = 600$ MeV at $\tau_0 = 0.15$ fm/c, {much smaller than the usual initial time of hydrodynamic calculations, since one of our purposes is to study the pre-equilibrium dynamics of the medium as well. Also for our two-component system} a forward {late-time} attractor for these moments is {observed when}  keeping the initialization time $\tau_0$ and effective temperature $T_0$ constant while varying the initial momentum-space anisotropy $\alpha_0$~\cite{Alalawi:2022pmg, Strickland:2018ayk}.

To compute the attractor, we use $\tau_{0, {\rm attr}} = 0.001$ fm/c and tune the initial anisotropy to $\alpha_0 \simeq 0.0025$, in order to obtain a slow-roll and non-diverging solution for $\Bar{\omega} \rightarrow 0$. These are the defining requirements for the attractor solution, according to Refs.~\cite{Strickland:2018ayk, Romatschke:2017vte, Strickland:2017kux, Jankowski:2023fdz}.  Furthermore, {for each of the above cases} we {explore} different values of the initial {quark fugacity} $\gamma_{q, 0}$. Since in the early stages the plasma is expected to be {gluon-dominated}~\cite{Kurkela:2018oqw, Lappi:2006fp, Gelis:2010nm}, we decided to {compare} the $\gamma_{q, 0} = \{0.1, 1\}$ cases, in order to appreciate discrepancies {due to deviations of quarks from chemical equilibrium.  Setting $\gamma_{q, 0} = 0.1$ provides a more realistic description of the QGP produced in HIC's~\cite{Strickland:1994rf, Martinez:2008di,Martinez:2007pjh}}.
{Accordingly}, the attractor solution, starting at very early times, should be initialized with an extremely small population of quarks. In our simulations, we use $\gamma_{q, 0} = 10^{-4}$ for the attractor, given that this solution, in practice, remains unchanged {even when} lower initial values of the fugacity are {employed}.
{Notice however that, because of Eq.~(\ref{gamma_q})}, $\gamma_{q,0} = 1$ {is a stable fixed point. Hence, in the chemical equilibrium case} we need to {set $\gamma_{q,0} = 1$} for the attractor solution as well. {In fact, our implementation of the kinetic equations} does not allow the {relative quark} abundance to change over time if quarks are taken in chemical equilibrium at the beginning of the longitudinal expansion.

In addition, for the {same} scaled moments, we also study the presence of a {possible} pullback early-time attractor by holding the initial anisotropy and temperature constant while adjusting the initialization time {to earlier and earlier values}~\cite{Alalawi:2022pmg, Jankowski:2023fdz, Strickland:2017kux}.

\begin{figure}[!hbt]
    \centering
    \includegraphics[width=0.325\textwidth] {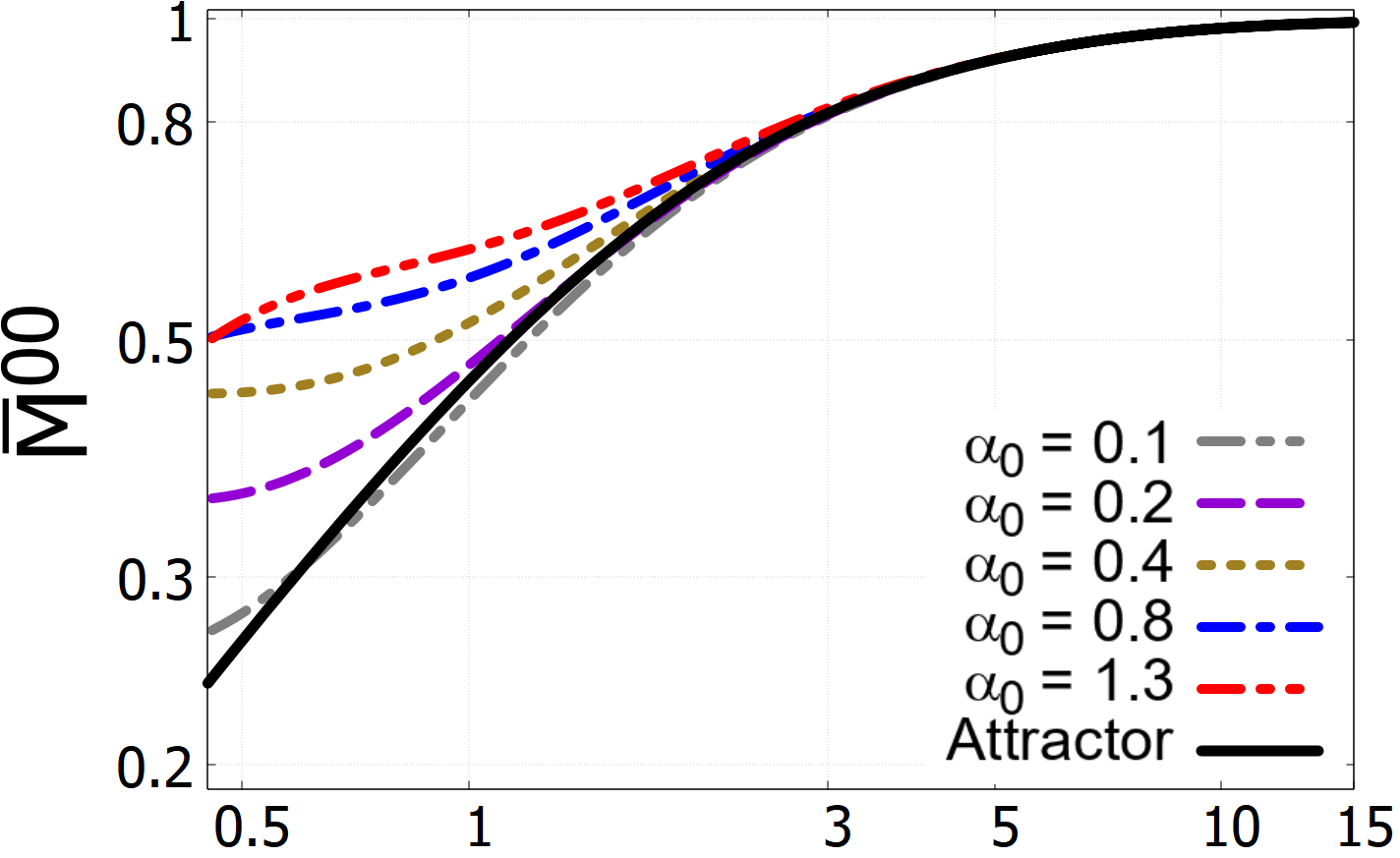} 
    \includegraphics[width=0.325\textwidth] {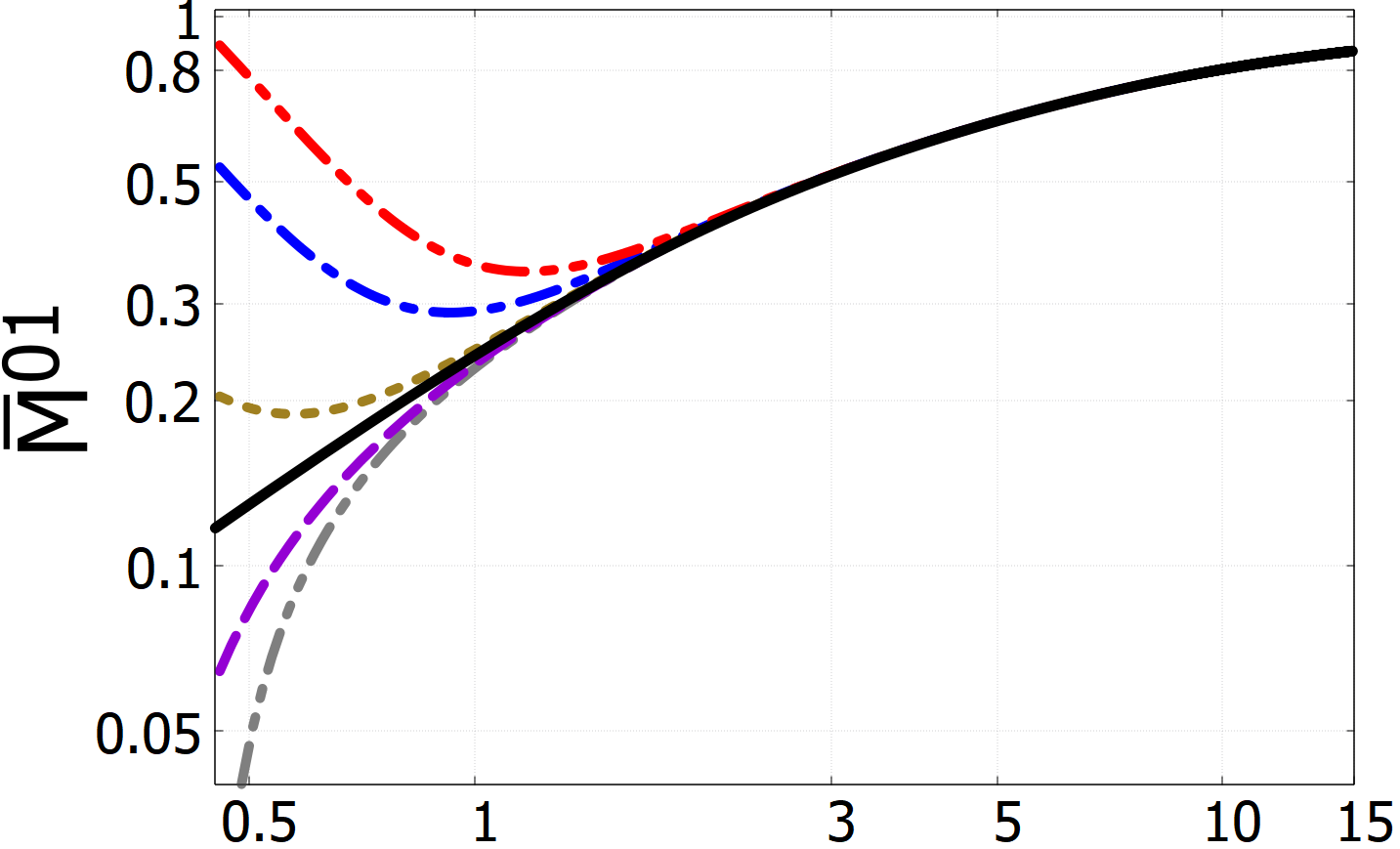}
    \includegraphics[width=0.325\textwidth] {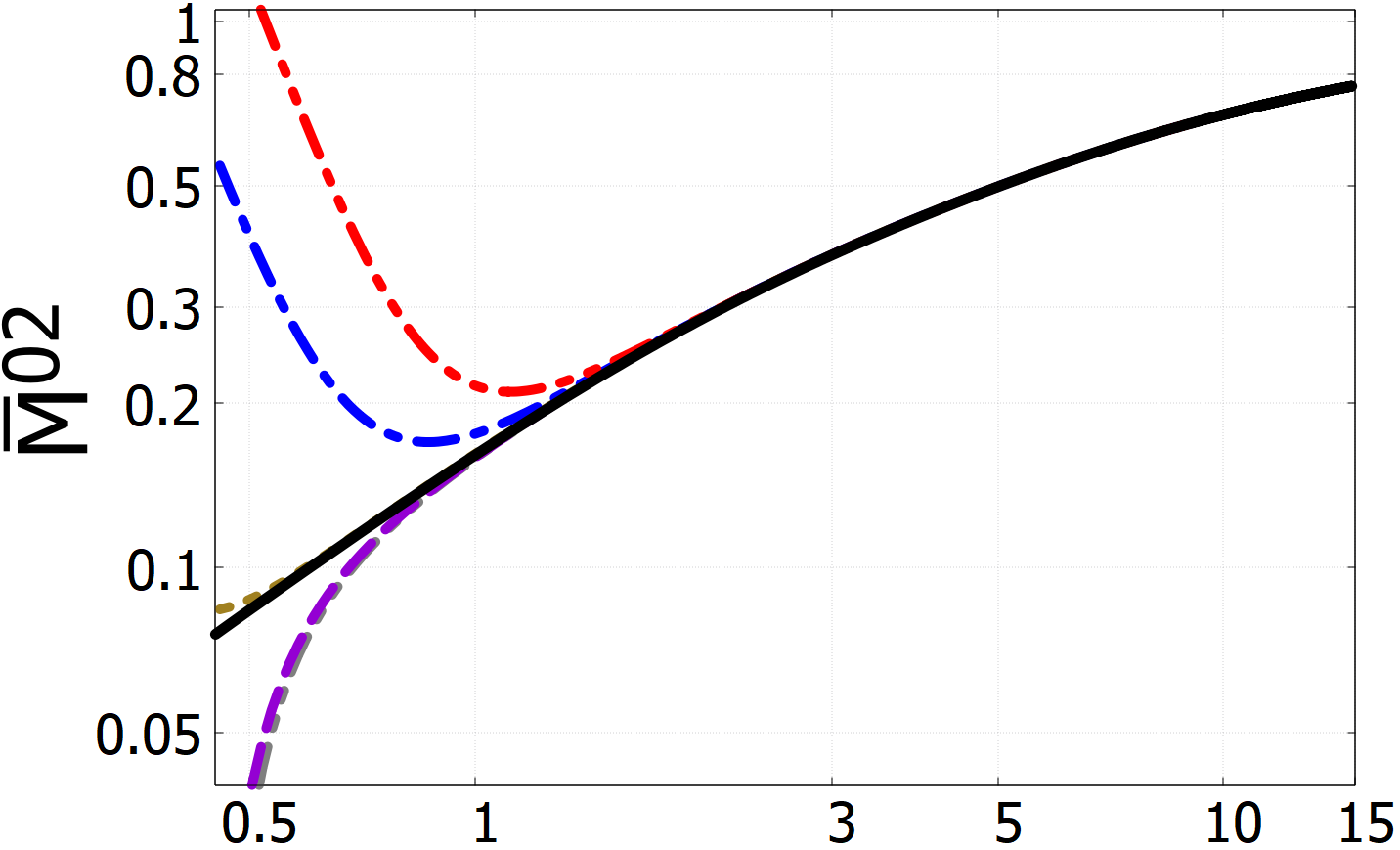} \\
    \includegraphics[width=0.325\textwidth] {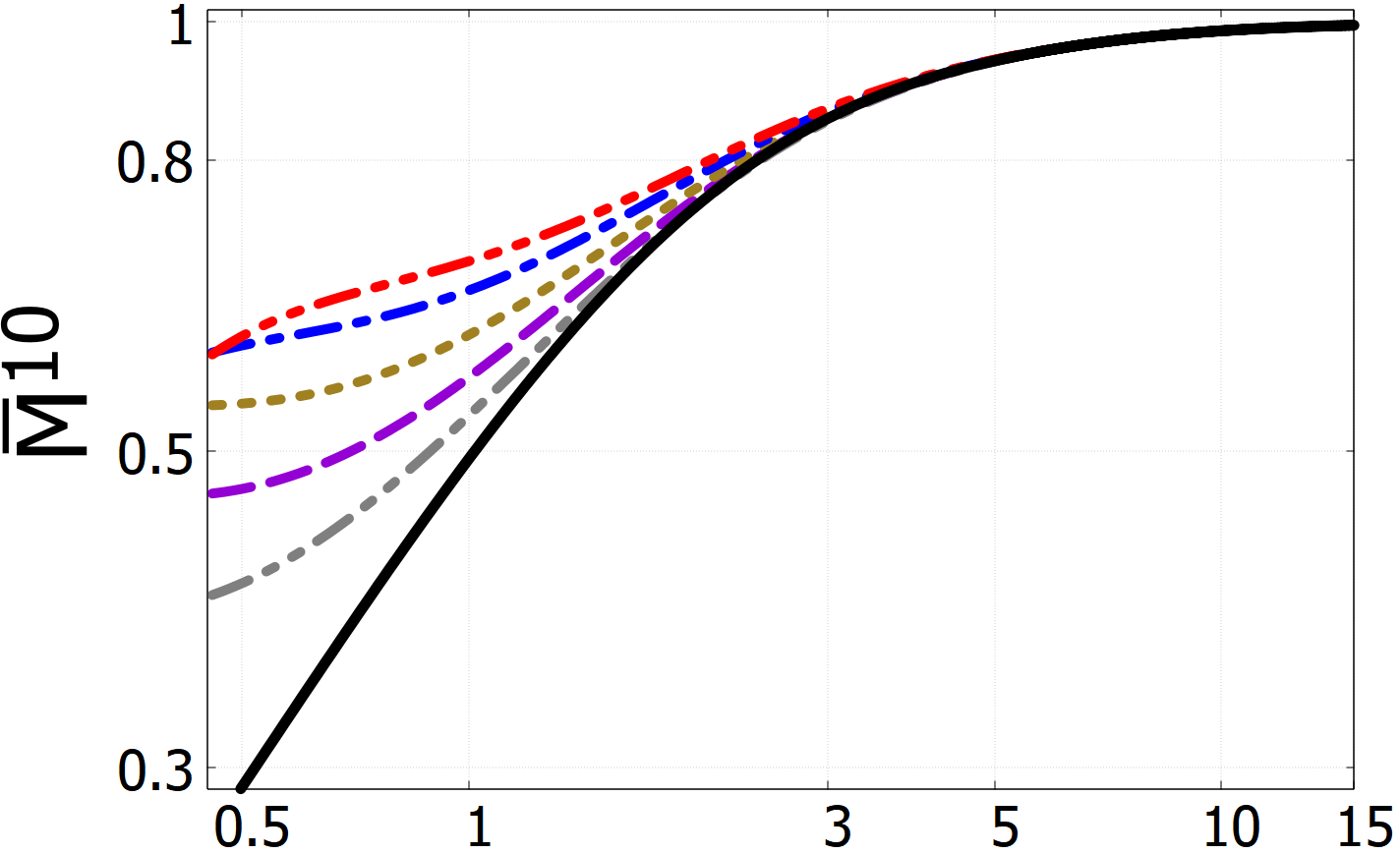} 
    \includegraphics[width=0.325\textwidth] {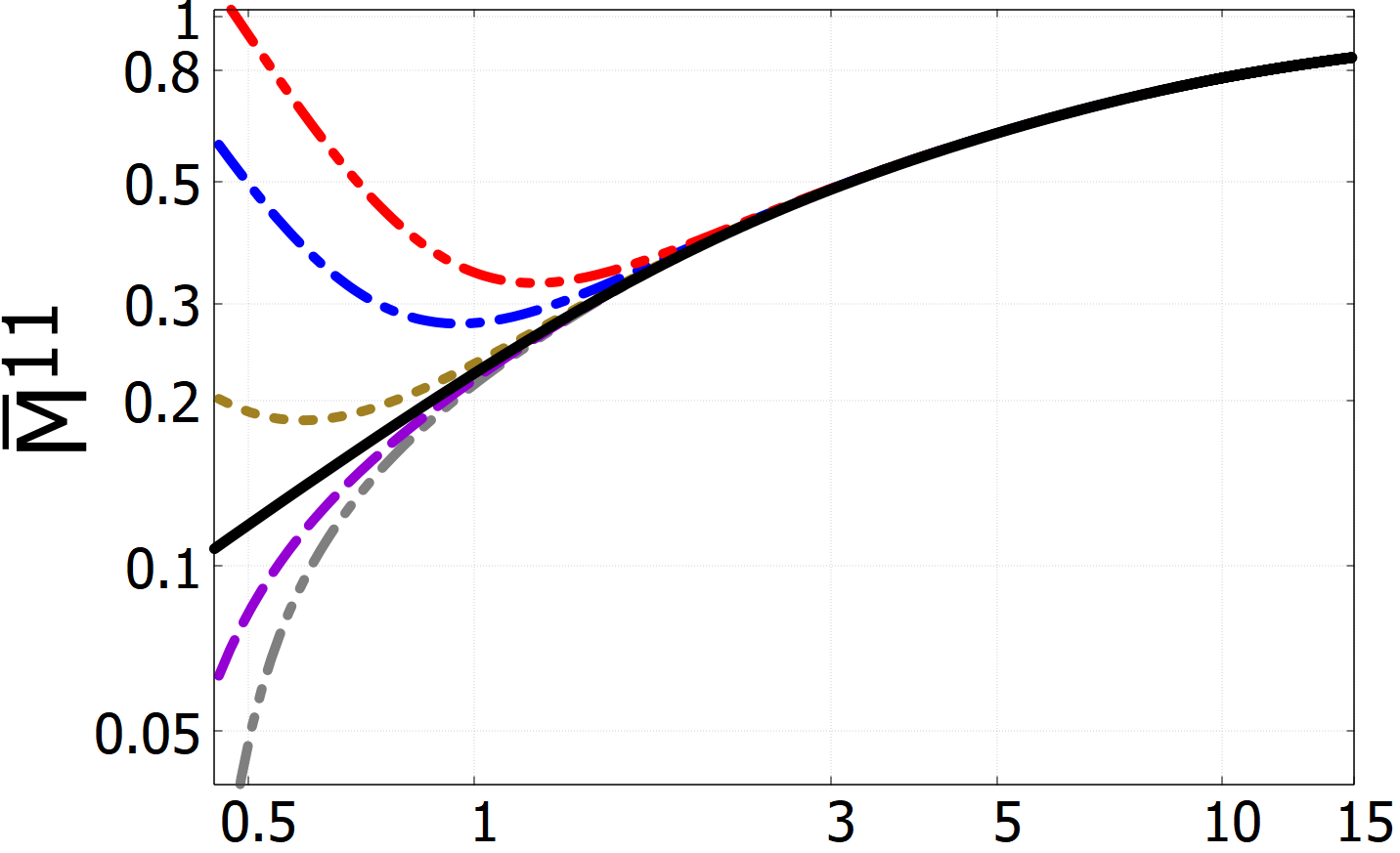}
    \includegraphics[width=0.325\textwidth] {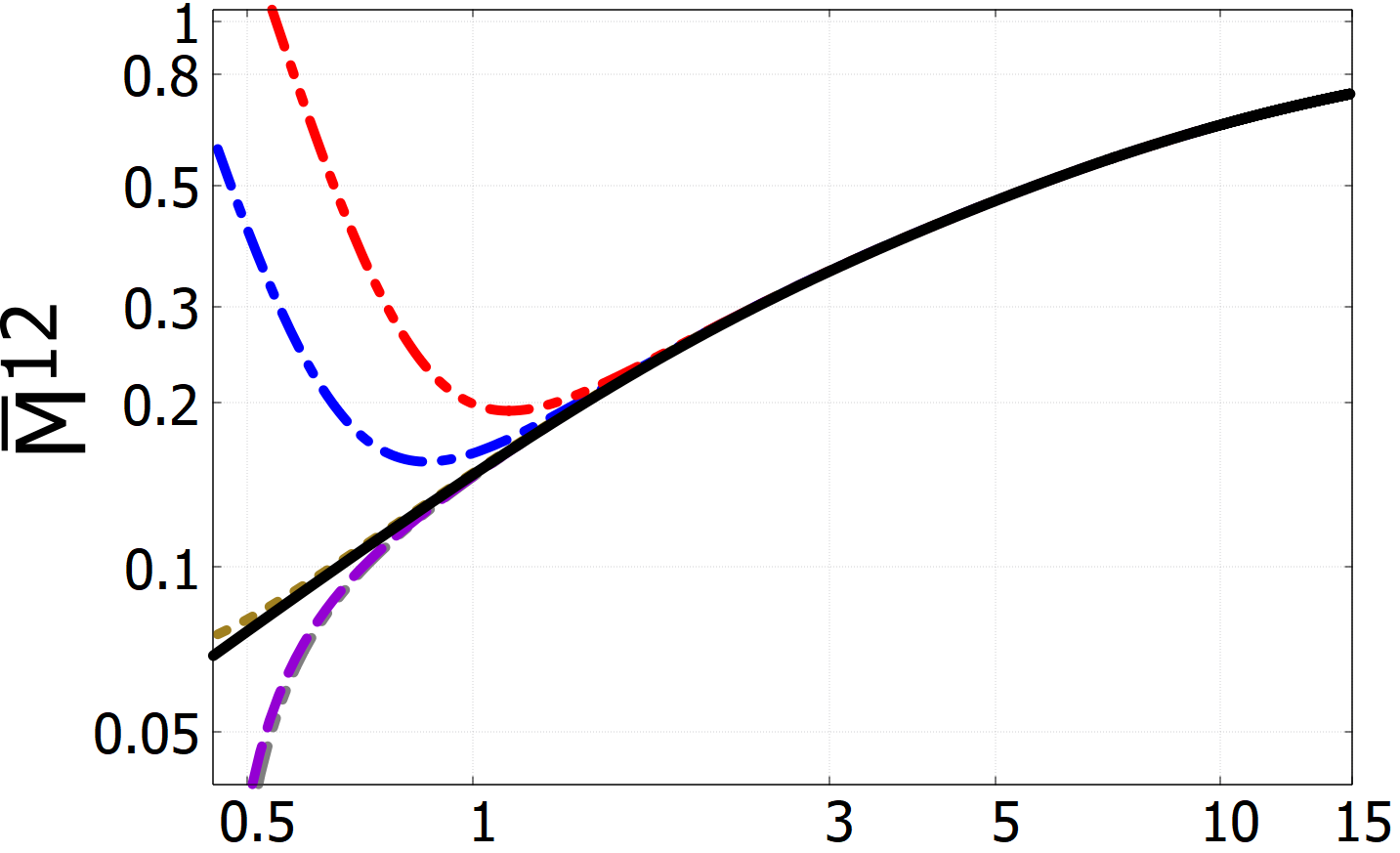}\\
    \includegraphics[width=0.325\textwidth] {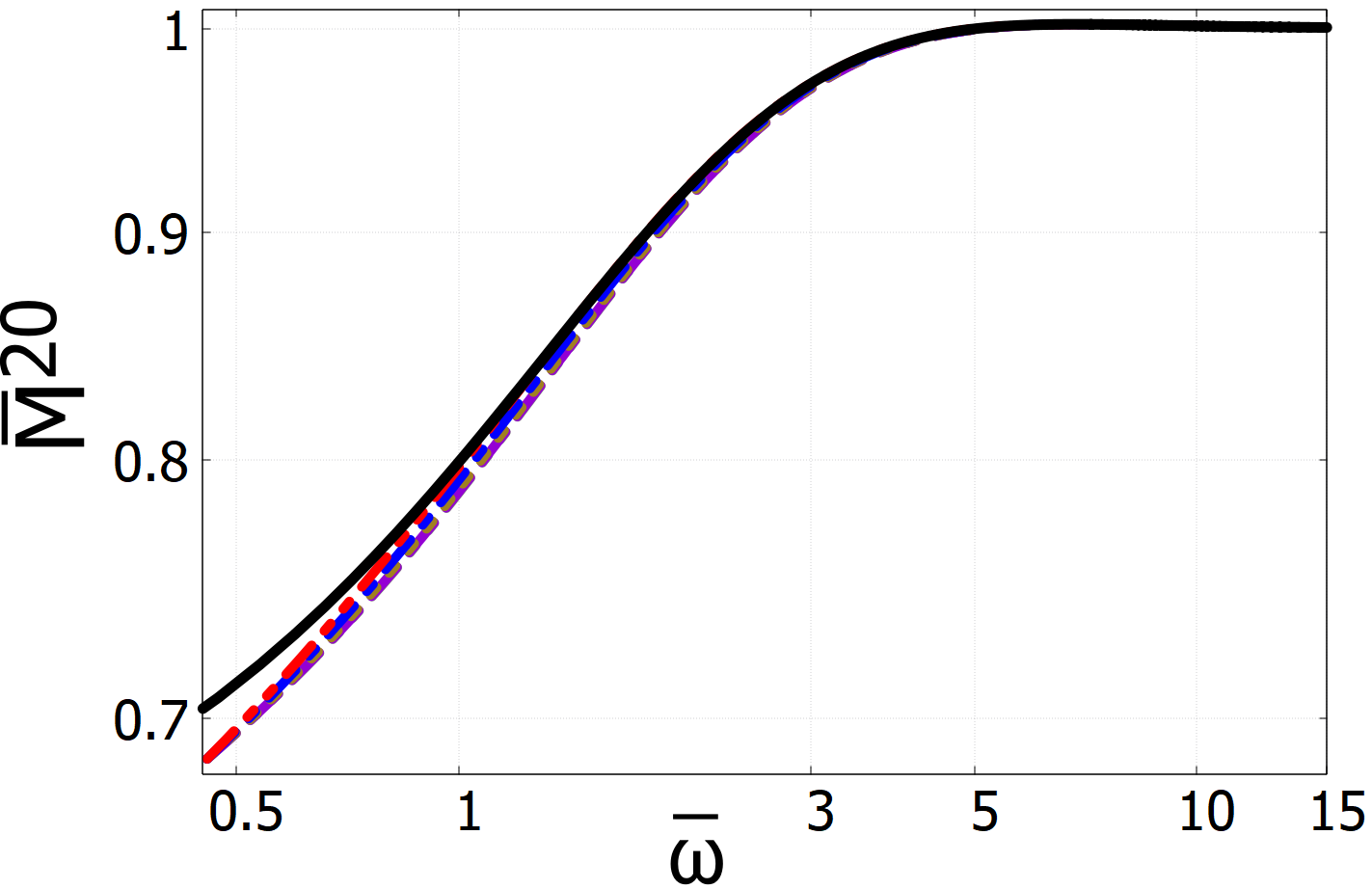} 
    \includegraphics[width=0.325\textwidth] {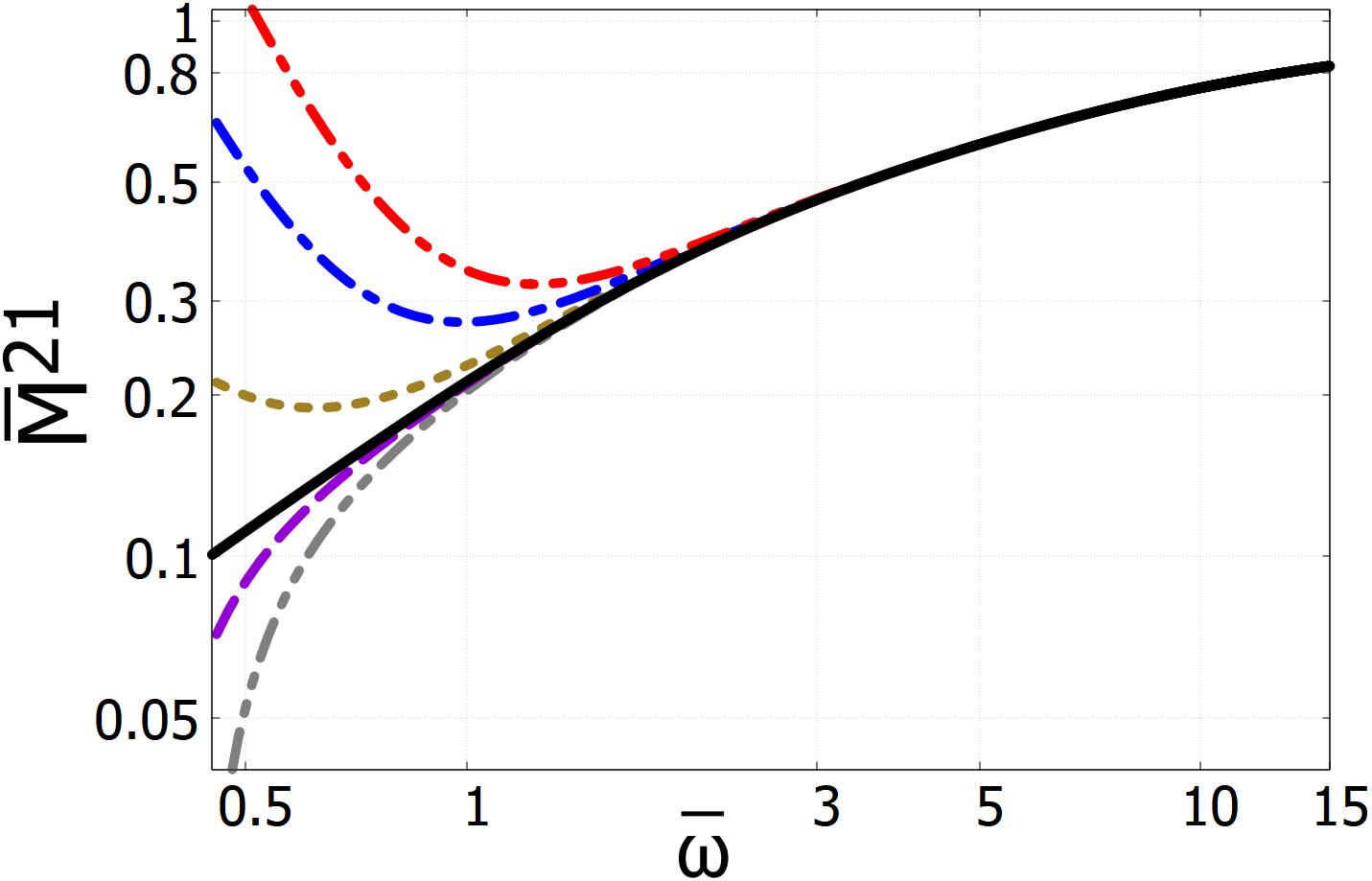}
    \includegraphics[width=0.325\textwidth] {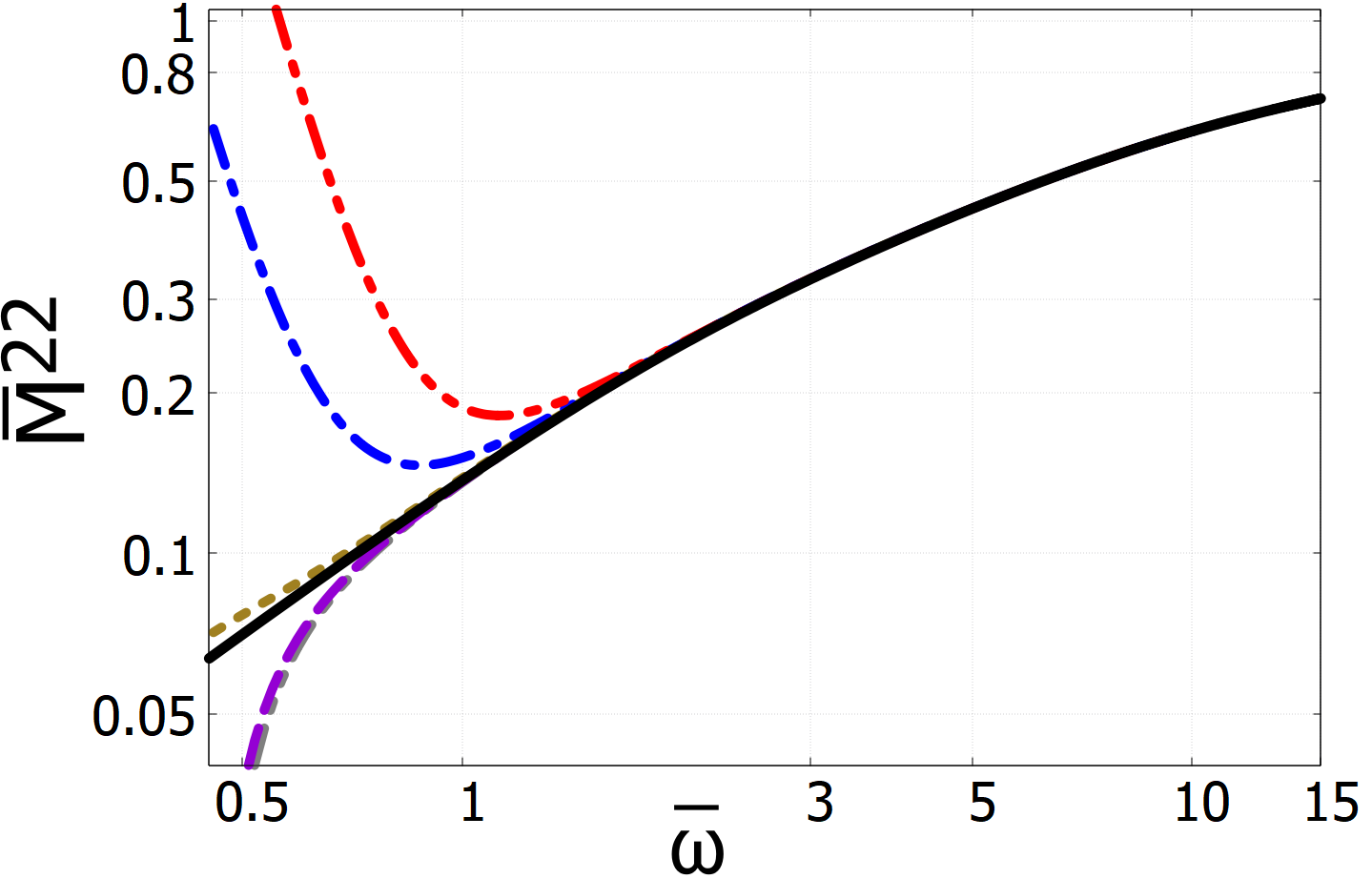}  
    \caption{Scaled moments $\overline{M}^{\hspace{0.05cm} nm}$ obtained from the attractor solution (solid black line) compared to a set of exact solutions {of the 2RTA-BE} (various dashed colored lines) initialized at $\tau_0 = 0.15$ fm/c and $T_0 = 600$ MeV, with varying initial anisotropy parameter $\alpha_0$ at {fixed specific gluon viscosity} $\eta/s = 0.2$. {All curves are plotted as functions of the scaling variable} $\Bar{\omega} = \tau/\tau_{\rm eq}$. The {initial quark} fugacity parameter is set to $\gamma_{q, 0} = 0.1$.}
    \label{fig:mom-late-under}
\end{figure}
{We start our analysis considering the hydrodynamization of our two-component system, here studied through the lowest-order} scaled moments $\overline{M}^{\hspace{0.05cm} nm}$, {focusing in particular on their convergence to the corresponding late-time attractor solutions. We first address the case of an initially gluon-dominated plasma, setting} $\gamma_{q, 0}= 0.1$ {for the quark fugacity. Our} numerical results are presented in Fig.~\ref{fig:mom-late-under}.
{First of all, we notice that, at variance with the 1RTA case,} $\overline{M}^{\hspace{0.05cm} 20}\!\equiv\!\varepsilon/\varepsilon_{\rm eq}\!\ne\! 1$ {during most of the system's expansion}. {This follows from the more complex Landau matching condition in the 2RTA case given in Eq.~(\ref{eq:LandauMC}), as shown in Ref.~\cite{Florkowski:2012as}. On the contrary, with a single relaxation time, one would have $\varepsilon=\varepsilon_{\rm eq}$ throughout the entire evolution~\cite{Strickland:2018ayk, Florkowski:2013lya, Maksymiuk:2017szr}.}
{Notice that, in light of Eqs.~(\ref{eq:split-fs}) and~(\ref{eq:split-coll})} all the solutions for the {rescaled} energy density start from the same value
\begin{equation*}
\overline{M}^{\hspace{0.05cm} 20}(\tau_0) = \frac{2 \hspace{0.07cm} \gamma_{q, 0} + r}{2 + r} \hspace{0.07cm} \frac{2 + \Bar{r}}{2 \hspace{0.07cm} \gamma_{q, 0} + \Bar{r}}\,.   
\end{equation*}
{On the other hand, for $\tau>\tau_0$ the collisional term in Eq.~(\ref{eq:split-coll}) starts to play a role and the system} evolution depends on the chosen initial condition for the anisotropy coefficient $\alpha_0$.
{As a numerical consistency check of our code we can define the quantity $\varepsilon_c \equiv \varepsilon_q + \varepsilon_g/C_R$. From the generalized Landau matching condition in Eq.~(\ref{eq:LandauMC})} the quantity $\Bar{\varepsilon}_c \equiv \varepsilon_c/\varepsilon_{c, {\rm eq}}$ should be equal to one at all times. Deviations from this value {can only be attributed to errors or lack of numerical precision in the code.}
Such an inspection has been performed for all the {initial momentum-anisotropy and fugacity coefficients employed in our numerical calculations, resulting in solutions always satisfying this consistency condition for all values of $\Bar{\omega}$}.
As can be seen from Fig.~\ref{fig:mom-late-under}, {for all the considered moments} our solutions approach the {corresponding} attractor in a finite amount of time, {after which the} information about the initial configuration of the system is completely lost~\cite{Strickland:2018ayk, Soloviev:2021lhs, Florkowski:2013lya, Romatschke:2017ejr}. Moreover, the hydrodynamization process occurs well before the system reaches {local thermodynamic} equilibrium (where $\overline{M}^{\hspace{0.05cm} nm} \simeq 1$). The {approach to} the late-time attractor is governed by the exponential decay of linearized non-hydrodynamic modes~\cite{Kurkela:2019set, Jankowski:2023fdz} and this is a consequence of the collisions among particles in the plasma, {whose} effect is contained in Eq.~(\ref{eq:split-coll}), which becomes dominant, {with respect to the free-streaming term}, at late times ($\Bar{\omega} > 1$)~\cite{Soloviev:2021lhs, Strickland:2017kux, Broniowski:2008qk}. {In the following, as a criterium to quantify the convergence of a generic solution for the scaled moment $\overline{M}^{\hspace{0.05cm} nm}$ to the corresponding late-time attractor, we will request a relative error smaller than $10^{-4}$}.
The slowest convergence seems to be the one for moments with $m = 0$, which appear in the leftmost column. Considering for example $\overline{M}^{\hspace{0.05cm} 10}$, the generic solutions approach the attractor only after $\Bar{\omega} \gtrsim 4$. For $m \neq 0$, however, one sees that all the moments computed from {any initialization} visibly merge with the forward attractor after a shorter scaled time: $\Bar{\omega} \gtrsim 3$ if $m = 1$, $\Bar{\omega} \gtrsim 2$ if $m = 2$. Hence, the greater the index $m$, the faster the approach to the attractor at late times. This {looks} in agreement with the considerations provided in Refs.~\cite{Strickland:2018ayk, Strickland:2019hff, Alalawi:2022pmg}. The convergence scaled-time seems also weakly dependent on the value of $n$, in {the explored cases}. {All these issues are} discussed in {deeper} detail in Refs.~\cite{Strickland:2018ayk, Strickland:2019hff}, for instance, where a single relaxation time (1RTA) is assumed.

 \begin{figure}[!hbt]
    \centering
    \includegraphics[width=0.325\textwidth] {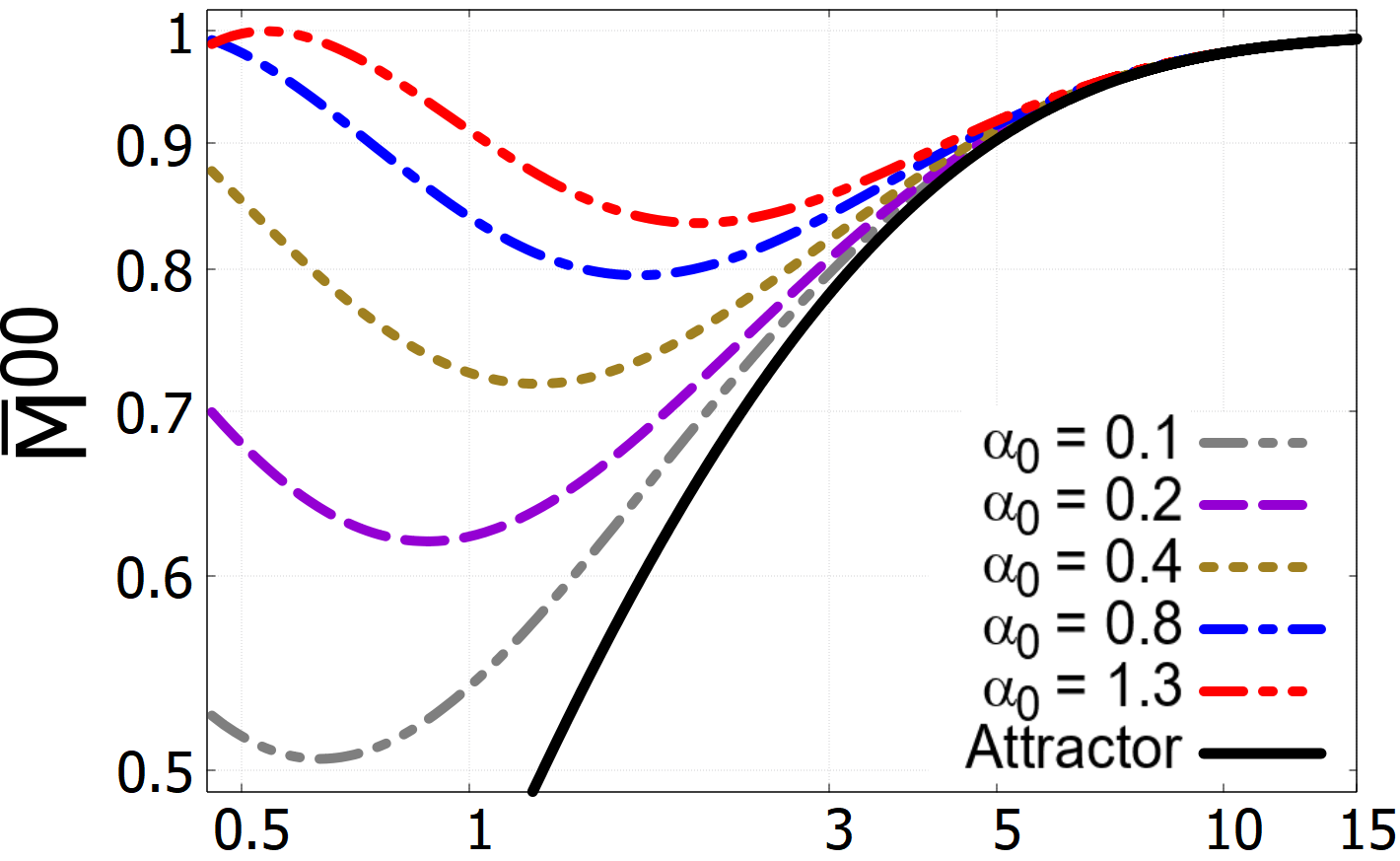} 
    \includegraphics[width=0.325\textwidth] {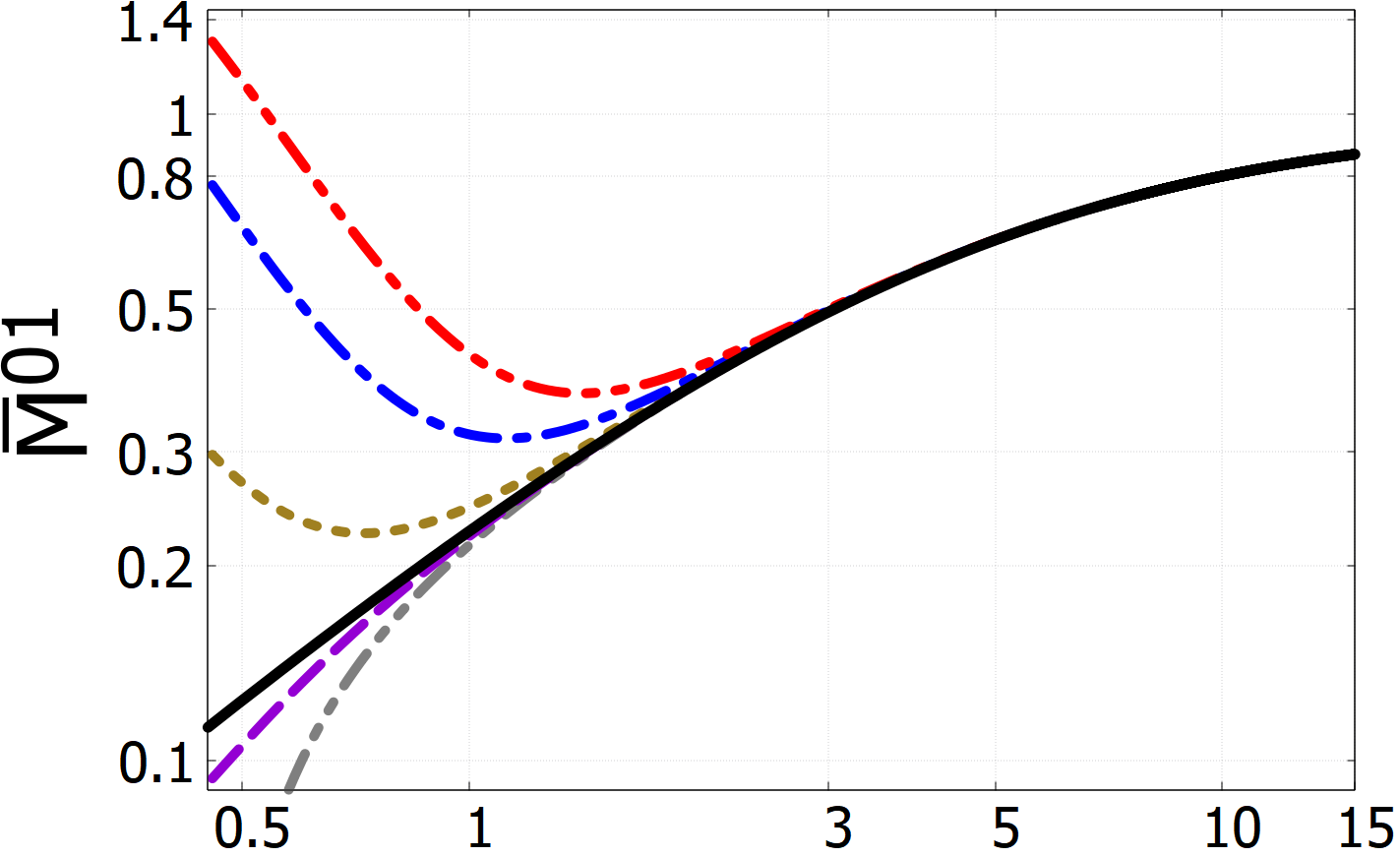}
    \includegraphics[width=0.325\textwidth] {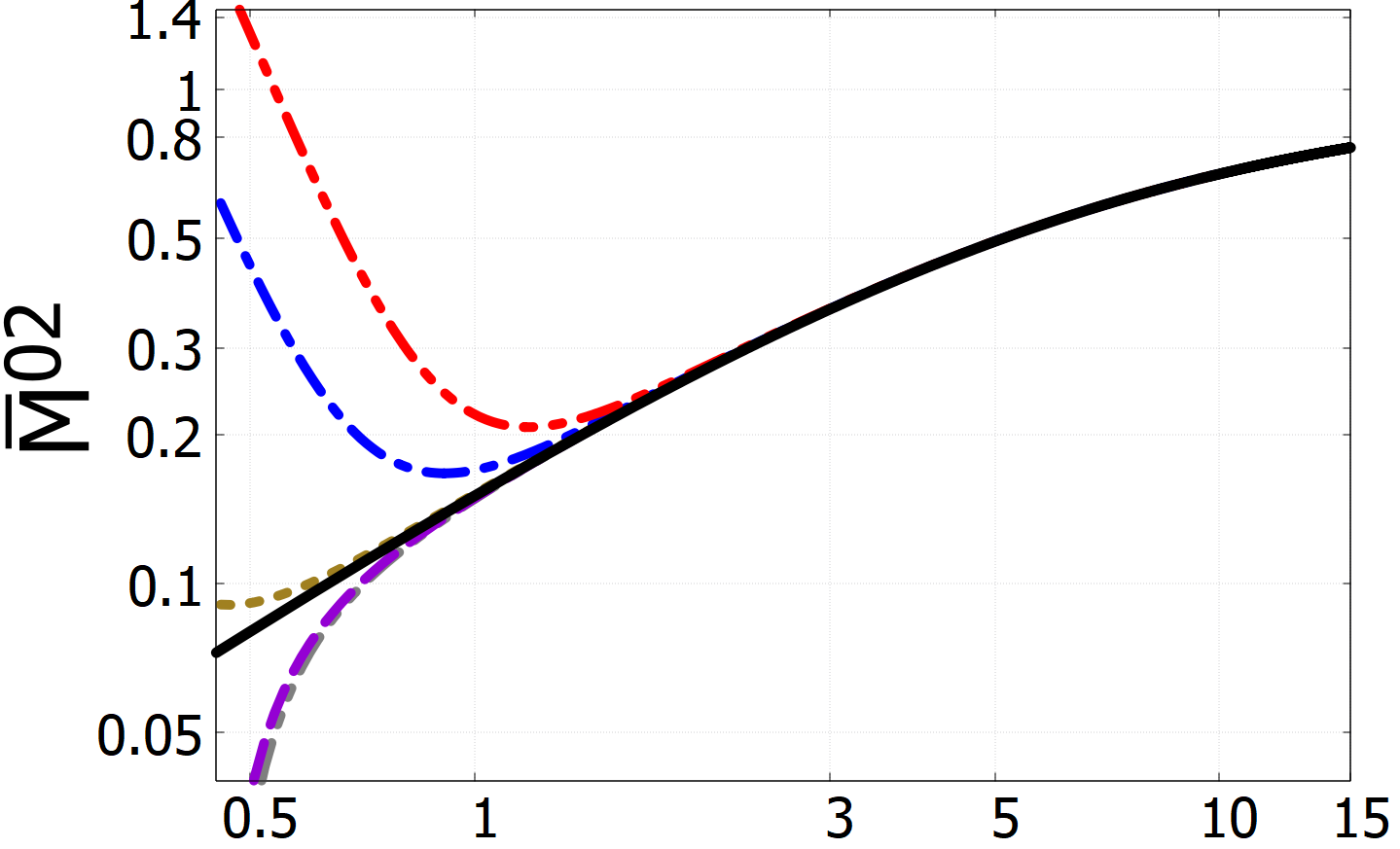} \\
    \includegraphics[width=0.325\textwidth] {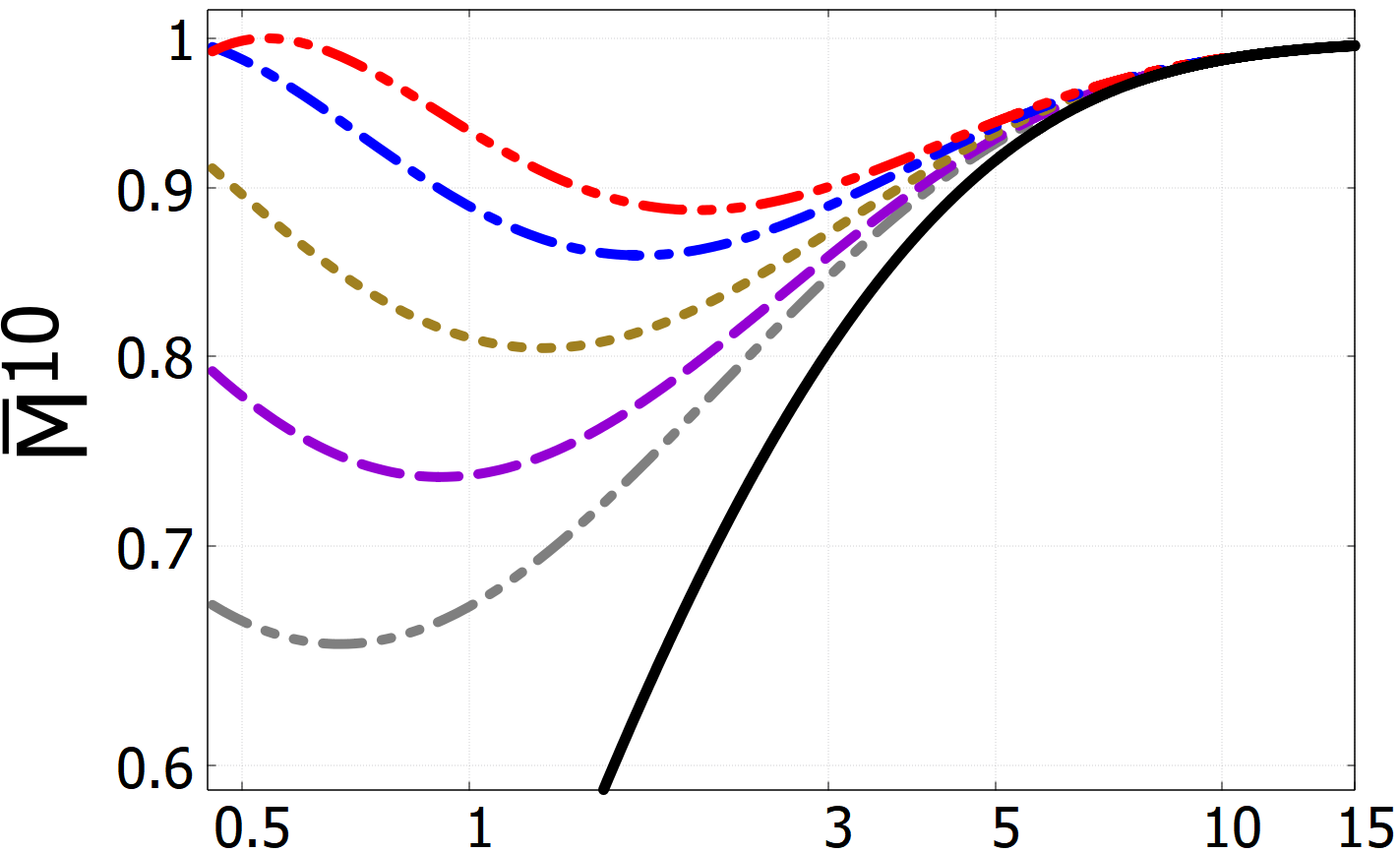} 
    \includegraphics[width=0.325\textwidth] {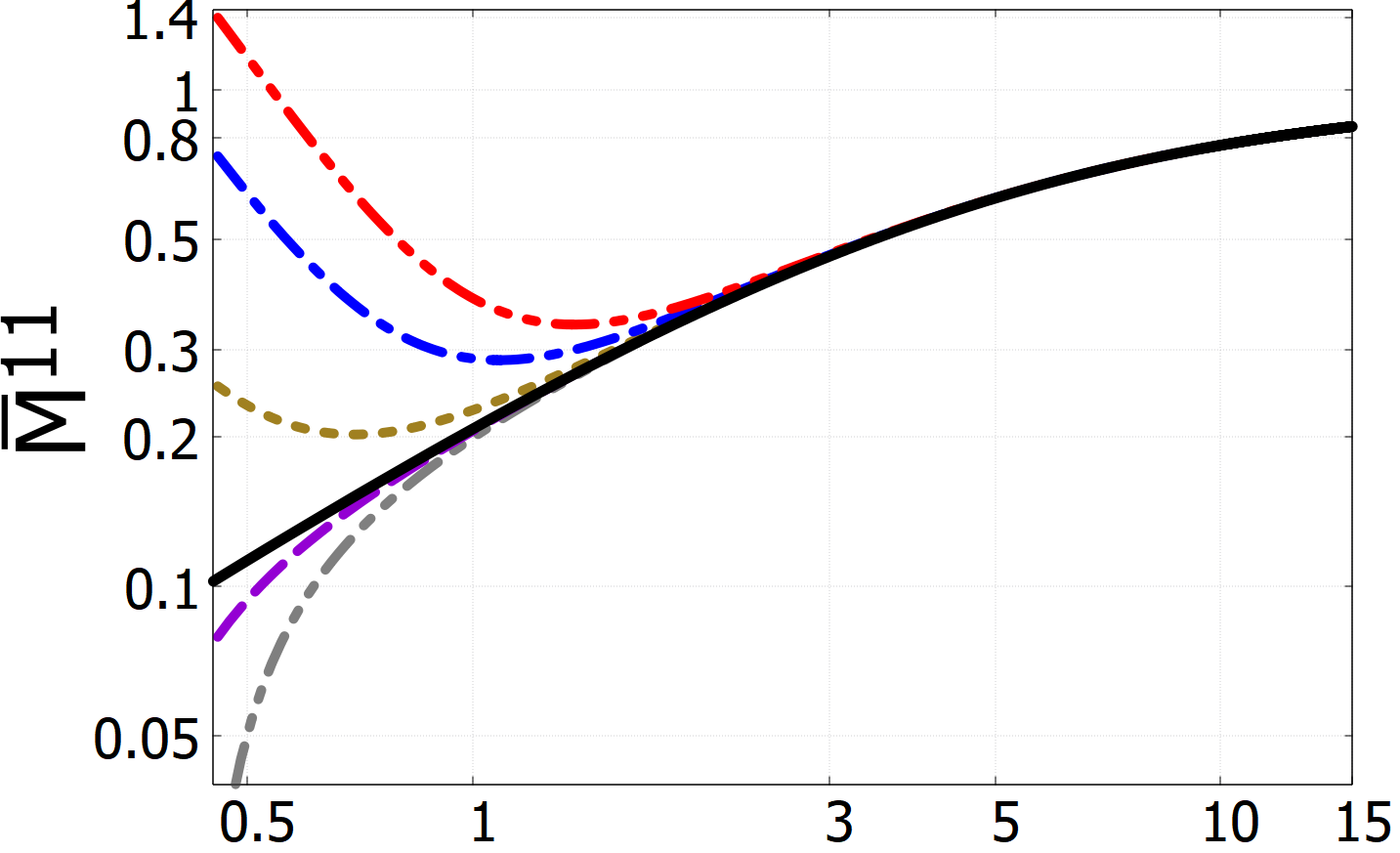}
    \includegraphics[width=0.325\textwidth] {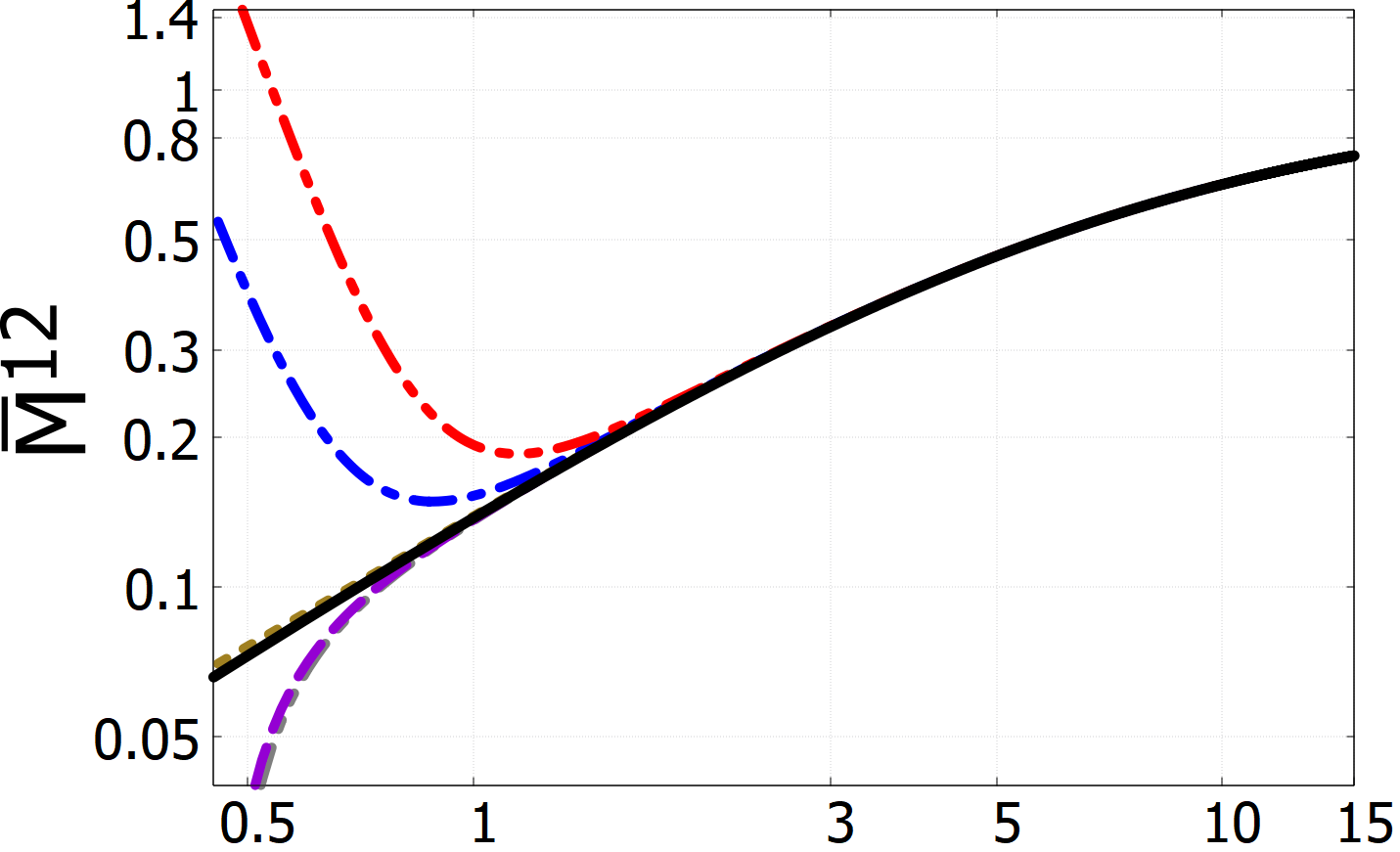}\\
    \includegraphics[width=0.325\textwidth] {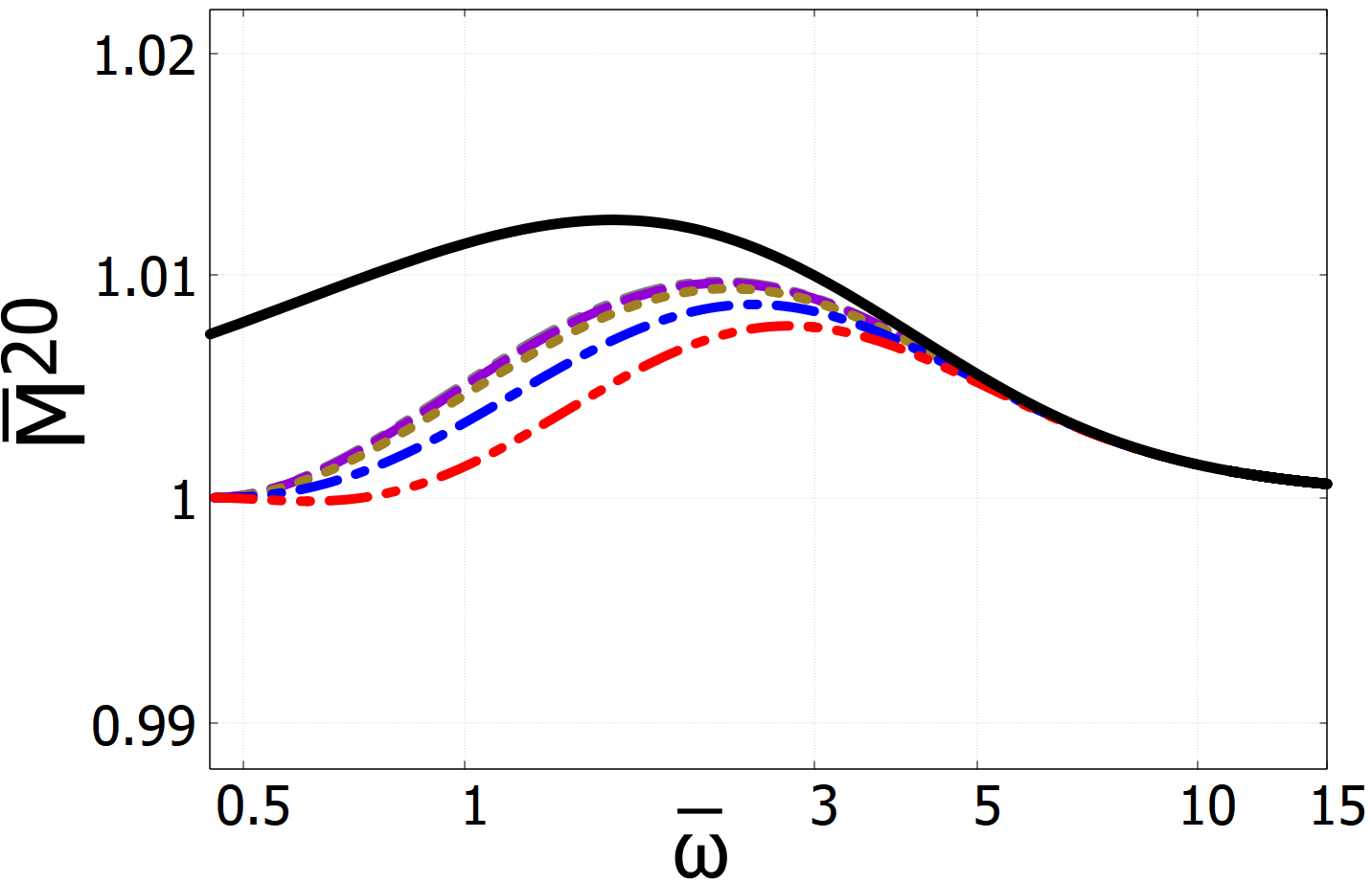} 
    \includegraphics[width=0.325\textwidth] {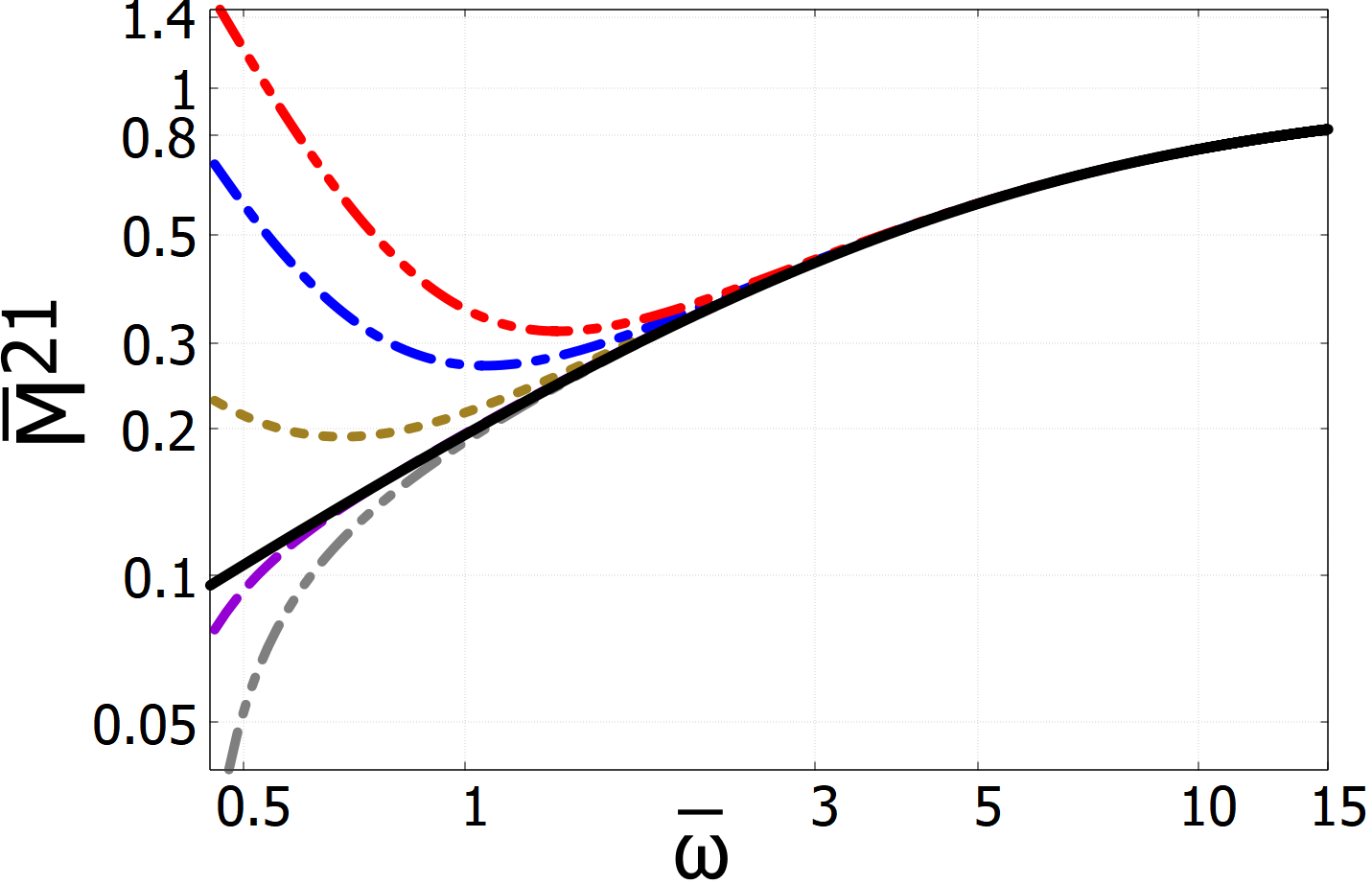}
    \includegraphics[width=0.325\textwidth] {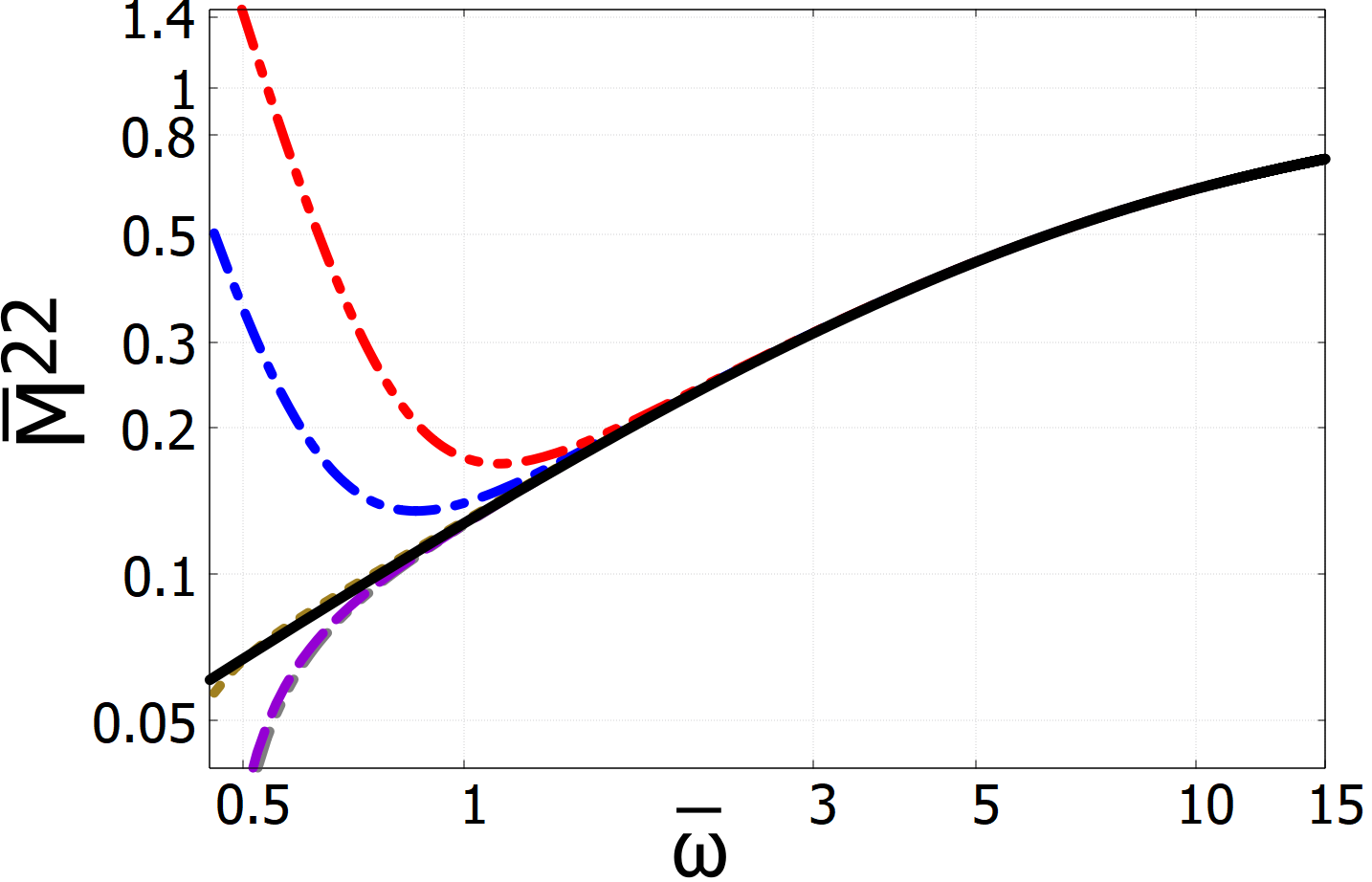}  
    \caption{The same as in Fig.~\ref{fig:mom-late-under}, but for the chemical equilibrium case $\gamma_{q, 0} = 1$.}
    \label{faccia2}
\end{figure}
In Fig.~\ref{faccia2}, we {address} the case in which quarks are initially in chemical equilibrium, i.e. $\gamma_{q, 0} = 1$ ({condition which, as discussed above, is preserved during the evolution of the system}), {their relative abundance with respect to gluons being simply given by the ratio of their respective degrees of freedom.}
Also in this situation, the initial ``free-streaming" expansion drives the system away from equilibrium until collisions start to be important~\cite{Strickland:2018ayk, Alalawi:2020zbx}. Shortly afterward, all the solutions converge to the attractor represented by the solid black line, aligning along the same curve~\cite{Soloviev:2021lhs, Jankowski:2023fdz}.

Comparing Figs.~\ref{fig:mom-late-under} and \ref{faccia2}, we note that the {time necessary to reach} the forward attractor {depends on the} initial {quark fugacity}. {For each scaled moment, as a criterium to establish the convergence to the attractor,} in table~\ref{conv_tau}, we adopt the choice:
\begin{equation*}
\frac{\Bigl|\overline{M}^{\hspace{0.05cm} nm}_{\rm attr}(\Bar{\omega}_c) - \overline{M}^{\hspace{0.05cm} nm}_{\alpha_0 = 1.3}(\Bar{\omega}_c)\Bigr|}{\overline{M}^{\hspace{0.05cm} nm}_{\alpha_0 = 1.3}(\Bar{\omega}_c)} < 10^{-4} \,,
\end{equation*}
{since the $\alpha_0 = 1.3$ case, corresponding to an initial prolate momentum distribution, is the most distant from the expected realistic evolution of the sytem, as} evident from Figs.~\ref{fig:mom-late-under} and \ref{faccia2}. {For each scaled moment this allows us to define the dimensionless convergence time $\Bar{\omega}_c$, whose values are quoted in table~\ref{conv_tau}}.
{One can notice that} the hydrodynamization process is anticipated a bit when one takes a {quark under-population} at the beginning of the system's expansion. Moments with $m = 0$ are the most influenced by the initial value of the fugacity parameter $\gamma_{q, 0}$.
\begin{center}
\begin{tabular}{ | m{4em} | m{1cm}| m{1cm} | m{1cm} |}
  \hline
  $\boldsymbol{\Bar{\omega}_c}$ & m = 0 & m = 1 & m = 2\\ 
  \hline
  n = 0 & 4.0 & 2.3 & 1.9\\ 
  \hline
  n = 1 & 4.6 & 2.5 & 2.0\\ 
  \hline
  n = 2 & 3.7 & 2.7 & 2.1\\
  \hline
\end{tabular}
\hspace{0.4cm}
\begin{tabular}{ | m{4em} | m{1cm}| m{1cm} | m{1cm} |}
  \hline
  $\boldsymbol{\Bar{\omega}_c}$ & m = 0 & m = 1 & m = 2\\ 
  \hline
  n = 0 & 7.1 & 2.5 & 2.0\\ 
  \hline
  n = 1 & 7.3 & 2.7 & 2.1\\ 
  \hline
  n = 2 & 5.4 & 2.8 & 2.2\\
  \hline
\end{tabular}
\captionof{table}{Values of the convergence scaled time $\Bar{\omega}_c$ for the moments in Figs.~\ref{fig:mom-late-under} and~\ref{faccia2}. {The two displayed tables correspond to the initial quark fugacity} $\gamma_{q, 0} = 0.1$ (left {panel}) and $\gamma_{q, 0} = 1.0$ (right {panel}).}
\label{conv_tau}
\end{center}

{As already discussed, two of the above scaled moments have a direct physical interpretation}: $\overline{M}^{\hspace{0.05cm} 10} \!\equiv\! n/n_{\rm eq}$, {where $n\!\equiv\! n_g+n_q$, and $\overline{M}^{\hspace{0.05cm} 20}\! \equiv\! \varepsilon/\varepsilon_{\rm eq}$}, {where $\varepsilon\!\equiv\! \varepsilon_g+\varepsilon_q$.}
The numerical results for $\overline{M}^{\hspace{0.05cm} 10}$, shown in Figs.~\ref{fig:mom-late-under} and~\ref{faccia2}, {are consistent with the expectations from Eq.~(\ref{rta6})}: {particle number} is not conserved during the system's expansion, {since $q\overline q$ pairs can be created/annihilated and gluons can be radiated/absorbed. Otherwise, in the case of particle-number conservation addressed in Ref.~\cite{Strickland:2019hff}, one would have $\overline{M}^{\hspace{0.05cm} 10}=1$ during the whole system's evolution. Concerning the scaled energy density, solutions for $\overline{M}^{\hspace{0.05cm} 20}$ corresponding to different initializations approach the attractor only quite late, as the other $m=0$ moments, when $\overline{M}^{\hspace{0.05cm} 20}$ is already very close to 1 and deviations from local thermodynamic equilibrium are small: hence, the attractor solution of kinetic equations is less relevant in this case.}

\begin{figure}[!hbt]
    \centering
    \includegraphics[width=0.495\textwidth] {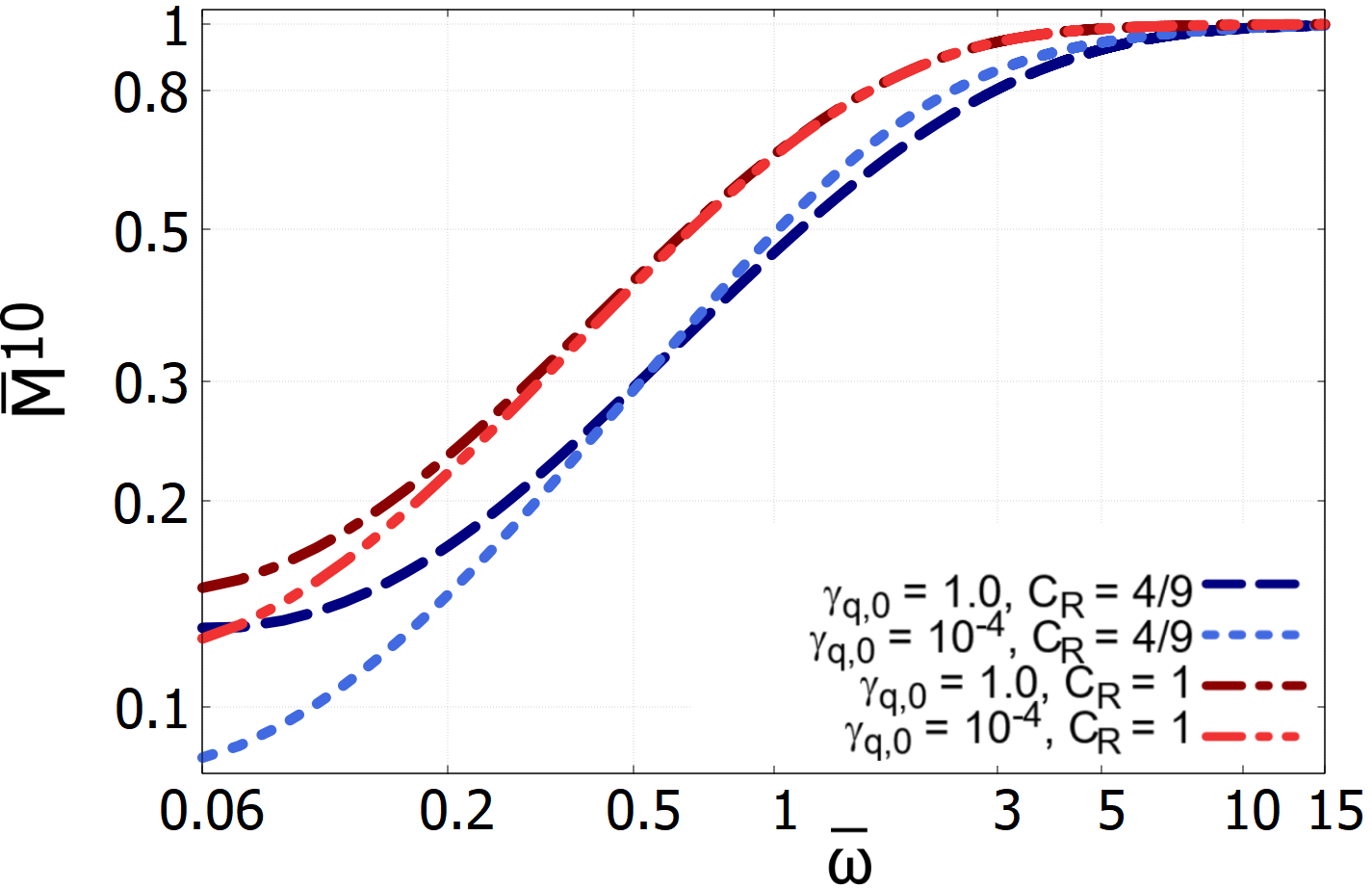} 
    \includegraphics[width=0.495\textwidth] {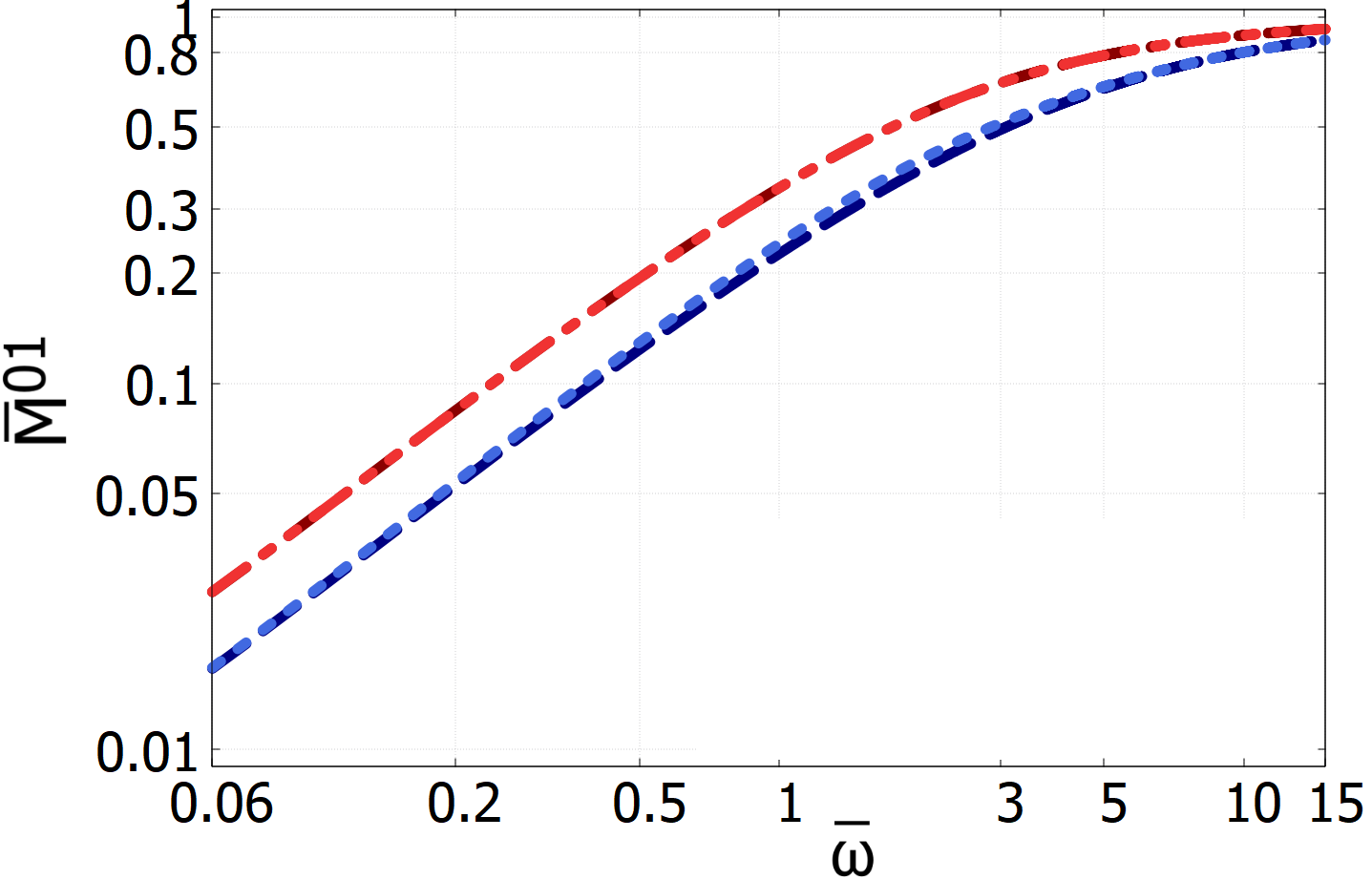} \\ 
    \caption{{Attractor solutions for the scaled moments $\overline{M}^{\hspace{0.05cm} 10}$ (particle number, left panel) and $\overline{M}^{\hspace{0.05cm} 01}$ (longitudinal pressure, right panel) for the case of two different relaxation times ($C_R = 4/9$) and of a single relaxation time ($C_R = 1$) for quarks and gluons. Both the scenarios of chemical equilibrium ($\gamma_{q,0}=1$) and of an initial under-population of quarks ($\gamma_{q,0}=10^{-4}$) are explored.} }
    \label{1rta}
\end{figure}
{We now wish to compare how the approach to equilibrium changes if the two components of the fluid mixture have different relaxation times ($C_R = 4/9$) or the same relaxation time ($C_R = 1$), taken to be the one of gluons. As a representative example, in Fig.~\ref{1rta} we focus on two scaled moments, $\overline{M}^{\hspace{0.05cm} 10}$ and $\overline{M}^{\hspace{0.05cm} 01}\!\equiv\! P_L/P_{\rm eq}$ and consider their attractor solutions (notice the very small initialization time) of the 2RTA and 1RTA-BE, both in the case of chemical equilibrium ($\gamma_{q,0}=1$) and of an initial under-population of quarks ($\gamma_{q,0}=10^{-4}$).}
One can observe that the approach to equilibrium (i.e. $\overline{M}^{\hspace{0.05cm} nm} \rightarrow 1$), in the 2RTA case is systematically retarded with respect to the case of a common relaxation time for quarks and gluons, {consistently with the quarks being characterized by $\tau_{{\rm eq}, q} = 9/4 \hspace{0.07cm} \tau_{\rm eq}$}, meaning that the quark component of the plasma takes more time to relax to the equilibrium configuration~\cite{Florkowski:2012as}, as will be also shown in more detail in Sec.~\ref{distrib_num} starting from the distribution function itself.

\begin{figure}[!hbt]
    \centering
    \includegraphics[width=0.325\textwidth] {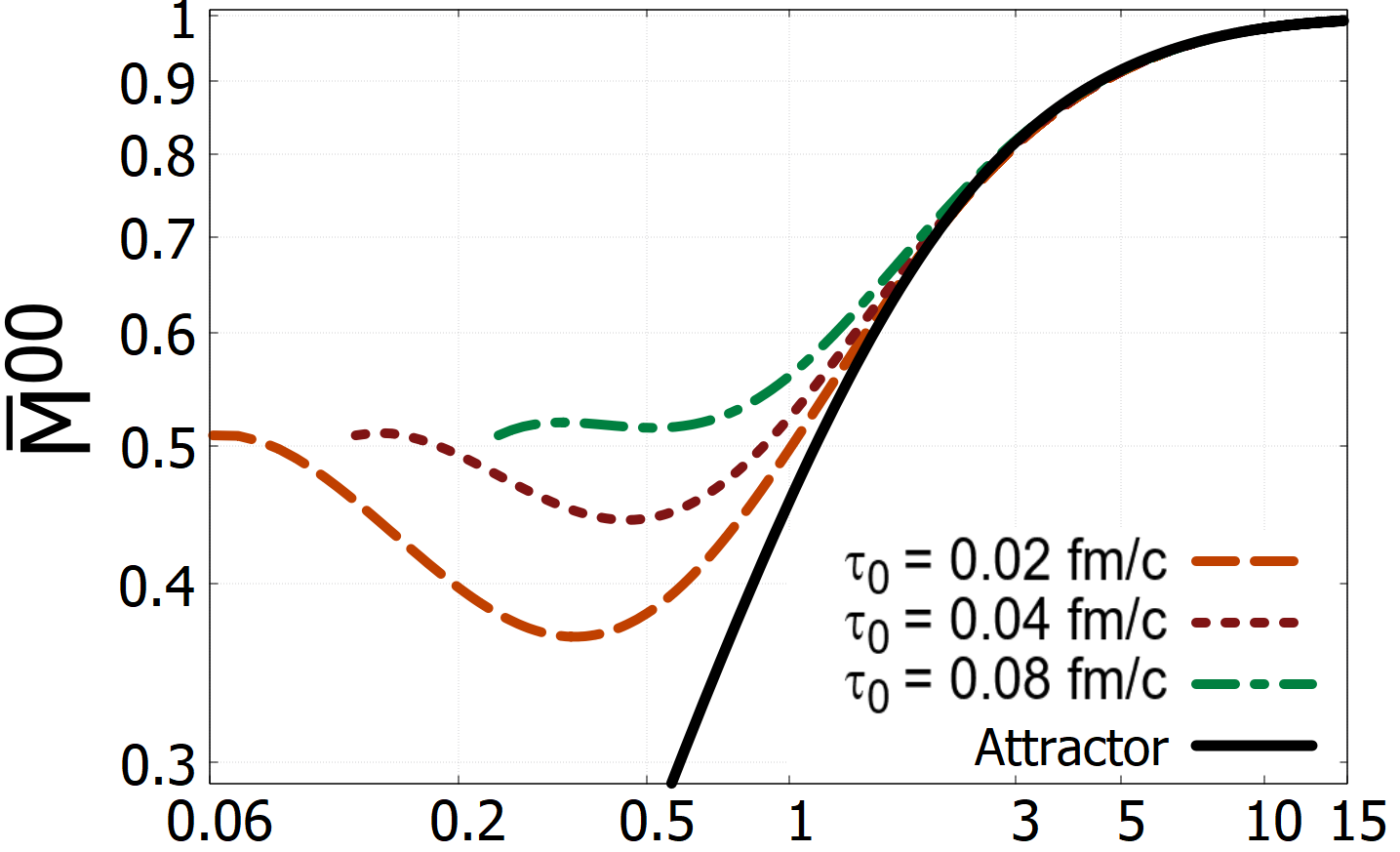} 
    \includegraphics[width=0.325\textwidth] {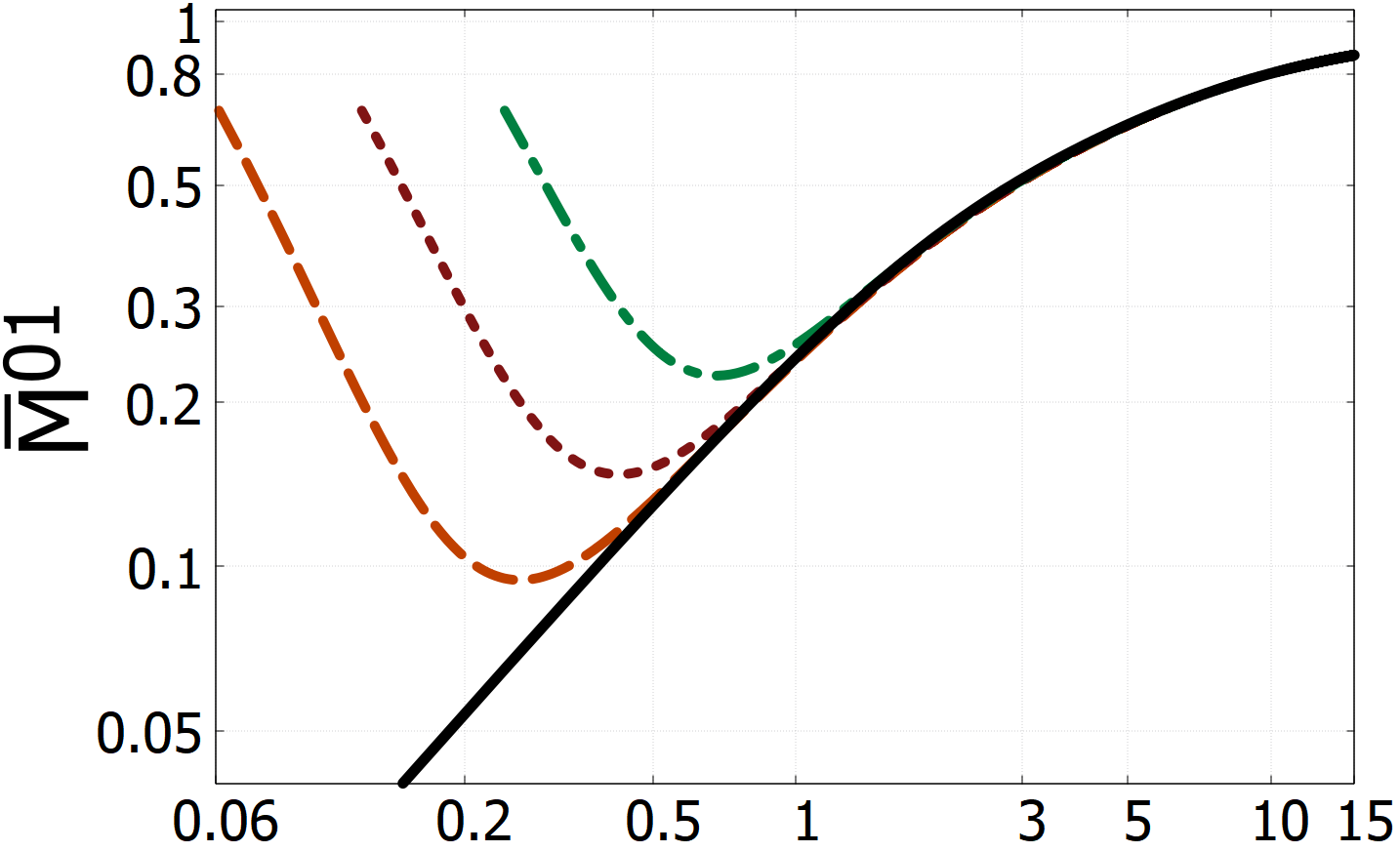}
    \includegraphics[width=0.325\textwidth] {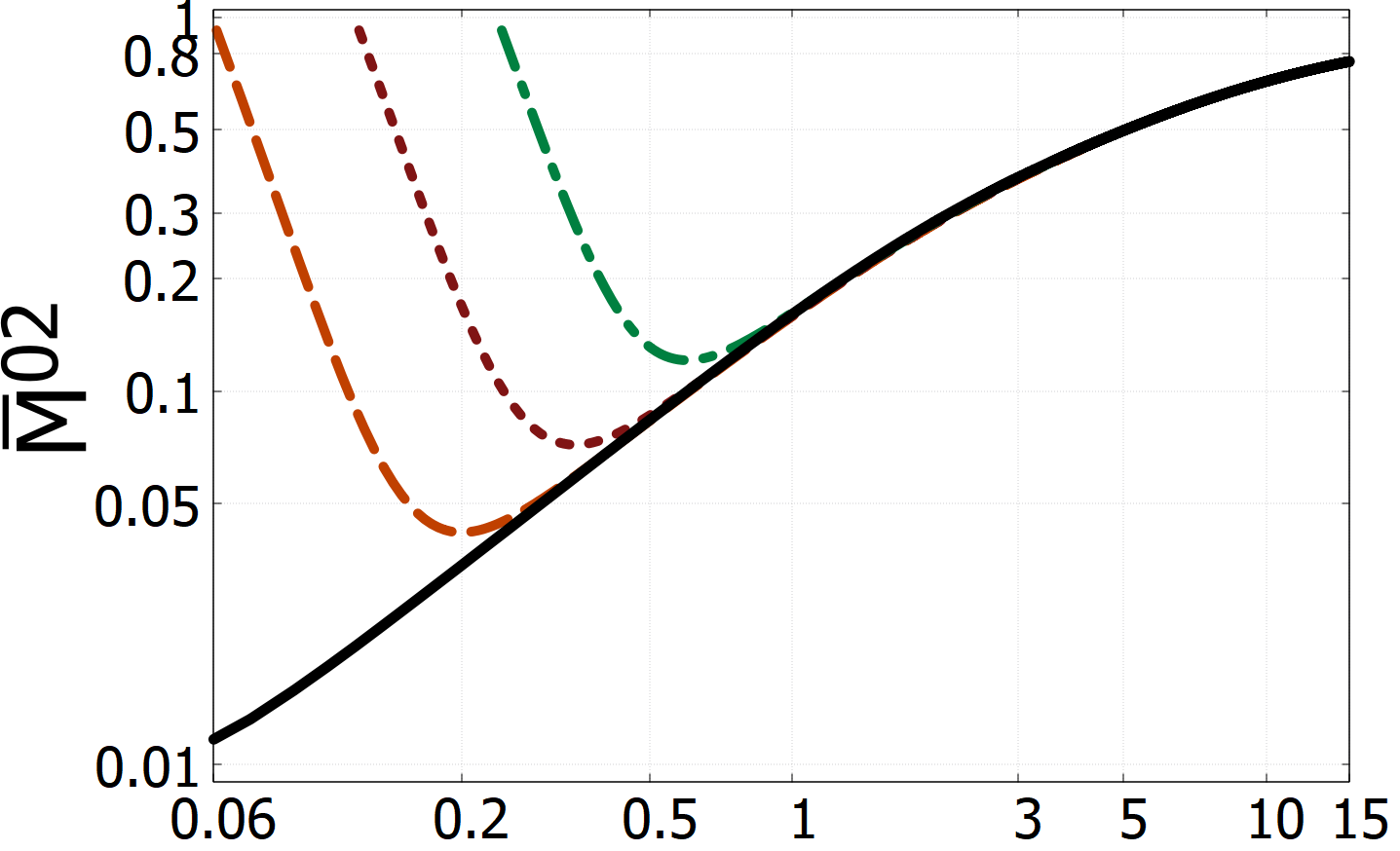} \\
    \includegraphics[width=0.325\textwidth] {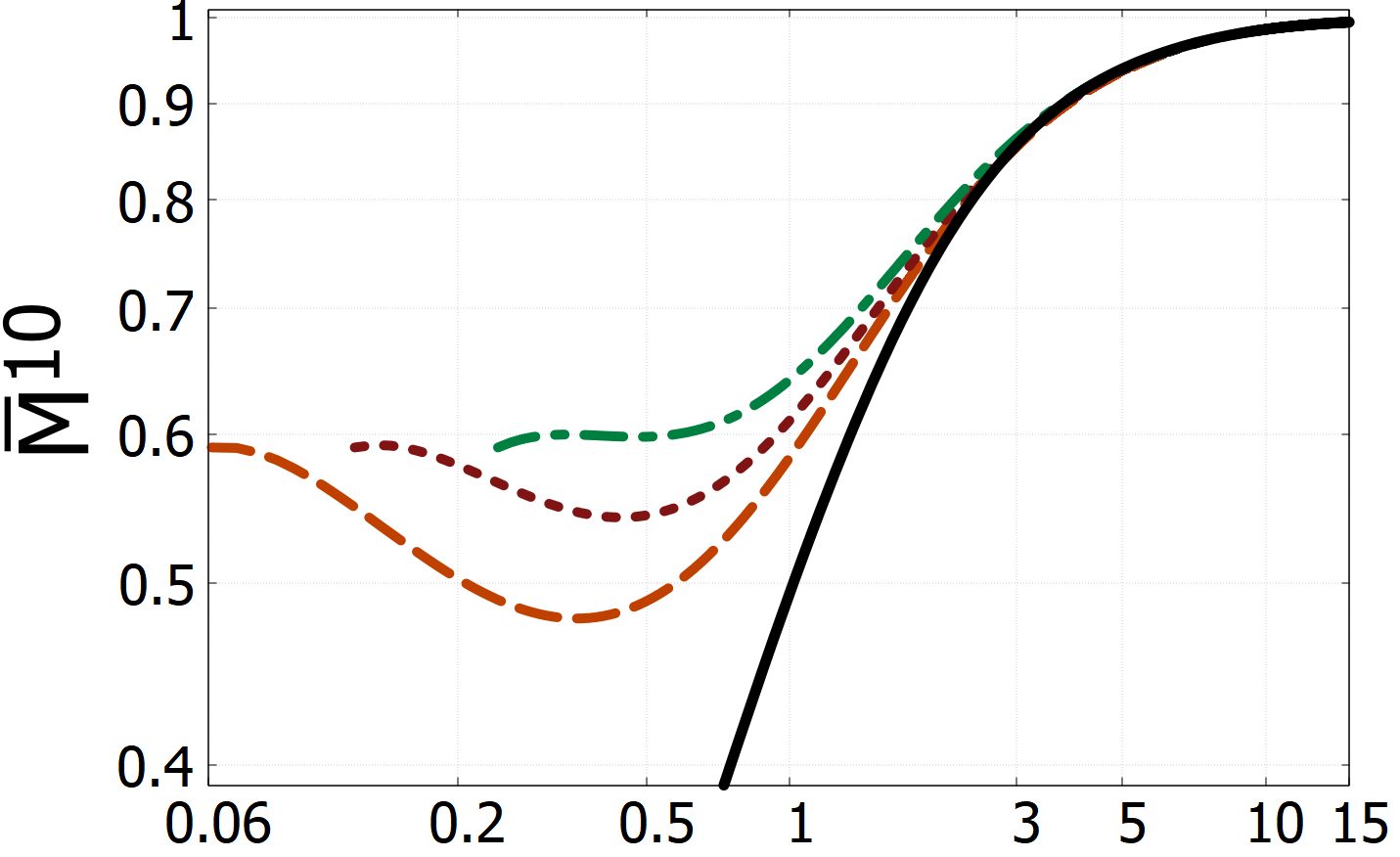} 
    \includegraphics[width=0.325\textwidth] {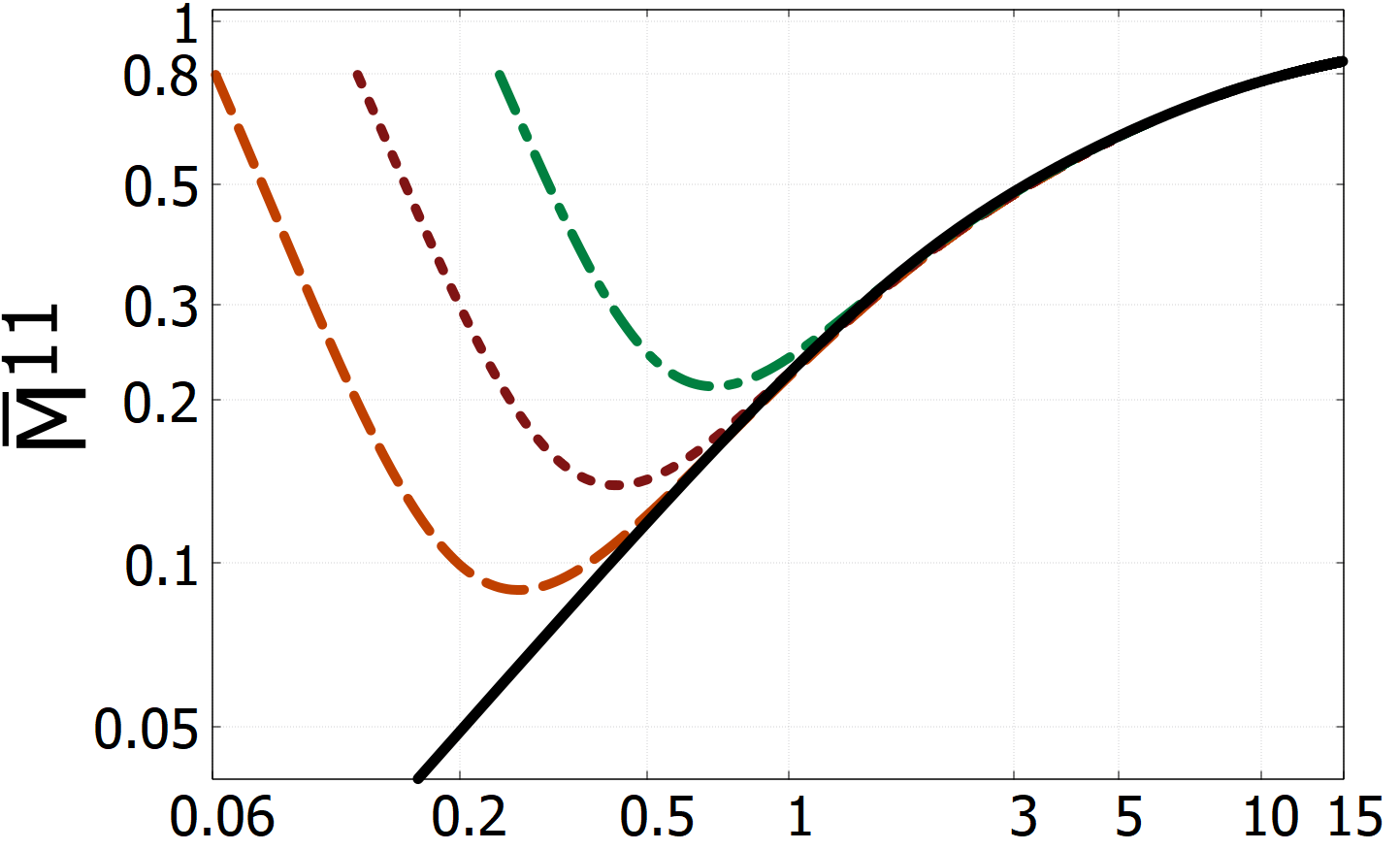}
    \includegraphics[width=0.325\textwidth] {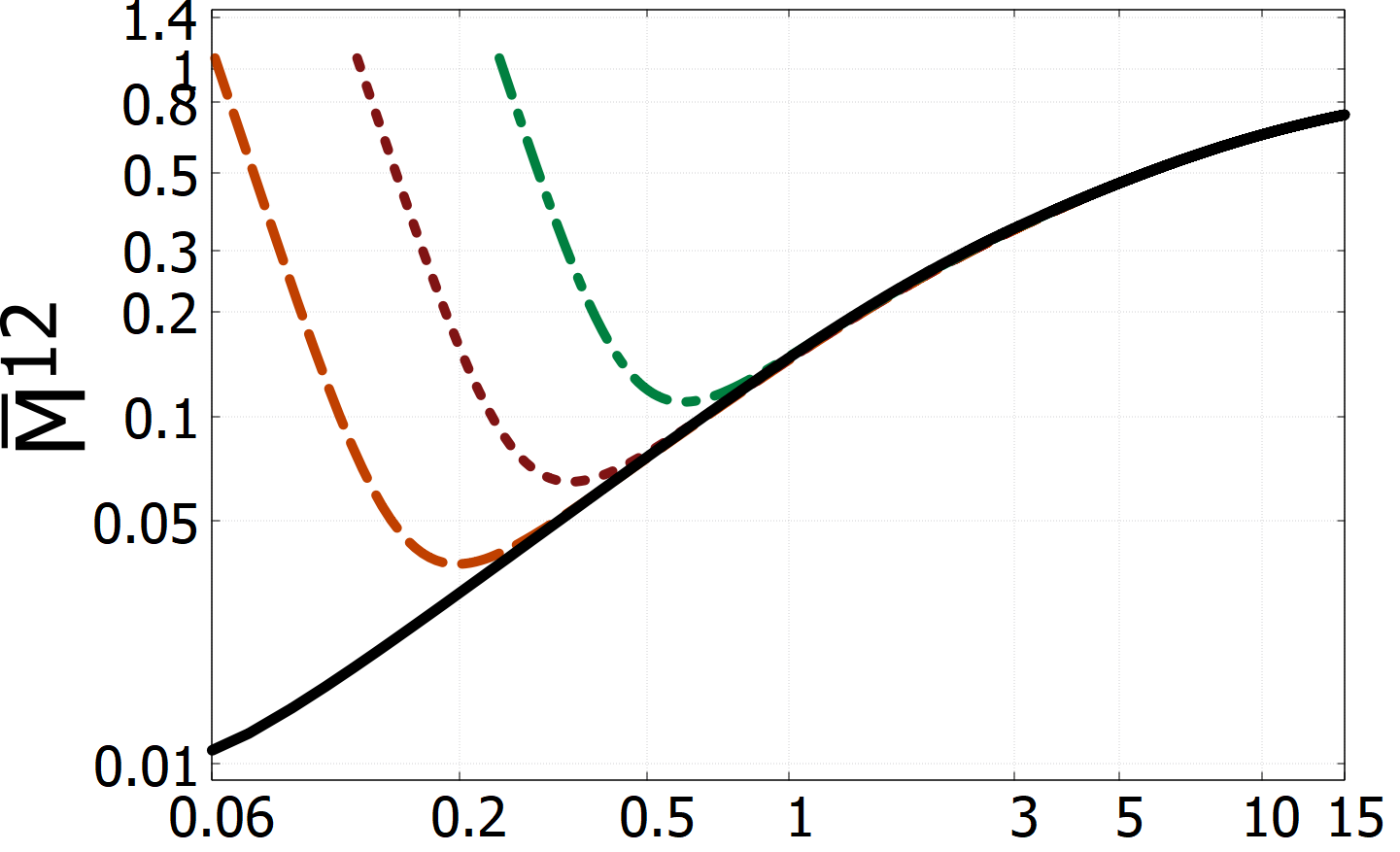}\\
    \includegraphics[width=0.325\textwidth] {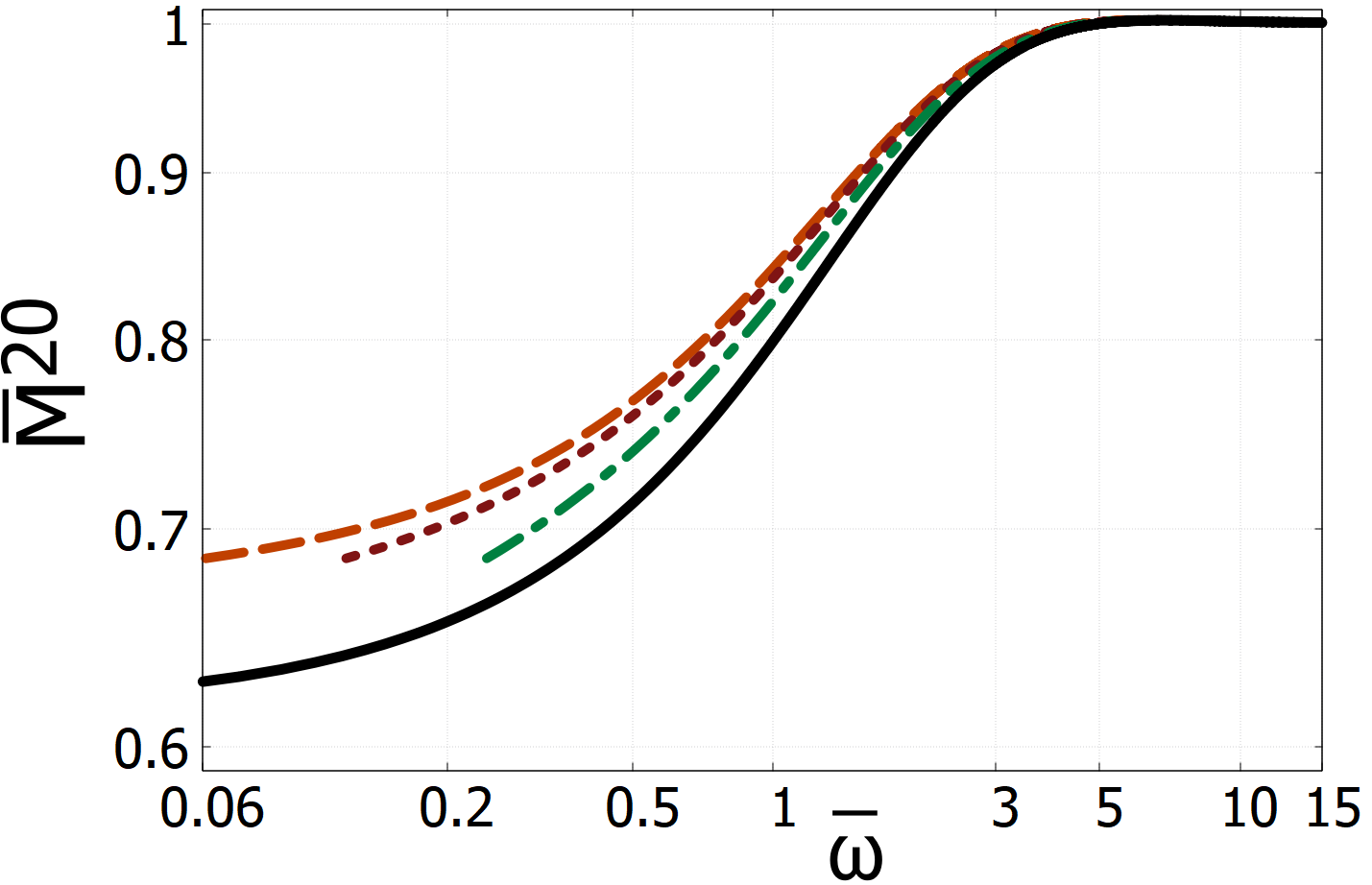} 
    \includegraphics[width=0.325\textwidth] {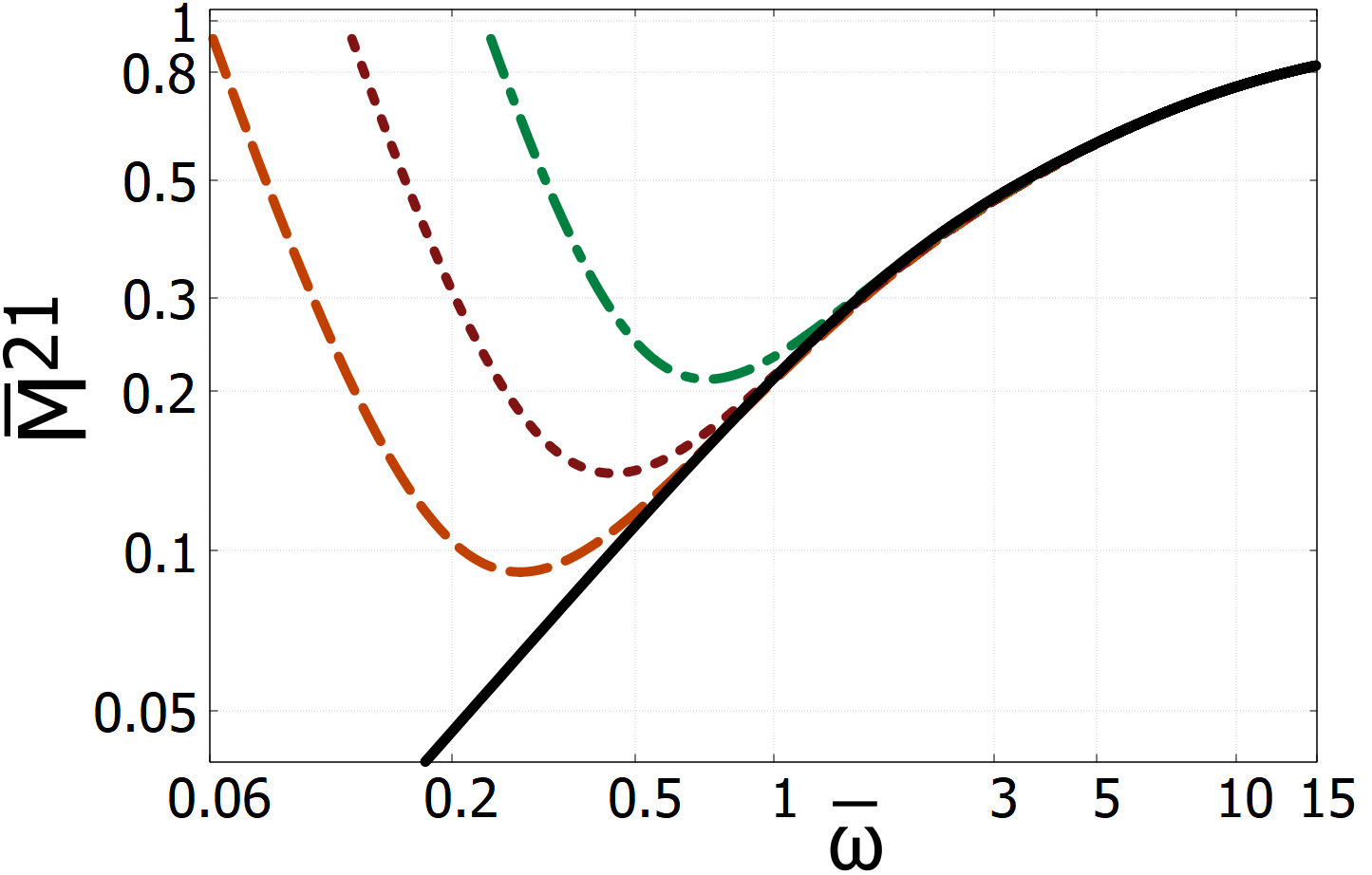}
    \includegraphics[width=0.325\textwidth] {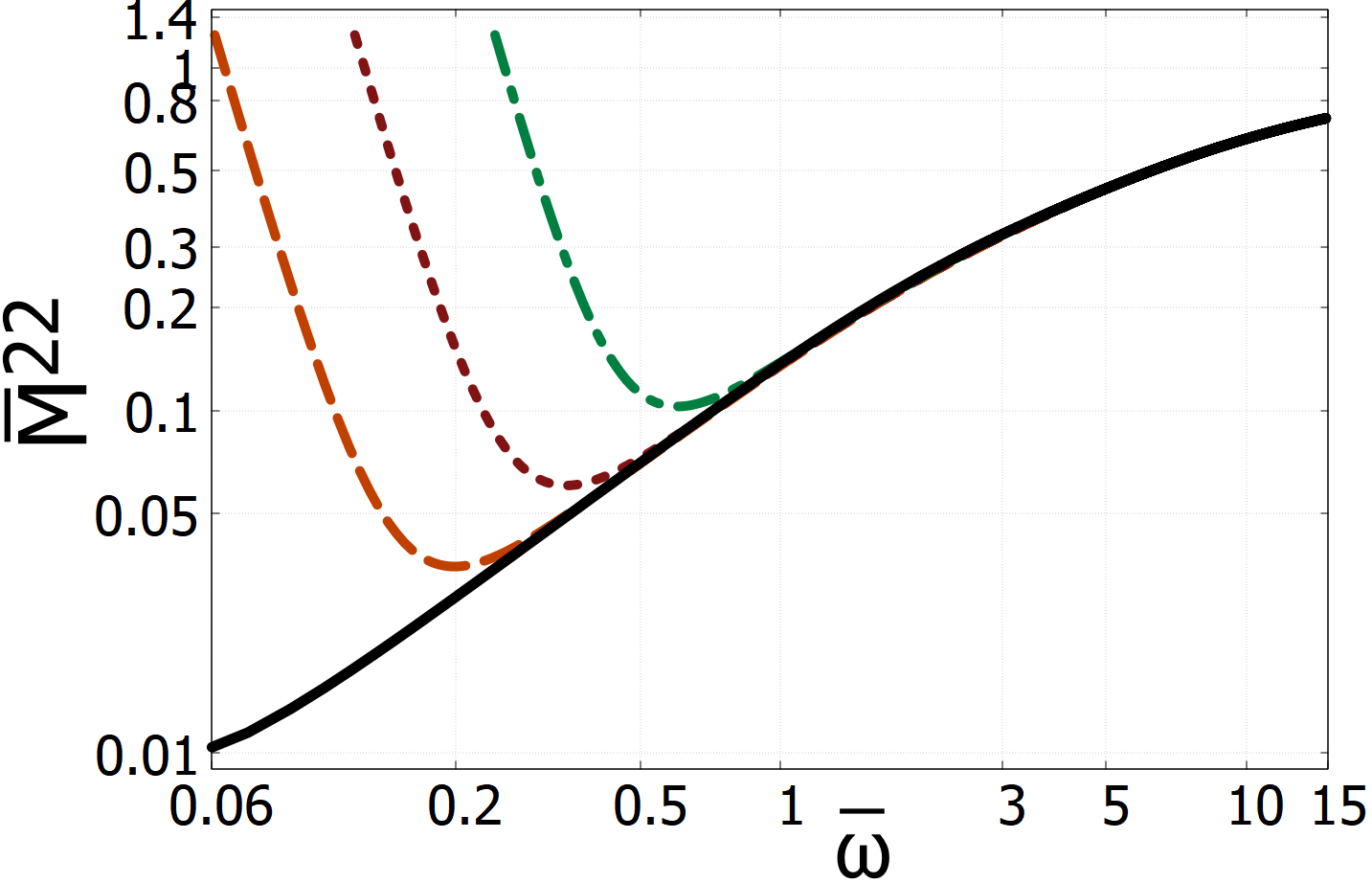}  
    \caption{Study of the convergence to the early-time attractor (solid black line) of the scaled moments $\overline{M}^{\hspace{0.05cm} nm}$ (various dashed colored lines) obtained from the exact solution of the BE in 2RTA. Here we initialize {the system setting} $T_0 = 600$ MeV and $\alpha_0 = 1$ {at different} initial longitudinal proper-times $\tau_0$. The gluon specific viscosity is taken to be $\eta/s = 0.2$. The time-variable is given by $\Bar{\omega} = \tau/\tau_{\rm eq}$. Different panels refer to different values of the indexes $n$ and $m$. The displayed curves refer to {an initial quark fugacity} $\gamma_{q, 0} = 0.1$, i.e. to a gluon-dominated plasma.}
    \label{earl1}
\end{figure}
\begin{figure}[!hbt]
    \centering
    \includegraphics[width=0.325\textwidth] {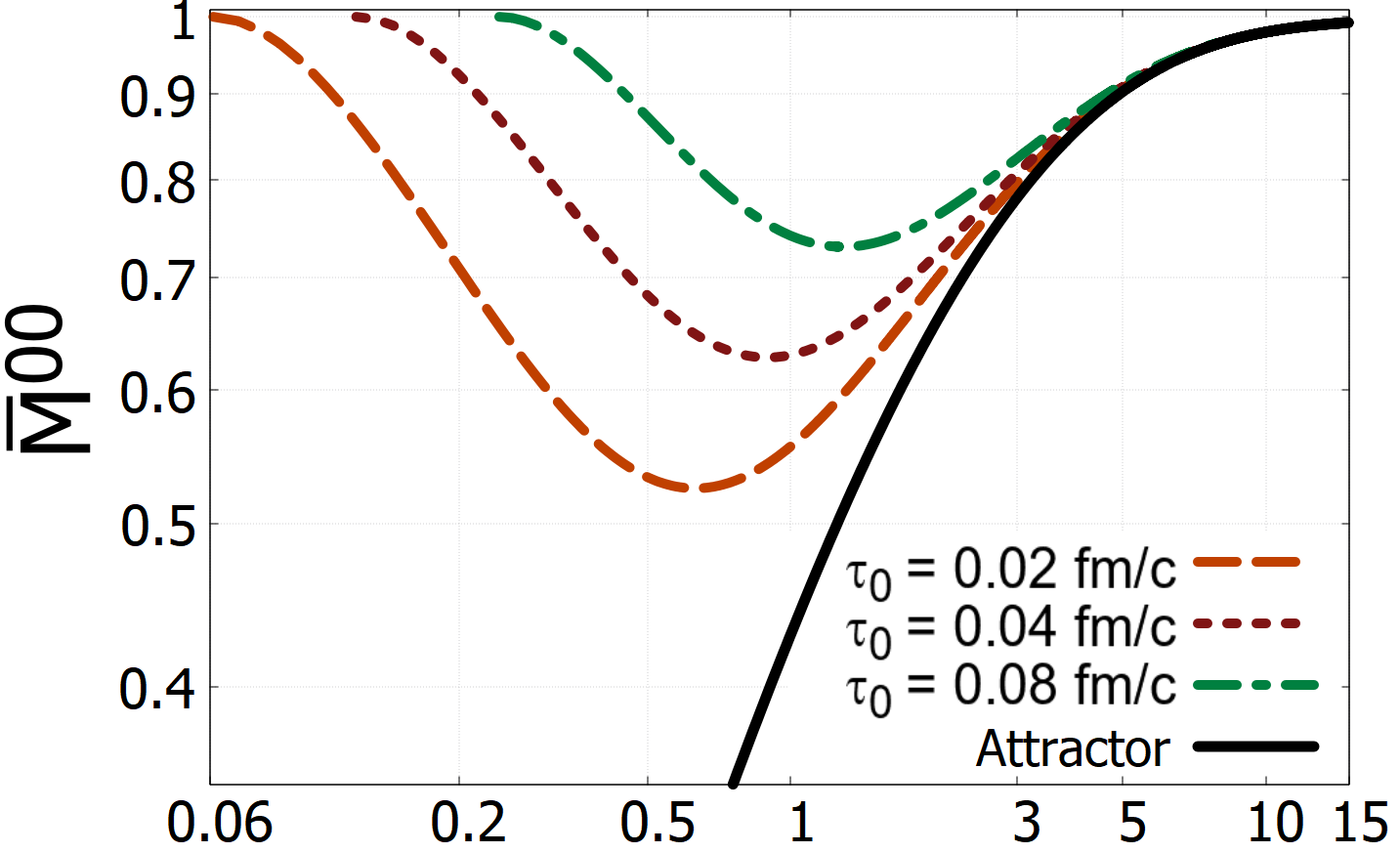} 
    \includegraphics[width=0.325\textwidth] {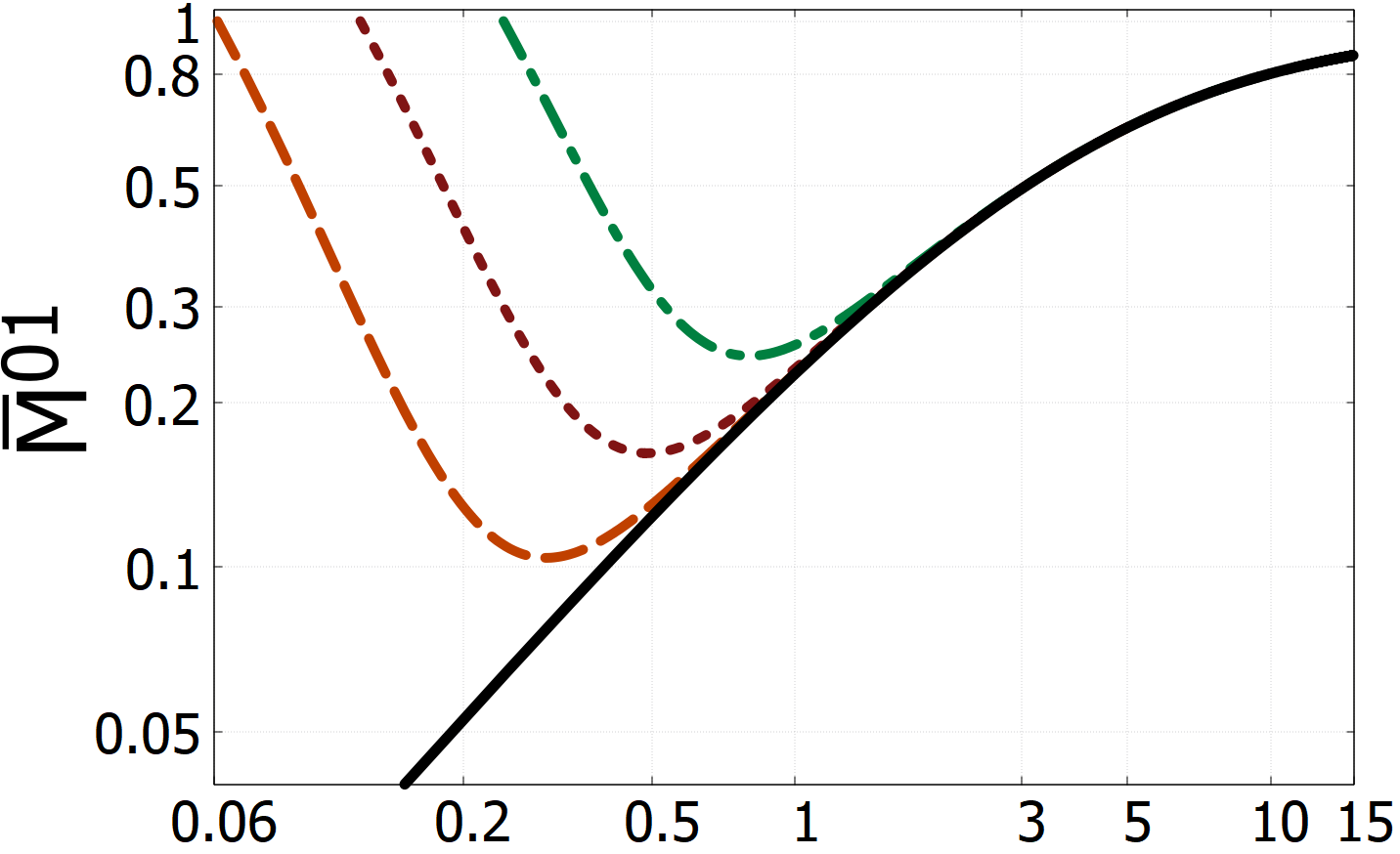}
    \includegraphics[width=0.325\textwidth] {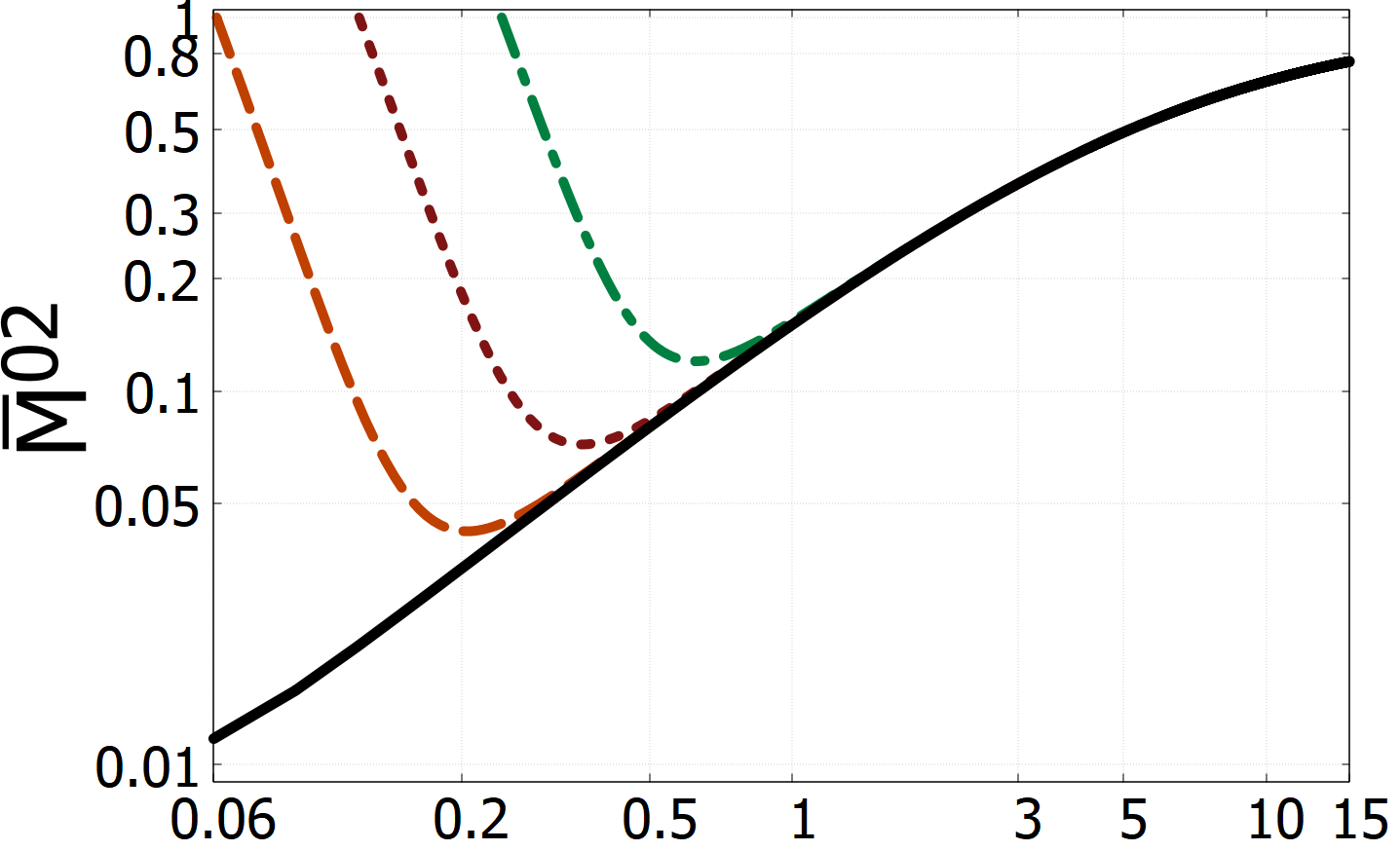} \\
    \includegraphics[width=0.325\textwidth] {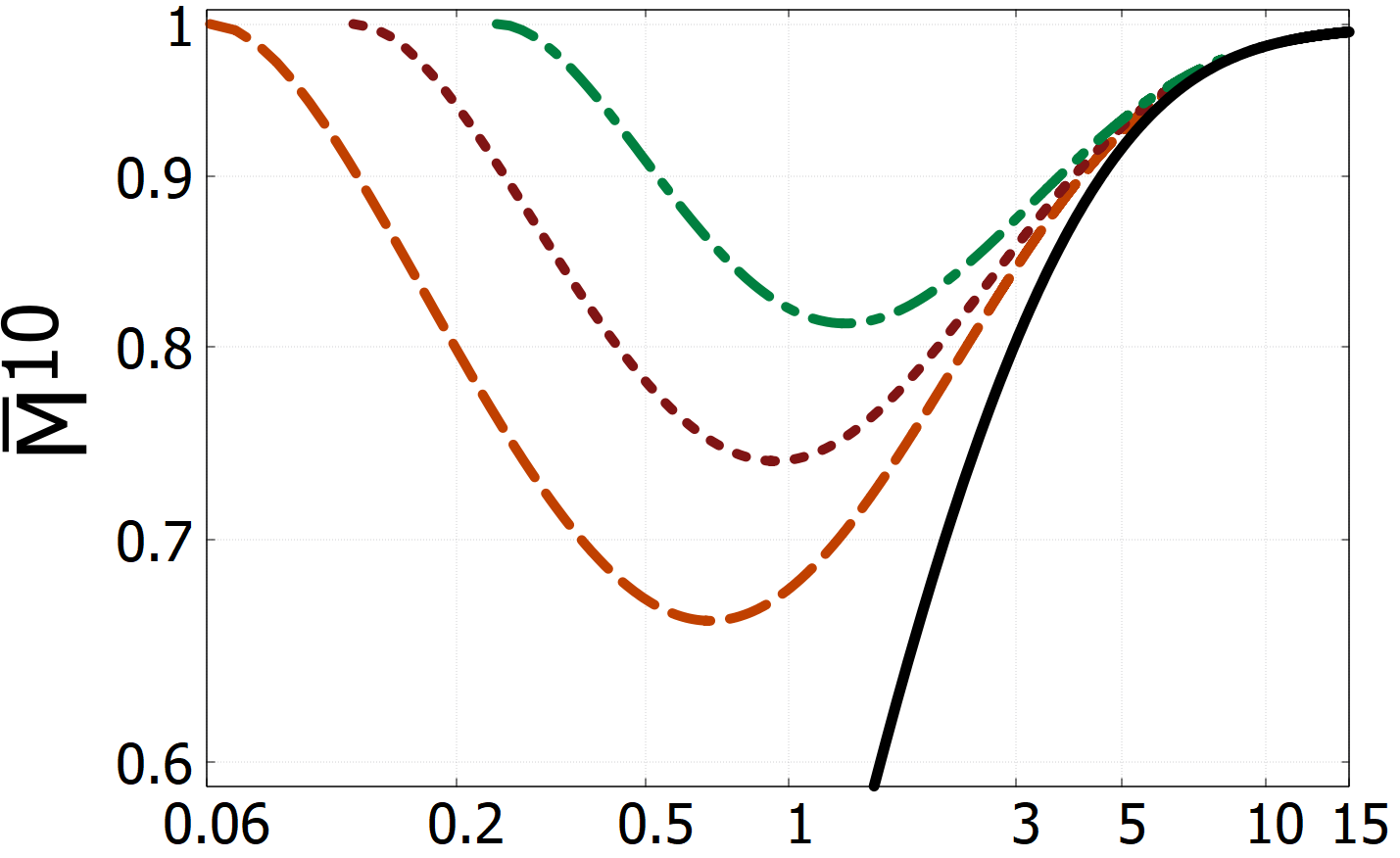} 
    \includegraphics[width=0.325\textwidth] {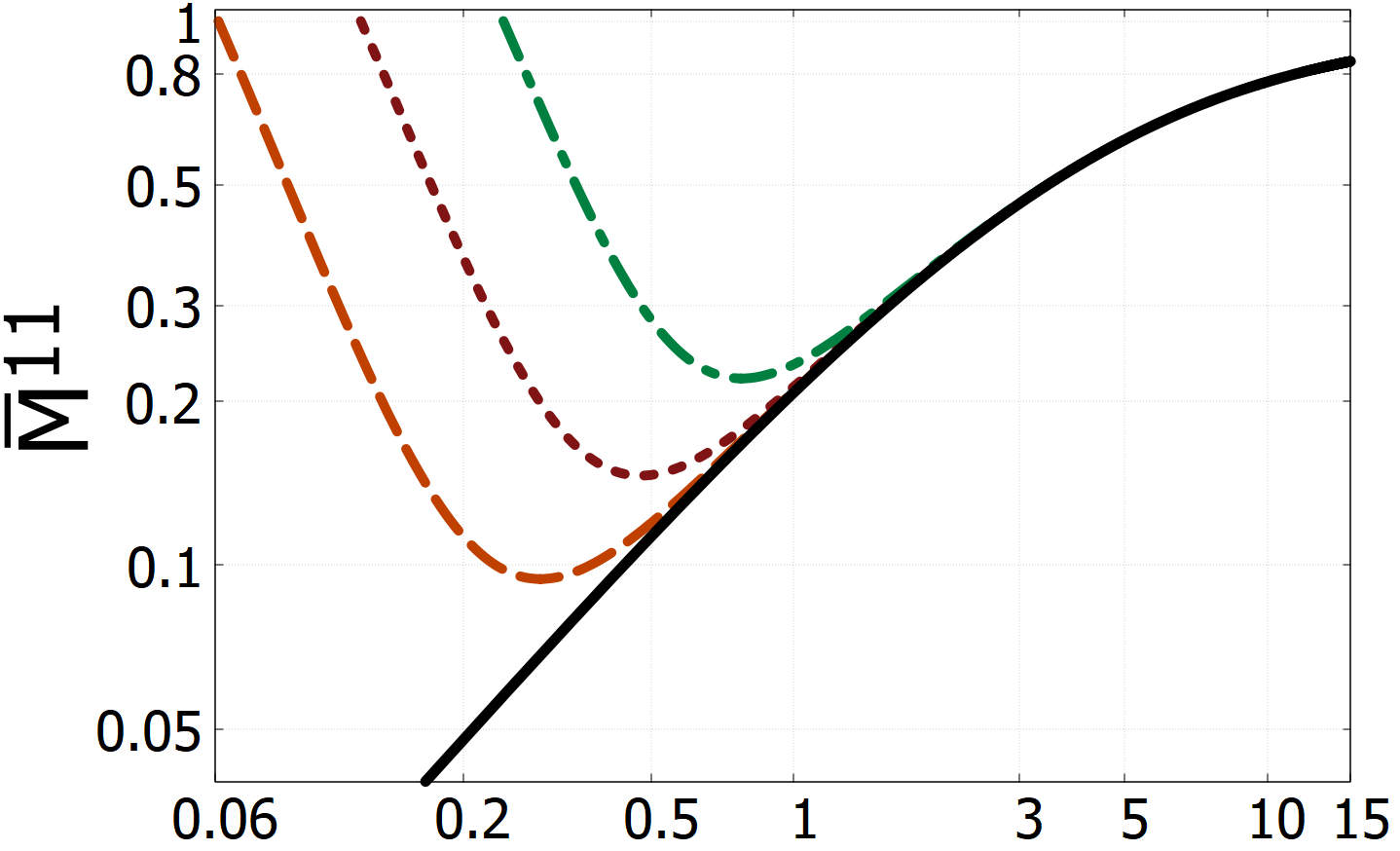}
    \includegraphics[width=0.325\textwidth] {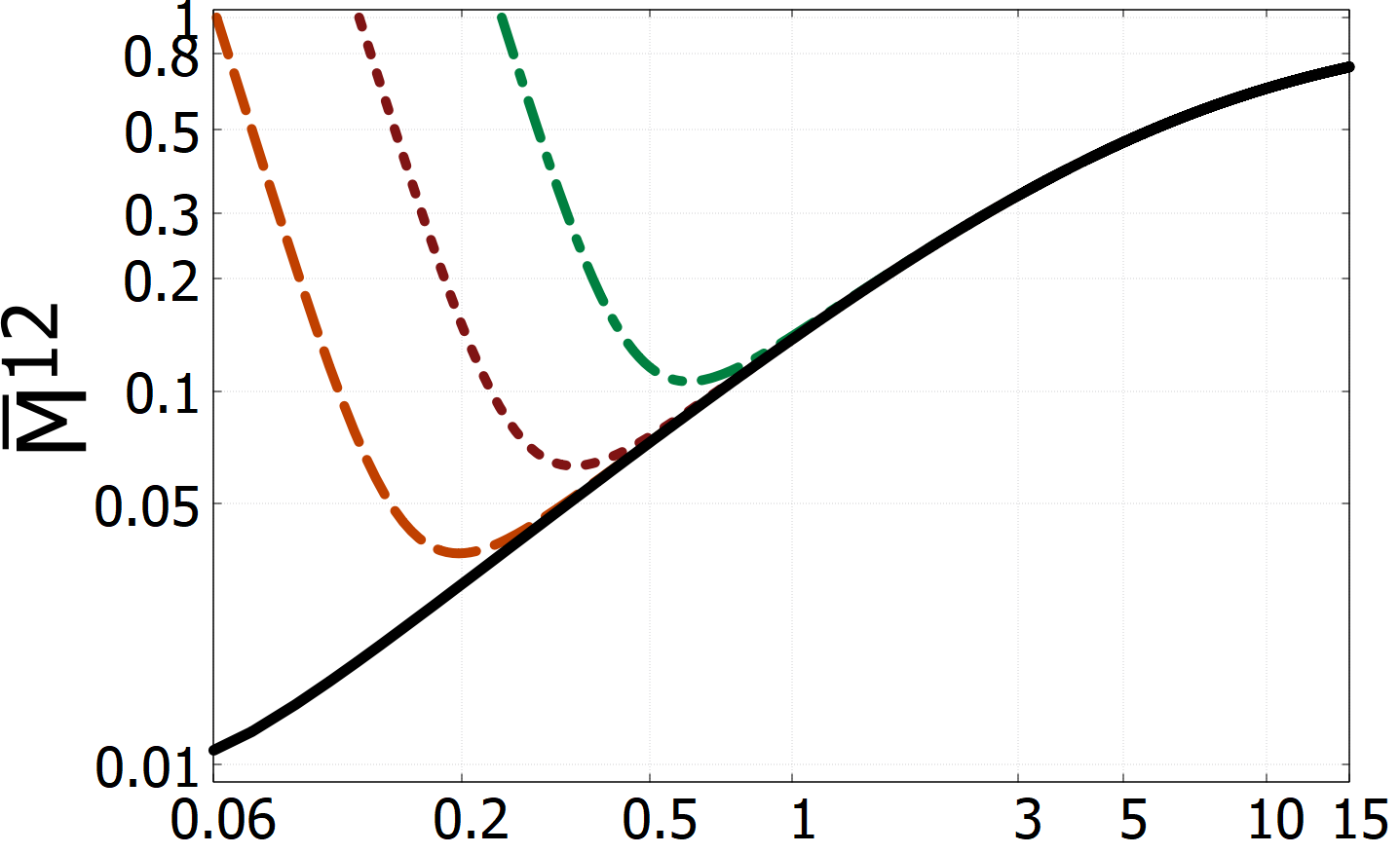}\\
    \includegraphics[width=0.325\textwidth] {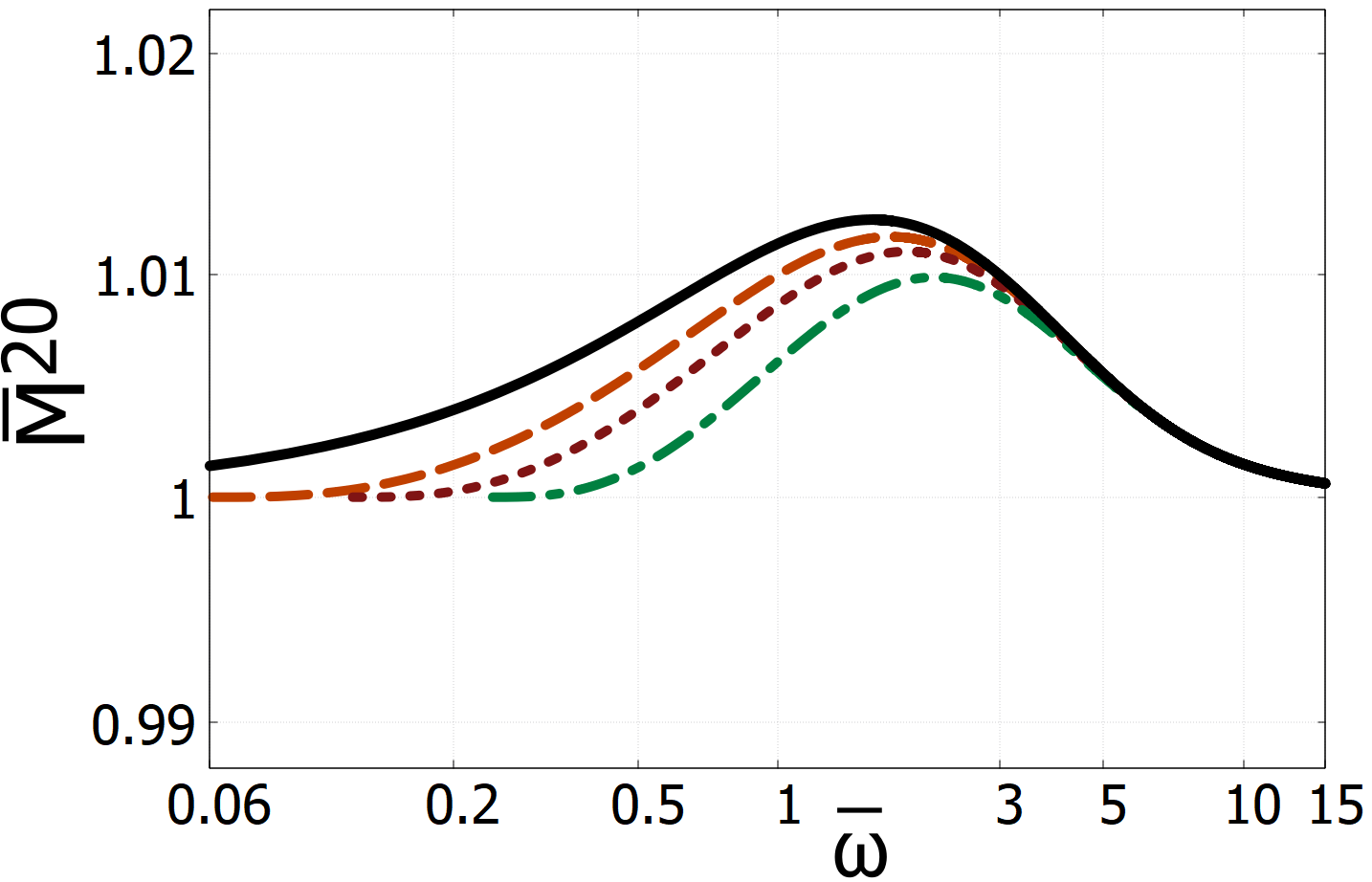} 
    \includegraphics[width=0.325\textwidth] {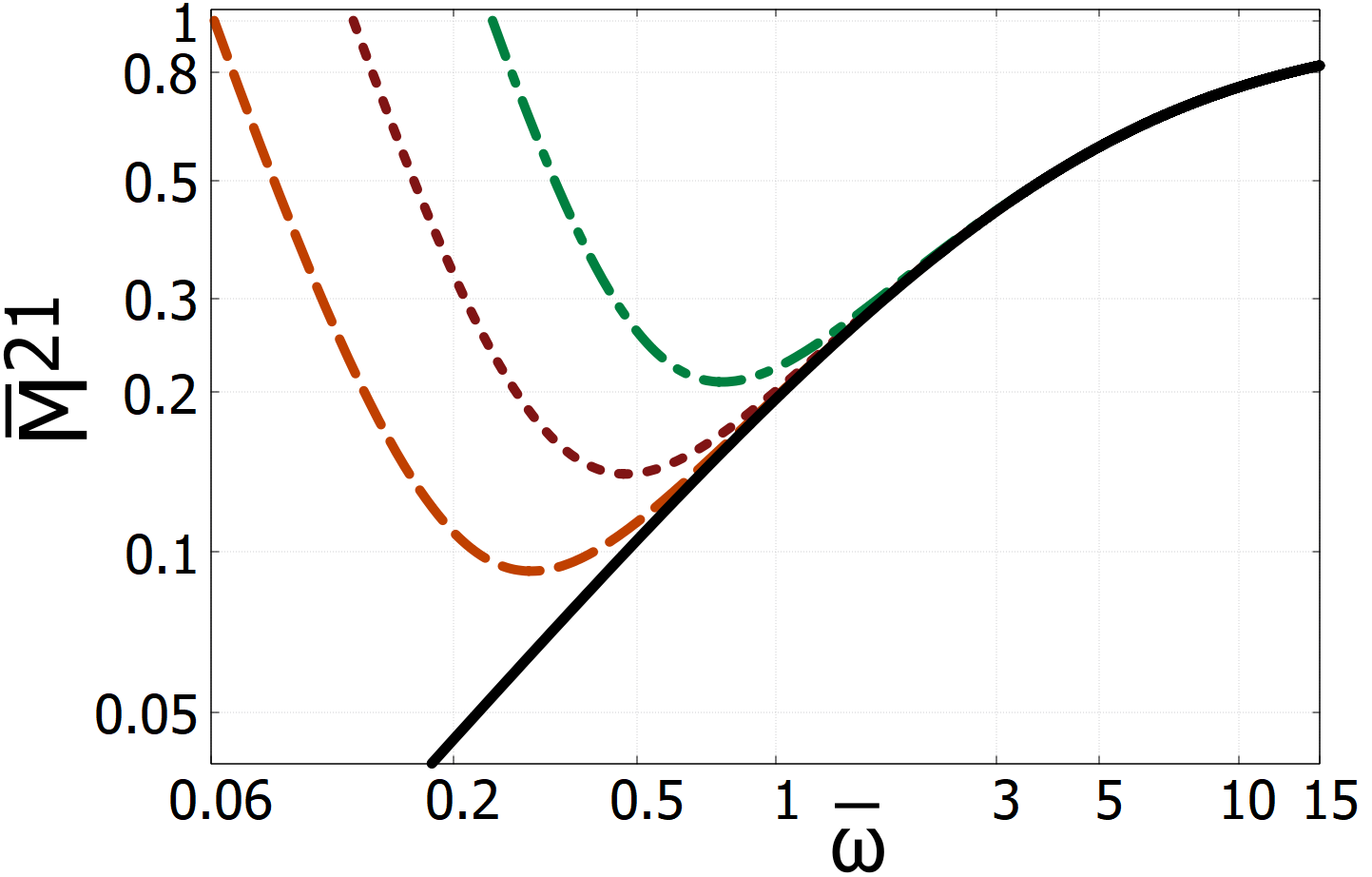}
    \includegraphics[width=0.325\textwidth] {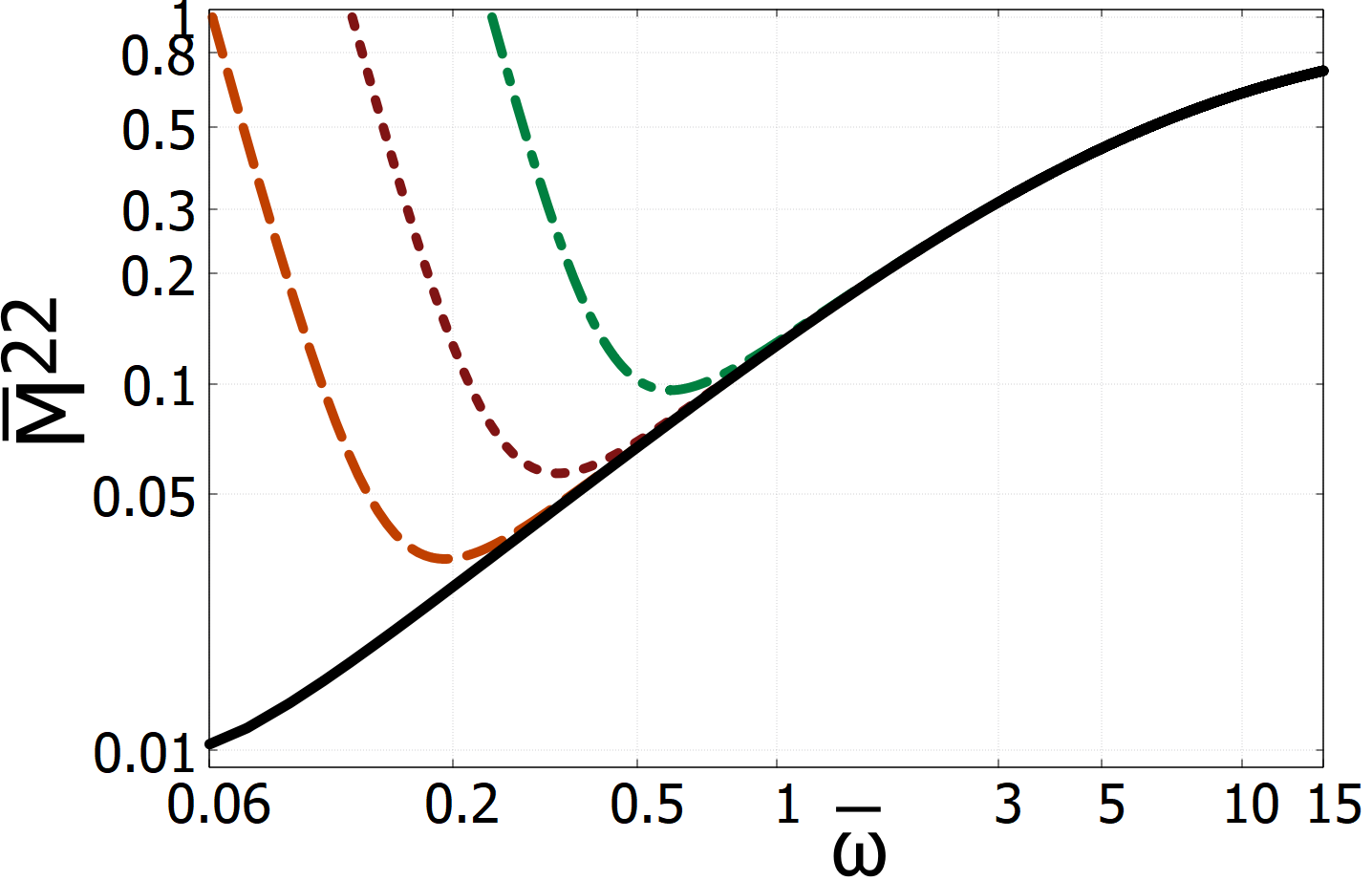}  
    \caption{{The same as in Fig.~\ref{earl1}, but for the chemical equilibrium case $\gamma_{q, 0} = 1$.}}
    \label{earl2}
\end{figure}
In order to {investigate the possible} presence of an early-time (pullback) attractor in this {2RTA} model, we {evaluate the same previously considered scaled moments $\overline M^{nm}$ keeping} the initial anisotropy and temperature fixed while manipulating the initialization time $\tau_0$~\cite{Alalawi:2022pmg, Jankowski:2023fdz, Strickland:2017kux}.
{Moving the latter to smaller and smaller values} reveals that, for all scaled moments with $m > 0$, {the same previously discussed} non-equilibrium {solution which acts as a late-time ($\Bar\omega > 1$) attractor actually} extends to very early times, {with a convergence of the different solutions driven by the fast expansion of the fireball for $\Bar\omega\lesssim 1$, when collisions cannot have played a major role yet}. However, for moments with $m = 0$, although the solutions tend towards the attractor, the convergence appears slower, and the solutions do not collapse {on a universal curve} in the first stages of the expansion, apparently. This situation is depicted in Figs.~\ref{earl1} and~\ref{earl2}, where we used $\gamma_{q, 0} = \{0.1, 1\}$ as the initial quark {fugacity}, respectively, and we took $\tau_0 = \{0.02, 0.04, 0.08\}$ fm/c for the initialization time. In all these plots we fix $\alpha_0 = 1$, which corresponds to an initially isotropic momentum distribution. {Clearly, this is not a physically realistic scenario, but our purpose is exactly to show how solutions corresponding to arbitrary initial conditions rapidly relax towards the attractor.}
It is noteworthy that higher-order moments exhibit a rapid collapse to their respective attractors, suggesting a fast convergence of the high-momentum region of the single-particle distribution function itself to a universal form~\cite{Alalawi:2022pmg, Soloviev:2021lhs, Strickland:2018ayk}. This behavior closely resembles the observations {performed for a conformal fluid in the 1RTA} case in Refs.~\cite{Strickland:2018ayk, Strickland:2017kux}.
The presence of a pullback attractor for scaled moments $\overline{M}^{\hspace{0.05cm} nm}$ with $m > 0$ is actually typical of {systems described by} kinetic theory but is not observed in strongly-coupled systems like the ones governed by the gauge-gravity duality~\cite{Kurkela:2019set, Janik:2006gp, Spalinski:2018mqg, Romatschke:2017vte}. {The above} observations are also consistent with the results obtained for non-conformal systems~\cite{Alalawi:2022pmg, Chattopadhyay:2021ive, Jaiswal:2021uvv}. {Furthermore}, a similar behavior has been identified also in the case in which the collision kernel of the BE is not approximated through the RTA~\cite{Nugara:2023eku}.

{We now wish to better understand} the slow hydrodynamization of the $m = 0$ scaled moments, {together with the early-time attractor observed for $m>0$. First of all, notice that moments of order $m$, in the local rest frame of the fluid, involve a convolution of $p_z^{2m}$ with the single-particle momentum distribution.}
{The latter} includes a ``free-streaming" (FS) and a ``thermalizing" component - as we can see in Eq.~(\ref{BF12}) - with the former being highly squeezed along the $p_{x,y}$-axes ({corresponding to} extremely small $\langle p_z^{2m}\rangle_{\rm FS}$) {due to the violent longitudinal expansion of the medium, which induces a strong red-shift of the $z$-component of the particle momenta. Hence, all $m>0$ moments rapidly converge to a universal solution, losing memory of the initial condition which only enters in the FS contribution to the particle distribution}. {On the other hand the $m=0$ moments, not sensitive to the suppression of the high-$p_z$ tail of the particle distributions, are less affected by the fast longitudinal expansion of the medium}. {As long as $\Bar\omega\lesssim 1$ the dynamics is simply governed by the longitudinal expansion. However, for larger values of  $\Bar{\omega}$, the relative weight of this free-streaming contribution rapidly decreases, reflecting the smaller and smaller fraction of particles which have not undergone any interaction}~\cite{Broniowski:2008qk, Alalawi:2020zbx, Strickland:2018ayk, Florkowski:2013lya}.  {Hence, also the $m=0$ moments can converge to the attractor solution, this time due to the effect of collisions which tend to isotropize the system which is said to experience ``hydrodynamization'', since all the scaled moments of the particle distributions are now described by the same attractor solution, independently of the initial condition.}

The moments following both a late- and an early-time attractor {are said to display a universal behavior}. The property of universality suggests that the complexity of the initial evolution rapidly reduces within a short {proper-time interval}. However, we noticed that moments with $m = 0$ do not exhibit an attractor behavior at early times. Despite this, the results are still semi-universal after a short time, meaning that the convergence to the forward attractor solution occurs at fairly early times ($\Bar{\omega}\gtrsim 1$), leading to the system's hydrodynamization~\cite{Soloviev:2021lhs,Alalawi:2022pmg, Kurkela:2019set}.
In {the absence of} a distinct early-time attractor, the behavior of the system in the initial stages ($\Bar{\omega} \lesssim 1$) {is not simply described by} a single, stable trajectory {to which} different initial conditions converge. Instead, the system may display a complex behavior (depending on the chosen initial condition) during its early evolution~\cite{Heller:2018qvh, Spalinski:2022cgj, Jankowski:2023fdz}.

To summarize, the results presented herein for the 2RTA-{scheme look} similar to those obtained from conformal kinetic theory in 1RTA. In fact, even the $m = 0$ moments exhibit a universal forward attractor. This finding, {according to} Refs.~\cite{Strickland:2018ayk, Strickland:2017kux, Strickland:2019hff}, implies the existence of an attractor for the complete single-particle distribution function itself, with a slower convergence at very low longitudinal momenta.

%%%%%%%%%%%%%%%%%%%%%%%%%%%%%%%%%%%%%%%%%%%%%%%%%%%%%%%%%%%%%%%%%%%%%%%%%
\subsection{Numerical solutions for the scaled shear corrections}
%%%%%%%%%%%%%%%%%%%%%%%%%%%%%%%%%%%%%%%%%%%%%%%%%%%%%%%%%%%%%%%%%%%%%%%%%
\label{visc_numerical}
In this section we address the time-evolution of the shear viscous correction $\overline{\phi}$ {to the energy-momentum tensor}, given in Eq.~(\ref{visc}), using the {previously obtained} results for the scaled moments. {Starting from different initializations for the momentum anisotropy and quark abundance} we are going to see {if also $\overline{\phi}$ displays} an attractor behavior. In order to discuss the universality of this quantity, we use the same {parameter set} as in Figs.~\ref{fig:mom-late-under} and~\ref{faccia2} (forward attractor) and in Figs.~\ref{earl1} and~\ref{earl2} (pullback attractor).
\begin{figure}[!hbt]
    \centering
    \includegraphics[width=0.495\textwidth] {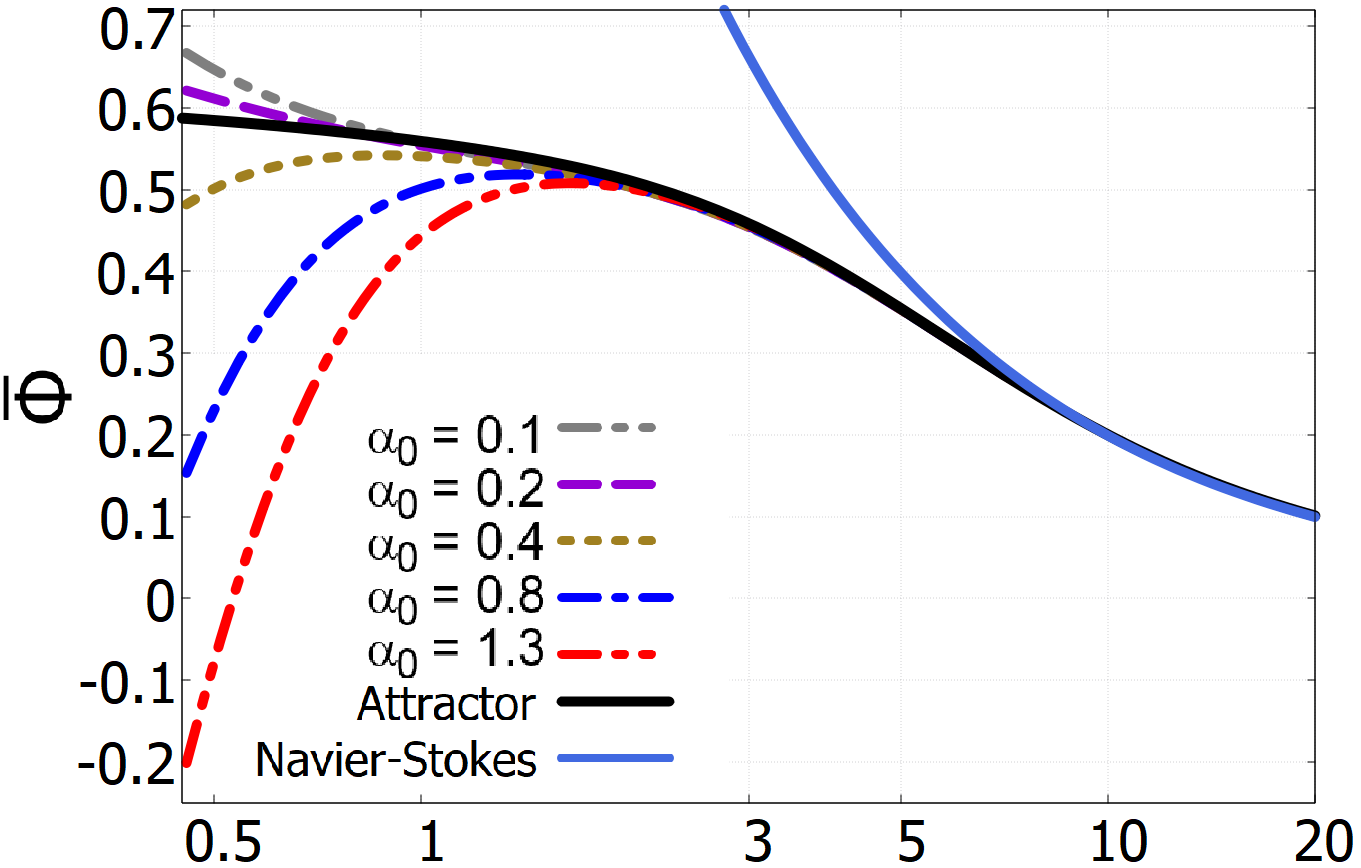}
    \includegraphics[width=0.495\textwidth] {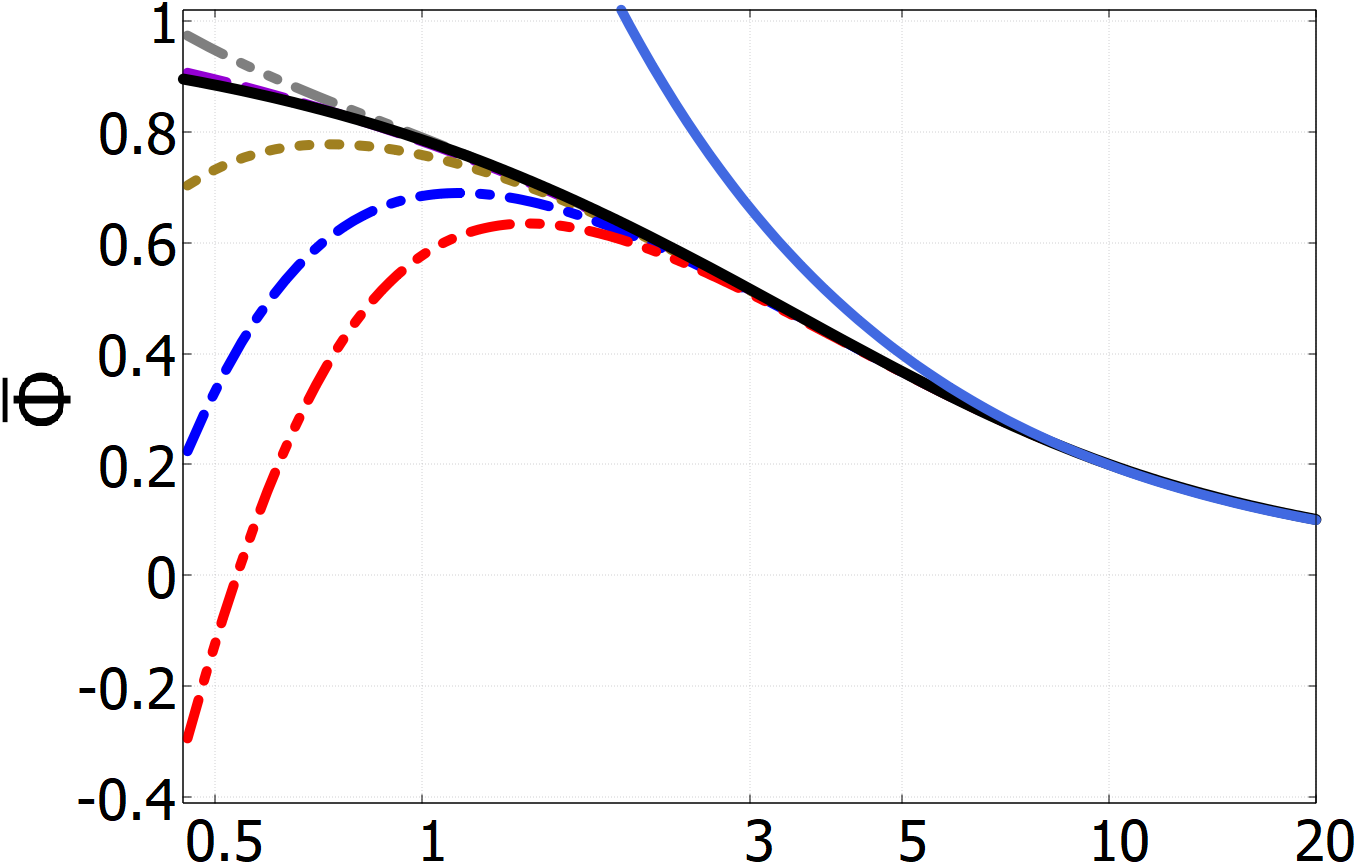} \\ 
    \vspace{0.2cm}
    \includegraphics[width=0.495\textwidth] {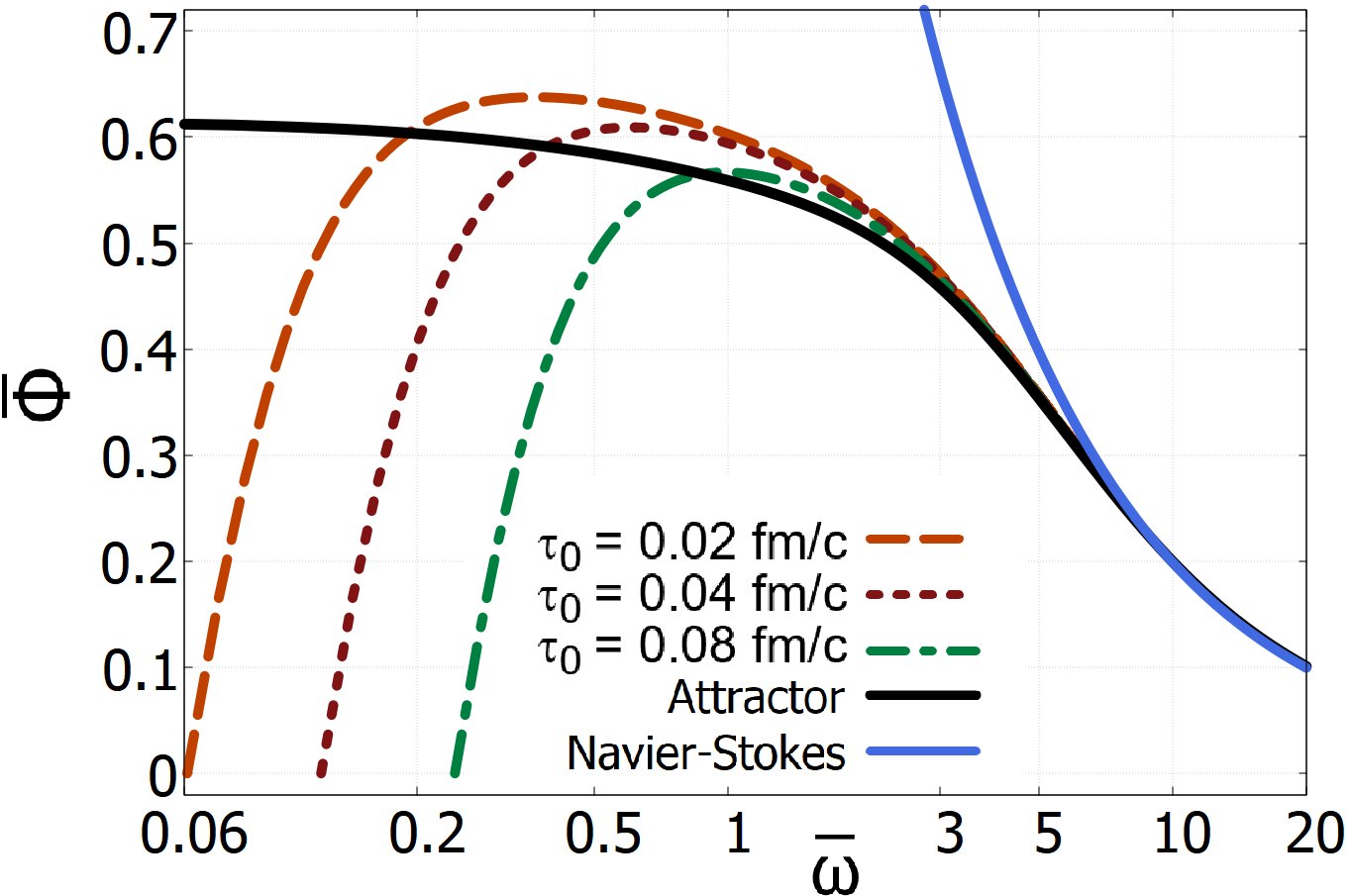}
    \includegraphics[width=0.495\textwidth] {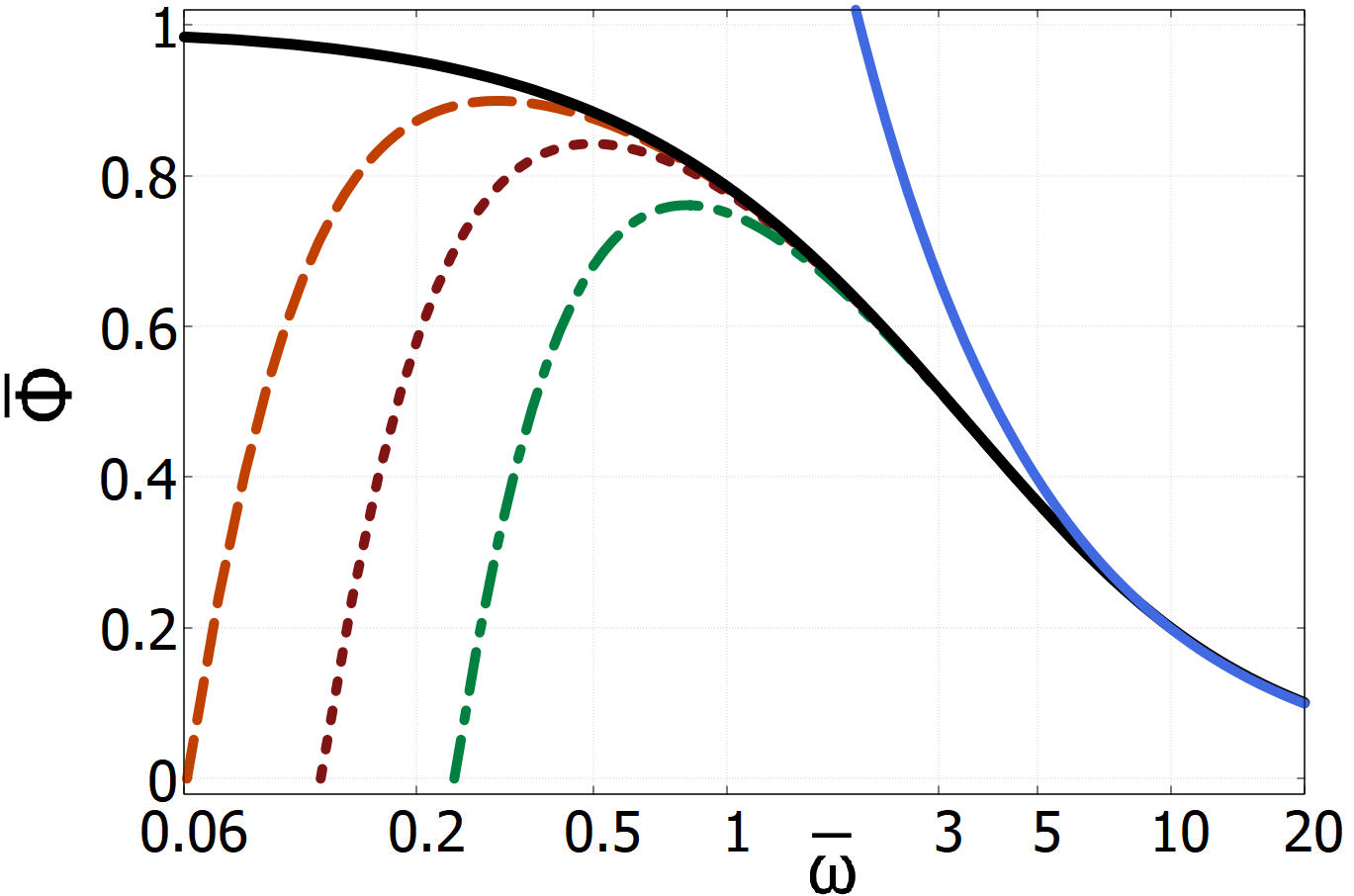}  
    \caption{Scaled shear correction $\overline{\phi}$ (various dashed colored lines) {to the energy-momentum tensor}. Exact solutions {of the 2RTA-BE} are compared to the Navier-Stokes {result (cyan solid line) and to the attractor solution (continuous black curve)}. The time-variable is given by $\Bar{\omega} = \tau/\tau_{\rm eq}$ {and logarithmic units are employed to cover both the pre-equilibrium and the universal asymptotic regime}. Columns from left to right show the cases $\gamma_{q, 0} = \{0.1, 1\}$, respectively. In the upper row {the convergence to the} forward attractor {is displayed}, while in the lower row the early-time behavior is considered.}
    \label{shear}
\end{figure}
{Notice that}, due to its definition in Eq.~(\ref{visc}), the quite slow convergence to a universal curve of the solutions for $\overline{M}^{20}$ {(see bottom-left panels of Figs.~\ref{earl1} and~\ref{earl2}) may prevent the emergence of an attractor behavior for $\overline{\phi}$}.
{However, even in the pre-hydrodynamic regime, when not collapsed onto a single curve, considering their absolute value, the different solutions for $\overline{M}^{20}$ display only small deviations among each other. Hence},
in practice, {as can be seen in Fig.~\ref{shear}}, a sort of universal late-time result (solid black curve) is already present for $\overline{\phi}$ {also in a pre-asymptotic regime (see the different values of $\Bar{\omega}_c$ in table \ref{conv_tau}), when $\overline{M}^{01}$ has already reached convergence to its non-equilibrium attractor} -- which occurs for $\Bar{\omega} \sim 2$ ($\gamma_{q,0} = 0.1$) or $\Bar{\omega} \sim 3$ ($\gamma_{q,0} = 1$) -- while $\overline{M}^{20}$ not.
In Fig.~\ref{shear} we also note that $\overline{\phi}$ can assume negative values {during the first stages of the medium evolution} if the initial distribution function has a prolate shape ($\xi_0 < 0$ and hence $\alpha_0 > 1$); {physically, this amounts to have $P_T<P_L$. Notice however that, as already stressed, due to the fast longitudinal expansion of the system in HIC's, the physical range for the initial anisotropy coefficient is} $\xi_0 \in (0, +\infty)$~\cite{Strickland:2014pga, Alqahtani:2017mhy}.

As expected, in Fig.~\ref{shear} we can see that at late times the scaled shear correction decreases  and asymptotically tends to zero (although very slowly), when the system reaches {local thermodynamic equilibrium}. On the other hand, it is evident that for most of the longitudinal expansion, the viscous correction $\phi$ remains always comparable to the {equilibrium} fluid pressure $P_{\rm eq}$.
{This highligths the importance of adopting, in the first stages of the evolution, a kinetic description of the medium, capable of accounting for the initial} large anisotropies in momentum space, unlike standard hydrodynamic approaches {which rely on some form of perturbative expansion around an equilibrium configuration}~\cite{Denicol:2010xn,Jaiswal:2013npa, Jaiswal:2015mxa, Romatschke:2017ejr}. An alternative strategy to circumvent this difficulty is the one represented by anisotropic hydrodynamics~\cite{Martinez:2010sc}.

{In Fig.~\ref{shear} we also display the} Navier-Stokes result (solid cyan curve) from Eq.~(\ref{navier}). {As one can see, all curves collapse on top of a universal forward attractor before reaching the Navier-Stokes limit (which is recovered when $\Bar{\omega} \gtrsim 9$) and well before thermodynamic equilibrium. The approach to the late-time attractor occurs approximately for $\Bar{\omega} \gtrsim 4$ (if $\gamma_{q,0} = 0.1$) or $\Bar{\omega} \gtrsim 5$ (if $\gamma_{q,0} = 1$). This fact is again a signal of the relatively fast hydrodynamization of the system which holds also in the case of different relaxation times for the two components of the fluid and could justify the} success {of hydrodynamics} in describing {several soft observables measured in HIC's} at RHIC and at LHC.

To complete our discussion, in the bottom row of Fig.~\ref{shear} {we investigate the possible presence of an early-time attractor for $\Bar{\phi}$. According to Figs.~\ref{earl1} and~\ref{earl2}, we start from an isotropic momentum distribution ($\alpha_0 = 1$), very far from what one expects for a realistic initial condition, and we check whether, initializing the system at earlier and earlier times, all the solutions of the 2RTA-BE rapidly converge (when $\Bar{\omega} \lesssim 1$) to the same universal curve. One clearly sees that this is not the case: at very early times the different solutions can even cross the attractor (when quarks are initially under-populated) and depart from it. Hence no early-time attractor is present for the scaled viscous correction $\Bar{\phi}$ and this has to be attributed to the absence of a pullback attractor for $\overline{M}^{20}$ in the 2RTA case. This is particularly evident in the case of initial departure from chemical equilibrium (bottom-left panel).}

%%%%%%%%%%%%%%%%%%%%%%%%%%%%%%%%%%%%%%%%%%%%%%%%%%%%%%%%%%%%%%%%%%%%%%%%
\subsection{The numerical exact distribution function for partons}
%%%%%%%%%%%%%%%%%%%%%%%%%%%%%%%%%%%%%%%%%%%%%%%%%%%%%%%%%%%%%%%%%%%%%%%%

\label{distrib_num}
{We devote this section to present our numerical results for the quark and gluon distribution functions arising from the exact solution of the 2RTA-BE given} in Eq.~(\ref{BF12}).
In order to appreciate the isotropization of $f_a$ ($a = q, g$), we put ourselves into the LRF of the fluid, where {at late times -- as a consequence of reaching equilibrium --} the distribution is expected to be isotropic. {In full generality}, due to longitudinal boost invariance, we can {focus on the center of the fireball, setting} $z = 0$.  {Since for Bjorken flow $v^z=z/t$, the results obtained in the cell with $z=0$ at time $t$ describe the fluid at longitudinal proper time $\tau$ everywhere else.} Within this assumption the relativistic invarant $w$ in Eq.~(\ref{eq:w-def}) turns out to be simply
\begin{equation*}
    w = \tau \hspace{0.07cm} p_z\,,
\end{equation*}
{since in the LRF ($z = 0$) one can identify the relative time with the proper time $t=\tau$.}
\begin{figure}[!hbt]
    \centering
    \includegraphics[width=0.345\textwidth] {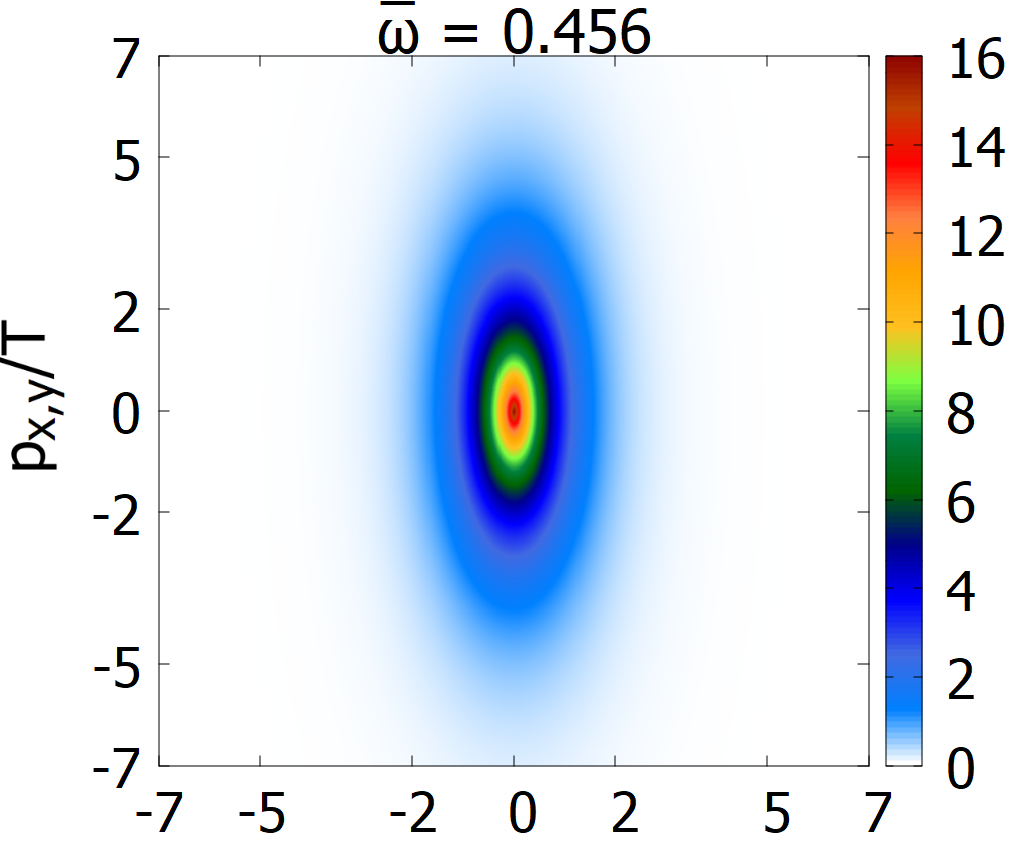} 
    \includegraphics[width=0.315\textwidth] {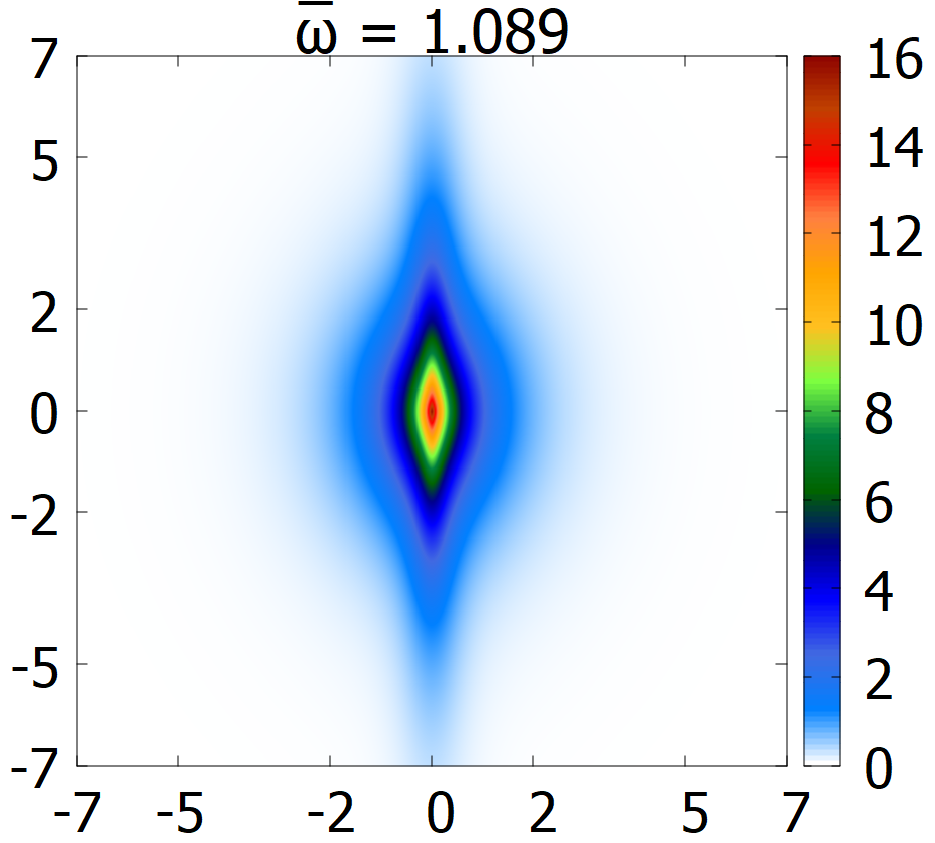}
    \includegraphics[width=0.315\textwidth] {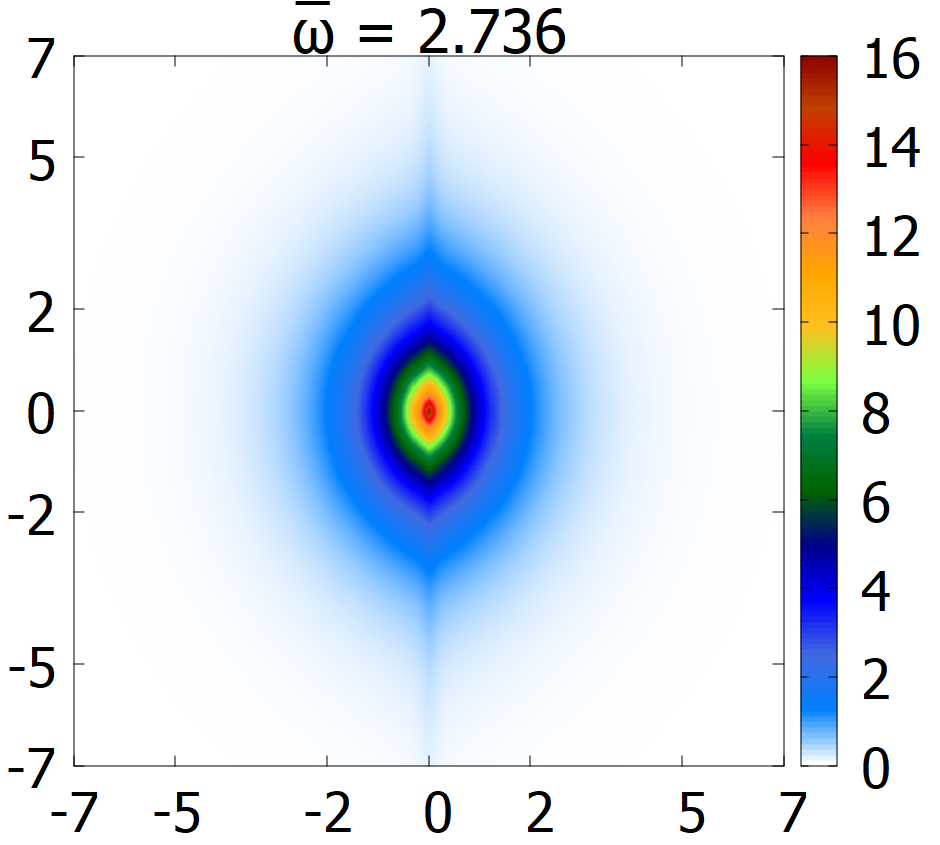} \\
    \vspace{0.1cm}
    \includegraphics[width=0.345\textwidth] {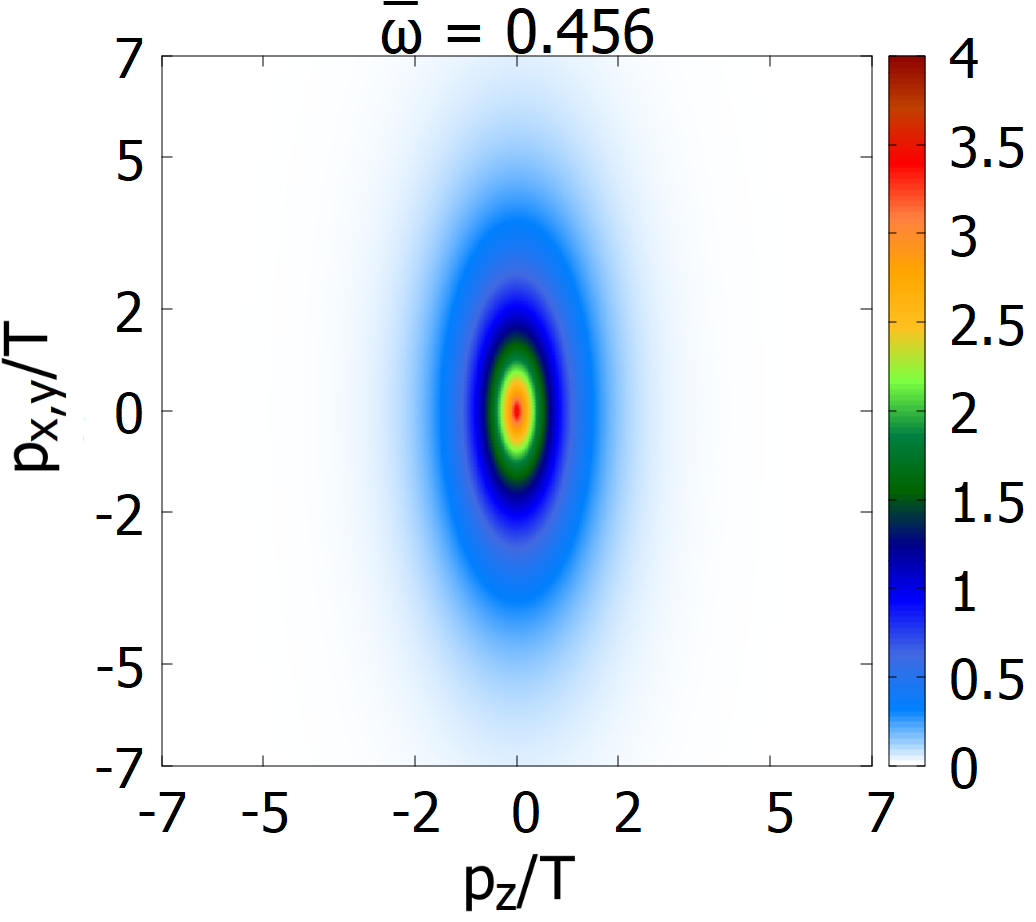} 
    \includegraphics[width=0.315\textwidth] {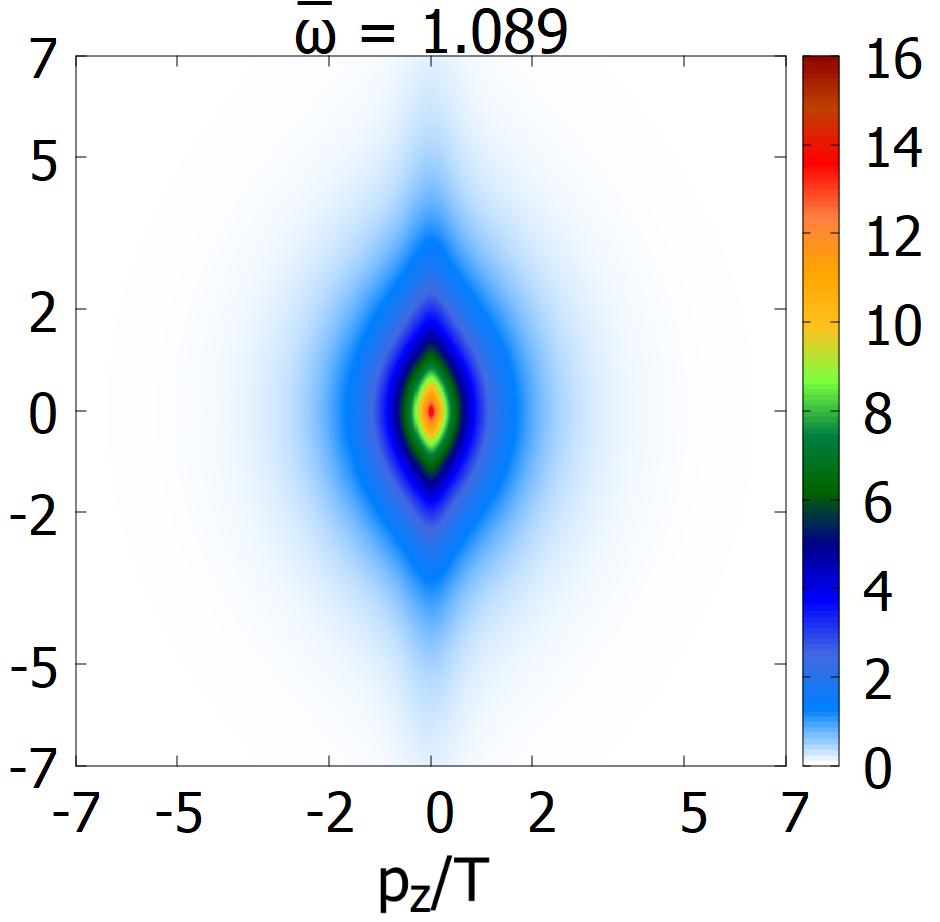}
    \includegraphics[width=0.315\textwidth] {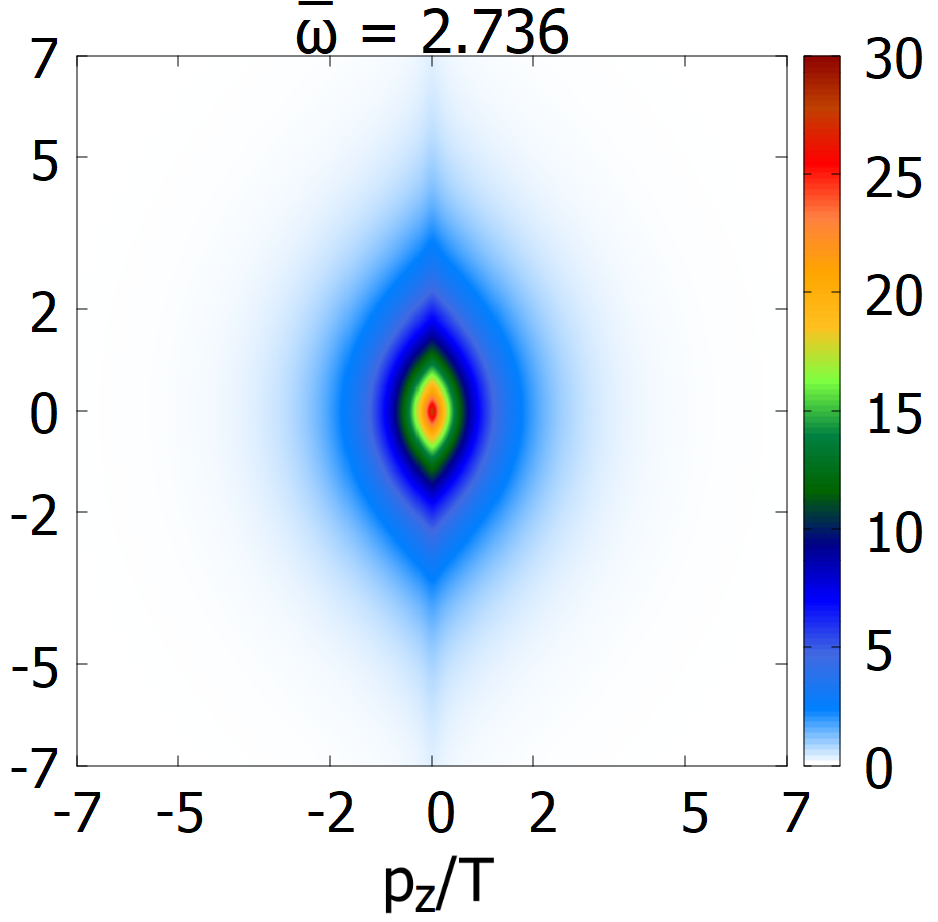}  
    \caption{Snapshots of the {gluon (upper panels) and quark (lower panels)} {momentum} distribution for different
scaled times $\Bar{\omega}$ {arising from the solution of the 2RTA-BE starting from} an oblate initial condition, {with $\alpha_0=0.4$}, for a gluon-dominated plasma ($\gamma_{q,0}=0.1$).}
    \label{f_gq}
\end{figure}
Hence we plot a section -- for one vanishing component of the {transverse momentum} $p_T \equiv \sqrt{p^2_x + p^2_y}$ -- of the quark (bottom row) and gluon (top row) {momentum distributions} at three different scaled times $\Bar{\omega} = \tau/\tau_{\rm eq}$. The {single-particle} distributions are initialized {according to} the Strickland-Romatschke form in Eqs.~(\ref{rta3}) and (\ref{rta4}). {We first address the case,} in Fig.~\ref{f_gq}, {of an initial condition with {an oblate momentum distribution ($\alpha_0=0.4$)} and quark fugacity} $\gamma_{q, 0} = 0.1$. {The other parareters -- } $\tau_0$, $T_0$ and specific viscosity $\eta/s$ -- are the same as in Figs.~\ref{fig:mom-late-under} and~\ref{faccia2}.
As depicted in Fig.~\ref{f_gq}, the exact solution for the distribution function typically contains two visually separated components, {corresponding to the two different terms in Eq.~(\ref{BF12})}. The first component, {arising from the fraction of initial particles which have not undergone any interaction}, exhibits an anisotropic nature, progressively {getting more and more squeezed around} $p_z \sim 0$. This {free-streaming contribution corresponds to the first term} in Eq.~(\ref{BF12}). By itself this contribution {would lead to a stronger and stronger anisotropy}, but its amplitude exhibits an exponential decay due to the damping function $D_a$ - defined in Eq.~(\ref{damp}) - causing the loss of information on the initial state. This consideration is qualitatively in agreement with the results presented in Refs.~\cite{Strickland:2018ayk, Nugara:2023eku, Florkowski:2013lya, Jaiswal:2021uvv, Kurkela:2018vqr}. {Due to the initial under-population of quarks, suppressed by a $\gamma_{q,0} = 0.1$ fugacity factor, the contribution from the initial free-streaming is more relevant for the gluon distribution.} However, we also note that the amplitude of this anisotropic component decays {more slowly} for quarks than for gluons, {due to the longer quark relaxation time $\tau_{{\rm eq}, q} = \frac{9}{4} \hspace{0.07cm} \tau_{{\rm eq}, g}$.} The other component, which tends to isotropize the system, comes from the second term in Eq.~(\ref{BF12}), {usually referred to as} ``evolution term". This contribution becomes increasingly dominant for $\Bar{\omega} > 1$, when the interaction rate between particles is larger than the expansion rate of the plasma, {and it is responsible for the asymptotic approach to local thermodynamic equilibrium.}

\begin{figure}[!hbt]
    \centering
    \includegraphics[width=0.345\textwidth] {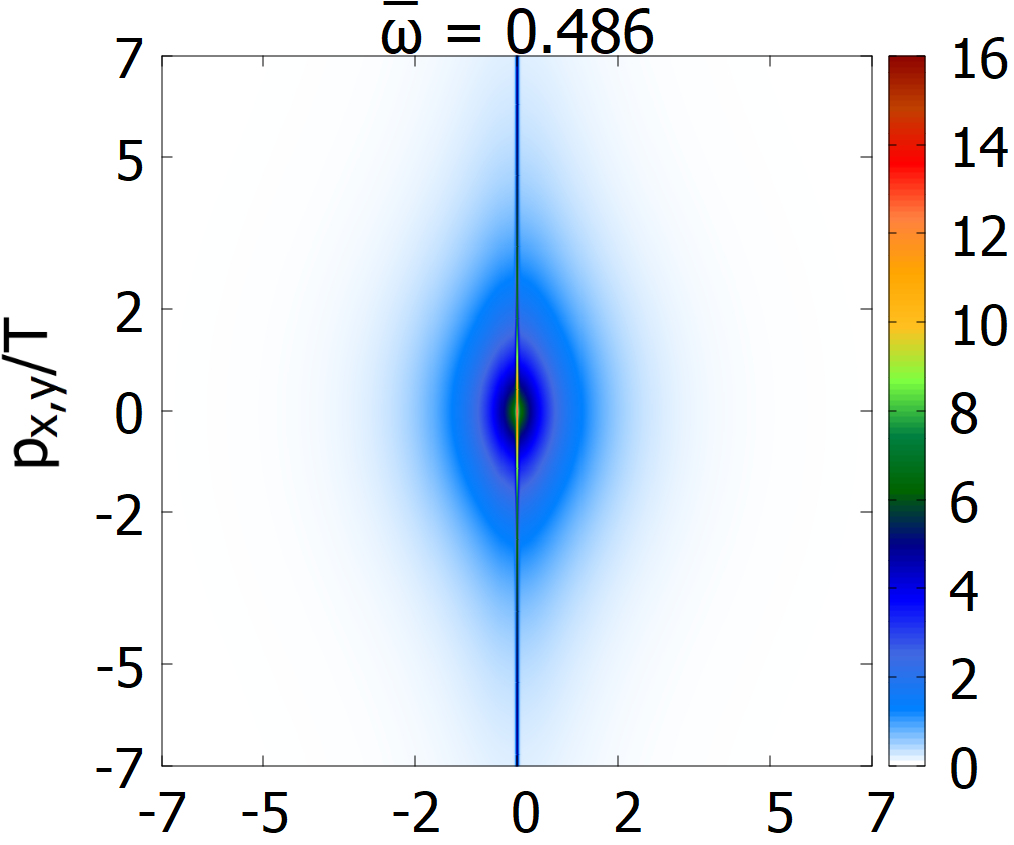} 
    \includegraphics[width=0.315\textwidth] {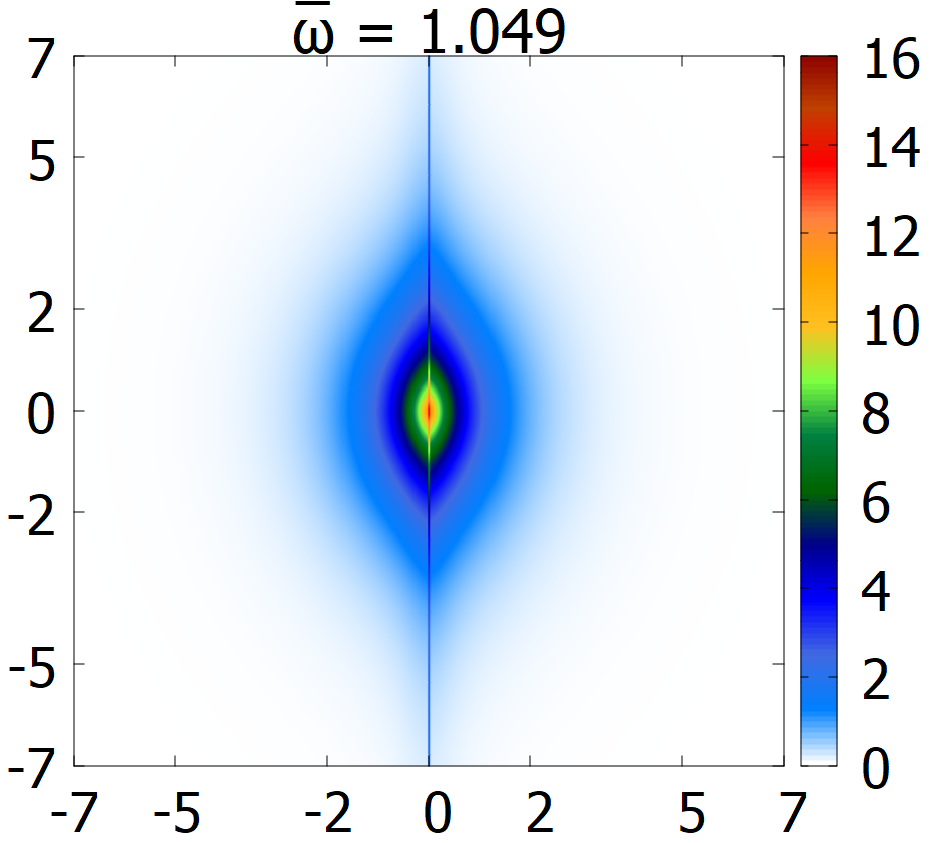}
    \includegraphics[width=0.315\textwidth] {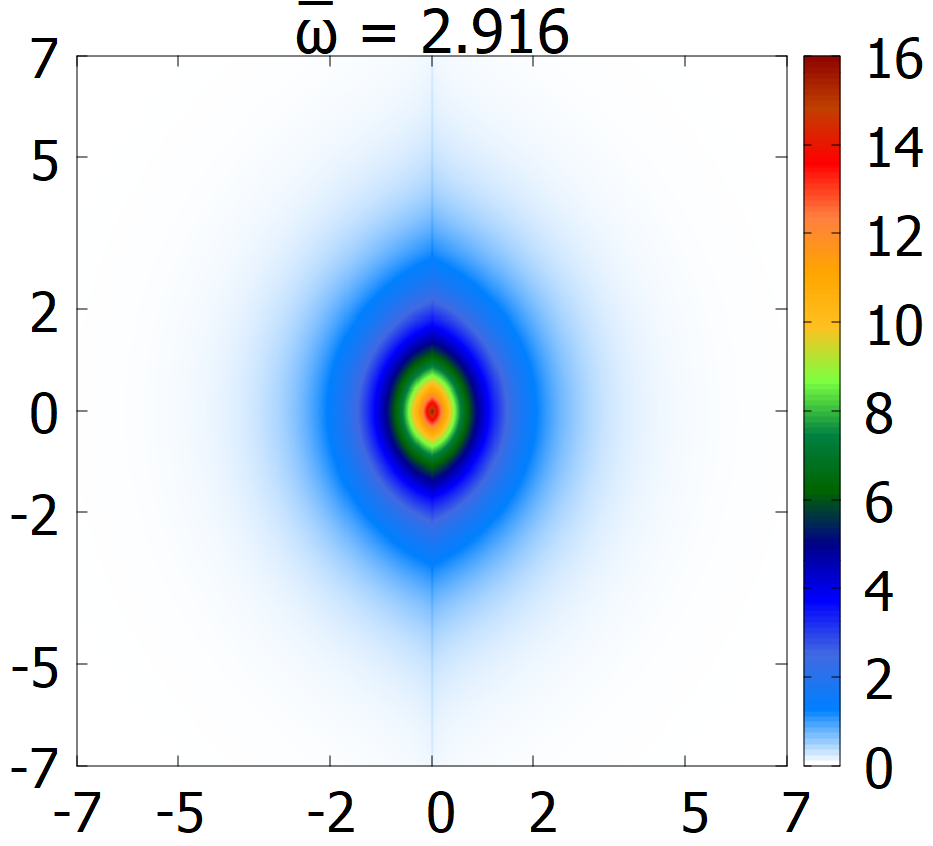} \\
    \vspace{0.1cm}
    \includegraphics[width=0.345\textwidth] {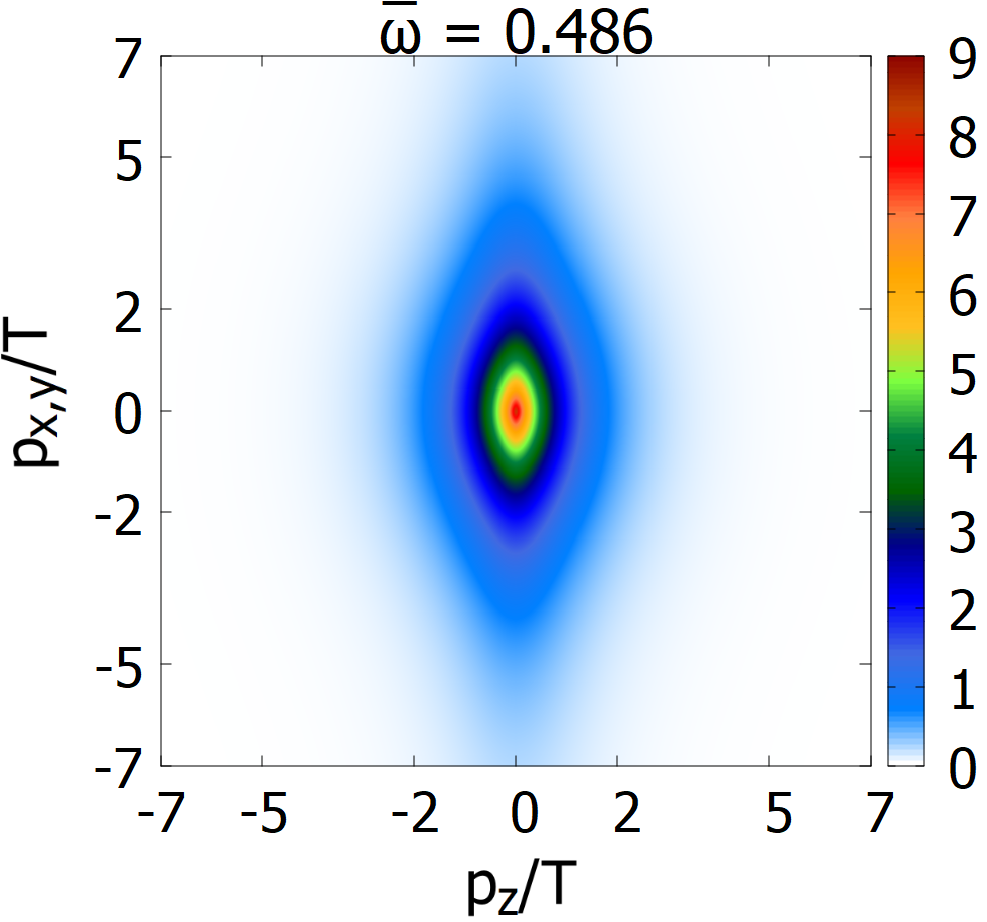} 
    \includegraphics[width=0.315\textwidth] {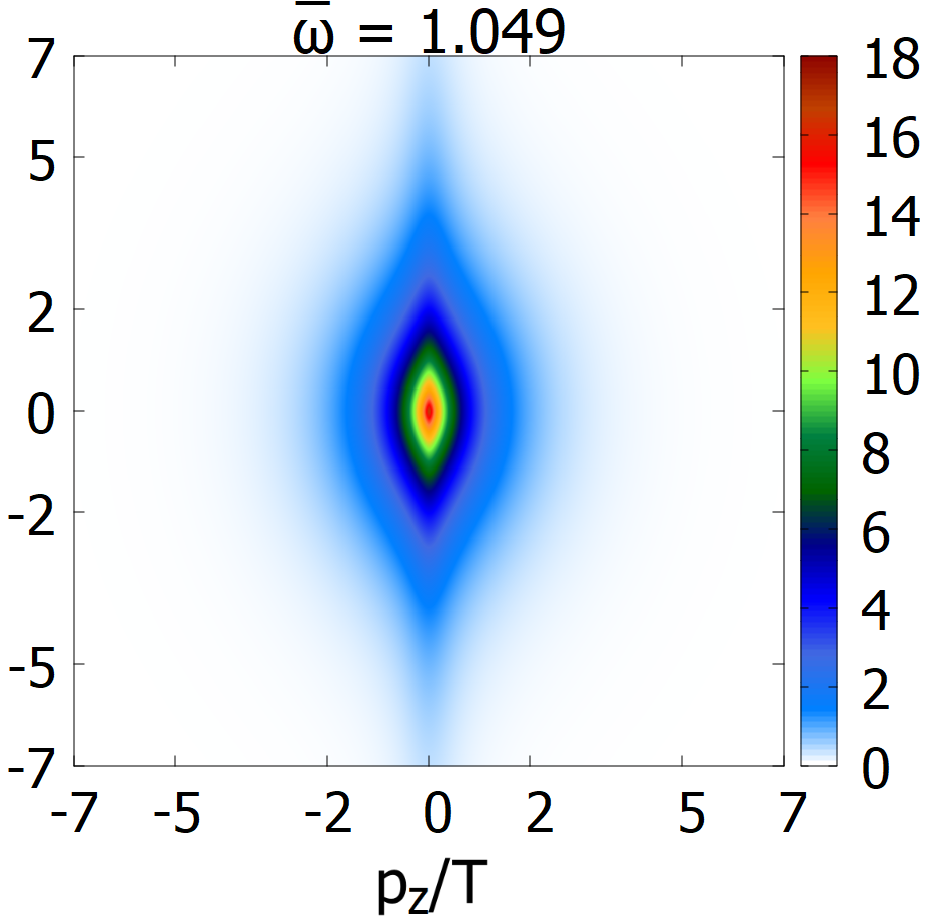}
    \includegraphics[width=0.315\textwidth] {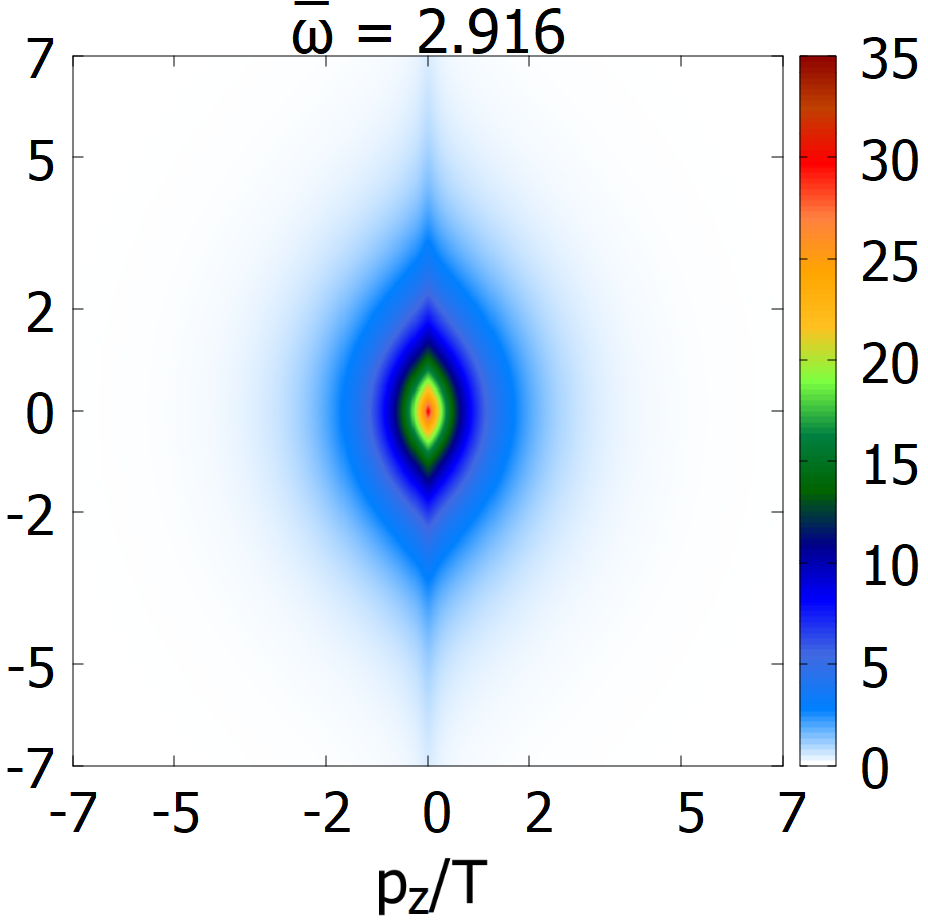}  
    \caption{{Snapshots of the {gluon (upper panels) and quark (lower panels)} {attractor for the single-particle momentum} distribution at different
scaled times $\Bar{\omega}$. Notice that the attractor solution is initialized at a much earlier time than the ones shown in the above panels. Here an initial gluon-dominated plasma, with $\gamma_{q,0}=10^{-4}$ is considered.}}
    \label{f_gq_attr}
\end{figure}
 Now we focus on Fig.~\ref{f_gq_attr}, which shows analogous snapshots of the single-particle distribution function for quarks and gluons. However, in this case, the temperature evolution {arising} from Eq.~(\ref{T_ex}) {refers to} the attractor solution. Similar characteristics to the previously discussed individual solution are observed, but with the notable distinction that the {free-streaming term starting from the attractor} initial condition induces a collection of highly-squeezed modes with $p_z \sim 0$ for the gluon component, as emphasized in Ref.~\cite{Strickland:2018ayk}. These $p_z\sim 0$ modes characterize also the initial quark distribution, but, due to the very small initial quark fugacity ($\gamma_{q,0}=10^{-4}$) and the very early initialization of the system ($\tau_0=0.001$ fm/c) when searching for the attractor solution, they are not visible in the lower panels of Fig.~\ref{f_gq_attr}, which refer to much later times.
In Refs.~\cite{Strickland:2018ayk, Strickland:2017kux, Florkowski:2013lya}, it was shown (in the 1RTA limit) that the slower approach to the late-time attractor for scaled moments $\overline{M}^{\hspace{0.05cm} nm}$ with $m = 0$ is a consequence of these squeezed modes. In fact, if $m \neq 0$, powers of $p_z$ within the phase-space integral naturally mitigate the influence of these highly squeezed modes, see Eq.~(\ref{robo}). However, when $m = 0$, {these} squeezed modes {strongly affect the moments of the single-particle distributions}, particularly during the early stages of the longitudinal expansion, resulting in a slower decay of the non-hydrodynamic modes - see Fig.~\ref{fig:mom-late-under}.

From Figs.~\ref{f_gq} and~\ref{f_gq_attr}, we also note that the {the quark population grows during the system evolution}: more and more {$q\bar q$ pairs } are produced through processes like $g + g \rightarrow q + \Bar{q}$. {Eventually, the occupancy of zero-momentum modes reflects the counting of the internal degrees of freedom (16 for gluons and 36 for quarks, assuming 3 massless flavors), signalling the attainment of chemical equilibrium.}

\begin{figure}[!hbt]
    \centering
    \includegraphics[width=0.345\textwidth] {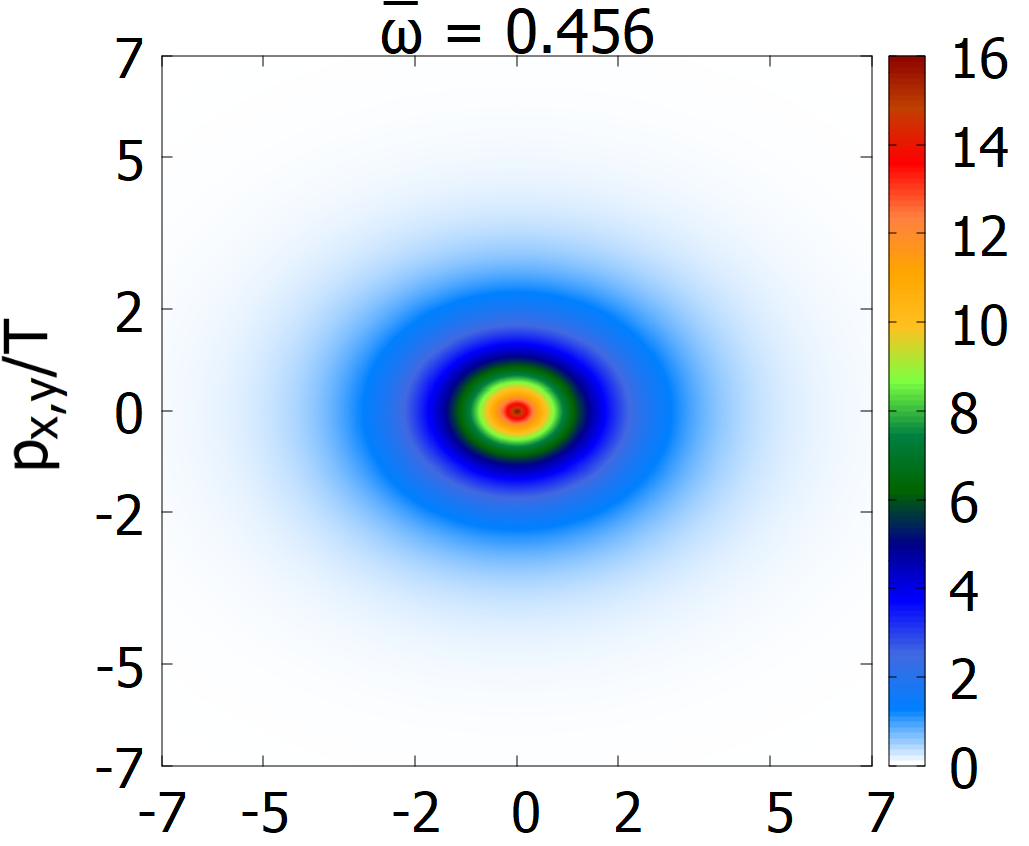} 
    \includegraphics[width=0.315\textwidth] {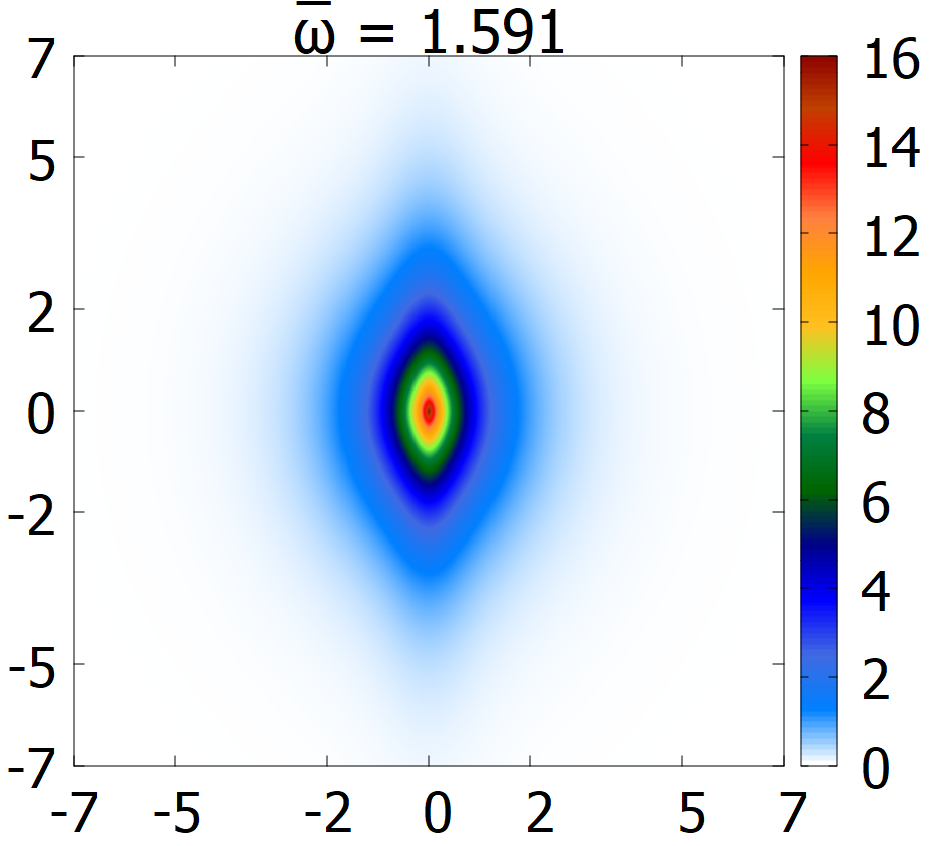}
    \includegraphics[width=0.315\textwidth] {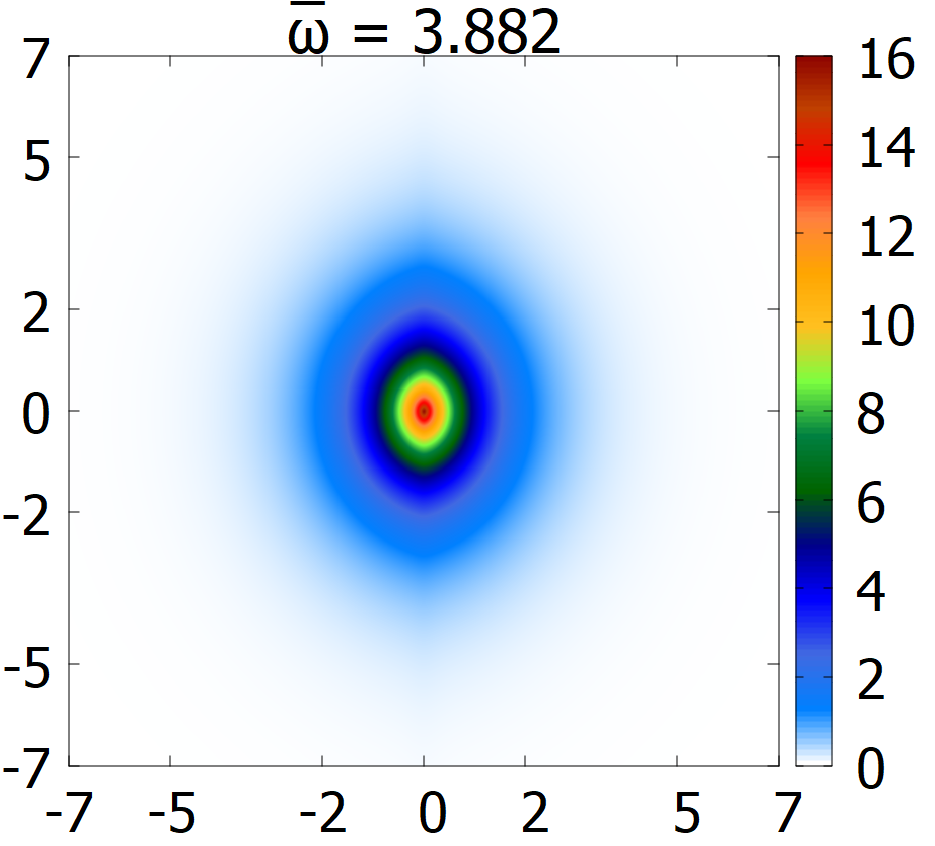} \\
    \vspace{0.1cm}
    \includegraphics[width=0.345\textwidth] {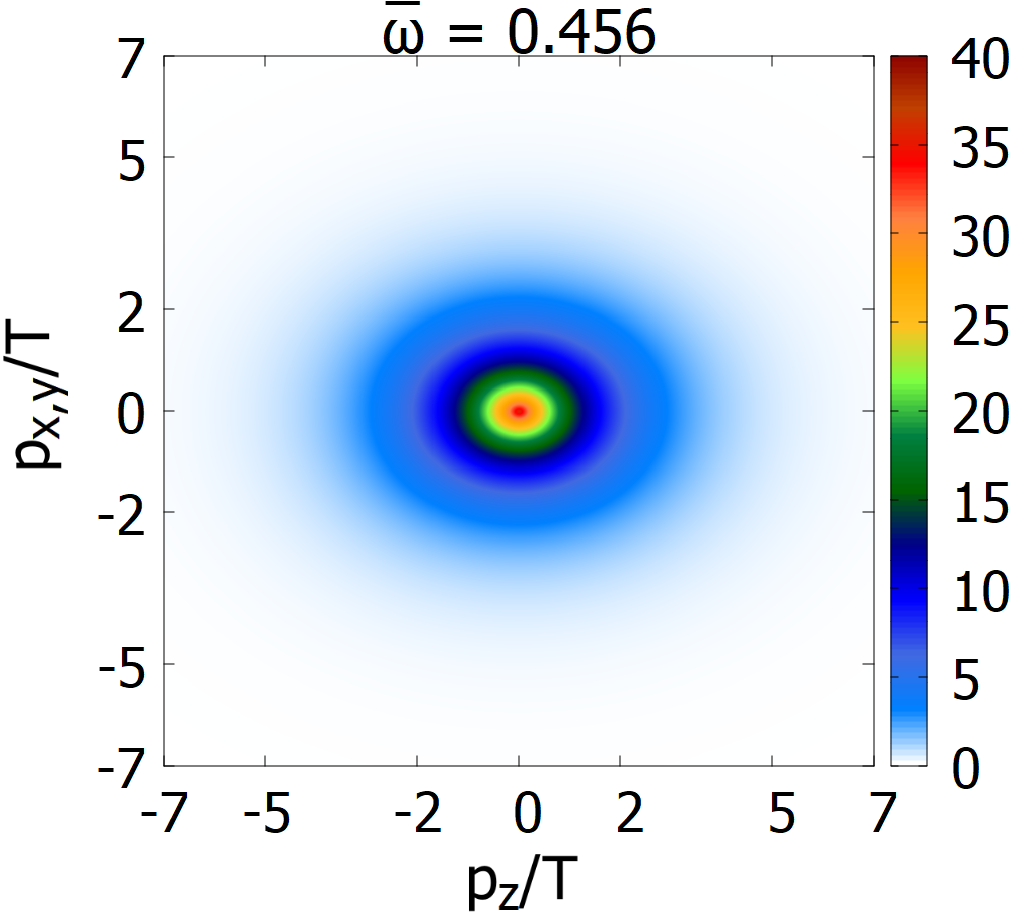} 
    \includegraphics[width=0.315\textwidth] {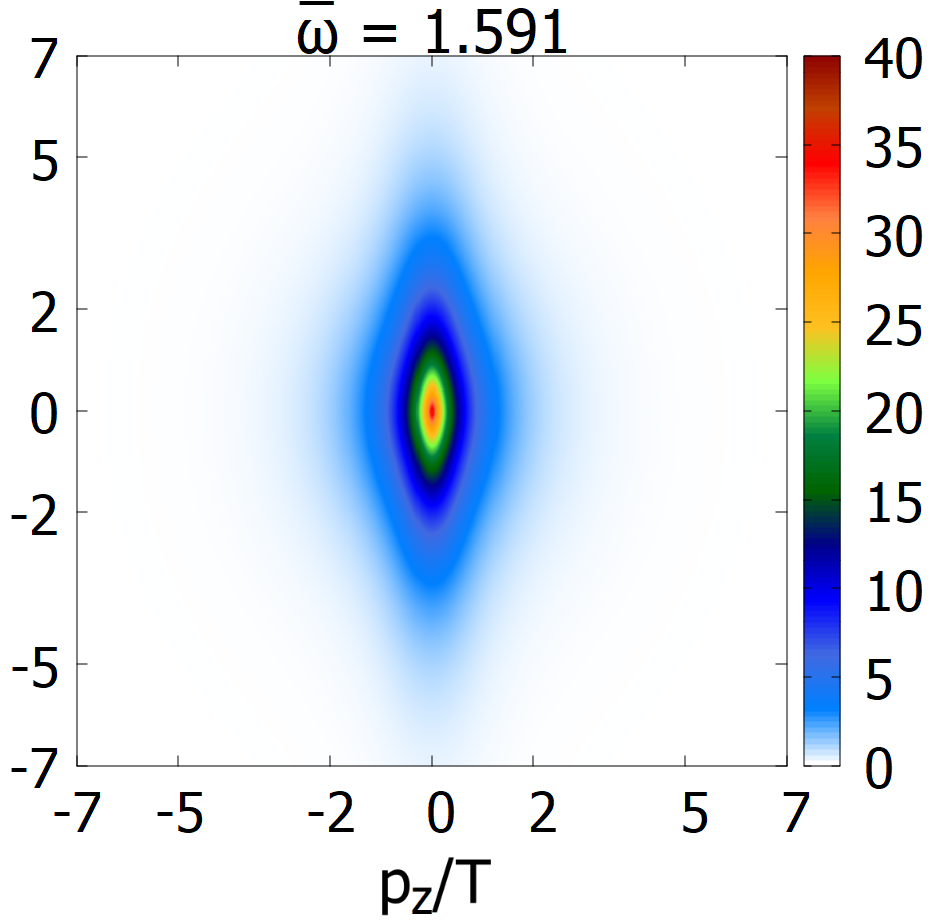}
    \includegraphics[width=0.315\textwidth] {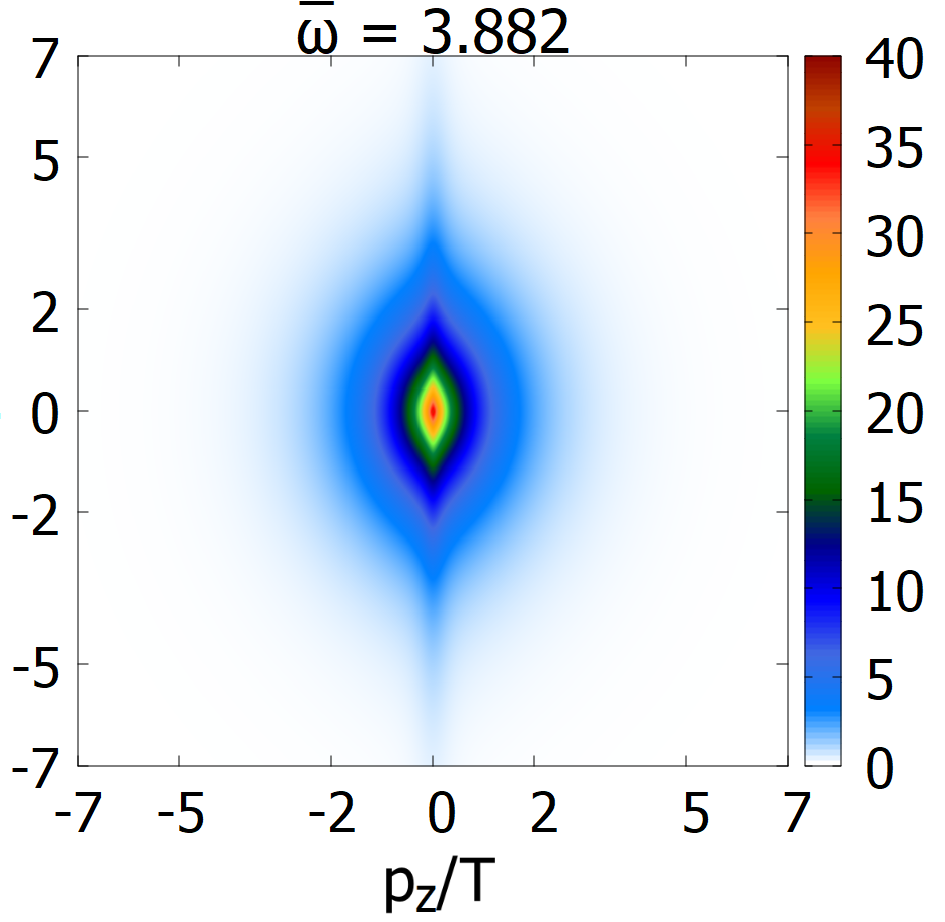}  
    \caption{{The same as in Fig.~\ref{f_gq}, but for a chemically equilibrated plasma ($\gamma_{q,0}=1$) and an initial prolate momentum distribution ($\alpha_0=1.3$).}}
    \label{f_gq_re}
\end{figure}
\begin{figure}[!hbt]
    \centering
    \includegraphics[width=0.345\textwidth] {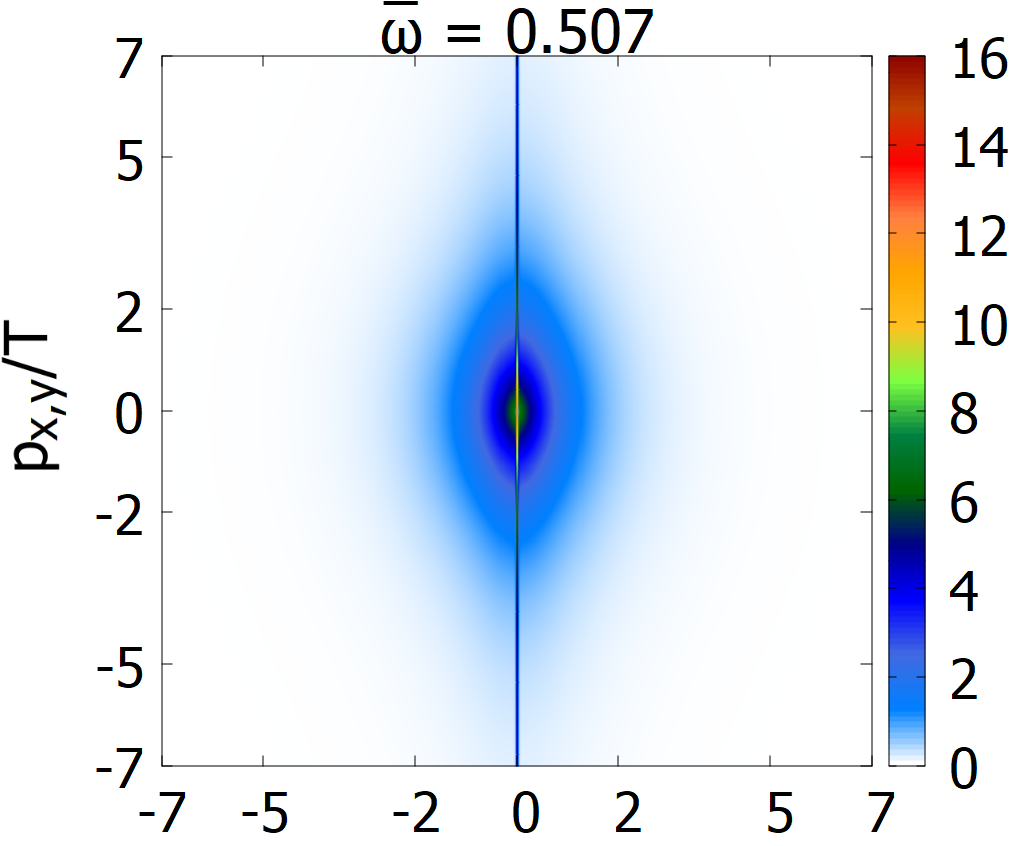} 
    \includegraphics[width=0.315\textwidth] {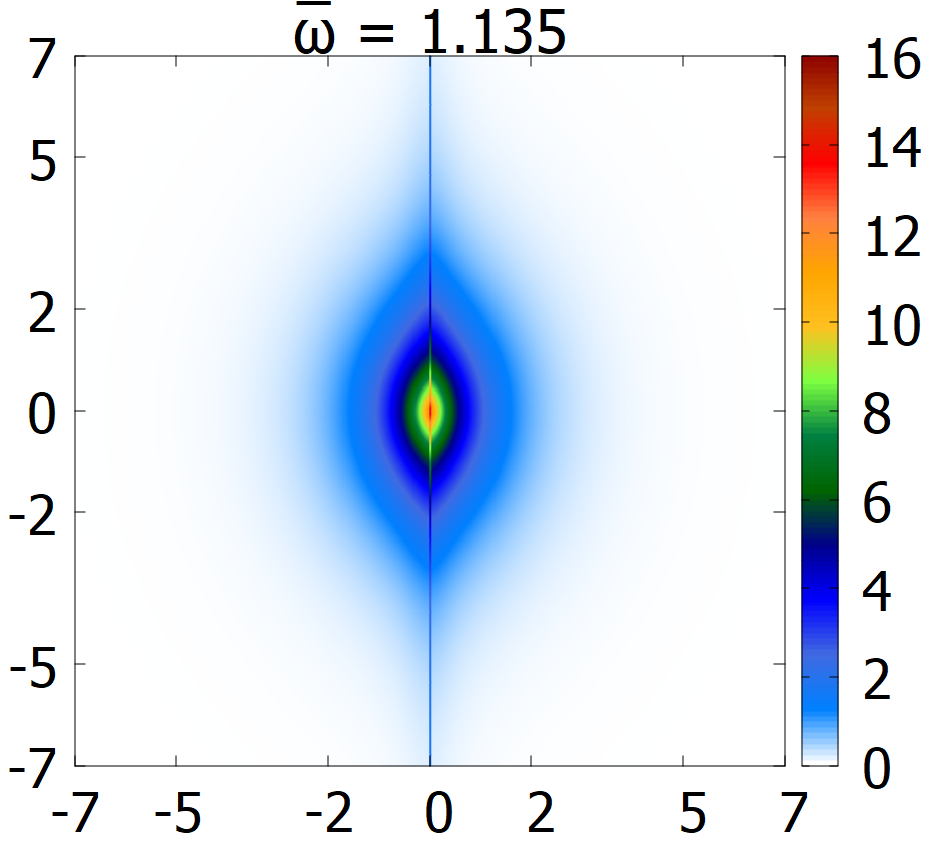}
    \includegraphics[width=0.315\textwidth] {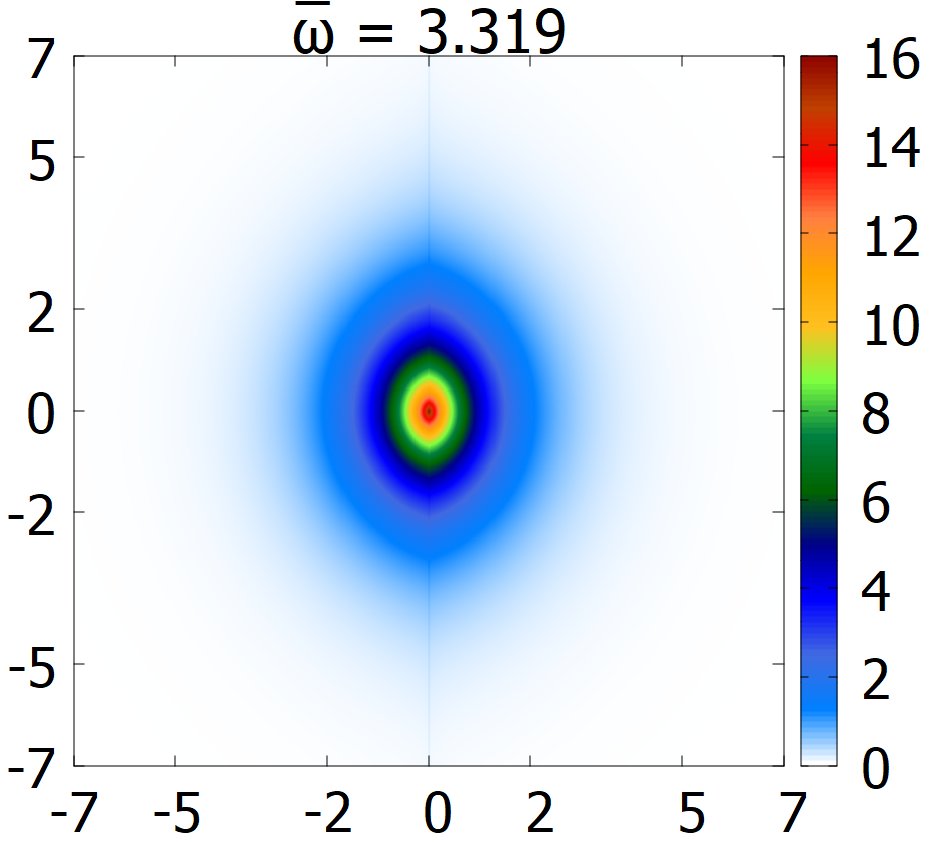} \\
    \vspace{0.1cm}
    \includegraphics[width=0.345\textwidth] {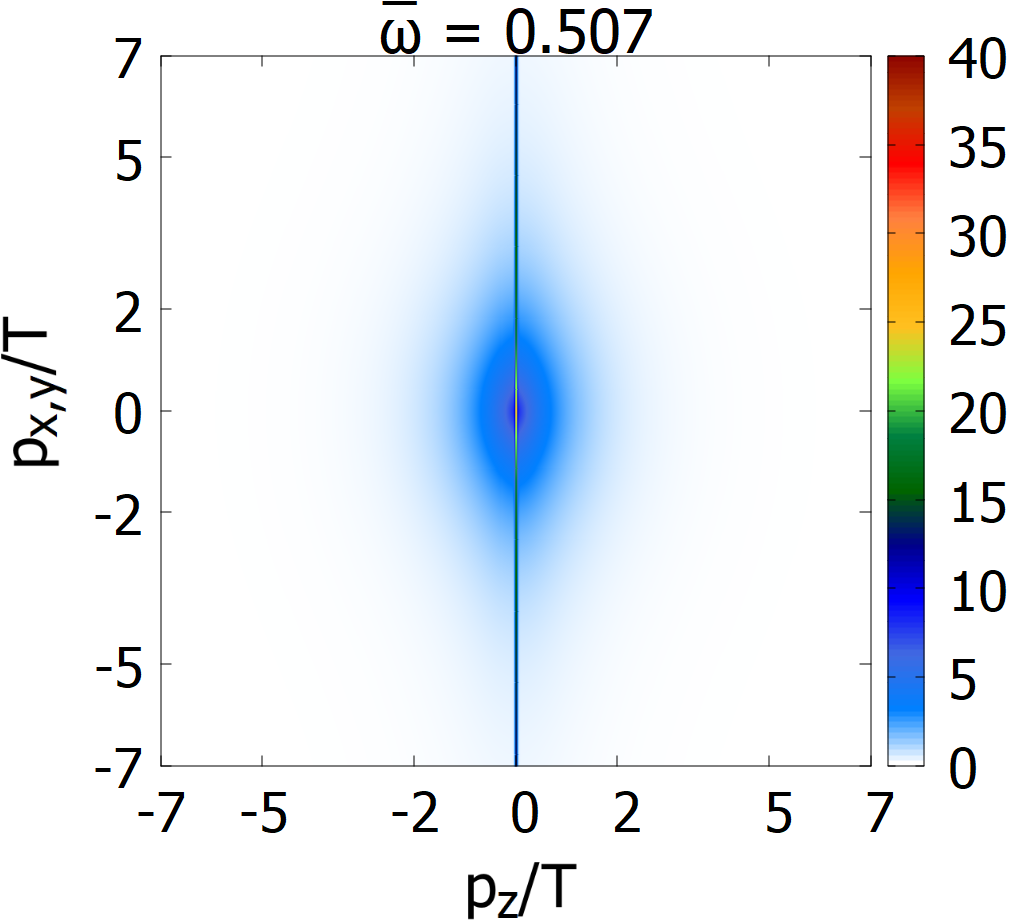} 
    \includegraphics[width=0.315\textwidth] {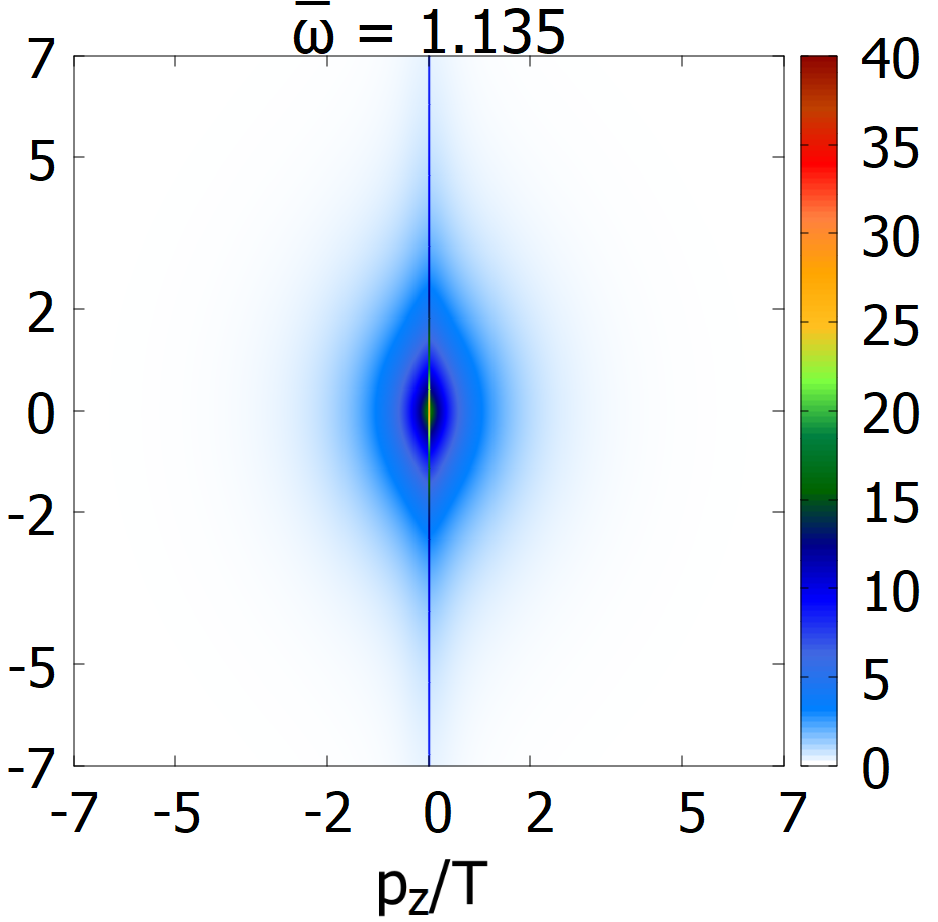}
    \includegraphics[width=0.315\textwidth] {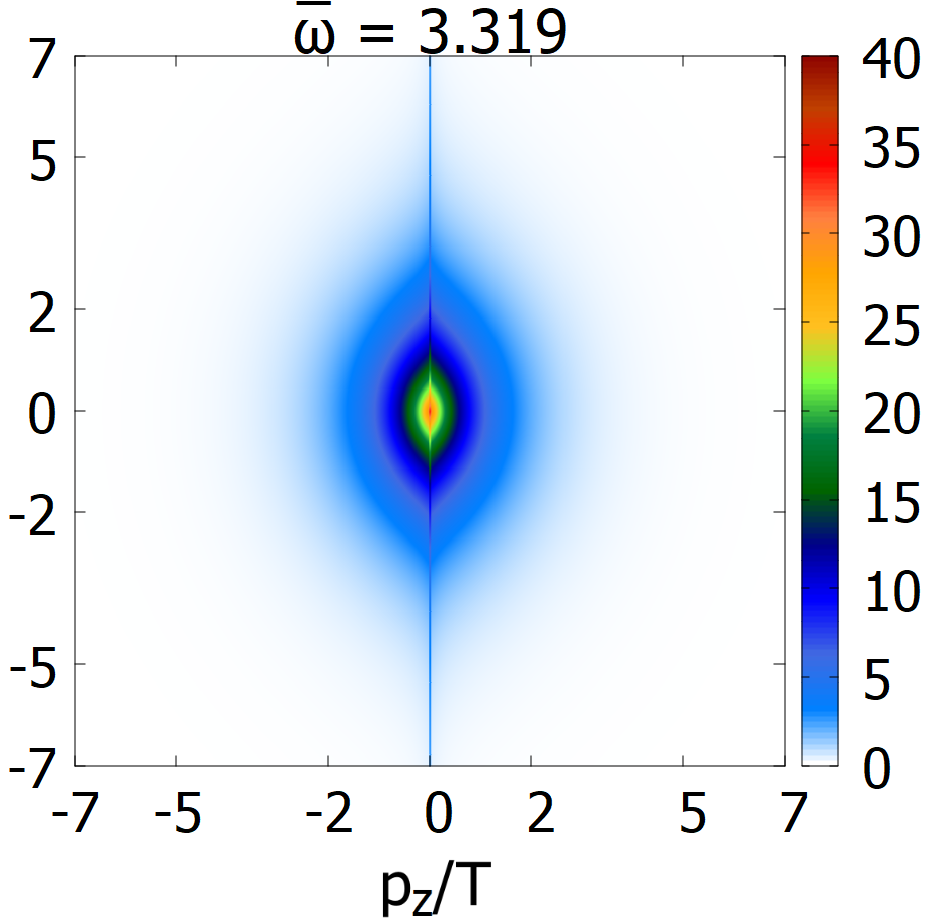}  
    \caption{{The same as in Fig.~\ref{f_gq_attr}, but for a chemically equilibrated plasma, with $\gamma_{q,0}=1$}.}
    \label{f_gq_attr1}
\end{figure}
In Figs.~\ref{f_gq_re} and~\ref{f_gq_attr1} we {address the case of a chemically equilibrated plasma, setting} $\gamma_{q,0} = 1.0$ {for the initial quark fugacity}. Fig.~\ref{f_gq_re} refers to an initial prolate distribution, with $\alpha_0 = 1.3$.  The interpretation of these snapshots is the same as for Fig.~\ref{f_gq}. The only thing that changes is the shape of the momentum distribution function, which from prolate ($P_L > P_T$) quickly becomes oblate ($P_L < P_T$) in the early stages of the evolution, see Fig.~\ref{faccia2}. This is due to the violent longitudinal expansion, which occurs at early times, causing the dominance of the free-streaming term in Eq.~(\ref{BF12}) and consequently the appearence of $p_z \sim 0$ modes. {Fig.~\ref{f_gq_attr1} refers to the attractor solution for a plasma in chemical equilibrium. Notice that in this case, at variance with Fig.~\ref{f_gq_attr}, the $p_z \sim 0$ modes in the quark distribution are clearly visible, since their initial population is not suppressed by the very small fugacity factor employed in Fig.~\ref{f_gq_attr}.}

\begin{figure}[!hbt]
    \centering
    \includegraphics[width=0.495\textwidth] {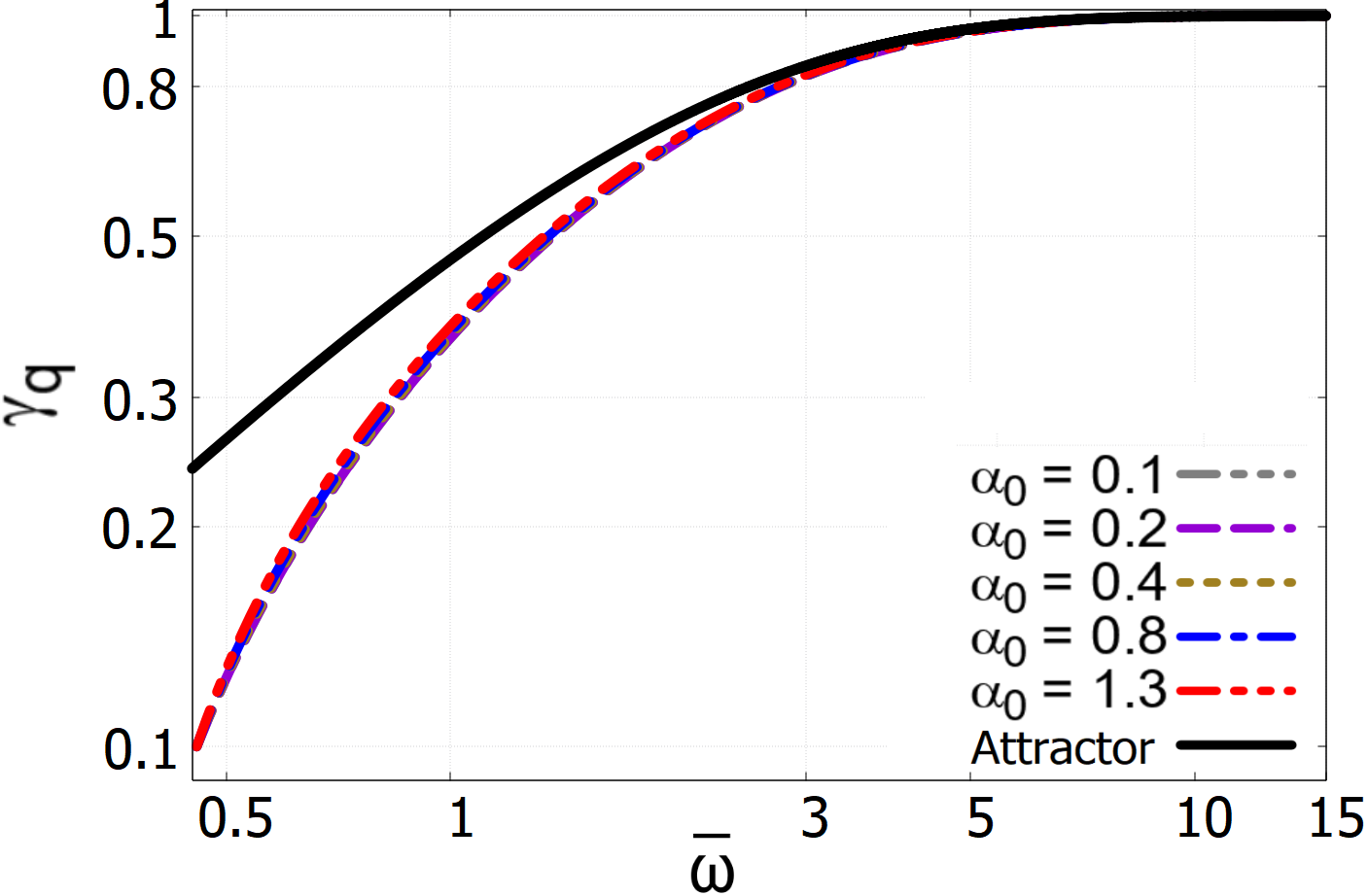}
    \includegraphics[width=0.495\textwidth] {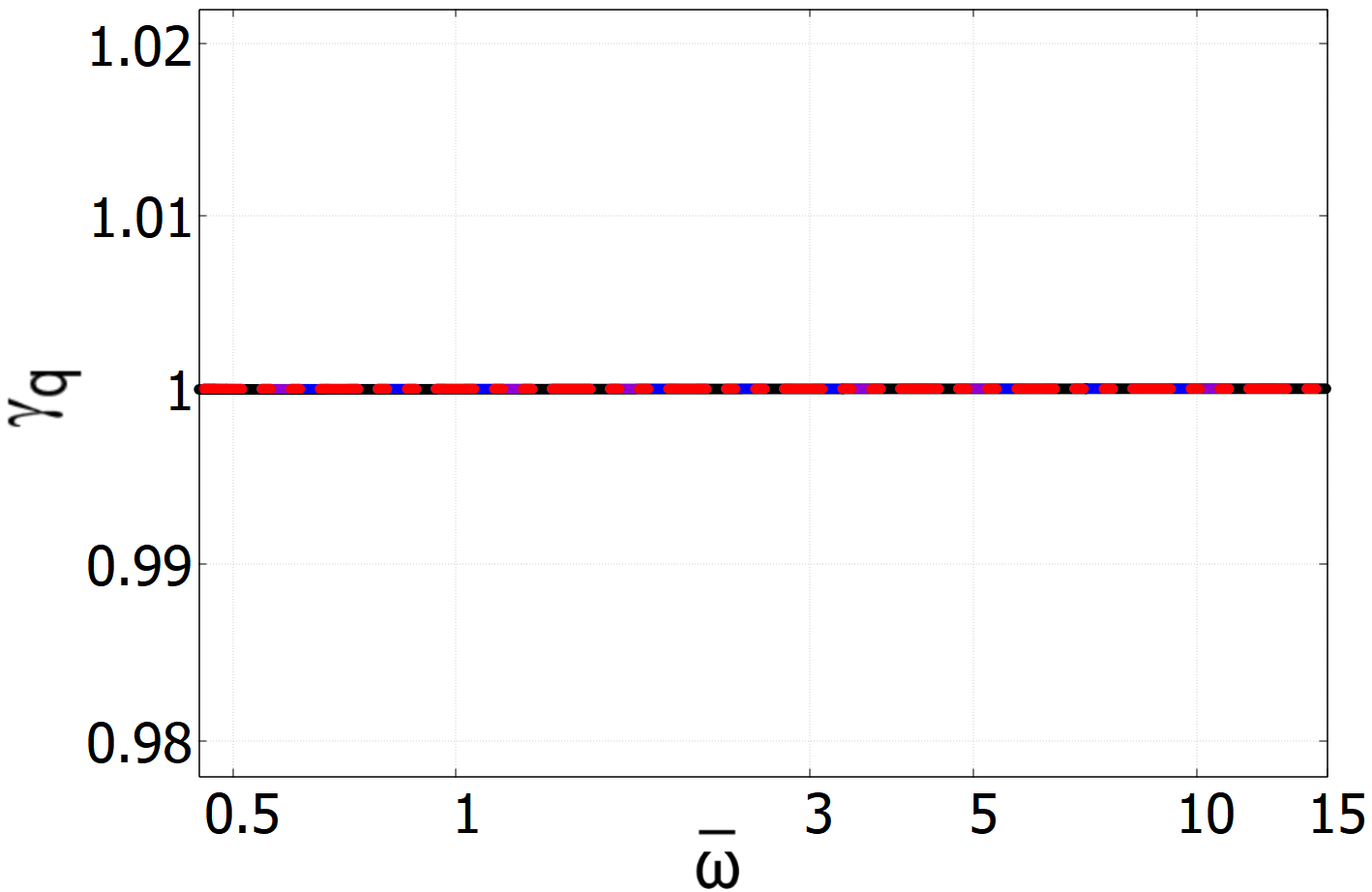}    
    \caption{{Evolution of the quark fugacity}, obtained from the exact solution of the 2RTA-BE, as a function of the scaled time $\Bar{\omega}$. {The two figures refers to a gluon-dominated ($\gamma_{q,0}=0.1$, left panel) and to a chemically equilibrated ($\gamma_{q,0}=1$, right panel) plasma at $\tau_0$, respectively}.}
    \label{dof_q}
\end{figure}
{Deviations of the quark abundance from its chemical equilibrium value, quantified by Eq.~(\ref{gamma_q}) for the fugacity, are displayed in Fig.~\ref{dof_q}}, where we used the same parameters and initializations as in Figs.~\ref{fig:mom-late-under} and \ref{faccia2}. {As one can see, starting from an initial quark under-population (left panel)}, {as appropriate for the plasma produced in HIC's~\cite{Kurkela:2018oqw, Lappi:2006fp, Strickland:1994rf}}, the quark abundance eventually approaches its equilibrium value $\gamma_{q, {\rm eq}} = 1$ at very late times.
More quantitatively, taking as a criterium for chemical equilibration $\gamma_q=0.9$, from the left panel of Fig.~\ref{dof_q} one can see that the latter is reached at the scaled time $\overline\omega\approx 3.8$, which is of the same order as the one found for the occurrence of hydrodynamization based on the analysis of the lowest-order moments of the particle distributions (see Table~\ref{conv_tau}). This is a direct consequence of our RTA approach, where, for a given parton species, the same relaxation time describes both elastic and inelastic interactions. For a more rigorous treatment, relying on a microscopic evaluation of $2\leftrightarrow2$ and $1\leftrightarrow 2$ scattering matrix elements we refer the reader to the EKT approach~\cite{Kurkela:2018xxd}, where a longer timescale for chemical equilibration compared to hydrodynamization was found.

As already discussed for the scaled energy density $\overline{M}^{\hspace{0.05cm} 20}$, also in this case it is not clear whether the concept of an off-equilibrium attractor is relevant for the quark abundance, because the convergence of the various solutions occurs when the fugacity has almost reached its equilibrium value (see the left panel of Fig.~\ref{dof_q}). 
{On the other hand, in the right panel of Fig.~\ref{dof_q} one can see that, within our approach, if the system starts at chemical equilibrium no deviations from the latter can develop during its expansion.} 

{The behavior of $\gamma_q(\Bar{\omega})$ in the left panel of Fig.~\ref{dof_q} allows one to better understand the evolution of the scaled energy density $\overline{M}^{\hspace{0.05cm} 20}$ in Fig.~\ref{fig:mom-late-under}, referring to the case $\gamma_{q, 0} = 0.1$. Deviations of $\overline{M}^{\hspace{0.05cm} 20}$ from unity are first of all due to the quenching of the quark abundance during the early stages of the evolution of the medium. Notice, in fact, that the analogous results for the chemical-equilibrium case shown in Fig.~\ref{faccia2} are much closer to unity, since in this case deviations are only due to the $C_R$ factor appearing in the} 2RTA Landau matching condition - see Eq.~(\ref{eq:LandauMC}).
%%%%%%%%%%%%%%%%%%%%%%%%%%%%%%%%%%%%%%%%%%%%%%%%%%%%%%%%%%%%%%%%%%%%
\subsection{Evolution of the entropy density for a mixture of quarks and gluons}
%%%%%%%%%%%%%%%%%%%%%%%%%%%%%%%%%%%%%%%%%%%%%%%%%%%%%%%%%%%%%%%%%%%%
\label{entropy_num}
{In this section we address the evolution of the entropy density of our out-of-equilibrium system. This represents, first of all, a consistency check of our numerical calculations, since by definition the entropy density must asymptotically approach its equilibrium value from below as the system thermalizes; furthermore, one must have a positive rate of entropy production during the entire system evolution. At the same time, knowing how such a rate evolves is important in order to fix through the final hadron yields the initial condition of hydrodynamic calculations, usually assuming the additional entropy generated by dissipative effects during the hydrodynamic stage to be small compared to the one arising from the pre-equilibrium dynamics of the system. In our simplified kinetic approach we will be able to check how solid this assumption is and also to estimate the contribution to entropy production from the progressive population of quark modes.}

\begin{figure}[!hbt]
    \centering
    \includegraphics[width=0.495\textwidth] {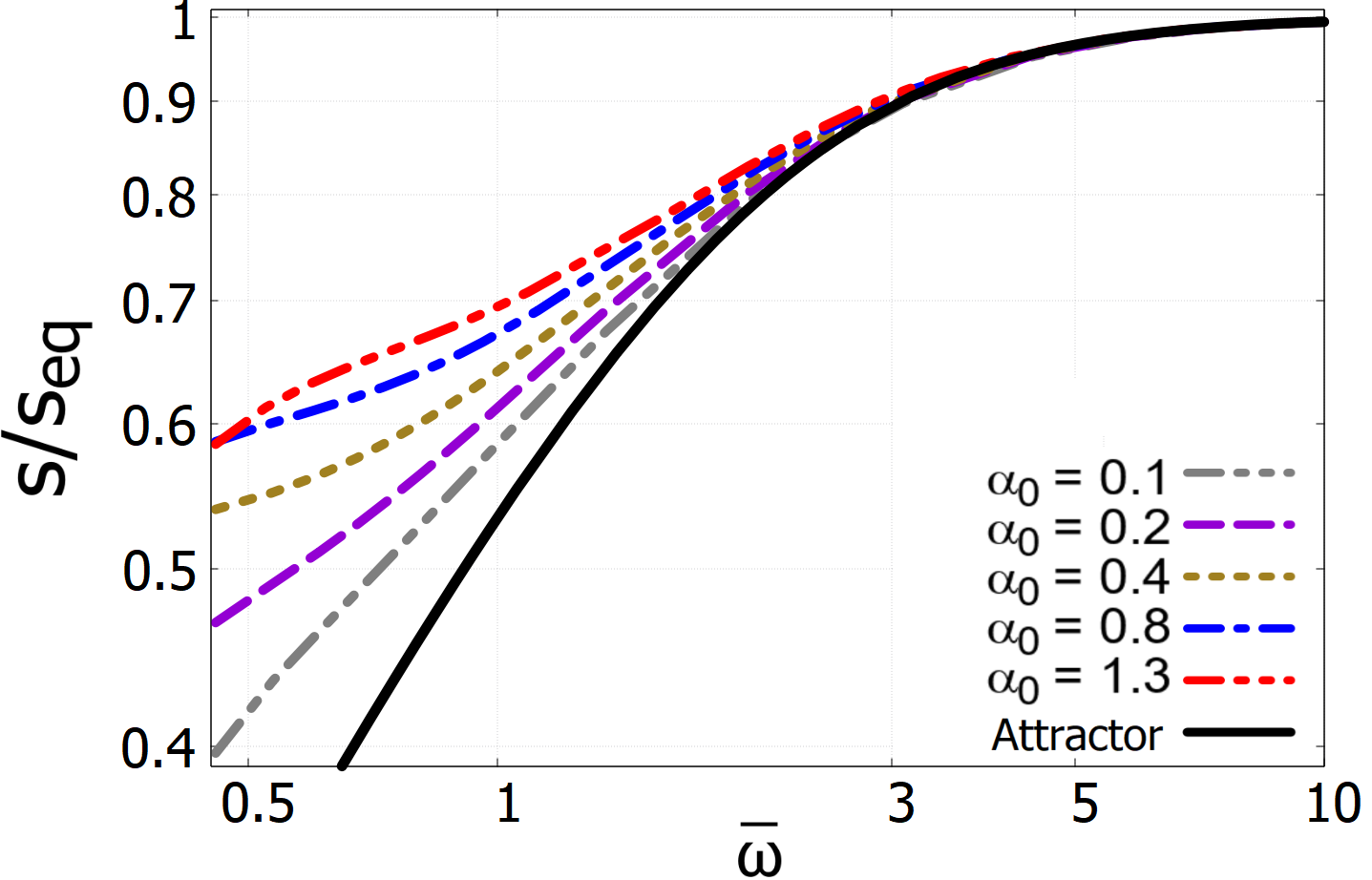}
    \includegraphics[width=0.495\textwidth] {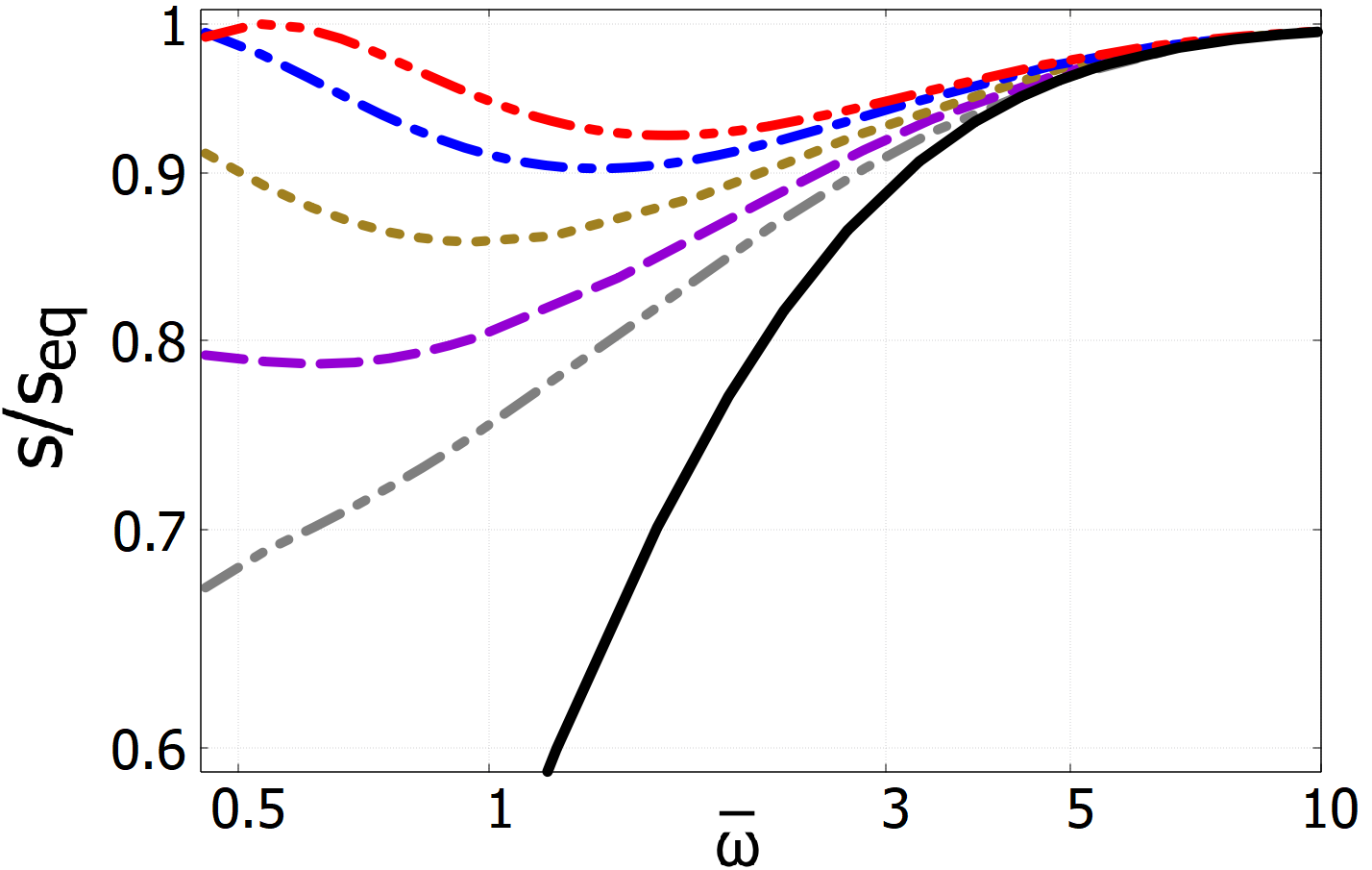}
        \caption{Late-time behavior of the scaled entropy density as a function of $\Bar{\omega}$, for $\tau_0 = 0.15$ fm/c, $T_0 = 600$ MeV and $\eta/s = 0.2$. {The left and right panels refers to the cases} $\gamma_{q, 0} = 0.1$ and $\gamma_{q, 0} = 1$, {respectively}.}
    \label{s}
\end{figure}

\begin{figure}[!hbt]
    \centering
    \includegraphics[width=0.495\textwidth] {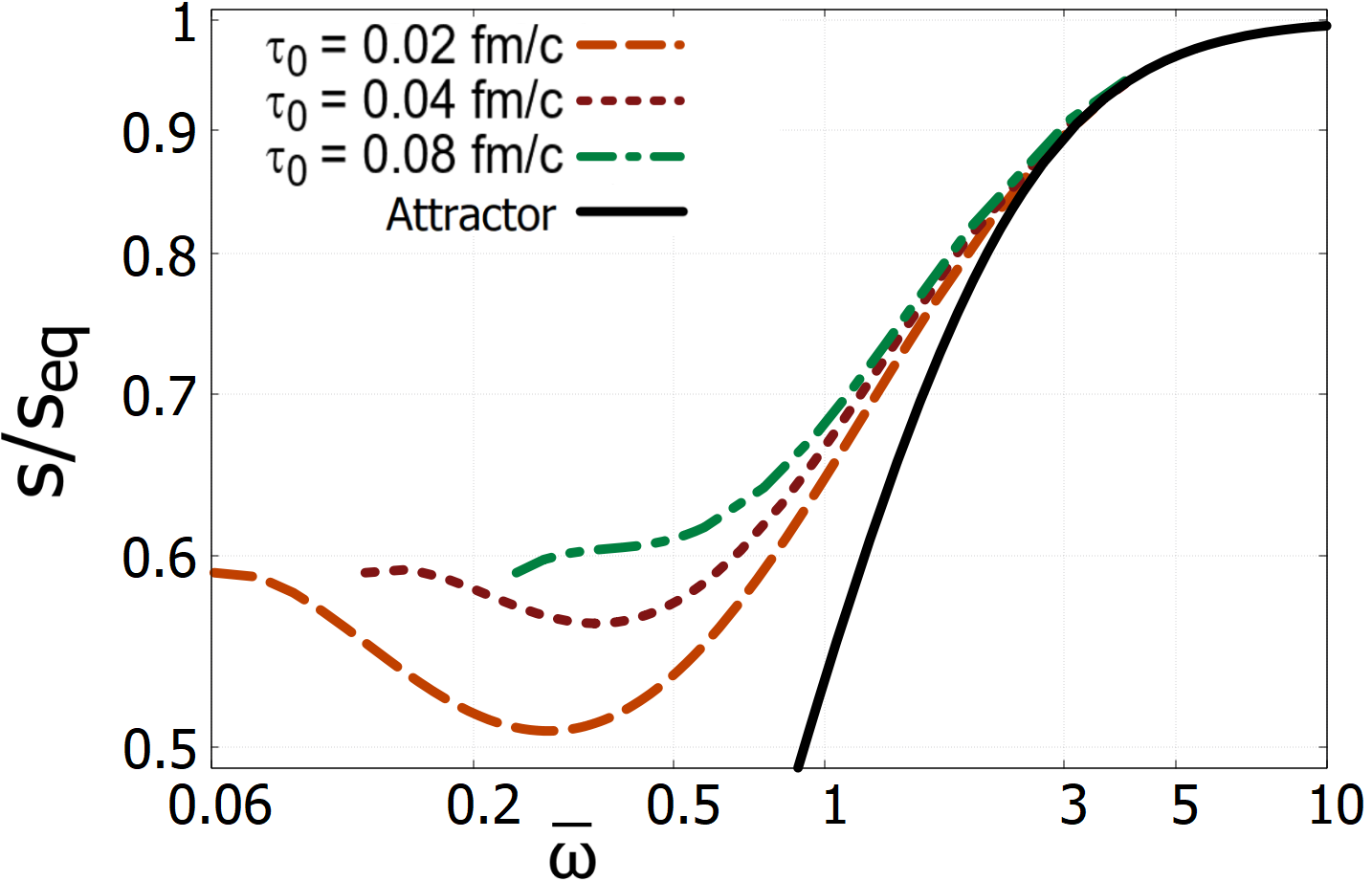}
    \includegraphics[width=0.495\textwidth] {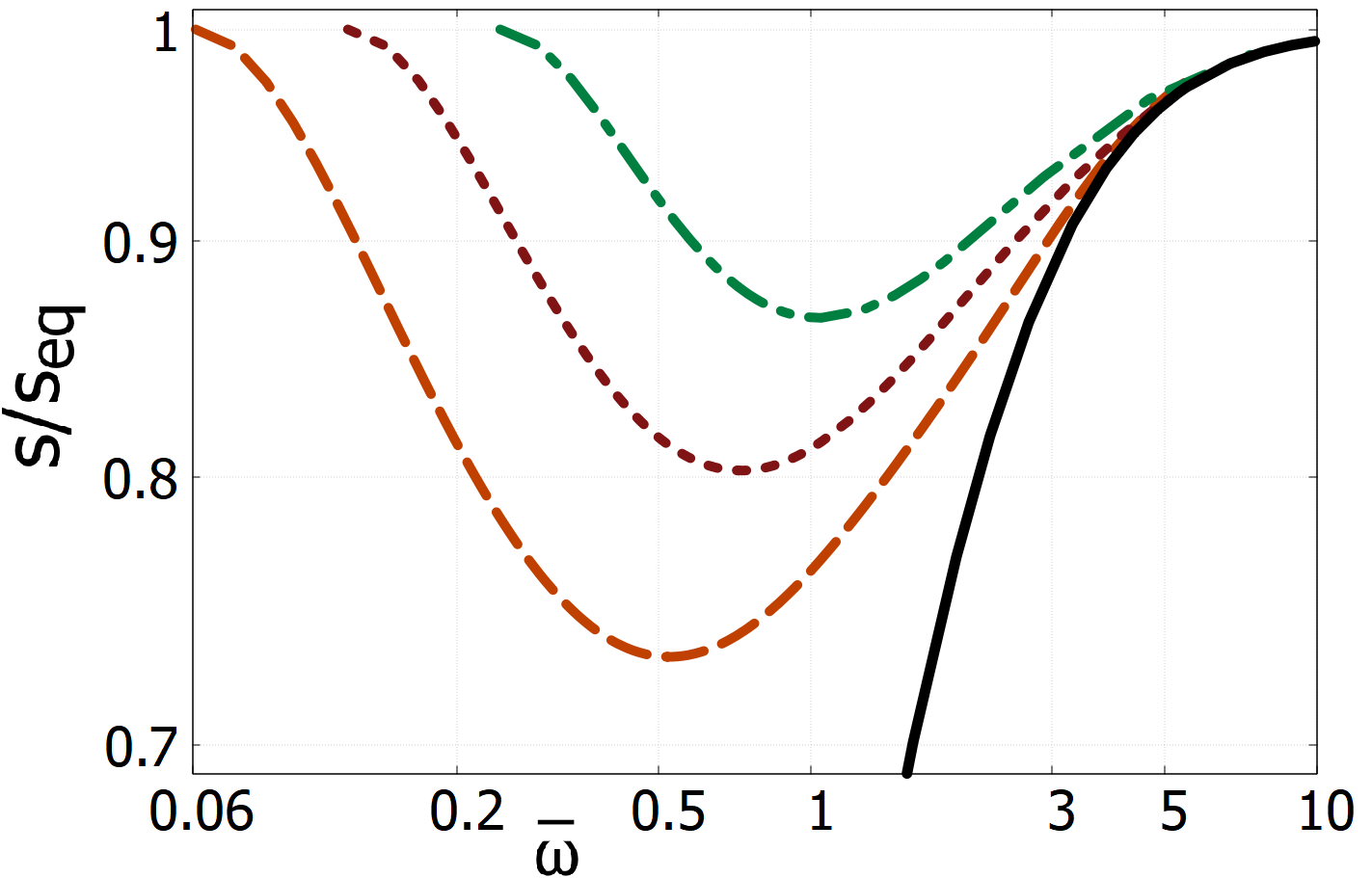}    
    \caption{Early-time behavior of the scaled entropy density as a function of $\Bar{\omega}$, for $\alpha_0 = 1.0$, $T_0 = 600$ MeV and $\eta/s = 0.2$. {The left and right panels refers to the cases} $\gamma_{q, 0} = 0.1$ and $\gamma_{q, 0} = 1$, {respectively}.}
    \label{s_re}
\end{figure}

{We first address the evolution of} the scaled entropy density $s/s_{\rm eq}$ in terms of the dimensionless time $\Bar{\omega} \equiv \tau/\tau_{\rm eq}$, {checking, as usual, the possible presence of a} late- and early-time attractor. {To initialize the system in the different cases, we employ the same set of parameters} as for the scaled moments $\overline{M}^{\hspace{0.05cm} nm}$ in Sec.~\ref{mom_num}. In order to {determine} $s(\tau)$, we compute the integral in Eq.~(\ref{entropy}) by {dividing the integration domain in several sub-intervals within which we} adopt a gaussian quadrature rule (9 gaussian points for each direction in momentum space). To obtain the attractor solution, {which displays an} extremely high anisotropy at early times (see Figs.~\ref{f_gq_attr} and~\ref{f_gq_attr1}), a Simpson algorithm is preferred. Notice, {first of all}, that in all curves in Figs.~\ref{s} and~\ref{s_re} {the $s/s_{\rm eq}$ ratio approaches unity from below} at late times. This is consistent with the fact that {the equilibrium particle distribution is the one maximizing entropy}.

In Fig.~\ref{s}, {the existence of a} forward attractor {for $s/s_{\rm eq}$ is scrutinized comparing the late-time behavior of the various solutions of the 2RTA-BE corresponding to the initial anisotropy coefficient} $\alpha_0 = \{0.1,0.2, 0.4, 0.8, 1.3\}$. {As one can see, the various solutions converge to a universal curve before the limit $s=s_{\rm eq}$ is reached.}

Moreover, in Fig.~\ref{s_re}, the early-time behavior is also addressed by varying the initial proper time $\tau_0 = \{0.02, 0.04, 0.08\}$ fm/c while keeping $\alpha_0 = 1$ fixed, {corresponding to an initial isotropic momentum distribution}. As can be observed, the scaled entropy density $s/s_{\rm eq}$ does not possess a pullback attractor, exactly as what happened for the moments with $m = 0$ (see Sec.~\ref{mom_num}).
{This can be undestood recalling that in the definition of the entropy density - see Eq.~(\ref{entropy}) - powers of $p_z$ do not appear factorized. These powers of the longitudinal momentum  would crucial to suppress the effect of the squeezed $p_z \sim 0$ modes, getting a faster approach to the attractor solution~\cite{Strickland:2018ayk, Strickland:2017kux, Florkowski:2013lya}.} This fact explains why the convergence to the forward attractor is quite slow also for scaled entropy density.
{To summarize}, the scaled entropy density is characterized only by a universal late-time attractor, while showing a complex behavior in the early stages, depending on {the system initialization}.

\begin{figure}[!hbt]
    \centering 
    \includegraphics[width=0.45\textwidth] {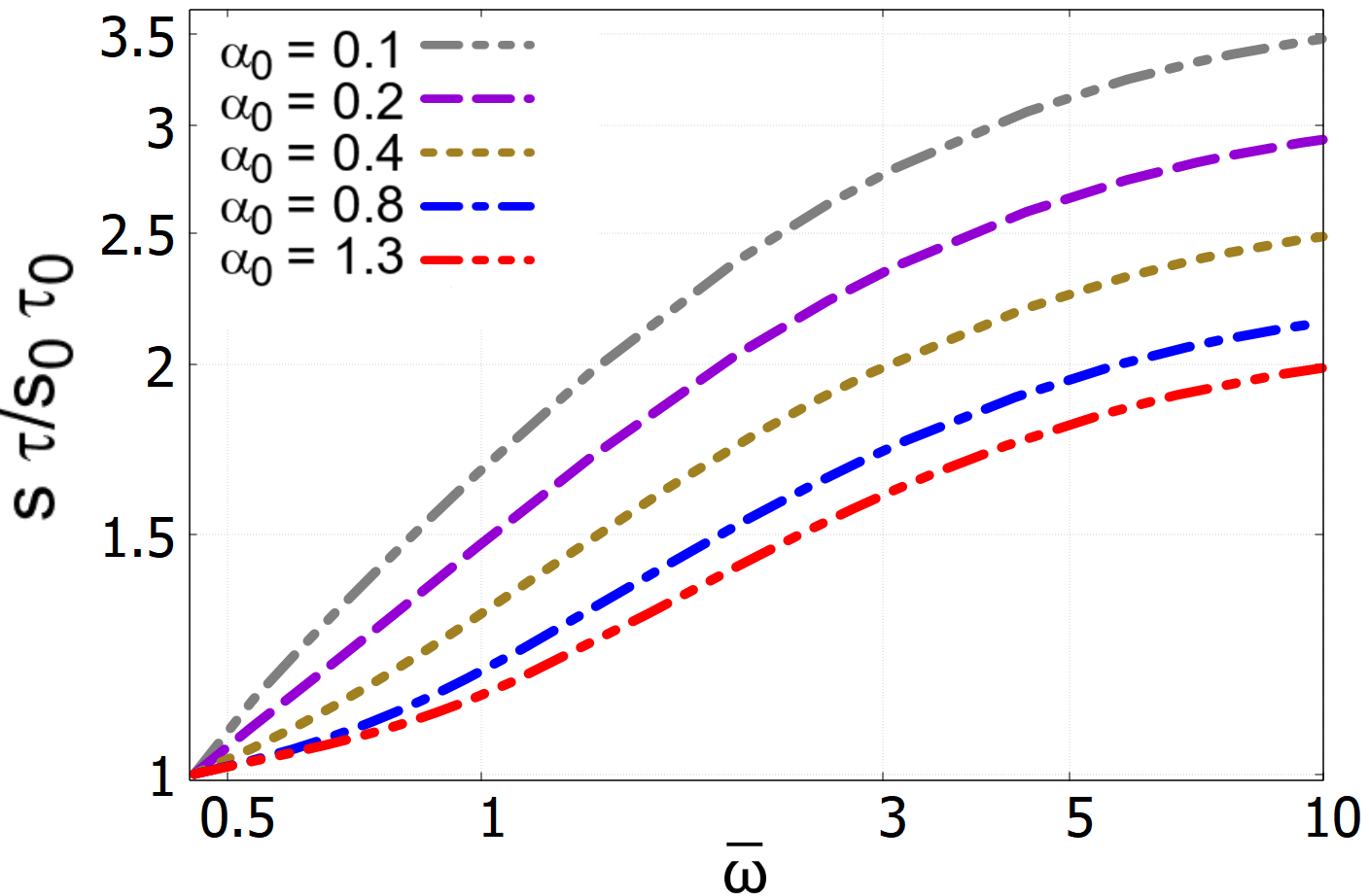}
    \includegraphics[width=0.45\textwidth] {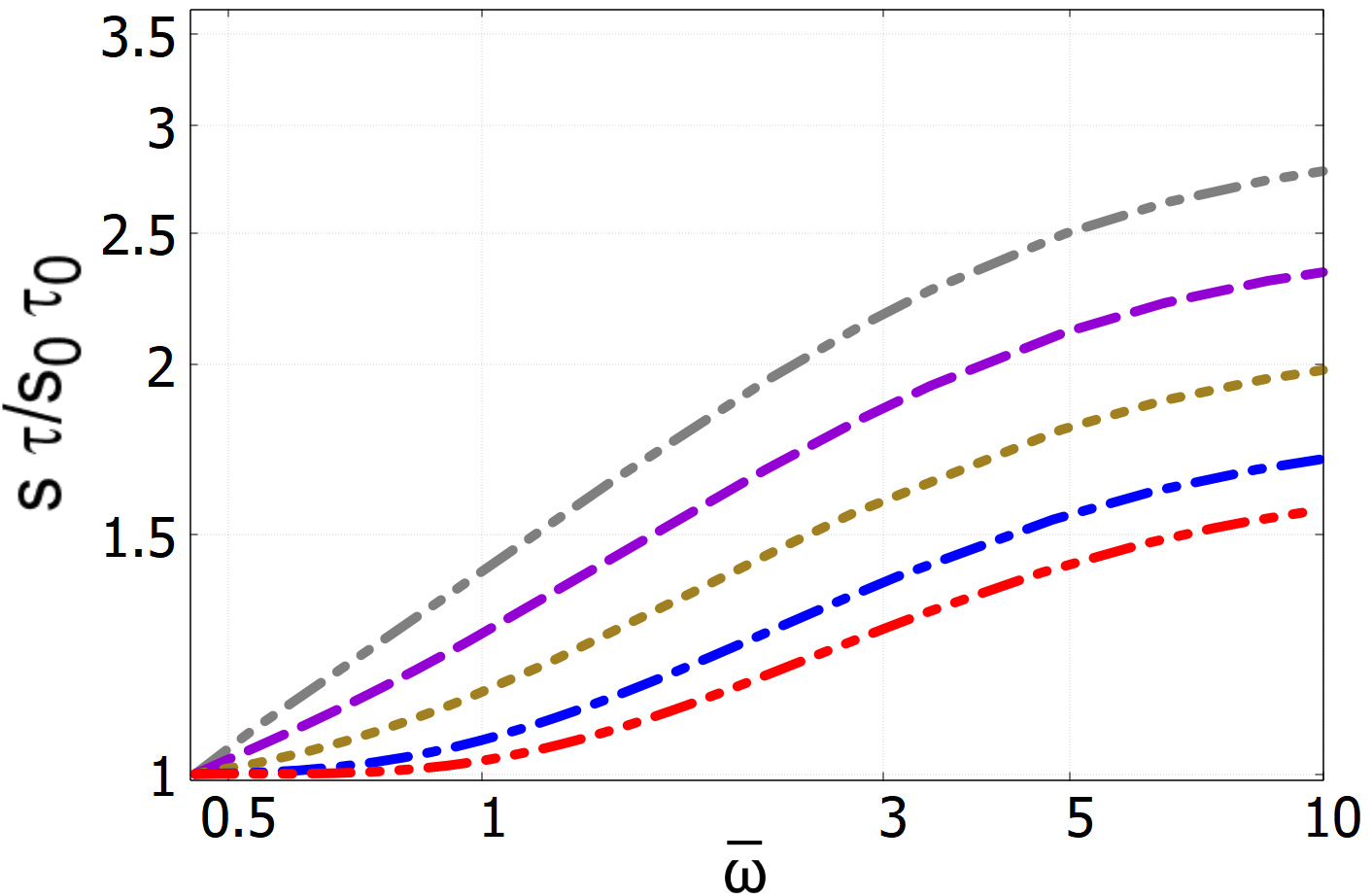}    \caption{Entropy production {displayed as a function} of the scaled time $\Bar{\omega}$. Here we set $\tau_0 = 0.15$ fm/c, $T_0 = 600$ MeV and $\eta/s = 0.2$. {The left and right panels refer to the cases} $\gamma_{q, 0} = 0.1$ and $\gamma_{q, 0} = 1$, {respectively}.} 
    \label{s_prod}
\end{figure}
{We now address the important issue of entropy production. This is done taking into account the ratio
\begin{equation}
\frac{s\!\cdot\!\tau}{s_0\!\cdot\!\tau_0}\,,\label{eq:stau}    
\end{equation}
which allows one to quantify entropy production with respect to the case of an ideal fluid, where} the entropy density simply gets diluted due to the longitudinal expansion {of the system, leading to
\begin{equation*}
s_{\rm id}(\tau) = s_0 \hspace{0.07cm} \frac{\tau_0}{\tau}\,.   
\end{equation*}}
Because of the dissipative {processes} occurring during the evolution of the system, {the ratio in Eq.~(\ref{eq:stau})} should remain greater than 1 and display a monotonous increase. {For a conformal fluid}, these (irreversible) processes are {due to} shear viscosity and, in the case of an initially gluon-dominated plasma, to the gradual population of quark modes.
From Fig.~\ref{s_prod} it is evident that the more the initial momentum-anisotropy parameter $\alpha_0$ differs from 1 ({corresponding to an isotropic distribution}) the greater the {amount of entropy production is,} as expected. A highly anisotropic momentum distribution corresponds, in fact, to a more ordered (hence with lower initial entropy) configuration than the isotropic case towards which the system evolution is driven. {Notice also that, in the case of an initial quark under-population (left panel), the amount of produced entropy is larger than in a chemically equilibrated plasma (right panel). Having most of the energy carried by gluons rather than equipartitioned among all degrees of freedom also corresponds to a more ordered configuration. As the system gets more isotropic and the quark fugacity closer to unity the rate of entropy production decreases, as can be seen from the flattening of the slope of the curves in Fig.~\ref{s_prod}.}

Note that, {in our kinetic calculations, we have taken an initial time $\tau_0=0.15$ fm/c much smaller than the one at which hydrodynamic codes are usually initialized and that} most of the entropy production takes place when the system is in the pre-hydrodynamic stage. Thus, when $\Bar{\omega} \gtrsim 5$ (for $\gamma_{q, 0} = 0.1$) or $\Bar{\omega} \gtrsim 7$ (for $\gamma_{q, 0} = 1$) -- value at which the hydrodynamization (see Fig.~\ref{s}) is expected to occur for $s/s_{\rm eq}$ -- the rate of entropy production is very small. This finding is actually in good agreement with the results presented in Ref.~\cite{Kurkela:2018vqr}, which, however, were derived in the context of EKT~\cite{Arnold:2002zm, Kurkela:2015qoa}.

\begin{figure}[!hbt]
    \centering 
    \includegraphics[width=0.55\textwidth] {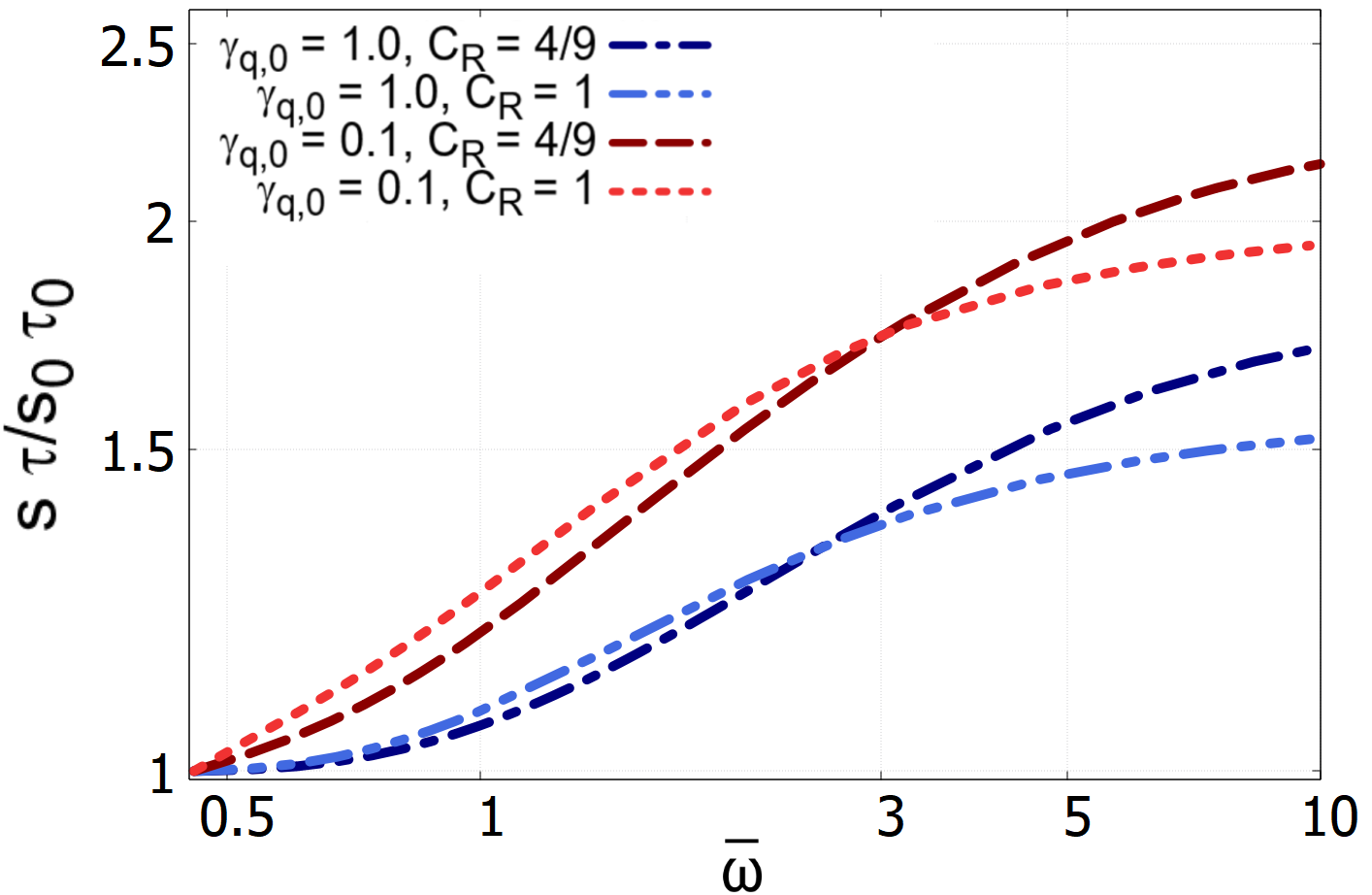}    
    \caption{{Dependence of the entropy production on the initial deviation from chemical equilibrium ($\gamma_{q,0}=0.1$ vs $\gamma_{q,0}=1$) and on the different relaxation time of quarks and gluons ($C_R=4/9$ vs $C_R=1$)}. The depicted curves correspond to an initial anisotropy coefficient $\alpha_0 = 0.8$.}
    \label{comp}
\end{figure}
Finally, in Fig.~\ref{comp}, we can appreciate how much the entropy production is influenced by the initial fugacity parameter $\gamma_{q, 0}$, {quantifying the deviation of quarks from chemical equilibrium}, and by the value assigned to the ratio $C_R$ {defined in Eq.~(\ref{rta17})}, {quantifying the different relaxation times of quarks and gluons ($C_R=1$ corresponding to a unique relaxation time).}
We note, from Fig.~\ref{comp}, that having two relaxation times results only in a relatively small increase of the entropy production during the longitudinal expansion ($+ \, 15 \%$, at the end of the simulation), while it is evident that {there is a much stronger dependence} on the initial abundance of quarks ($+ \,40 \%$ at $\Bar{\omega} = 10$ in the case of an initial quark under-population).

%%%%%%%%%%%%%%%%%%%%%%%%%%%%%%%%%%%%%%%%%%%%%%%%%%%%%%%%%%%%%
\section{Conclusions and outlook}
\label{conclusions}
%%%%%%%%%%%%%%%%%%%%%%%%%%%%%%%%%%%%%%%%%%%%%%%%%%%%%%%%%%%%%
In the present work we studied the (0+1)D evolution of a system of quarks and gluons undergoing a purely longitudinal boost-invariant expansion (Bjorken flow). In doing so we accounted for the different relaxation times of the two species appearing in the Boltzmann equations for the respective single-particle distributions.

One of the motivations of providing a description based on kinetic theory of the first stages of the medium evolution in HIC's is to employ a tool capable of addressing a regime -- unaccessible to hydrodynamics -- in which the expansion rate of the system is greater than the interaction rate among its constituents, here assumed to behave as on-shell particles.

At the same time, in order to provide phenomenological predictions for the realistic case of a (3+1)D expansion (furthermore, valid also in a strongly-coupled regime) with a more limited allocation of computing resources, relativistic hydrodynamics (embedding dissipative corrections) represents a much more convenient tool, which was found to work unreasonably well even at times $\lesssim 1$ fm/c and for very small systems like the ones produced in proton-proton collisions. These are situations in which neither the inverse Reynolds number -- see Eq.~(\ref{eq:Reynolds}) -- nor the Knudsen number ($\propto 1/\overline{\omega}$) are small (dissipative corrections to the energy-momentum tensor are large, the particle interaction-rate can be comparable to the expansion rate of the system and the mean-free-path comparable to the system size), lying beyond what one would think to be the limit of applicability of hydrodynamics. 

A major recent achievement sheding light on the above issue was the discovery of the attractor behavior displayed by the solutions of hydrodynamic~\cite{Heller:2015dha, Strickland:2017kux} -- within different approximation schemes -- and kinetic-theory~\cite{Strickland:2018ayk, Florkowski:2013lya} equations: different initializations of the system, after a transient phase, can lead to a universal behavior well before pressure isotropy and local thermal equilibrium are achieved. This process is known as ``hydrodynamization" and, so far, had been mainly investigated in the case of a single relaxation time. However, the medium formed in HIC's is made of particles characterized by different interaction cross-sections due to their different color charge and, furthermore, in the early stages its constituents are neither in kinetic nor in {chemical} equilibrium, the system being initially gluon dominated~\cite{Strickland:1994rf, Kurkela:2018xxd, Kurkela:2018oqw}. This led us to develop a simplified setup capable however of dealing with this situation: the description of the system evolution was still based on the relativistic Boltzmann equation within the relaxation-time approximation, but different relaxation times for quarks and gluons were employed (2RTA-BE) and a possible departure of the system from chemical equilibrium, quantified by the quark fugacity $\gamma_q$ was allowed. The final goal was clearly to verify the existence of a universal behavior of the system also in this more general, but closer to reality, situation.

To accomplish the above task, as commonly done in the literature, we studied the moments of the single-particle phase-space distributions normalized to their equilibrium value, as defined in Eq.~(\ref{eq:red-mom}). Similarly to the 1RTA case, we found that all moments converge to an attractor solution at late times and this occurs {before} the system reaches isotropy and thermalization in its local rest frame. This finding provides further support to the idea of a quite fast hydrodynamization of the system, also in the more realistic case of a medium made of particles with different interaction rates.  The main effect of having two different interaction rates is twofold. First of all, due to the non-trivial Landau matching condition~\cite{Florkowski:2012as} in Eq.~(\ref{eq:LandauMC}), the $\overline{M}^{\hspace{0.05cm} 20}$ scaled moment, associated to the energy density, is no longer identically 1. Secondly, the isotropization of the system is retarded compared to the 1RTA case, as can be observed in Fig.~\ref{1rta}, because quarks relax more slowly than gluons.

Concerning the early-time dynamics, as in the situation of one common relaxation time, only the scaled moments with $m > 0$, involving positive even powers $p_z^{2m}$ of the longitudinal momentum, were found to display a universal behavior even at early times (pullback attractor). This has to be attributed to the high population of $p_z\sim 0$ modes arising from the fast longitudinal expansion of the system.
In particular, no early-time attractor for the scaled energy density given by $\overline{M}^{\hspace{0.05cm} 20}$, no longer fixed by the Landau matching condition as in the 1RTA case, was found. Accordingly, within the 2RTA setup, the scaled viscous correction $\overline{\phi}$ to the energy-momentum tensor becomes independent of its initial condition only at late times, but not during the pre-hydrodynamic regime, as can be noticed in Fig.~\ref{shear}. This is due to the absence of an early-time attractor for the $\overline{M}^{\hspace{0.05cm} 20}$ moment, which enters into the definition of $\overline{\phi}$ in Eq.~(\ref{visc}). On the contrary, for a unique relaxation time $\overline{M}^{\hspace{0.05cm} 20}\!\equiv\! 1$, hence $\overline{\phi}$ displays a universal dynamics also at early times: the rapid expansion makes the system forget about the initial conditions well before collisions can play a major role.

Beside scrutinizing the existence of a late and early-time attractor for the various scaled moments of the particle phase-space distributions we also addressed the issue of the chemical composition of the system, in which quark modes are initially extremely underpopulated, their deviation from the chemical-equilibrium abundance being quantified by the fugacity factor $\gamma_q<1$ -- see Eq.~(\ref{gamma_q}). We found that the system approaches chemical equilibrium within approximately the same time-scale necessary for the hydrodynamization of the lowest-order moments ($m = 0$), as can be observed from a comparison between Fig.~\ref{dof_q} and the left panel of Table~\ref{conv_tau}.
%{\em This is not perfectly in agreement with the main result of Ref.~\cite{Kurkela:2018xxd}, claiming that, through an analysis of the energy density ($\overline{M}^{\hspace{0.05cm} 20}$ in our notations), hydrodynamization takes place faster than chemical equilibration. However, it is not surprising that kinetic theory in RTA cannot lead exactly to the same statement as that of Ref.~\cite{Kurkela:2018xxd}, since the latter is based on EKT, which represents a more realistic model to address chemical equilibration in strongly-interacting systems, whereas our approach, despite requiring less computational resources, is actually suitable for describing only weakly-interacting systems. Having clarified this point, our finding in 2RTA is still a step forward compared to the case of one single relaxation time for quarks and gluons, because in this situation the procedure followed in Ref.~\cite{Kurkela:2018xxd} is not even well defined, due to the standard Landau matching condition $\varepsilon = \varepsilon_{\rm eq}$, which implies a completely flat profile for the scaled energy density that compromises the analysis mentioned above, as shown in Refs.~\cite{Strickland:2018ayk,Strickland:2019hff,Florkowski:2013lya}, for instance.}

Both out-of and in chemical equilibrium the scaled entropy density $s/s_{\rm eq}$ was found to display a late-time attractor, but not a universal early-time dynamics. We found a sizable entropy production during the system's evolution, but mostly concentrated in the early stages, before hydrodynamization (see Fig.~\ref{s_prod}). Notice that in hydrodynamic calculations the final hadron multiplicity is generally used as a proxy for the initial entropy of the fireball at the beginning of its fluid dynamic expansion, assuming that subsequent dissipative effects provide only a minor correction in comparison to the initial parton production: this is confirmed by our analysis, where we found that, after hydrodynamization, the entropy of the mixture increases at most by 20 percent. As expected, entropy production turned out to be larger when starting from a more ordered configuration corresponding to a larger momentum-space anisotropy and/or quark under-population. This should be taken into account when using final hadronic observables to fix the initial condition of the system.

To summarize, also in the case of two species of particles the system displays a universal late-time dynamics, although the different relaxation time of quarks and gluons makes the approach to full equilibrium slower. However, for some of the observables considered in this paper (e.g. the inverse Reynolds number and the quark abundance), the presence of species with different interaction cross-sections prevents the development of an early-time attractor, the medium dynamics remaining sensitive to the initialization of the system for longer time. Furthermore quarks, beside carrying a smaller color charge and hence interacting more weakly than gluons, are also extremely under-populated immediately after the collision of the two nuclei. The gradual population of quark modes represents an important mechanism of entropy production at work in HIC's, as shown in Fig.~\ref{comp}.   

The initial deviation of quarks from isotropy in momentum-space and chemical equilibrium can be of relevance to correctly describe observables particularly sensitive to the early stage of the fireball evolution, such as the dilepton production-rate~\cite{Martinez:2007pjh,Martinez:2008di}. We leave this study for a possible future publication, together with the comparison of our 2RTA-BE results with viscous and anisotropic hydrodynamic calculations generalized to the case of a fluid mixture with two different relaxation times~\cite{Florkowski:2012as,Fotakis:2022usk,Hu:2022vph}. 

\section*{Acknowledgements}
F.F. and A.B. acknowledge financial support by MUR within the Prin$\_$2022sm5yas project.

%%%%%%%%%%%%%%%%%%%%%%%%%%%%%%%%%%%%%%%%%%%%%%%%%%%%%%%%%%%%%%%%
\appendix
%%%%%%%%%%%%%%%%%%%%%%%%%%%%%%%%%%%%%%%%%%%%%%%%%%%%%%%%%%%%%%%%

%%%%%%%%%%%%%%%%%%%%%%%%%%%%%%%%%%%%%%%%%%%%%%%%%%%%%%%%%%%%%
\section{Kinetic equations in RTA for a Bjorken-expanding fluid}
\label{Milne}
%%%%%%%%%%%%%%%%%%%%%%%%%%%%%%%%%%%%%%%%%%%%%%%%%%%%%%%%%%%%%
{The} ($0 + 1$)D Bjorken expansion {represents the easiest non-trivial flow field used in the literature to describe the evolution of the matter produced in HIC's~\cite{Bjorken:1982qr, Simeoni:2022hjh, Romatschke:2009im}. For the sake of consistency, here we summarize the main equations necessary to derive the results presented in the body of the text. In Minkowski coordinates, with metric tensor} $\eta_{\mu \nu} = \text{diag}(+1, -1, -1, -1)$, the fluid four-velocity is given by:  
\begin{equation}
    u^{\mu} \equiv \left( \frac{t}{\tau}, 0, 0, \frac{z}{\tau}  \right)^T \,.\label{eq:umu}
\end{equation}
{All scalar functions only depends}
on the longitudinal proper time $\tau \equiv \sqrt{t^2 - z^2}$. {The spacetime rapidity
\begin{equation}
    \zeta \equiv \arctanh \left( \frac{z}{t} \right) \,,
\end{equation}
is also introduced.
The inverse transformations are provided by
\begin{equation}
    t = \tau \hspace{0.07cm} \cosh \zeta \hspace{0.3cm},\hspace{0.3cm} z = \tau \hspace{0.07cm} \sinh \zeta\label{eq:inverse}
\end{equation}
and allow one to rewrite the fluid four-velocity as 
\begin{equation}
    u^\mu=\left(\cosh\zeta,0,0,\sinh\zeta\right)^T\,.
\end{equation}
Hence, the matrix (unessential transverse coordinates are neglected)
\begin{equation}
    \begin{pmatrix}
        \cosh\zeta &-\sinh\zeta\\
        -\sinh\zeta &\cosh\zeta
    \end{pmatrix}\label{eq:boost}    
\end{equation}
allows one to perform the boost to the LRF of the fluid, where $u^\mu_{\rm LRF}=(1,0,0,0)^T$. Consistently with Eqs.~(\ref{eq:umu}) and~(\ref{eq:inverse}) this corresponds to the spacetime point $t'=\tau$ and $z'=0$.
}

{Let us now consider the on-shell phase-space distribution $f(x,\vec p)$, which is more conveniently written in terms of the spacetime and particle rapidities $\zeta$ and $y$ rapidities:
\begin{equation}
\label{func}
    f(t,\vec x_T,z;\vec p_T, p_z)=f(\tau\cosh\zeta,\tau\sinh\zeta;p_T,p_T\sinh y)\,,
\end{equation}
where the homogeneity and isotropy of the system in the transverse plane is exploited and we focus on the case of massless particles. In Eq.~(\ref{func}), the particle-rapidity is defined as
\begin{equation}
 y \equiv \frac{1}{2}\,\ln{\left(\frac{E + p_z}{E - p_z}\right)}\,.   
\end{equation}
The phase-space distribution is a Lorentz scalar, thus in the case of a Bjorken expansion its functional form must depend only on the longitudinal proper-time $\tau$. This corresponds to perform a boost of rapidity $-\zeta$ to the origin. Hence, eventually, one is left with the following functional dependence
\begin{equation}
    f=f(\tau;p_T,p_T\sinh(y-\zeta))\,.
\end{equation}
The dependence on the longitudinal coordinate is more conveniently expressed through the boost-invariant variable
\begin{equation}
    \label{BF8}
    w \equiv t \hspace{0.07cm} p_z - z \hspace{0.07cm} E=\tau\, p_T\sinh(y-\zeta)\,.
\end{equation}
}{The requirement of longitudinal boost-invariance implies then that the phase-space distribution $f(x;\vec p)$ should be considered a function of only the three independent variables $\tau$ , $w$ and ${p}_T$~\cite{Strickland:2018ayk, Soloviev:2021lhs}, eventually getting
\begin{equation}
    f=f(\tau;w,p_T)\,.
\end{equation}
In order to perform the phase-space integration it is useful to introduced a second boost-invariant quantity}~\cite{Bialas:1984wv, Bialas:1987en}
\begin{equation}
\label{BF9}
    v(\tau, w, p_T) \equiv E \hspace{0.07cm} t - z \hspace{0.07cm} p_z = \sqrt{w^2 + p_T^2 \hspace{0.07cm} \tau^2}=\tau \,p_T\cosh(y-\zeta)\,,
\end{equation}
which holds for a conformal system of massless particles.
{In terms of these new variables the invariant integration measure over the particle momentum can be written as}~\cite{Florkowski:2013lya}:
\begin{equation}
    \label{BF10}
    d \chi \equiv \frac{d p_z \hspace{0.07cm} d^2 p_T}{(2 \pi)^3 \hspace{0.07cm} E} = \frac{d w \hspace{0.07cm} d^2 p_T}{(2 \pi)^3 \hspace{0.07cm} v}\,.
\end{equation}
{Notice that in defining the boost-invariant moments of the phase-space distribution in Eq.~(\ref{robo}) another four-vector
\begin{equation}
    z^{\mu} \equiv \left( \frac{z}{\tau}, 0, 0, \frac{t}{\tau} \right)^T=(\sinh\zeta,0,0,\cosh\zeta)^T \,.
\end{equation}
is introduced. The latter is a space-like vector ($z^2=-1$) and corresponds to the $z$-axis in the LRF of the fluid, $z^\mu_{\rm LRF}=(0,0,0,1)^T$~\cite{Soloviev:2021lhs, Strickland:2018ayk}.}

{In the case of a (0+1)D Bjorken expansion, using the above variables, the} Boltzmann equation in Eq.~(\ref{rta5}) {assumes a much simpler form, since all the spacetime dependence enters through the longitudinal proper-time $\tau$. Thus, one has
\begin{equation}
   p^{\mu}\partial_{\mu} = E\,\frac{\partial\tau}{\partial t}\frac{\partial}{\partial\tau}+p_z\, \frac{\partial\tau}{\partial z}\frac{\partial}{\partial\tau}= \frac{v}{\tau}\,\frac{\partial}{\partial\tau}
\end{equation}
and, with the flow-field in Eq.~(\ref{eq:umu})},
\begin{equation*}
    p \cdot u = \frac{v}{\tau}\,. 
\end{equation*}
One finally obtains~\cite{Denicol:2014xca}:
\begin{equation}
\label{BF11}
    \partial_{\tau} f_a = -\frac{f_a-f_{{\rm eq}, a}}{\tau_{{\rm eq}, a}} \,,
\end{equation}
where $f_a = f_a(\tau; w, {p}_T)$ and the index $a$ refers to the particle species. In Sec.~\ref{ex_2rta} we implicitly assumed that $f_a$ is an even function of $w$, namely $f_a (\tau; w, p_T) \equiv f_a (\tau; - w, p_T)$~\cite{Strickland:2018ayk}.

%%%%%%%%%%%%%%%%%%%%%%%%%%%%%%%%%%%%%%%%%%%%%%%%%%%%%%%%%%%%%
\section{Proof of the integral equation for general moments}
\label{strick_proof}
%%%%%%%%%%%%%%%%%%%%%%%%%%%%%%%%%%%%%%%%%%%%%%%%%%%%%%%%%%%%%

Also in the case of two different relaxation times for quarks and gluons ({in the following} labelled with the index $a$), in the ($0 + 1$)D Bjorken-flow regime, we are able to find an exact formal solution of the BE, which reads:
\begin{equation}
\label{exact}
    f_a(\tau; w, p_T) = D_a(\tau, \tau_0) \hspace{0.07cm} f_{0, a}(w, p_T) + \int_{\tau_0}^{\tau} \frac{d \tau'}{\tau_{{\rm eq}, a}(\tau')} \hspace{0.07cm} D_a(\tau, \tau') \hspace{0.07cm} f_{{\rm eq}, a}(\tau'; w, p_T) \,.
\end{equation}
Essentially, it has the same mathematical form of the exact solution for a single relaxation time (1RTA) - see Refs.~\cite{Soloviev:2021lhs, Strickland:2018ayk}.

At the beginning of the evolution, we consider for both quarks and gluons a momentum distribution of the Strickland-Romatschke form, {obtained through a deformation of an} isotropic Boltzmann distribution~\cite{Romatschke:2003ms}, i.e.
\begin{align}
\begin{split}
\label{f_0 strick}
    &f_{0, q}(w, p_T) \equiv 2 g_q \, \gamma_{q, 0}\, \exp \left[- \frac{\sqrt{(p \cdot u)^2 + \xi_0 (p \cdot z)^2}}{\Lambda_0}  \right] = 2 g_q \hspace{0.07cm} \gamma_{q, 0} \hspace{0.07cm} \exp \left[- \frac{\sqrt{(1 + \xi_0) w^2 + p_{T}^2 \hspace{0.07cm} \tau_{0}^2}}{\Lambda_0 \hspace{0.07cm} \tau_0} \right] \,, \\
    &f_{0, g}(w, p_T) \equiv g_g \hspace{0.07cm} \exp \left[- \frac{\sqrt{(p \cdot u)^2 + \xi_0 (p \cdot z)^2}}{\Lambda_0}  \right] = g_g \hspace{0.07cm} \exp \left[- \frac{\sqrt{(1 + \xi_0) w^2 + p_{T}^2 \hspace{0.07cm} \tau_{0}^2}}{\Lambda_0 \hspace{0.07cm} \tau_0} \right] \,,
\end{split}
\end{align}
where we used the relations $p \cdot  u = v/\tau$ and $p \cdot  z = -  w/\tau$ (already discussed in \ref{Milne}). The quantity $\xi_0$ is known as the initial anisotropy parameter and $\Lambda_0$ {provides the typical} transverse-momentum scale of the particles. We note that these distribution functions reduce to the equilibrium one if $\xi_0 = 0$ (isotropic fluid) and $\gamma_{q, 0} = 1$ (chemical equilibrium): {in this case} $\Lambda_0$ is equal to the initial temperature $T_0$ of the quark-gluon mixture. 

Now we recall that the general moment of the one-particle distribution function $f_a$ is defined as follows~\cite{Strickland:2018ayk}:
\begin{equation}
    M_a^{nm} \equiv \int d \chi \hspace{0.07cm} (p \cdot u)^n \hspace{0.07cm} (p \cdot z)^{2m} \hspace{0.07cm} f_a (\tau; w, p_T) \,, \,\text{ \small{with}} \,\, a = q, g \,.
    \label{eq:Mnmdef}
\end{equation}
In the above integral, the momentum-space integration measure is given {by Eq.~(\ref{BF10})}, expressed in terms of the boost-invariant variables, $w$ and $v = \sqrt{w^2 + p_{T}^2 \hspace{0.07cm} \tau^2}$ (see \ref{Milne}). This allows us to recast the general moment in Eq.~(\ref{eq:Mnmdef}) in the form:
\begin{align}
\begin{split}
\label{genmom}
    M_a^{nm} & =  \int \frac{d w \hspace{0.07cm} d^2 p_T}{(2 \pi)^3 \hspace{0.07cm} v} \hspace{0.07cm} \left( \frac{v}{\tau} \right)^n \hspace{0.07cm} \left( \frac{w}{\tau} \right)^{2m} \hspace{0.07cm} f_a(\tau; w, p_T) =\\
    & = \frac{1}{(2 \pi)^3 \hspace{0.07cm} \tau^{\hspace{0.05cm}n + 2m}} \hspace{0.07cm} \int d w \hspace{0.07cm} d^2 p_T \hspace{0.1cm} v^{n - 1} \hspace{0.07cm} w^{2m} \hspace{0.07cm} f_a(\tau; w, p_T) \,.
    \end{split}
\end{align}
If we substitute the exact solution for the distribution function {of the species $a$ given} in Eq.~(\ref{exact}) into Eq.~(\ref{genmom}) we find an integral equation of the form 
\begin{equation}
\label{strick}
    M_a^{nm}(\tau) = D_a(\tau, \tau_0) \hspace{0.07cm} M^{nm}_{0, a}(\tau) + \int_{\tau_0}^{\tau} \frac{d \tau'}{\tau_{{\rm eq}, a}(\tau')} \hspace{0.07cm} D_a(\tau, \tau') \hspace{0.07cm} M^{nm}_{{\rm eq}, a}(\tau') \,,
\end{equation}
{where the expressions for $M^{nm}_{0, a}(\tau)$ and $M^{nm}_{{\rm eq}, a}(\tau')$ are specified in the following}.

%------------------------------------------------------------
\subsection{Free-streaming term}
%------------------------------------------------------------
$M^{nm}_{0, a}(\tau)$ {is the so-called ``free-streaming'' term, the only one present in the absence of interactions. It does not coincide with $M^{nm}_{0, a}(\tau_0)$ due to the longitudinal expansion of the system. One has:}
\begin{align*}
\begin{split}
M^{nm}_{0, a}(\tau) & \equiv \frac{1}{(2 \pi)^3 \hspace{0.07cm} \tau^{\hspace{0.05cm}n + 2m}} \hspace{0.07cm} \int d w \hspace{0.07cm} d^2 p_T \hspace{0.1cm} v^{n - 1} \hspace{0.07cm} w^{2m} \hspace{0.07cm} f_{0, a}(w, p_T) =\\
& = \frac{G_{0,a}}{(2 \pi)^3 \hspace{0.07cm} \tau^{\hspace{0.05cm}n + 2m}} \hspace{0.07cm} \int d w \hspace{0.07cm} d^2 p_T \hspace{0.1cm} (w^2 + p_{T}^2 \hspace{0.07cm} \tau^2)^{\frac{n -1}{2}} \hspace{0.07cm} w^{2m} \hspace{0.07cm} \exp \left[ - \sqrt{\left( \frac{w}{w_0}\right)^2 + \left( \frac{p_T}{\Lambda_0}\right)^2} \right] \,,
\end{split}    
\end{align*}
where we have introduced the new quantity $w_0 \equiv \alpha_0 \hspace{0.07cm} \Lambda_0 \hspace{0.07cm} \tau_0$, with $\alpha_0 = (1 + \xi_0)^{-\frac{1}{2}}$ which corresponds to another way of rewriting the initial anisotropy parameter in momentum space. 
{In the above}, the following compact notation
\begin{equation}
\label{G0}
G_{0, a} \equiv
    \begin{cases}
    g_g \hspace{0.5cm}, \text{ \small{if}} \hspace{0.5cm} a = g \\
    2 g_q \, \gamma_{q, 0} \hspace{0.5cm}, \text{ \small{if}} \hspace{0.5cm} a = q
    \end{cases}
\end{equation}
{is adopted} in order to {count the number of quark and gluon} degrees of freedom (including the particle-antiparticle degeneracy) {present} in the plasma at the beginning of its evolution.
{One conveniently introduces the following dimensionless variables:}
\begin{equation*}
    \begin{cases}
    \hat{w} \equiv \frac{w}{w_0} \\[0.2cm]
    \hat{p}_T \equiv \frac{p_T}{\Lambda_0}
    \end{cases}
    \hspace{0.1cm} \longrightarrow \hspace{0.3cm} d w = w_0 \hspace{0.07cm} d \hat{w} \hspace{0.3cm},\hspace{0.3cm} d^2 p_T = \Lambda_{0}^2 \hspace{0.07cm}d^2\hat{p}_T \,.
\end{equation*}
This substitution allows one to obtain:
\begin{align*}
\begin{split}
M^{nm}_{0, a}(\tau) & = \frac{G_{0, a} \hspace{0.07cm} w_{0}^{2m +1} \hspace{0.07cm} \Lambda_{0}^2}{(2 \pi)^3 \hspace{0.07cm} \tau^{\hspace{0.05cm}n + 2m}} \hspace{0.07cm} \int d \hat{w} \hspace{0.07cm} d^2 \hat{p}_T \hspace{0.1cm} (w_{0}^2 \hspace{0.07cm} \hat{w}^2 + \Lambda_{0}^2 \hspace{0.07cm} \tau^2 \hspace{0.07cm} \hat{p}_{T}^2)^{\frac{n -1}{2}} \hspace{0.07cm} \hat{w}^{2m} \hspace{0.07cm} \exp \Bigl[ - \sqrt{\hat{w}^2 + \hat{p}_T^2} \Bigr] =\\
& = \frac{G_{0, a} \hspace{0.07cm} w_{0}^{2m +1} \hspace{0.07cm} \Lambda_{0}^{n + 1}}{(2 \pi)^3 \hspace{0.07cm} \tau^{\hspace{0.05cm}2m + 1}} \hspace{0.07cm} \int d \hat{w} \hspace{0.07cm} d^2 \hat{p}_T \hspace{0.1cm} \left( \frac{w_{0}^2 }{\Lambda_0^2 \hspace{0.07cm} \tau^2}\hspace{0.07cm} \hat{w}^2 + \hat{p}_{T}^2 \right)^{\frac{n -1}{2}} \hspace{0.07cm} \hat{w}^{2m} \hspace{0.07cm} \exp \Bigl[ - \sqrt{\hat{w}^2 + \hat{p}_T^2} \Bigr] \,.
\end{split}    
\end{align*}
With the introduction of the new parameter $y \equiv \frac{w_0}{\Lambda_0 \hspace{0.07cm} \tau} = \alpha_0 \hspace{0.07cm} \frac{\tau_0}{\tau}$, one gets
\begin{equation*}
    M^{nm}_{0, a}(\tau) = \frac{G_{0, a} \hspace{0.07cm} \Lambda_0^{n + 2m +2}}{(2 \pi)^3} \hspace{0.07cm} y^{2m + 1} \hspace{0.07cm} \int d \hat{w} \hspace{0.07cm} d^2 \hat{p}_T \hspace{0.1cm} \left( y^2 \hspace{0.07cm} \hat{w}^2 + \hat{p}_{T}^2 \right)^{\frac{n -1}{2}} \hspace{0.07cm} \hat{w}^{2m} \hspace{0.07cm} \exp \Bigl[ - \sqrt{\hat{w}^2 + \hat{p}_T^2} \Bigr] \,.
\end{equation*}
{The integration can be more easily performed introducing} spherical coordinates:
\begin{equation*}
    \begin{cases}
        \hat{p}_x = \hat{p} \hspace{0.07cm} \sin \theta \hspace{0.07cm} \cos \varphi \\
         \hat{p}_y = \hat{p} \hspace{0.07cm} \sin \theta \hspace{0.07cm} \sin \varphi \\
         \hat{w} = \hat{p} \hspace{0.07cm} \cos \theta
    \end{cases}
    \quad\longrightarrow \quad
    d \hat{w} \hspace{0.07cm} d^2 \hat{p}_T = (2 \pi)\,\hat{p}^2 d \hat{p}\,\sin\theta\,d\theta\,,
\end{equation*}
where {isotropy in the transverse plane allows one to trivially perform the azimuthal intregration}. {One is left with}
\begin{align*}
\begin{split}
M^{nm}_{0, a}(\tau) & = \frac{G_{0, a} \hspace{0.07cm} \Lambda_{0}^{n + 2m + 2}}{(2 \pi)^2} \hspace{0.07cm} y^{2m + 1} \hspace{0.07cm} \int_{0}^{+ \infty} d \hat{p} \hspace{0.07cm} \hat{p}^{n + 2m +1} \hspace{0.07cm} e^{- \hat{p}} \hspace{0.1cm} \int_{0}^{\pi} d\theta\,\sin\theta \hspace{0.07cm} (y^2 \hspace{0.07cm} \cos^2 \theta + \sin^2 \theta)^{\frac{n - 1}{2}} \hspace{0.07cm} (\cos\theta)^{2m} \,,
\end{split}    
\end{align*}
where one recognizes the integral representation of the Euler Gamma-function:
\begin{equation*}
    \Gamma(z) \equiv \int_0^{+ \infty} dx \hspace{0.1cm} x^{z - 1} \hspace{0.07cm} e^{-x} \hspace{0.3cm},\hspace{0.3cm} \mathfrak{Re}(z) > 0 \,.
\end{equation*}
Furthermore, one can conveniently define:
\begin{equation}
\label{H}
    H^{nm}(y) \equiv y^{2m + 1} \hspace{0.07cm} \int_{0}^{\pi} d \theta \hspace{0.07cm} \sin \theta \hspace{0.07cm} (y^2 \hspace{0.07cm} \cos^2 \theta + \sin^2 \theta)^{\frac{n - 1}{2}} \hspace{0.07cm} (\cos \theta)^{2m} \,.
\end{equation}
Then, our first important result is given by
\begin{equation}
\label{babbo}
    M_{0, a}^{nm} (\tau) = G_{0, a} \hspace{0.07cm} \frac{\Gamma(n + 2m + 2)}{(2 \pi)^2} \hspace{0.07cm} \Lambda_0^{n + 2m +2} \hspace{0.07cm} H^{nm}(y) \,.
\end{equation}
{Considering} Eq.~(\ref{H}), {one can} recast integral in a different way:
\begin{equation*}
    H^{nm}(y) = y^{2m + 1} \hspace{0.07cm} \int_{-1}^1 d u \hspace{0.07cm} u^{2m} \hspace{0.07cm} \Bigl[ 1 - (1 - y^2) \, u^2 \Bigr]^{\frac{n - 1}{2}} \,,
\end{equation*}
where $u \equiv \cos \theta$. Since the integrand is even in $u$, it can be manipulated as follows:
\begin{align*}
    \begin{split}
         H^{nm}(y) = 2 \hspace{0.07cm} y^{2m + 1} \hspace{0.07cm} \int_{0}^1 d u \hspace{0.07cm} u^{2m} \hspace{0.07cm} \Bigl[ 1 - (1 - y^2) u^2 \Bigr]^{\frac{n - 1}{2}} 
        = y^{2m + 1} \hspace{0.07cm} \int_{0}^1 d t \hspace{0.07cm} t^{m - \frac{1}{2}} \hspace{0.07cm} \Bigl[ 1 - (1 - y^2) t \Bigr]^{\frac{n - 1}{2}} \,,
    \end{split}
\end{align*}
where the last integral comes from the change of variable $t \equiv u^2$.

It is well known that the Gaussian hypergeometric function admits the following integral representation
\begin{equation*}
     _2F_1(a,b,c;z) \equiv \frac{\Gamma(c)}{\Gamma(b) \hspace{0.07cm} \Gamma(c - b)} \hspace{0.07cm} \int_0^1 dt \hspace{0.07cm} t^{b - 1} \hspace{0.07cm} (1 - t)^{c - b - 1} \hspace{0.07cm} (1 - tz)^{-a} \hspace{0.07cm} \,,
\end{equation*}
whose conditions of existence are: $\mathfrak{Re}(c) > \mathfrak{Re}(b) > 0$ and $|\text{arg}(1 - z)| < \pi$, since this special function exhibits a cut on the real semi-axis [$1 , + \infty$).
By comparing the two previous integrals, one concludes that {in order to express} $H^{nm}$ {in terms of a Gaussian hypergeometric function one should identify}
\begin{equation*}
    a = \frac{1 - n}{2} \hspace{0.3cm},\hspace{0.3cm} b = m + \frac{1}{2} \hspace{0.3cm},\hspace{0.3cm} c = b + 1 \,.
\end{equation*}
Using the fundamental gamma property $\Gamma(z + 1) = z \hspace{0.07cm} \Gamma(z)$ and the equality
\begin{equation*}
    _2F_1(a,b,c;z) = \hspace{0.07cm} _2F_1(b,a,c;z) \,,
\end{equation*}
one eventually finds that:
\begin{equation}
\label{Hnm}
    H^{nm}(y) = \frac{2}{2m + 1} \hspace{0.07cm} y^{2m + 1} \hspace{0.07cm} _2F_1 \left(m + \frac{1}{2}, \frac{1 - n}{2}, m + \frac{3}{2}; 1 - y^2 \right)\,.
\end{equation}
Looking at Eq.~(\ref{Hnm}), it is evident that, in our case of interest, the integral definition of the Gaussian hypergeometric function is valid, since the conditions of existence are always fulfilled. 

Some brief cross-checks can be done. First of all, taking $n = 2$ and $m = 0$, one finds from Eqs.~(\ref{H}) and (\ref{Hnm}) that:
\begin{equation}
\label{fono}
    H^{20}(y) = 2 \hspace{0.07cm} y \hspace{0.07cm} _2F_1 \left(\frac{1}{2}, -\frac{1}{2} , \frac{3}{2}; 1 - y^2 \right) = y \int_0^{\pi} d \theta \hspace{0.07cm} \sin \theta \hspace{0.07cm} \sqrt{y^2 \hspace{0.07cm} \cos^2 \theta + \sin^2 \theta} \,.
\end{equation}
This is simply the $H(y)$ function that appears in the integral equation for the effective temperature in Ref.~\cite{Florkowski:2013lya} for the case of a conformal fluid.

{If one consider the case of a fluid mixture of two particle species with different relaxation times,}
using the Landau matching condition in Eq.~(\ref{eq:LandauMC}) and applying this relation at the initial longitudinal proper time $\tau_0$, one gets
\begin{equation*}
    M^{20}_{0, q}(\tau_0) + \frac{M^{20}_{0, g}(\tau_0)}{C_R} \equiv \varepsilon_{{\rm eq}, q} + \frac{\varepsilon_{{\rm eq}, g}}{C_R} \hspace{0.07cm} \Bigg|_{\tau = \tau_0} = \frac{3}{\pi^2} \hspace{0.07cm} \biggl( 2 \hspace{0.07cm} g_q + \frac{g_g}{C_R} \biggr) \hspace{0.07cm} T_0^4 \,,
\end{equation*}
where the equilibrium energy densities $\varepsilon_{{\rm eq},q}$ and $\varepsilon_{{\rm eq},g}$ are written in Eq.~(\ref{eq:LandauMC2}).
{In order to find the relation between the initial effective temperature $T_0$, the transverse-momentum scale $\Lambda_0$ and the anisotropy parameter $\alpha_0$ one must set} $n = 2$ and $m = 0$ in Eq.~(\ref{babbo}) {at the initial time} $\tau = \tau_0$, obtaining
\begin{equation*}
    \frac{3}{2 \hspace{0.07cm} \pi^2} \hspace{0.07cm} \Lambda_0^4 \hspace{0.07cm} H(\alpha_0) \hspace{0.07cm} \biggl( 2 g_q \hspace{0.07cm} \gamma_{q, 0} + \frac{g_g}{C_R} \biggr) = \frac{3}{\pi^2} \hspace{0.07cm} \biggl( 2 \hspace{0.07cm} g_q + \frac{g_g}{C_R} \biggr) \hspace{0.07cm} T_0^4 \,.
\end{equation*}
{The difference in the overall factor between the left- and right-hand side of the equation arises from the normalization $H(\alpha_0=1)=2$, referring to the case of an initial isotropic distribution.} One can immediately determine the relation between $\Lambda_0$ and the other initial parameters:
\begin{equation}
    \label{gogga}
    \Lambda_0^4 = \frac{2 \hspace{0.07cm} T_0^4}{H(\alpha_0)} \hspace{0.07cm} \frac{2 + \Bar{r}}{2 \hspace{0.07cm} \gamma_{q, 0} + \Bar{r}} \,,
\end{equation}
with the coefficient $\Bar r$ defined in Eq.~(\ref{nonna}).

Hence, one concludes that:
\begin{equation}
\label{M0}
    M_{0, a}^{nm}(\tau) = G_{0, a} \hspace{0.07cm} \frac{\Gamma(n + 2m + 2)}{(2 \pi)^2} \hspace{0.07cm} 2^{\frac{n + 2m +2}{4}} \hspace{0.07cm} T_0^{n + 2m +2} \hspace{0.07cm} \frac{H^{nm} \left( \alpha_0 \hspace{0.07cm} \frac{\tau_0}{\tau}\right)}{ \Bigl[ H(\alpha_0) \Bigr]^{\frac{n + 2m + 2}{4}}} \hspace{0.07cm} \biggl( \frac{2 + \Bar{r}}{2 \gamma_{q, 0} + \Bar{r}} \biggr)^{\frac{n + 2m +2}{4}} \,.
\end{equation}
{As a further cross-check}, considering the case in which $n = 0$ and $m = 1$, one gets
\begin{equation}
\label{bono}
    H^{01}(y) = \frac{2}{3} \hspace{0.07cm} y^3 \hspace{0.07cm} _2F_1 \left(\frac{3}{2}, \frac{1}{2} , \frac{5}{2}; 1 - y^2 \right) = y^3 \hspace{0.07cm} \int_0^{\pi} d \theta \hspace{0.07cm} \frac{\sin \theta \hspace{0.07cm} \cos^2 \theta}{\sqrt{y^2 \hspace{0.07cm} \cos^2 \theta + \sin^2 \theta}} \,.
\end{equation} 
This quantity corresponds to the integral $H_L(y)$ introduced in Ref.~\cite{Florkowski:2013lya}.

%------------------------------------------------------------
\subsection{Evolution term}
%------------------------------------------------------------

Now we focus on the second term of Eq.~(\ref{strick}) and, in particular, we consider
\begin{equation*}
    M^{nm}_{{\rm eq}, a}(\tau') \equiv \frac{1}{(2 \pi)^3 \hspace{0.07cm} \tau^{\hspace{0.05cm}n + 2m}} \hspace{0.07cm} \int d w \hspace{0.07cm} d^2 p_T \hspace{0.07cm} v^{n - 1} \hspace{0.07cm} w^{2m} \hspace{0.07cm} f_{{\rm eq}, a}(\tau'; w, p_T) \,,
\end{equation*}
{where the $\tau^{-n-2m}$ dependence follows from $p \cdot  u = v/\tau$ and $p \cdot  z = -  w/\tau$} and $f_{{\rm eq}, a}$ is the usual (relativistic) Boltzmann isotropic distribution
\begin{equation*}
    f_{{\rm eq}, a}(\tau'; w, p_T) \equiv G_{{\rm eq}, a} \hspace{0.07cm} \exp \left[ - \frac{p \cdot u}{T(\tau')} \right] = G_{{\rm eq}, a} \hspace{0.07cm} \exp \left[ - \frac{\sqrt{w^2 + p_T^2 \hspace{0.07cm} \tau'^{2}}}{T(\tau') \hspace{0.07cm} \tau'} \right] \,.
\end{equation*}
In the above expression one has $G_{{\rm eq},g} = g_g$ and $G_{{\rm eq}, q} = 2 g_q$, since {thermodynamic equilibrium entails $\gamma_{\rm eq} = 1$}.

Luckily, one can evaluate this integral directly from the free-streaming result, by noticing that $f_{0, a}$ becomes $f_{{\rm eq}, a}$ under the substitutions:
\begin{equation*}
    \alpha_0 \hspace{0.07cm} \tau_0 \rightarrow \tau' \hspace{0.3cm},\hspace{0.3cm} \Lambda_0 \rightarrow T(\tau') \hspace{0.3cm},\hspace{0.3cm} G_{0, a} \rightarrow G_{{\rm eq}, a} \,.
\end{equation*}
Using the fact that $H(1) = 2$, one {immediately} determines the final result from Eqs.~(\ref{babbo}) or~(\ref{M0}), equivalently:
\begin{equation}
    \label{Meq}
    M_{{\rm eq}, a}^{nm}(\tau') = G_{{\rm eq}, a} \hspace{0.07cm} \frac{\Gamma(n + 2m +2)}{(2 \pi)^2} \hspace{0.07cm} T^{n + 2m + 2}(\tau') \hspace{0.07cm} H^{nm} \left( \frac{\tau'}{\tau} \right)\,.
\end{equation}
Now, {for a given particle species, one has} all the ingredients to evaluate the $M_a^{nm}(\tau)$ moment in Eq.~(\ref{strick}) from the results shown in Eqs.~(\ref{M0}) and (\ref{Meq}).

However, we are interested in the total moments, which describe the collective behavior of the whole mixture of quarks and gluons during its longitudinal expansion, namely:
\begin{equation*}
    M^{nm} \equiv \sum_{a = q, g} M_a^{nm} \,\,.
\end{equation*}
{In summing over the particle species one has to account for their different relaxation time, which affects their respective damping functions $D_a(\tau_2,\tau_1)$ quantifying the no-interaction probability. For instance, one evaluates:}
\begin{align*}
    \begin{split}
        &\sum_a G_{0, a} \hspace{0.07cm} D_a(\tau, \tau_0) = g_g \hspace{0.07cm} D_g(\tau, \tau_0) + 2 g_q \hspace{0.07cm} \gamma_{q, 0} \hspace{0.07cm} D_q(\tau, \tau_0) = g_g \hspace{0.07cm} \exp \left[ - \int_{\tau_0}^\tau  \frac{d \tau''}{\tau_{\rm eq}(\tau'')} \right] +\\
        &+ 2 g_q \hspace{0.07cm} \gamma_{q, 0} \hspace{0.07cm} \exp \left[ - C_R \hspace{0.07cm} \int_{\tau_0}^\tau  \frac{d \tau''}{\tau_{\rm eq}(\tau'')} \right] = g_g \hspace{0.07cm} D(\tau, \tau_0) + 2 g_q \hspace{0.07cm} \gamma_{q, 0} \hspace{0.07cm} \Bigl[D(\tau, \tau_0)\Bigr]^{C_R} \,,
    \end{split}
\end{align*}
where the definition of the damping functions for quarks and gluons -- see Eq.~(\ref{damp}) -- and the Casimir-scaling assumption $\tau_{{\rm eq}, q} \equiv \tau_{\rm eq}/C_R$ {have been employed}.
{Furthermore, the simplified notation
$D_g(\tau, \tau_0) \equiv D(\tau, \tau_0)$ is adopted}. The previous relation can be further manipulated to obtain:
\begin{equation}
    \label{gaia}
    \sum_a G_{0, a} \hspace{0.07cm} D_a(\tau, \tau_0) = g_q \hspace{0.07cm} D(\tau, \tau_0) \hspace{0.07cm} \biggl\{ r + 2 \hspace{0.07cm} \gamma_{q, 0} \Bigl[ D(\tau, \tau_0) \Bigr]^{C_R - 1} \biggr\} \,.
\end{equation}
The second quantity {one needs} to calculate is given by:
\begin{equation*}
    \sum_a \frac{G_{{\rm eq}, a}}{\tau_{{\rm eq}, a}(\tau')} \hspace{0.07cm} D_a(\tau, \tau') = \frac{g_g}{\tau_{\rm eq}(\tau')} \hspace{0.07cm} D(\tau, \tau') + \frac{2 g_q \hspace{0.07cm} C_R}{\tau_{\rm eq}(\tau')} \hspace{0.07cm} \Bigl[ D(\tau, \tau') \Bigr]^{C_R} \,.
\end{equation*}
The latter {is more conveniently} rewritten as:
\begin{equation}
    \label{gaio}
    \sum_a \frac{G_{{\rm eq}, a}}{\tau_{{\rm eq}, a}(\tau')} \hspace{0.07cm} D_a(\tau, \tau') = \frac{g_q \hspace{0.07cm} C_R}{\tau_{\rm eq}(\tau')} \hspace{0.07cm} D(\tau, \tau') \hspace{0.07cm} \biggl\{ \Bar{r} + 2 \hspace{0.07cm} \Bigl[ D(\tau, \tau') \Bigr]^{C_R - 1} \biggr\} \,.    
\end{equation}
Putting together the results in Eqs.~(\ref{M0}), (\ref{Meq}), (\ref{gaia}) and (\ref{gaio}) and substituting them into Eq.~(\ref{strick}) one finally obtains:
\begin{align}
    \begin{split}
    \label{eq:mom-splitr}
        &M^{nm}(\tau) = \hspace{0.1cm} g_q \hspace{0.07cm} \frac{\Gamma(n + 2m +2)}{(2\pi)^2}\times\\
        &\times\Biggl\{ D(\tau, \tau_0) \hspace{0.07cm} \biggl[ r + 2 \hspace{0.07cm} \gamma_{q, 0} \hspace{0.07cm} \biggl( D(\tau,\tau_0)\biggr)^{C_R - 1} \biggr] \hspace{0.07cm} \biggl( \frac{2 \hspace{0.07cm} \left(2 + \Bar{r}\right)}{2 \hspace{0.07cm} \gamma_{q, 0} + \Bar{r}} \biggr)^\frac{n + 2m + 2}{4} \hspace{0.07cm} \hspace{0.07cm} T_{0}^{\hspace{0.07cm} n +2m +2} \hspace{0.1cm} \frac{H^{nm}\left( \alpha_0 \hspace{0.07cm} \frac{\tau_0}{\tau} \right)}{\Bigl[H(\alpha_0) \Bigr]^{\frac{n +2m +2}{4}}}+\\
        &+ C_R \hspace{0.07cm} \int_{\tau_0}^{\tau} \frac{d \tau'}{\tau_{\rm eq}(\tau')} \hspace{0.07cm} D(\tau, \tau') \hspace{0.07cm} \biggl[ \Bar{r} + 2 \hspace{0.07cm} \biggl( D(\tau,\tau')\biggr)^{C_R - 1} \biggr] \hspace{0.07cm} T^{\hspace{0.07cm} n +2m +2}(\tau') \hspace{0.07cm} H^{nm} \left( \frac{\tau'}{\tau} \right) \Biggr\} \,.
               \end{split}              
    \end{align}
This represents the main result we wanted to prove. The integral equation (\ref{eq:mom-splitr}) allows one to compute, through numerical integration, the exact general moment of the phase-space distribution function, in the case of two different relaxation times for quarks and gluons.

Finally, {one has to} find an evolution equation for the effective temperature of the system. To this end, {one first explicitly writes} the integral equation for the moments of {the phase-space distribution of} particle $a$:
\begin{align}
    \begin{split}
    \label{eq:mom-splitff}
        M_a^{nm}(\tau) = \, &\frac{\Gamma(n + 2m +2)}{(2\pi)^2} \hspace{0.07cm} \Biggl\{ G_{0, a} \hspace{0.07cm} D_a(\tau, \tau_0) \hspace{0.07cm} \biggl( \frac{2 + \Bar{r}}{2 \hspace{0.07cm} \gamma_{q, 0} + \Bar{r}} \biggr)^\frac{n + 2m + 2}{4} \hspace{0.07cm} 2^{\frac{n + 2m +2}{4}} \hspace{0.07cm} T_{0}^{\hspace{0.07cm} n +2m +2} \hspace{0.1cm} \times\\ 
        &\times \frac{H^{nm}\left( \alpha_0 \hspace{0.07cm} \frac{\tau_0}{\tau} \right)}{\Bigl[H(\alpha_0) \Bigr]^{\frac{n +2m +2}{4}}} 
        + G_{{\rm eq}, a} \hspace{0.07cm} \int_{\tau_0}^{\tau} \frac{d \tau'}{\tau_{{\rm eq}, a}(\tau')} \hspace{0.07cm} D_a(\tau, \tau') \hspace{0.07cm} T^{\hspace{0.07cm} n +2m +2}(\tau') \hspace{0.07cm} H^{nm} \left( \frac{\tau'}{\tau} \right) \Biggr\}\,.
    \end{split}
\end{align}
{In order to determine the effective temperature of the fluid mixture through}
the Landau matching condition in Eq.~(\ref{eq:LandauMC}) {one has to set $n = 2$ and $m = 0$}. Consequently, one gets:
\begin{align}
\label{xeno}
    \begin{split}
        \varepsilon_q(\tau) + \frac{\varepsilon_g(\tau)}{C_R} &= \frac{3}{2 \hspace{0.07cm} \pi^2} \hspace{0.07cm} \biggl\{ 2 \hspace{0.07cm} \Biggl[ D_q(\tau, \tau_0) \hspace{0.07cm} G_{0, q} + D_g(\tau, \tau_0) \hspace{0.07cm} \frac{G_{0, g}}{C_R} \biggr] \hspace{0.07cm} T_0^4 \hspace{0.07cm} \biggl( \frac{2 + \Bar{r}}{2 \hspace{0.07cm} \gamma_{q, 0} + \Bar{r}} \biggr) \hspace{0.07cm} \frac{H\left( \alpha_0 \hspace{0.07cm} \frac{\tau_0}{\tau} \right)}{H(\alpha_0)} +\\
        &+ \int_{\tau_0}^\tau \frac{d \tau'}{\tau_{\rm eq}(\tau')} \biggl[ C_R \hspace{0.07cm} G_{{\rm eq}, q} \hspace{0.07cm} D_q(\tau, \tau') + \frac{G_{{\rm eq}, g}}{C_R} \hspace{0.07cm} D_g(\tau, \tau') \biggr] \hspace{0.07cm} T^4(\tau') \hspace{0.07cm} H\left(\frac{\tau'}{\tau}\right)\Biggr\} =\\
        &= \frac{3 \hspace{0.07cm} g_q}{2 \hspace{0.07cm} \pi^2} \hspace{0.07cm} \Biggl\{ 2 \hspace{0.07cm} D(\tau, \tau_0) \hspace{0.07cm} \biggl[ 2 \hspace{0.07cm} \gamma_{q, 0} \hspace{0.07cm} \Bigl( D(\tau, \tau_0) \Bigr)^{C_R - 1}  + \Bar{r} \biggr] \hspace{0.07cm} T_0^4 \hspace{0.07cm} \biggl( \frac{2 + \Bar{r}}{2 \hspace{0.07cm} \gamma_{q, 0} + \Bar{r}} \biggr) \hspace{0.07cm} \frac{H\left( \alpha_0 \hspace{0.07cm} \frac{\tau_0}{\tau} \right)}{H(\alpha_0)} +\\
        &+ C_R \hspace{0.07cm} \int_{\tau_0}^\tau \frac{d \tau'}{\tau_{\rm eq}(\tau')} \hspace{0.07cm} D(\tau, \tau') \hspace{0.07cm} \biggl[ 2 \hspace{0.07cm} \Bigl( D(\tau, \tau') \Bigr)^{C_R - 1}  + \frac{\Bar{r}}{C_R} \biggr] \hspace{0.07cm} T^4(\tau') \hspace{0.07cm} H\left(\frac{\tau'}{\tau}\right)\Biggr\} \equiv\\
        &\equiv \varepsilon_{{\rm eq}, q}(\tau) + \frac{\varepsilon_{{\rm eq}, g}(\tau)}{C_R} = \frac{3 \hspace{0.07cm} g_q}{\pi^2} \hspace{0.07cm} \bigl( 2 + \Bar{r} \bigr) \hspace{0.07cm} T^4(\tau) \,,
    \end{split}
\end{align}
After some algebra, one eventually obtains:
\begin{align}
    \begin{split}
    \label{T_exff}
        T^4(\tau) &=  D(\tau, \tau_0) \hspace{0.07cm} \biggl[ \Bar{r} + 2 \hspace{0.07cm} \gamma_{q, 0} \hspace{0.07cm} \biggl( D(\tau,\tau_0)\biggr)^{C_R - 1} \biggr] \hspace{0.07cm} \bigl( 2 \hspace{0.07cm} \gamma_{q,0} + \Bar{r} \bigr)^{-1} \hspace{0.07cm} T_0^4 \hspace{0.07cm} \frac{H \left( \alpha_0 \hspace{0.07cm} \frac{\tau_0}{\tau} \right)}{H(\alpha_0)} \hspace{0.1cm}+ \\
        &+ \frac{C_R}{2 + \Bar{r}} \hspace{0.07cm} \int_{\tau_0}^{\tau} \frac{d \tau'}{2 \hspace{0.07cm} \tau_{\rm eq}(\tau')} \hspace{0.07cm} D(\tau, \tau') \hspace{0.07cm} \biggl[ \frac{\Bar{r}}{C_R} + 2 \hspace{0.07cm} \biggl( D(\tau,\tau')\biggr)^{C_R - 1} \biggr] \hspace{0.07cm} T^4(\tau') \hspace{0.07cm} H \left( \frac{\tau'}{\tau} \right) \,.
    \end{split}
\end{align}
{In the above the factor $(1/2)$ within the integral arises from the normalization $H(1)=2$.}
This integral equation enables one to study the cooling of {the quark-gluon mixture within the 2RTA-BE}.
{After getting the temperature evolution from Eq.~(\ref{T_exff}) one can calculate any general moment of the momentum distributions through Eq.~(\ref{eq:mom-splitr}).} 

%%%%%%%%%%%%%%%%%%%%%%%%%%%%%%%%%%%%%%%%%%%%%%%%%%
\bibliographystyle{elsarticle-num} 
\bibliography{main}
%%%%%%%%%%%%%%%%%%%%%%%%%%%%%%%%%%%%%%%%%%%%%%%%%%

\end{document}